\documentclass[twocolumn]{aastex631}

\newcommand\aastex{AAS\TeX}

\received{\today}
\submitjournal{AAS}

\shorttitle{\aastex\ ULIRG 0.5 kpc nuclear dense molecular gas}
\shortauthors{Imanishi et al.}

\begin{document}

\title{ALMA 0.5 kpc Resolution Spatially Resolved Investigations of 
Nuclear Dense Molecular Gas Properties in Nearby Ultraluminous
Infrared Galaxies Based on HCN and HCO$^{+}$ Three Transition Line Data}

\correspondingauthor{Masatoshi Imanishi, Shunsuke Baba}
\email{masa.imanishi@nao.ac.jp, shunsuke.baba.astro@gmail.com}

\author[0000-0001-6186-8792]{Masatoshi Imanishi}
\affil{National Astronomical Observatory of Japan, National Institutes 
of Natural Sciences (NINS), 2-21-1 Osawa, Mitaka, Tokyo 181-8588, Japan}
\affil{Department of Astronomy, School of Science, Graduate
University for Advanced Studies (SOKENDAI), Mitaka, Tokyo 181-8588,
Japan} 
\affil{Toyo University, 5-28-20, Hakusan, Bunkyo-ku, Tokyo 112-8606, 
Japan}

\author[0000-0002-9850-6290]{Shunsuke Baba}
\affil{Kagoshima University, Graduate School of Science and Engineering, 
Kagoshima 890-0065, Japan}
\affil{National Astronomical Observatory of Japan, National Institutes 
of Natural Sciences (NINS), 2-21-1 Osawa, Mitaka, Tokyo 181-8588, Japan}

\author[0000-0002-6939-0372]{Kouichiro Nakanishi}
\affil{National Astronomical Observatory of Japan, National Institutes 
of Natural Sciences (NINS), 2-21-1 Osawa, Mitaka, Tokyo 181-8588, Japan}
\affil{Department of Astronomy, School of Science, Graduate
University for Advanced Studies (SOKENDAI), Mitaka, Tokyo 181-8588,
Japan} 

\author[0000-0001-9452-0813]{Takuma Izumi}
\affil{National Astronomical Observatory of Japan, National Institutes 
of Natural Sciences (NINS), 2-21-1 Osawa, Mitaka, Tokyo 181-8588, Japan}
\affil{Department of Astronomy, School of Science, Graduate
University for Advanced Studies (SOKENDAI), Mitaka, Tokyo 181-8588,
Japan} 
\affil{Department of Physics, Graduate School of Science, Tokyo
Metropolitan University, 1-1 Minami-Osawa, Hachioji-shi, Tokyo
192-0397, Japan}



\begin{abstract}
We present the results of our ALMA $\lesssim$0.5 kpc-resolution dense 
molecular line (HCN and HCO$^{+}$ J=2--1, J=3--2, and J=4--3)
observations of 12 nearby (ultra)luminous infrared galaxies
([U]LIRGs). 
After matching beam sizes of all molecular line data to the same
values in all (U)LIRGs, we derive molecular line flux ratios, by
extracting spectra in the central 0.5, 1, 2 kpc circular
regions, and 0.5--1 and 1--2 kpc annular regions. 
Based on non-LTE model calculations, we quantitatively confirm
that the innermost ($\lesssim$0.5 kpc) molecular gas is very dense
($\gtrsim$10$^{5}$ cm$^{-3}$) and warm ($\gtrsim$300 K) in ULIRGs, and
that in one LIRG is also modestly dense (10$^{4-5}$ cm$^{-3}$) and
warm ($\sim$100 K).   
We then investigate the spatial variation of the HCN-to-HCO$^{+}$ flux
ratios and high-J to low-J flux ratios of HCN and HCO$^{+}$.
A subtle sign of decreasing trend of these ratios from the innermost
($\lesssim$0.5 kpc) to outer nuclear (0.5--2 kpc) region is
discernible in a significant fraction of the observed ULIRGs.
For two AGN-hosting ULIRGs which display the trend most clearly, we
find based on a Bayesian approach that the HCN-to-HCO$^{+}$ abundance
ratio and gas kinetic temperature systematically increase from 
the outer nuclear to the innermost region. 
We suggest that this trend comes from potential AGN effects,
because no such spatial variation is found in a starburst-dominated LIRG.
\end{abstract}




\section{Introduction} 

Ultraluminous infrared galaxies (ULIRGs) radiate very strong infrared 
emission with luminosity L$_{\rm IR}$ $\gtrsim$ 10$^{12}$L$_{\odot}$
and are usually seen as gas-rich galaxy major mergers in the nearby 
universe at $z <$ 0.3 \citep[e.g.,][]{sam96}. 
The observed infrared luminosity is much higher than UV--optical
luminosity in most cases, suggesting that the bulk of UV--optical
emission from luminous, but hidden, energy sources is absorbed by dust
which re-emits the absorbed energy as infrared thermal radiation.
Through galaxy merger processes, a large amount of molecular gas and
dust can concentrate into nuclear regions ($\lesssim$1--2 kpc).
Star-formation (= starburst) activity and mass-accretion onto existing 
supermassive black holes (SMBHs), the so-called active galactic
nucleus (AGN) activity, can occur there. 
Both the starburst and AGN activity can be the luminous hidden energy
sources of ULIRGs, but distinguishing the relative energetic
contribution of these two kinds of activity is not an easy task,
because of huge dust extinction toward the hidden energy sources.   
Observations at wavelengths of strong penetrating power against dust
are indispensable to scrutinize what is happening at nearby ULIRGs'
nuclei.  

In the (sub)millimeter wavelength range at 0.3--3.5 mm where dust 
extinction effects are very small \citep{hil83}, many rotational
J-transition lines of abundant molecules are found. 
At nearby merging ULIRGs' nuclei, the bulk of molecular gas is thought to be
in a dense form ($\gtrsim$10$^{4}$ cm$^{-3}$) \citep[e.g.,][]{gao04}.
For this reason, observations of (sub)millimeter rotational
J-transition emission lines of dense molecular gas tracers with high
dipole moments and/or high critical density can provide important
information about the enigmatic nature of nearby ULIRGs' nuclei. 
In fact, a starburst (= energy release by nuclear fusion inside stars)
and an AGN (= radiative energy generated by a mass accreting SMBH)
can have different physical and chemical effects to surrounding dense
molecular gas, so that particular molecular emission lines can be
strong depending on energy sources.  
Regarding the widely used bright CO emission lines, high-J (J
$\gtrsim$ 4--5) ones can probe dense (and warm) molecular gas at
nearby ULIRGs' nuclei because of higher critical density (and excitation
energy) than low-J (J=1--2) ones.
Because it is theoretically predicted that high-J CO emission lines,
relative to low-J CO ones, can be stronger in an AGN than in a 
starburst \citep[e.g.,][]{mei07,spa08}, detection of significantly
stronger high-J CO emission than that explained by starburst activity,
has been used to argue for the presence of a luminous AGN
\citep[e.g.,][]{van10,spi12,per14,mas15,lu17,esp22}, with a caution
that mechanical heating by shocks could also produce strong high-J
CO emission lines \citep[e.g.,][]{hai12,mei13,pel13,ros15,kam16}.

Additionally, an AGN can enhance the abundance of some particular
molecules, compared to a starburst, because stronger X-ray emission,
when normalized to UV luminosity, and a larger amount of hot 
($>$100 K) dust in an AGN, can make chemistry significantly different
from a starburst \citep[e.g.,][]{mei07,har13}.
It is desirable to see how dense molecular emission line flux ratios
differ between known starburst-dominated and AGN-important galaxies.
HCN and HCO$^{+}$ rotational J-transition line observations have been
conducted before to probe dense molecular gas in nearby ULIRGs, because 
(1) the dipole moments of HCN and HCO$^{+}$ are much larger
than the widely used CO \citep{shi15} and    
(2) HCN and HCO$^{+}$ are one of the brightest lines among putative dense
molecular gas tracers \citep[e.g.,][]{mar11,ala11,ala15}. 
However, these observations were done with large beams 
($\gtrsim$5$''$ or $\gtrsim$5 kpc at $z \gtrsim$ 0.05) using single
dish (sub)millimeter telescopes 
\citep[e.g.,][]{gao04,baa08,gra08,kri08,gre09,cos11,pap14,pri15,ued21,zho22,isr23} 
or pre-ALMA interferometric facilities with limited angular resolution
of $\gtrsim$1$\farcs$5 \citep[e.g.,][]{ima06b,ima07b,ima09a}.
Physical and chemical conditions at the energetically dominant nearby
ULIRGs' nuclei ($\lesssim$1--2 kpc) \citep[e.g.,][]{soi00,dia10,ima11,per21} 
may not be best probed with previously taken large-beam-sized
observational data, because of possible contamination from spatially
extended ($\gtrsim$a few kpc) molecular gas emission in the host galaxies.

With the advent of ALMA, conducting sensitive high-angular-resolution
($\lesssim$1$''$) dense molecular J-transition line observations in
the (sub)millimeter has routinely become possible. 
Sub-arcsecond and sub-kpc-resolution HCN and HCO$^{+}$ line
observations of the two nearby well-studied ULIRGs Arp 220 ($z =$
0.018) and Mrk 231 ($z =$ 0.042) were conducted 
\citep[e.g.,][]{sco15,aal15a,aal15b,mar16,sak21}.
In Arp 220, dense molecular gas properties were investigated, using
multiple J-transition molecular line data and non-LTE modeling
\citep[e.g.,][]{tun15,sli17,man17}.  
However, our understanding of dense molecular gas properties in nearby
ULIRGs' nuclei in general is still highly incomplete.
Sub-arcsec ($\lesssim$a few kpc)-resolution HCN and HCO$^{+}$
observational results of multiple nearby ULIRGs at J=2--1,
J=3--2 and J=4--3 lines have been reported   
\citep[e.g.,][]{ima13a,ima13b,ima14,ima16a,ima16b,ima18,ima21,ima22}. 
By combining these multiple J-transition HCN and HCO$^{+}$ line data
and by applying non-LTE model calculations,  
\citet{ima23} derived nuclear dense molecular gas properties of
ten nearby (U)LIRGs at 1--2 kpc physical resolution. 
However, possible spatial variation of the properties within nearby
ULIRGs' nuclei cannot be investigated in detail
with this resolution.    

\citet{ima19} obtained $\lesssim$0$\farcs$2-resolution HCN and
HCO$^{+}$ J=3--2 observational data of $>$20 nearby ULIRGs at $z <$ 0.15.  
The corresponding physical scale is $\lesssim$0.5 kpc, which enables
us to investigate dense molecular gas properties at $\lesssim$0.5 kpc
spatial resolution within nearby ULIRGs' nuclei, if multiple
J-transition line data with similar resolution are available.  
By adding $\lesssim$0.5 kpc-resolution HCN and HCO$^{+}$ J=2--1 and
J=4--3 line data 
\footnote{
HCN and HCO$^{+}$ J=1--0 lines were not observable with ALMA before
2022 for sources at $z \gtrsim$ 0.06, because these lines are shifted
to longer wavelength (= lower frequency) beyond the band 3 coverage 
(2.6--3.6 mm or 84--116 GHz). 
Observations of J=5--4 or even higher J-transition lines of HCN and
HCO$^{+}$ are difficult for ULIRGs at $z <$ 0.15, because these lines
fall in band 8 (385--500 GHz) or even higher frequency band.  
}
to the existing J=3--2 data, we will be able to obtain three
independent J-transition line data. 
They can be used to better constrain the possible spatial variation of 
(1) physical properties of dense molecular gas, based on excitation
conditions (high-J to low-J flux ratios) of both HCN and HCO$^{+}$, and  
(2) chemical properties, by comparing HCN and HCO$^{+}$ emission line
fluxes at the same J-transition.

In this paper, we present our new $\lesssim$0$\farcs$2 
($\lesssim$0.5 kpc)-resolution HCN and HCO$^{+}$ J=2--1 and J=4--3
observational results of nearby ULIRGs already observed at J=3--2 
with similarly high spatial resolution by \citet{ima19}. 
After matching beam sizes of multiple J-transition line data of both
HCN and HCO$^{+}$ to the same value, we attempt to investigate, 
with an aid of non-LTE calculations, 
(1) physical and chemical properties of nuclear dense molecular gas 
in a larger number of nearby ULIRGs, compared to the previous
study by \citet{ima23}, and 
(2) for the first time, how the properties spatially change between
the innermost ($\lesssim$0.5 kpc) and outer nuclear (0.5--2 kpc) regions. 
Throughout this paper, (1) we adopt the cosmological parameters, H$_{0}$
$=$ 71 km s$^{-1}$ Mpc$^{-1}$, $\Omega_{\rm M}$ = 0.27, and
$\Omega_{\rm \Lambda}$ = 0.73, (2) maps are shown in the ICRS coordinate,
and (3) flux ratios of HCN-to-HCO$^{+}$ and between different
J-transition lines are calculated in units of Jy km s$^{-1}$.
Density and temperature mean, respectively, H$_{2}$ volume number
density (n$_{\rm H_2}$) and kinetic temperature (T$_{\rm kin}$),
unless otherwise stated.

\section{Targets} 

Our targets are originally selected from nearby ULIRGs in the
well-studied IRAS 1 Jy sample \citep{kim98}. 
We limit our sample to ULIRGs which are (1) at $z <$ 0.15, (2) at
declination $<$ $+$20$^{\circ}$ (to be best observable from the ALMA
site in Chile), and (3) classified optically as LINERs or HII-regions  
(i.e., non-Seyferts; no obvious optical AGN signature), to investigate
optically elusive, but intrinsically luminous buried AGNs. 
\citet{ima19} observed 26 such ULIRGs   
(a complete sample with expected dense molecular line peak flux above
a certain threshold),  
at HCN and HCO$^{+}$ J=3--2, with $\lesssim$0.5 kpc resolution in most
cases, in ALMA Cycle 5.  
After excluding ULIRGs with too faint dense molecular
emission lines (HCN J=3--2 peak flux from the central
$\lesssim$0.5 kpc region is $\lesssim$1.5 mJy) and/or small
($\lesssim$1) HCN-to-HCO$^{+}$ J=3--2 flux ratios (i.e., no signature
of luminous buried AGNs; \citet{ima16b}), 16 ULIRGs were selected and
their HCN and HCO$^{+}$ J=2--1 and J=4--3 observations, at
$\lesssim$0.5 kpc resolution, were proposed in ALMA Cycle 7.    
Not all the proposed ULIRGs were observed, due to limited
observing time available, resulting in 11 observed ULIRGs. 
In addition to these ULIRGs, NGC 1614 (a starburst-dominated 
luminous infrared galaxy [LIRG] with L$_{\rm IR}$ =
10$^{11.7}$L$_{\odot}$ at $z =$ 0.016) is added for comparison, 
because J=2--1, J=3--2, and J=4--3 data of HCN and HCO$^{+}$ with 
$\lesssim$0.5 kpc resolution are available \citep{ima13a,ima16b,ima22}. 
Table \ref{tab:object} summarizes these (U)LIRGs studied in this paper. 
The 11 ULIRGs are not statistically complete and are possibly 
biased to AGN-important ULIRGs because sources with high ($\gtrsim$1)
HCN-to-HCO$^{+}$ J=3--2 flux ratios are selected \citep{ima16b}. 
However, we will be able to obtain valuable information on nuclear
dense molecular gas properties in an increased number of nearby
ULIRGs, because the observed sources are largely different from those
studied by \citet{ima23}. 
 

\begin{deluxetable*}{lcccrrrrcccc}[!hbt]
\tabletypesize{\scriptsize}
\tablecaption{Basic Properties of the Observed (Ultra)luminous 
Infrared Galaxies \label{tab:object}} 
\tablewidth{0pt}
\tablehead{
\colhead{Object} & \colhead{Redshift} & 
\colhead{d$_{\rm L}$} & \colhead{Scale} & 
\colhead{f$_{\rm 12}$} & 
\colhead{f$_{\rm 25}$} & 
\colhead{f$_{\rm 60}$} & 
\colhead{f$_{\rm 100}$} & 
\colhead{log L$_{\rm IR}$} & 
\colhead{Optical} & \colhead{AGN} &
\colhead{IR/submm/X} \\
\colhead{} & \colhead{} & \colhead{[Mpc]} & \colhead{[kpc/$''$]}  
& \colhead{[Jy]} & \colhead{[Jy]} & \colhead{[Jy]} & \colhead{[Jy]}  
& \colhead{[L$_{\odot}$]} & \colhead{Class} & \colhead{IR [\%]} 
& \colhead{AGN} \\  
\colhead{(1)} & \colhead{(2)} & \colhead{(3)} & \colhead{(4)} & 
\colhead{(5)} & \colhead{(6)} & \colhead{(7)} & \colhead{(8)} & 
\colhead{(9)} & \colhead{(10)} & \colhead{(11)} & \colhead{(12)} 
}
\startdata
IRAS 00091$-$0738 & 0.1180 & 543 & 2.1 & $<$0.07 & 0.22 & 2.63 & 2.52 
& 12.3 & HII & 58$\pm$6 & Y$^{b,c}$ \\
IRAS 00188$-$0856 & 0.1285 & 596 & 2.3 & $<$0.12 & 0.37 & 2.59 & 3.40 
& 12.4 & LINER & 35$\pm$4 & Y$^{b,c,d,e}$ \\    
IRAS 00456$-$2904 & 0.1100 & 504 & 2.0 & $<$0.08 & 0.14 & 2.60 & 3.38 
& 12.2 & HII & $<$0.05 & Y \\ 
IRAS 01166$-$0844 & 0.1172 & 539 & 2.1 & 0.07 & 0.17 & 1.74 
& 1.42 & 12.1 & HII & 88$^{+6}_{-10}$ & Y$^{b,c,e}$ \\ 
IRAS 01569$-$2939 & 0.1402 & 655 & 2.5 & $<$0.11 & 0.14 & 1.73 
& 1.51 & 12.3 & HII & 18$\pm$3 & Y$^{b}$ \\ 
IRAS 03250$+$1606 & 0.1286 & 596 & 2.3 & $<$0.10 & $<$0.15 & 1.38 
& 1.77 & 12.1 & LINER & $<$0.2 & Y$^{d}$ \\   
IRAS 10378$+$1108 & 0.1365 & 636 & 2.4 & $<$0.11 & 0.24 & 2.28 
& 1.82 & 12.3 & LINER & 14$\pm$2 & Y$^{b,c,d}$ \\   
IRAS 16090$-$0139 & 0.1334 & 621 & 2.4 & 0.09 & 0.26 & 3.61 & 4.87 
& 12.6 & LINER & 24$\pm$3 & Y$^{b,c,d,e}$ \\   
IRAS 22206$-$2715 & 0.1320 & 614 & 2.3 & $<$0.10 & $<$0.16 & 1.75 & 2.33 
& 12.2 & HII & $<$0.5 & Y \\ 
IRAS 22491$-$1808 & 0.0776  & 347 & 1.5 & 0.05 
& 0.55 & 5.44 & 4.45 & 12.2 & HII & $<$0.07 & Y$^{f}$ \\ \hline
IRAS 12112$+$0305 & 0.0730 & 326 & 1.4 & 0.12 & 0.51 
& 8.50 & 9.98 & 12.3 & LINER & $<$0.7 & Y$^{f,g}$ (NE nucleus) \\ \hline
NGC 1614\tablenotemark{a} & 0.0160 & 68 & 0.32 & 1.38 & 7.50 & 32.12 & 34.32 
& 11.7 & HII & $<$5 & \\ \hline 
\enddata

\tablenotetext{a}{Also known as IRAS 04315$-$0840. This is a LIRG
classified as starburst-dominated through various spectroscopic
observations \citep[e.g.,][]{bra06,ber09,ima10b,per15}.
}

\tablecomments{
Col.(1): Object name. 
IRAS 12112$+$0305 is listed separately, because we have only one
J-transition line data with $\lesssim$0.5 kpc resolution.
Col.(2): Redshift adopted from ALMA dense molecular line data 
\citep{ima16b}, which are slightly different from optically derived ones 
\citep{kim98} in some cases. 
Col.(3): Luminosity distance in Mpc. 
Col.(4): Physical scale in kpc arcsec$^{-1}$. 
Col.(5)--(8): f$_{12}$, f$_{25}$, f$_{60}$, and f$_{100}$ are 
{\it IRAS} fluxes at 12 $\mu$m, 25 $\mu$m, 60 $\mu$m, and 100 $\mu$m,
respectively, taken from \citet{kim98} or \citet{san03} or 
the IRAS Faint Source Catalog (FSC). 
Col.(9): Decimal logarithm of infrared (8$-$1000 $\mu$m) luminosity
in units of solar luminosity (L$_{\odot}$), calculated with
$L_{\rm IR} = 2.1 \times 10^{39} \times$ D(Mpc)$^{2}$
$\times$ (13.48 $\times$ $f_{12}$ + 5.16 $\times$ $f_{25}$ +
$2.58 \times f_{60} + f_{100}$) ergs s$^{-1}$ \citep{sam96}.
Col.(10): Optical spectroscopic classification by \citet{vei99} or
\citet{vei95}. 
``LINER'' and ``HII'' refer to LINER and HII-region, respectively. 
Col.(11): Infrared spectroscopically estimated bolometric contribution 
of AGN in \% by \citet{nar10} for all ULIRGs and by \citet{per15}
for the LIRG NGC 1614.
Col.(12): ``Y'' means the presence of the signatures of
optically elusive, but intrinsically luminous buried AGNs.
All ULIRGs show elevated ($\gtrsim$1) HCN-to-HCO$^{+}$ J=3--2 flux
ratios at $\sim$1.3 mm \citep{ima19}, possible signatures of luminous
AGNs \citep[e.g.,][]{ima16b}.
Other representative references for AGN signatures in the infrared
3--40 $\mu$m and/or (sub)millimeter spectra:  
$^{b}$: \citet{ima07a}.
$^{c}$: \citet{vei09}.
$^{d}$: \citet{ima06a}.
$^{e}$: \citet{ima10b}.
$^{f}$: \citet{ima18}.
$^{g}$: \citet{ima16b}.
}

\end{deluxetable*}

\section{Observations and Data Analysis} 

Our HCN and HCO$^{+}$ J=2--1 and J=4--3 observations of the 11 ULIRGs  
were conducted in our ALMA Cycle 7 program 2019.1.00027.S 
(PI = M. Imanishi). 
We employed the widest 1.875 GHz mode with 1920 channels for each
spectral window. 
HCN and HCO$^{+}$ lines were simultaneously observed in one sideband
(LSB or USB). 
For IRAS 22206$-$2715 and IRAS 22491$-$1808, although HCN and HCO$^{+}$
J=3--2 data were taken in ALMA Cycle 5 \citep{ima19}, the achieved
beam sizes were much larger than $\sim$0.5 kpc, unlike other ULIRGs. 
For these two ULIRGs, we thus took $\lesssim$0.5 kpc-resolution HCN and 
HCO$^{+}$ J=3--2 data as well. 
Table \ref{tab:obs} tabulates our ALMA Cycle 7 observation log. 
For nine ULIRGs except IRAS 10378$+$1108 and IRAS 12112$+$0305, both
J=2--1 and J=4--3 data of HCN and HCO$^{+}$ were obtained, and so after
combining with available or newly taken J=3--2 data, we have full
three J-transition HCN and HCO$^{+}$ data with $\lesssim$0.5 kpc
resolution.   
For IRAS 10378$+$1108, we obtained only J=2--1 data in our ALMA Cycle
7 program and so can combine $\lesssim$0.5 kpc-resolution J=2--1 and
J=3--2 data only. 
For IRAS 12112$+$0305, only J=4--3 data were taken in our ALMA Cycle 7
program. 
Because available J=3--2 data of IRAS 12112$+$0305 are not of
sufficiently small physical resolution ($>$0.5 kpc) \citep{ima19}, 
we will only display newly taken J=4--3 data of the primary
north-eastern (NE) nucleus (whose beam sizes are much smaller than
previously published J=4--3 data by \citet{ima18}), but will not
constrain nuclear dense molecular gas properties with $\lesssim$0.5
kpc resolution in detail.  

We started our analysis from pipeline-calibrated data, using 
the CASA version 6.1.1.15 \citep{CASA22}, provided by ALMA.
By choosing channels without showing obvious emission and absorption
lines, we determined the continuum level, and subtracted it using the 
CASA task ``uvcontsub''.
Then we applied the ``tclean'' task 
(Briggs-weighting; robust $=$ 0.5 and gain $=$ 0.1)
for the continuum-only and continuum-subtracted dense molecular line
data to create cleaned maps.
The final velocity resolution was $\sim$20 km s$^{-1}$ and the pixel
scale was 0$\farcs$02 pixel$^{-1}$.
According to the ALMA Cycle 7 Technical Handbook (equation 7.6) 
\footnote{https://almascience.eso.org/documents-and-tools/cycle7/alma-technical-handbook},  
the maximum recoverable scale (MRS) is $>$4$''$ at $\sim$0.85--2 mm
(i.e., the wavelength range of HCN and HCO$^{+}$ J=2--1, J=3--2,
and J=4--3) for the minimum baseline length of 15--29 m 
(Table \ref{tab:obs}).  
This MRS corresponds to $>$5 kpc for all the observed ULIRGs, so that
our targeting dense molecular line emission at ULIRGs' nuclei 
($\lesssim$1--2 kpc) should be fully recovered.  
This is also the case for the HCN and HCO$^{+}$ J=3--2 data of
ULIRGs taken in ALMA Cycle 5 \citep{ima19}. 
For the nearest LIRG NGC 1614 ($z =$ 0.016; 0.32 kpc arcsec$^{-1}$), 
molecular line emission with $\sim$1 kpc physical scale can be
safely recovered with our ALMA data taken before Cycle 5.
Because the J=2--1, J=3--2, and J=4--3 data were obtained at
different times, we will take into account the possible absolute flux
calibration uncertainty in individual ALMA observations, with
maximum $\sim$5\% for J=2--1 and $\sim$10\% for J=3--2 and J=4--3
(ALMA Cycle 5 and 7 Proposer's Guide) when we discuss molecular gas
properties based on the flux comparison at different J-transitions.    
However, because HCN and HCO$^{+}$ data at each J-transition were taken
simultaneously, the HCN-to-HCO$^{+}$ flux ratios at J=2--1, J=3--2, and
J=4--3 are not directly affected by this possible absolute flux
calibration uncertainty.
 
\begin{deluxetable*}{lllccc|ccc}[!hbt]
\tabletypesize{\scriptsize}
\tablecaption{ALMA Cycle 7 Observation Log \label{tab:obs}} 
\tablewidth{0pt}
\tablehead{
\colhead{Object} & \colhead{Line} & \colhead{Date} & \colhead{Antenna} & 
\colhead{Baseline} & \colhead{Integration} & \multicolumn{3}{c}{Calibrator} \\ 
\colhead{} & \colhead{} & \colhead{[UT]} & \colhead{Number} & \colhead{[m]} &
\colhead{[min]} & \colhead{Bandpass} & \colhead{Flux} & \colhead{Phase}  \\
\colhead{(1)} & \colhead{(2)} & \colhead{(3)} & \colhead{(4)} &
\colhead{(5)} & \colhead{(6)} & \colhead{(7)}  & \colhead{(8)} 
& \colhead{(9)}
}
\startdata 
IRAS 00091$-$0738 & J=2--1 & 2021 July 10 & 43 & 29--3396 & 26 &
J0006$-$0623 & J0006$-$0623 & J0017$-$0512 \\
 & J=4--3 & 2021 May 24 & 42 & 15--2375 & 13 & J0006$-$0623 &
J0006$-$0623 & J2358$-$1020 \\ \hline
IRAS 00188$-$0856 & J=2--1 & 2021 July 5 & 45 & 29--2996 & 25 &
J2258$-$2758 & J2258$-$2758 & J0017$-$0512 \\
 & J=4--3 & 2021 May 23 & 41 & 15--2375 & 14 & J0006$-$0623 &
J0006$-$0623 & J0051$-$0650 \\ \hline
IRAS 00456$-$2904 & J=2--1 & 2021 July 5 & 45 & 29--2996 & 26 &
J2357$-$5311 & J2357$-$5311 & J0106$-$2718 \\
 & J=4--3 & 2021 June 27 & 38 & 15--3396 & 18 & J2258$-$2758 &
J2258$-$2758 & J0038$-$2459 \\ \hline
IRAS 01166$-$0844 & J=2--1 & 2021 July 5 & 46 & 29--2996 & 16 &
J0238$+$1636 & J0238$+$1636 & J0110$-$0741 \\
 & J=4--3 & 2021 June 12 & 42 & 15--2386 & 18 & J0238$+$1636 &
J0238$+$1636 & J0116$-$1136 \\ \hline
IRAS 01569$-$2939 & J=2--1 & 2021 July 10 & 43 & 29--3396 & 38 &
J0334$-$4008 & J0334$-$4008 & J0145$-$2733 \\
 & J=4--3 & 2021 June 28 & 35 & 15--3638 & 16 & J2258$-$2758 &
J2258$-$2758 & J0145$-$2733 \\ \hline
IRAS 03250$+$1606 & J=2--1 & 2021 July 10 & 43 & 29--3396 & 23 &
J0238$+$1636 & J0238$+$1636 & J0325$+$2224 \\
 & J=4--3 & 2021 July 16 & 45 & 15--3638 & 36 & J0423$-$0120 & 
J0423$-$0120 & J0325$+$2224 \\ \hline
IRAS 10378$+$1108 & J=2--1 & 2021 July 19 & 36 & 15--3697 & 28 &
J1058$+$0133 & J1058$+$0133 & J1025$+$1253 \\ \hline
IRAS 16090$-$0139 & J=2--1 & 2021 July 12 & 45 & 29--3396 & 45 &
J1550$+$0527 & J1550$+$0527 & J1557$-$0001 \\
 & J=4--3 & 2021 June 10 & 40 & 15--2386 & 14 & J1517$-$2422 &
J1517$-$2422 & J1549$+$0237 \\ \hline
IRAS 22206$-$2715 & J=2--1 & 2021 July 9 & 45 & 29--3396 & 25 &
J2258$-$2758 & J2258$-$2758 & J2223$-$3137 \\
 & J=3--2 & 2021 May 20 & 46 & 15--2517 & 19 & J2258$-$2758 &
J2258$-$2758 & J2223$-$3137 \\ 
 & J=4--3 & 2021 May 15 & 44 & 15--2517 & 13 & J2258$-$2758 &
J2258$-$2758 & J2223$-$3137 \\ \hline
IRAS 22491$-$1808 & J=2--1 & 2021 July 9 & 43 & 29--3638 & 14 &
J2258$-$2758 & J2258$-$2758 & J2303$-$1841 \\
 & J=3--2 & 2021 May 15 & 41 & 15--2386 & 7 & J2258$-$2758 &
J2258$-$2758 & J2303$-$1841 \\ 
 & J=4--3 & 2021 June 10 & 40 & 15--2386 & 8 & J2258$-$2758 &
J2258$-$2758 & J2303$-$1841 \\ \hline
IRAS 12112$+$0305 & J=4--3 & 2021 July 10 & 45 & 29--3638 & 9 &
J1229$+$0203 & J1229$+$0203 & J1222$+$0413 \\ \hline
\enddata

\tablecomments{ 
Col.(1): Object name. 
Col.(2): Observed J-transition of HCN and HCO$^{+}$. 
Col.(3): Observation date in UT. 
Col.(4): Number of antennas used for observations. 
Col.(5): Baseline length in meters. Minimum and maximum baseline lengths are 
shown.  
Col.(6): Net on source integration time in minutes.
Cols.(7), (8), and (9): Bandpass, flux, and phase calibrator for the 
target source, respectively.
}

\end{deluxetable*}


\section{Results} 

\begin{figure*}
\begin{center}
\includegraphics[angle=0,scale=.42]{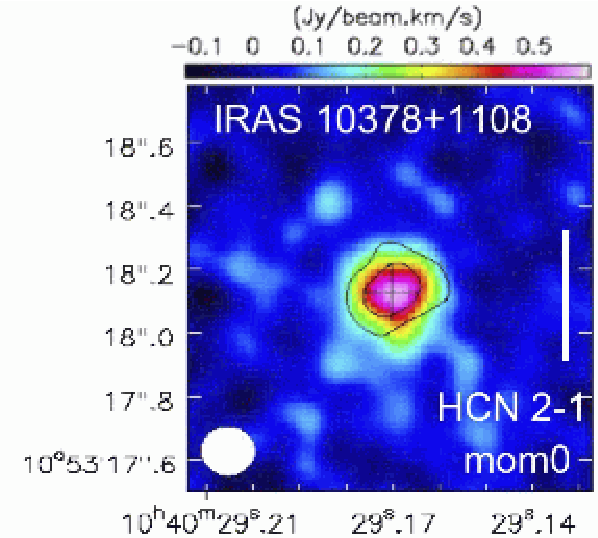} 
\includegraphics[angle=0,scale=.42]{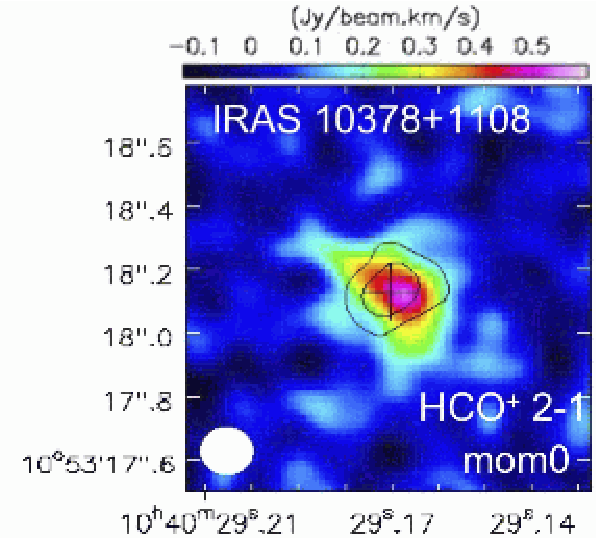} 
\includegraphics[angle=0,scale=.42]{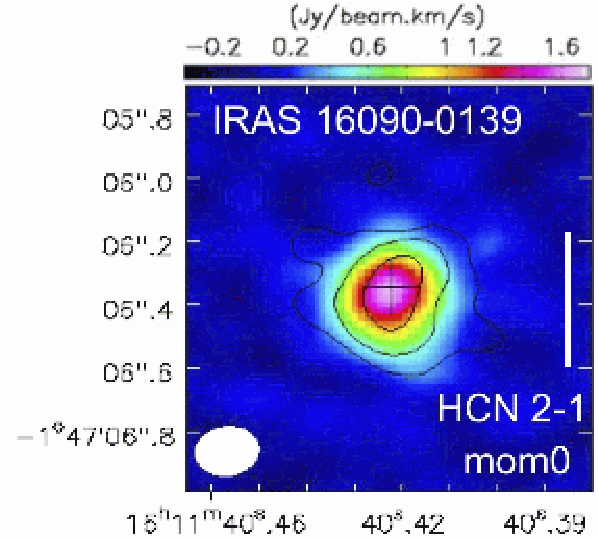} 
\includegraphics[angle=0,scale=.42]{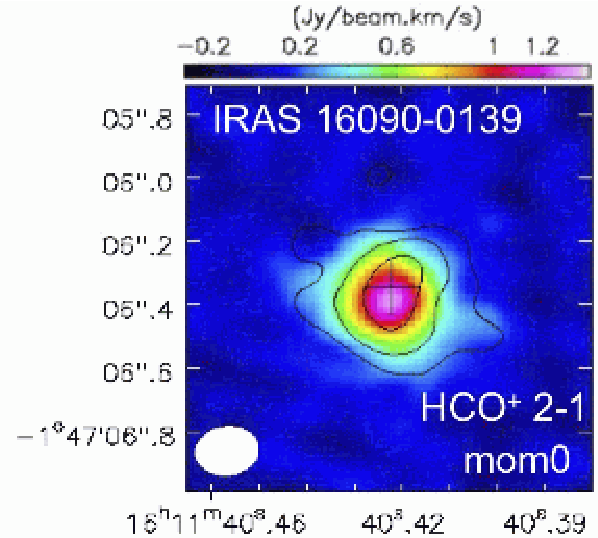} \\
\includegraphics[angle=0,scale=.42]{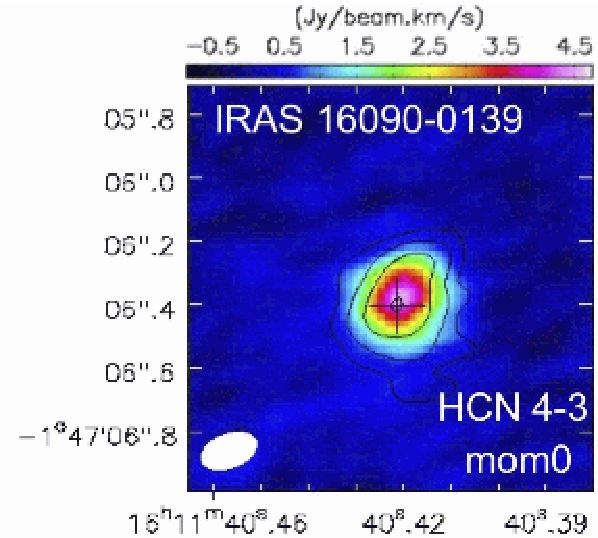} 
\includegraphics[angle=0,scale=.42]{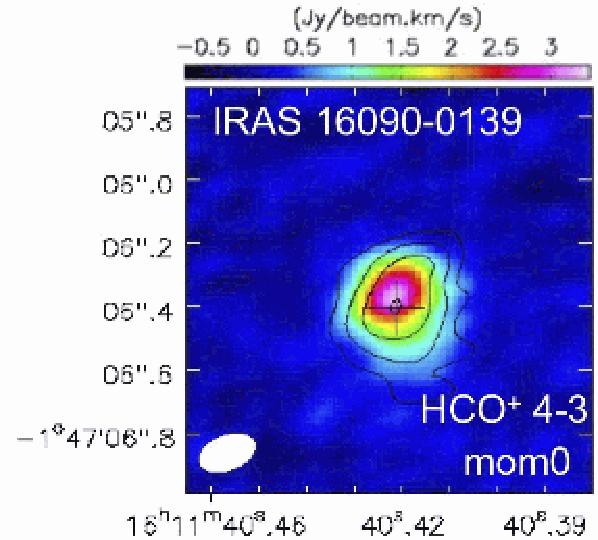} 
\includegraphics[angle=0,scale=.42]{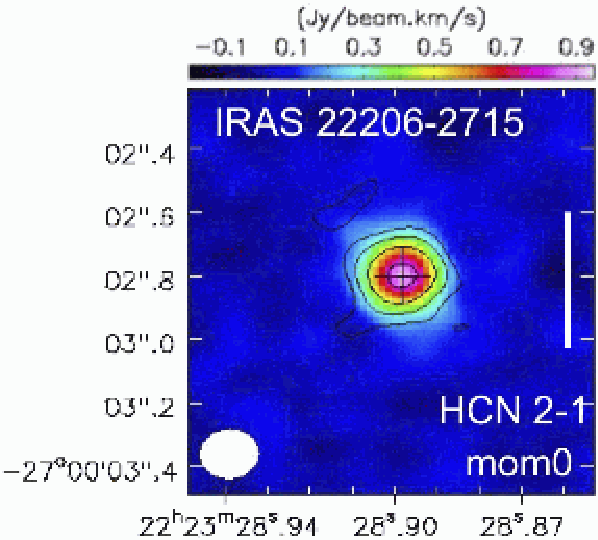} 
\includegraphics[angle=0,scale=.42]{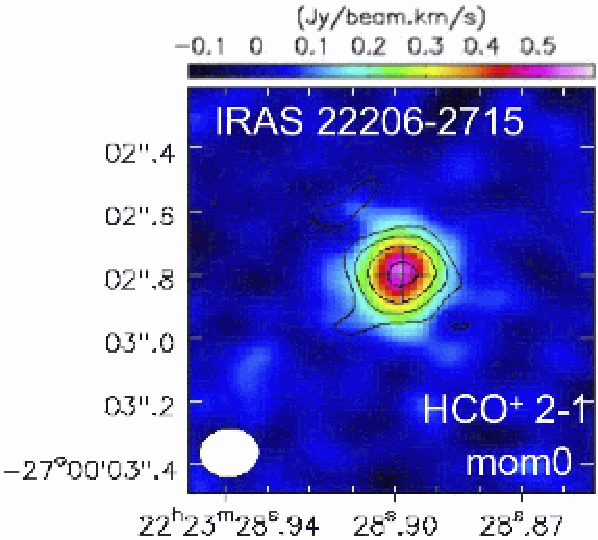} \\
\includegraphics[angle=0,scale=.42]{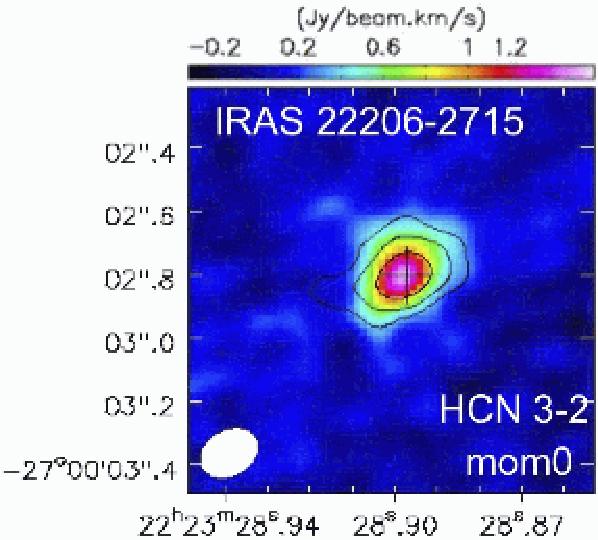} 
\includegraphics[angle=0,scale=.42]{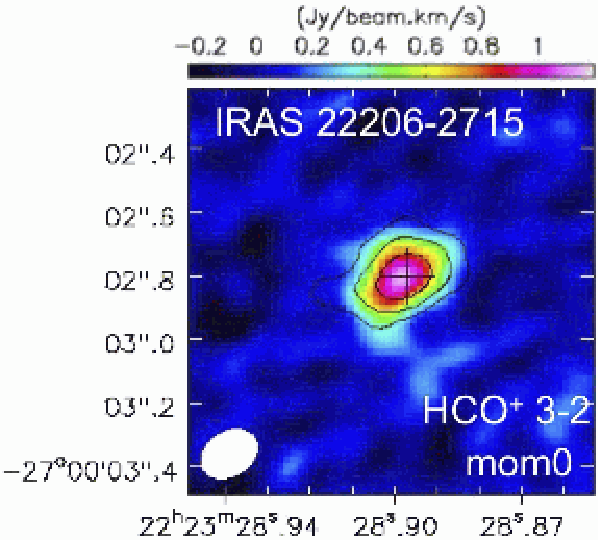} 
\includegraphics[angle=0,scale=.42]{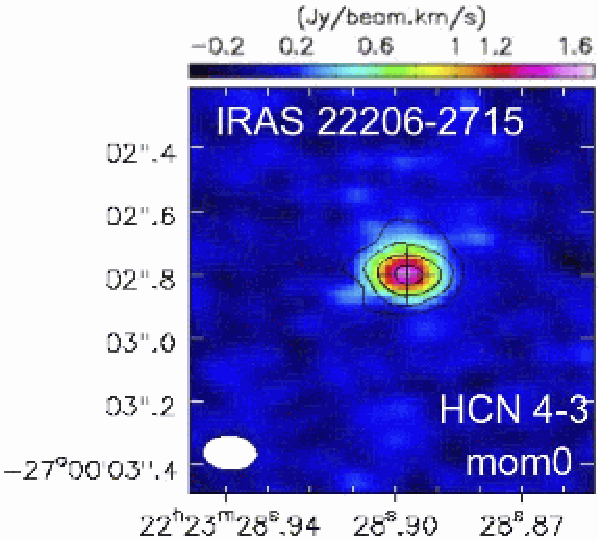} 
\includegraphics[angle=0,scale=.42]{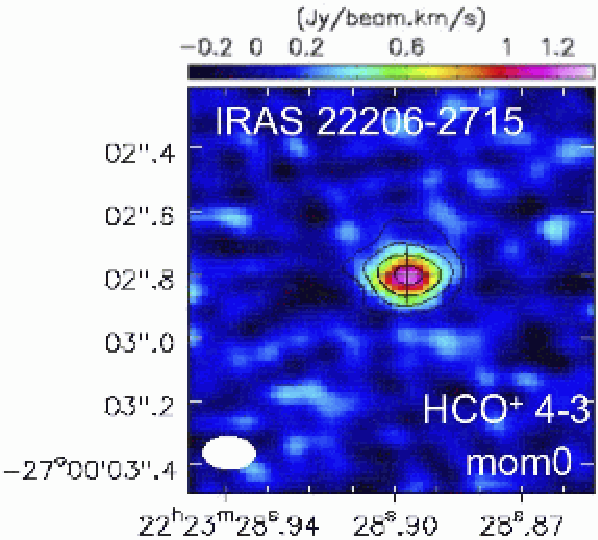} \\
\includegraphics[angle=0,scale=.42]{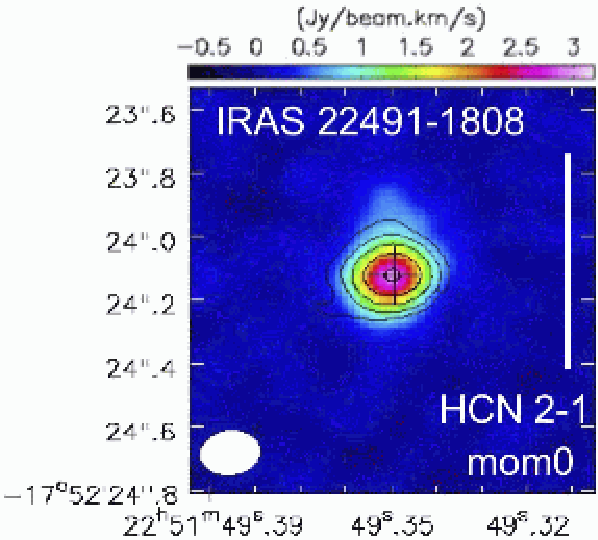} 
\includegraphics[angle=0,scale=.42]{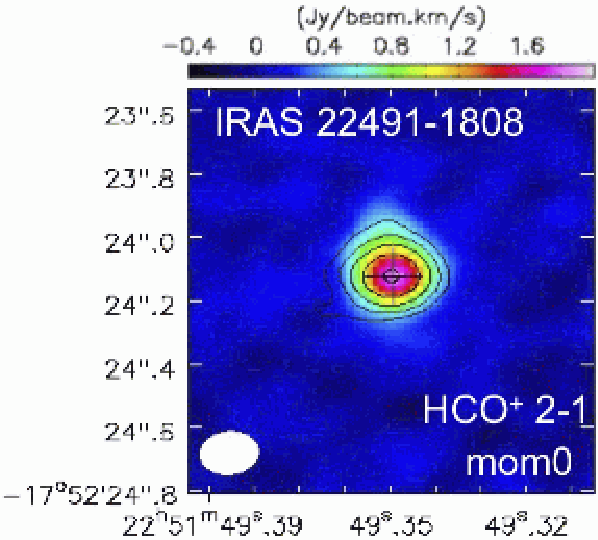} 
\includegraphics[angle=0,scale=.42]{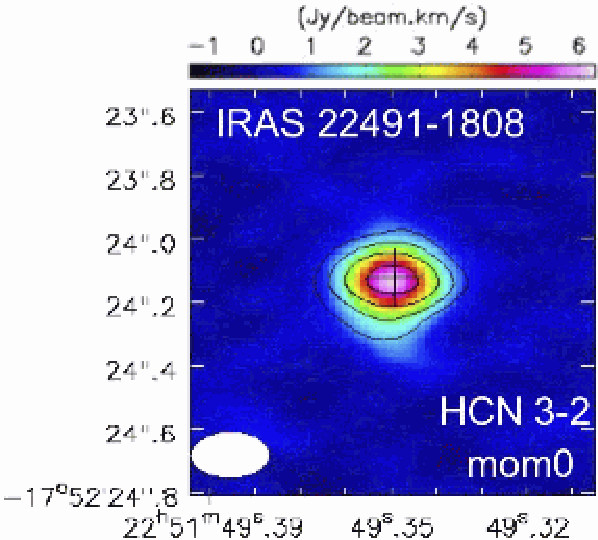} 
\includegraphics[angle=0,scale=.42]{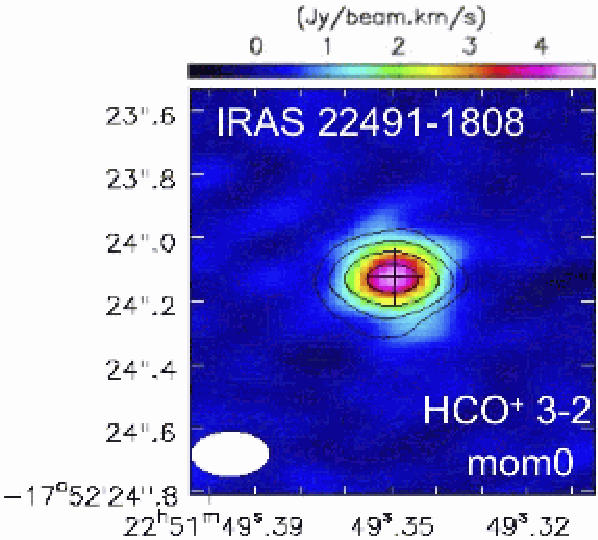} \\
\includegraphics[angle=0,scale=.42]{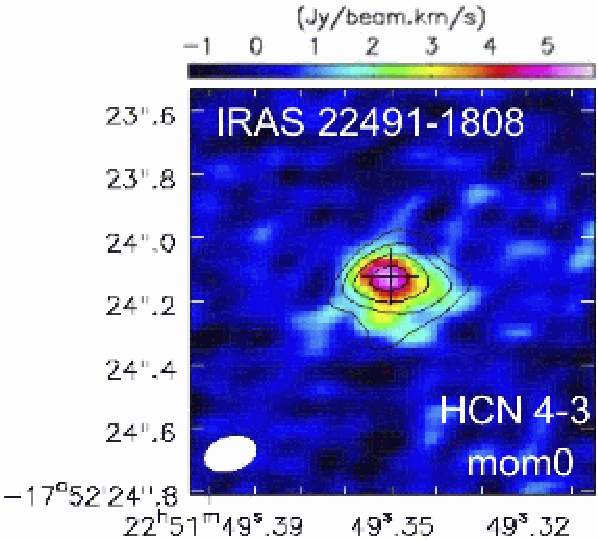} 
\includegraphics[angle=0,scale=.42]{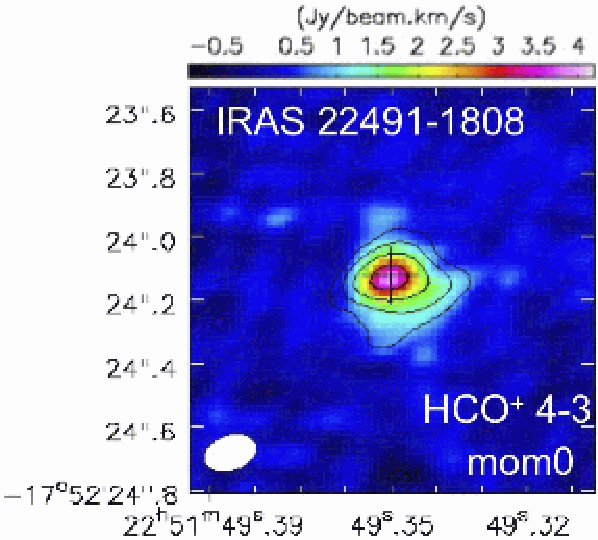} 
\includegraphics[angle=0,scale=.42]{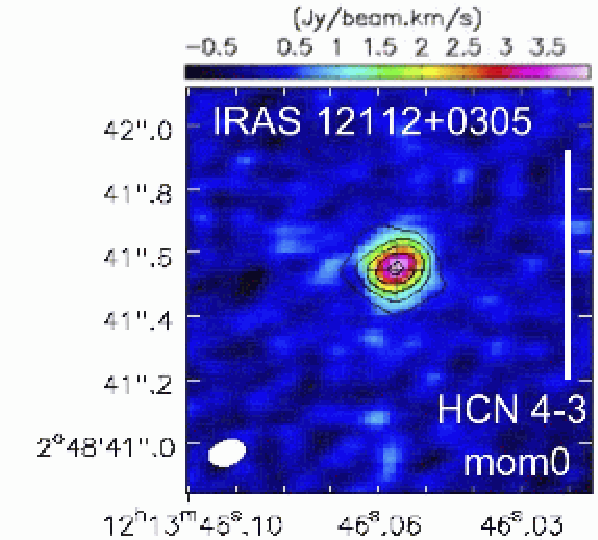} 
\includegraphics[angle=0,scale=.42]{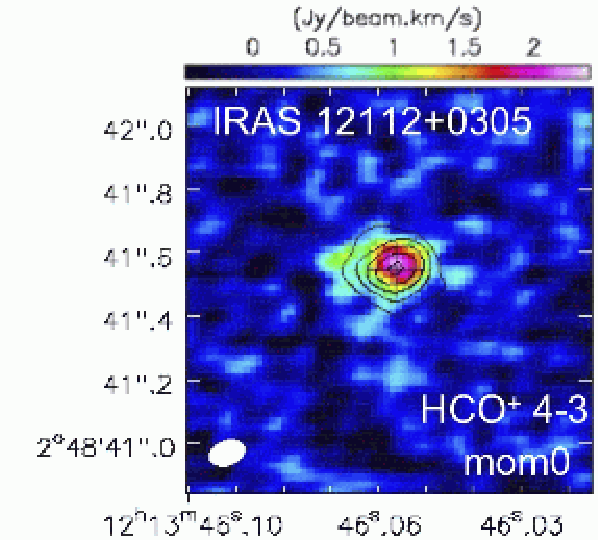} 
\end{center}
\caption{Integrated intensity (moment 0) map of HCN and
HCO$^{+}$ lines created from the original-beam-sized data (Table
\ref{tab:beam}, column 2--4) taken in ALMA Cycle 7.  
Simultaneously obtained continuum emission is overplotted as contours.
Continuum peak position is shown as a cross.
The continuum contours start from 4$\sigma$ and increase by a factor
of 2 (i.e., 8$\sigma$, 16$\sigma$, 32$\sigma$, and 64$\sigma$) for all 
sources.
The length of the vertical white solid bar in the first image of each
object corresponds to 1 kpc. 
Beam size for each moment 0 map is shown as a white filled circle 
in the lower-left region. Coordinates are in ICRS. 
\label{fig:mom0}
}
\end{figure*}

Table \ref{tab:beam} summarizes synthesized beam sizes in the cleaned maps.
Spatial resolution of $\lesssim$0.5 kpc is achieved for all data of
the observed (U)LIRGs. 
Figure \ref{fig:mom0} shows continuum (contours) and integrated-intensity 
(moment 0) maps of newly taken HCN and HCO$^{+}$ lines with 
the original beam size (Table \ref{tab:beam}, column 2--4).
HCN and HCO$^{+}$ emission lines are significantly detected at the
continuum peak positions.
Tables \ref{tab:cont} and \ref{tab:mom0} summarize, respectively,
continuum emission properties and dense molecular emission line properties
derived from the original-beam-sized moment 0 maps.   
Continuum spectral energy distributions from infrared 60 $\mu$m to
ALMA 0.85--2 mm for selected ULIRGs are presented in Appendix A.    
Intensity-weighted mean velocity (moment 1) maps of newly obtained HCN
and HCO$^{+}$ lines in ALMA Cycle 7, created from the original-beam-sized
data, are shown in Appendix B. 

\begin{deluxetable*}{lcccc}[!hbt]
\tabletypesize{\scriptsize}
\tablecaption{Summary of Synthesized Beam Sizes \label{tab:beam}} 
\tablewidth{0pt}
\tablehead{
\colhead{Object} & \multicolumn{3}{c}{Beam size (arcsec $\times$ arcsec)} 
& \colhead{arcsec} \\
\colhead{} & \colhead{J21 (J=2--1)} & 
\colhead{J32 (J=3--2)} & \colhead{J43 (J=4--3)} & 
\colhead{(for 0.5 kpc) } \\
\colhead{(1)} & \colhead{(2)} & \colhead{(3)} & \colhead{(4)} & \colhead{(5)} 
}
\startdata 
IRAS 00091$-$0738 & 0.19$\times$0.15 (Cy7) &
0.18$\times$0.13 (Cy5) & 0.17$\times$0.10 (Cy7) & 0.24 \\
IRAS 00188$-$0856 & 0.22$\times$0.17 (Cy7) &
0.18$\times$0.13 (Cy5) & 0.16$\times$0.10 (Cy7) & 0.22 \\
IRAS 00456$-$2904 & 0.22$\times$0.17 (Cy7) &
0.16$\times$0.12 (Cy5) & 0.19$\times$0.14 (Cy7) & 0.25 \\
IRAS 01166$-$0844 & 0.21$\times$0.17 (Cy7) &
0.12$\times$0.092 (Cy5) & 0.15$\times$0.090 (Cy7) & 0.24 \\
IRAS 01569$-$2939 & 0.21$\times$0.15 (Cy7) &
0.11$\times$0.11 (Cy5) & 0.21$\times$0.087 (Cy7) & 0.21 \\
IRAS 03250$+$1606 & 0.19$\times$0.18 (Cy7) &
0.13$\times$0.10 (Cy5) & 0.11$\times$0.084 (Cy7) & 0.22 \\
IRAS 10378$+$1108 & 0.17$\times$0.15 (Cy7) &
0.17$\times$0.15 (Cy5) & ---  & 0.21 \\
IRAS 16090$-$0139 & 0.20$\times$0.15 (Cy7) &
0.17$\times$0.15 (Cy5) & 0.19$\times$0.11 (Cy7) & 0.21 \\
IRAS 22206$-$2715  & 0.18$\times$0.15 (Cy7) &
0.18$\times$0.13 (Cy7) & 0.16$\times$0.099 (Cy7) & 0.22 \\
IRAS 22491$-$1808 & 0.17$\times$0.12 (Cy7) &
0.23$\times$0.13 (Cy7) & 0.16$\times$0.10 (Cy7) & 0.35 \\ \hline
IRAS 12112$+$0305 & --- & --- & 0.11$\times$0.071 (Cy7)
& 0.36 \\ \hline
NGC 1614 & 0.55$\times$0.37 (Cy5) & 1.1$\times$0.58
(Cy2) & 1.5$\times$1.3 (Cy0) & 1.56 \\ \hline
\enddata

\tablecomments{ 
Col.(1): Object name. 
Cols.(2)--(4): Synthesized beam size of continuum data in arcsec
$\times$ arcsec.  
ALMA Cycle of each data acquisition is also shown in parentheses.
Cy0, Cy2, Cy5, and Cy7 mean Cycle 0, Cycle 2, Cycle 5 and Cycle 7,
respectively.  
Col.(2): J21.
Col.(3): J32.
Col.(4): J43.
J21, J32, and J43 mean continuum data simultaneously taken during HCN
and HCO$^{+}$ J=2--1, J=3--2, and J=4--3 line observations, respectively.
The synthesized beam sizes are almost the same between each 
continuum and corresponding molecular line data.
Col.(5) Angular scale in arcsec corresponding to 0.5 kpc.
}

\end{deluxetable*}

\begin{figure*}[!hbt]
\begin{center}
\includegraphics[angle=0,scale=.42]{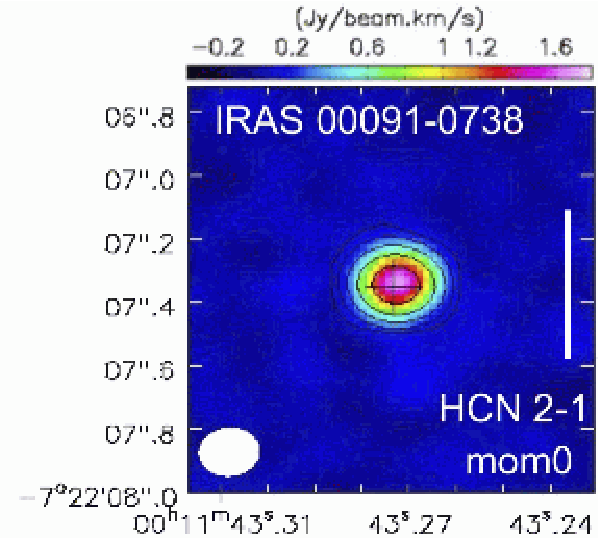} 
\includegraphics[angle=0,scale=.42]{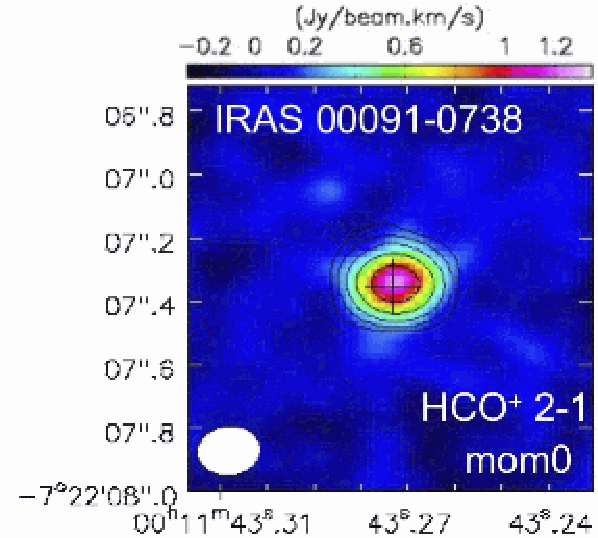} 
\includegraphics[angle=0,scale=.42]{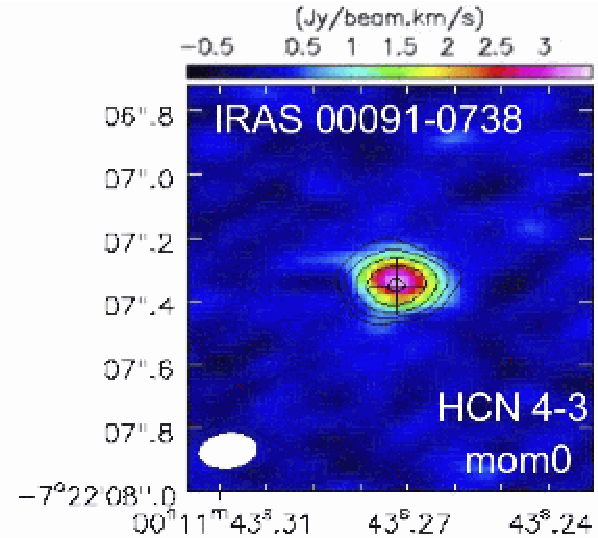} 
\includegraphics[angle=0,scale=.42]{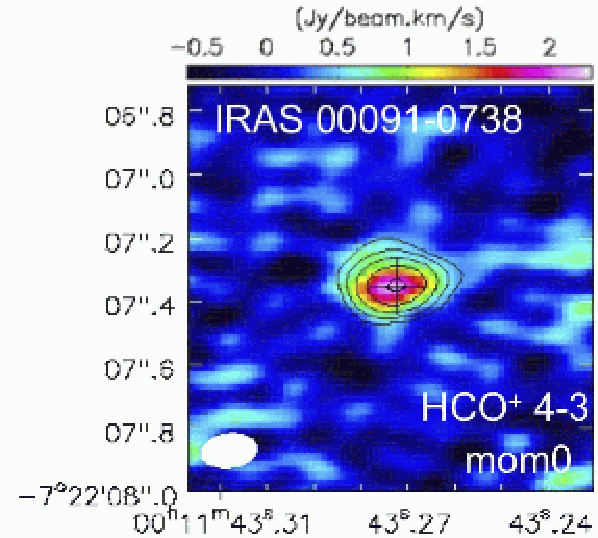} \\
\includegraphics[angle=0,scale=.42]{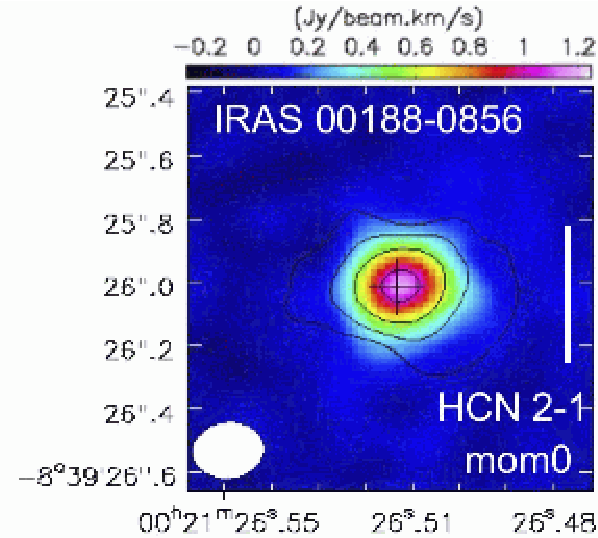} 
\includegraphics[angle=0,scale=.42]{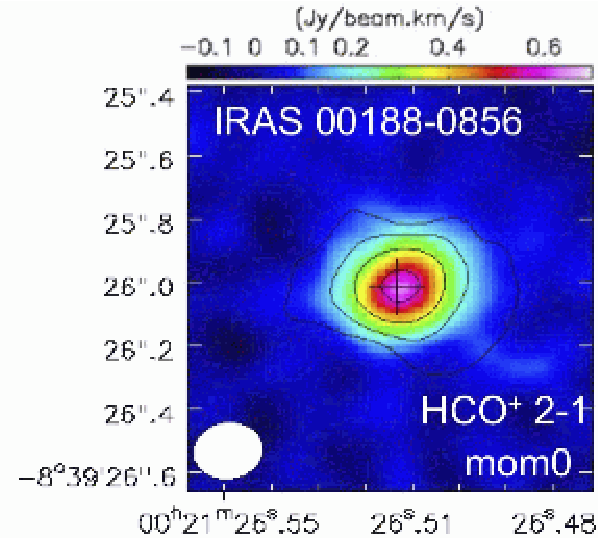} 
\includegraphics[angle=0,scale=.42]{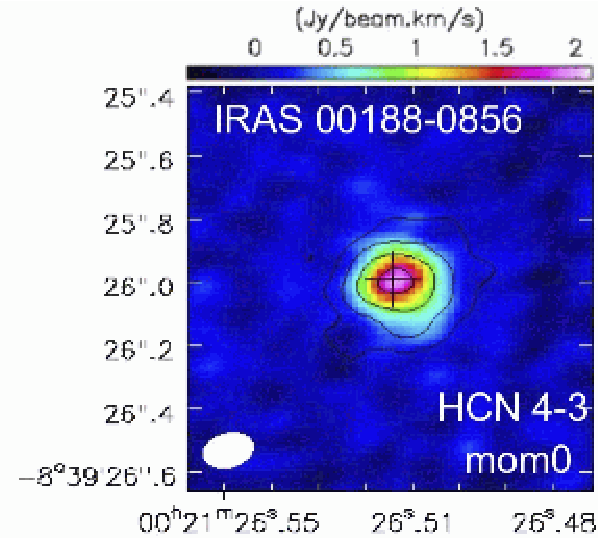} 
\includegraphics[angle=0,scale=.42]{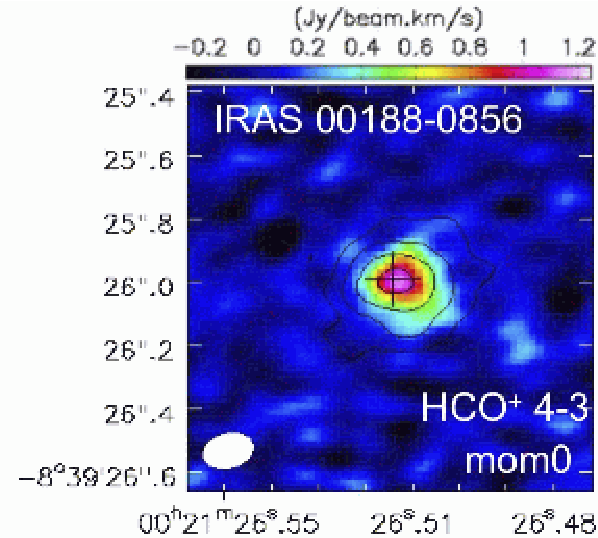} \\
\includegraphics[angle=0,scale=.42]{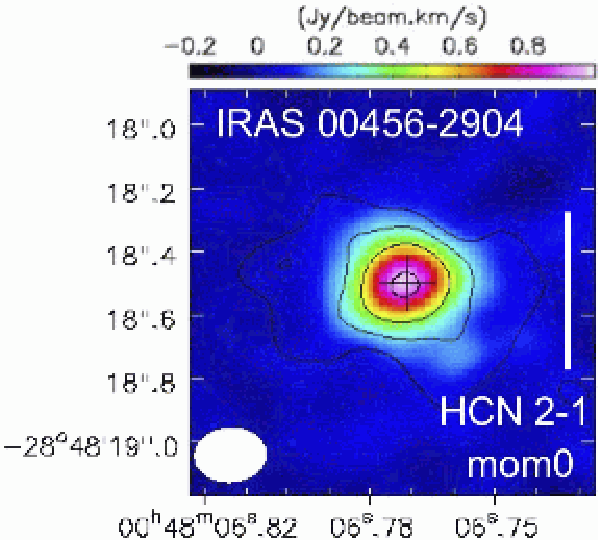} 
\includegraphics[angle=0,scale=.42]{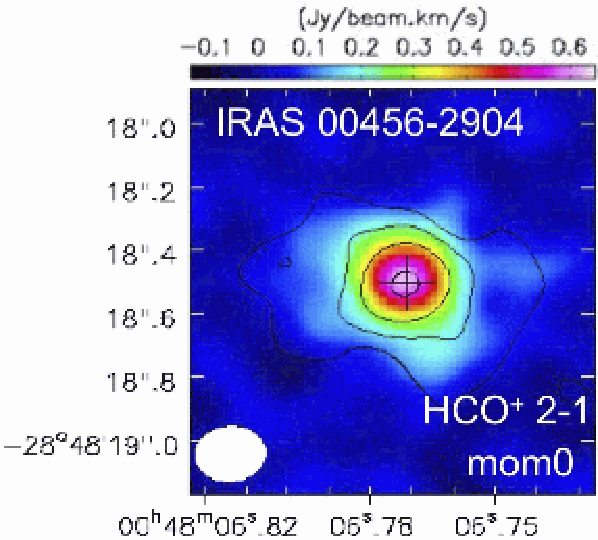} 
\includegraphics[angle=0,scale=.42]{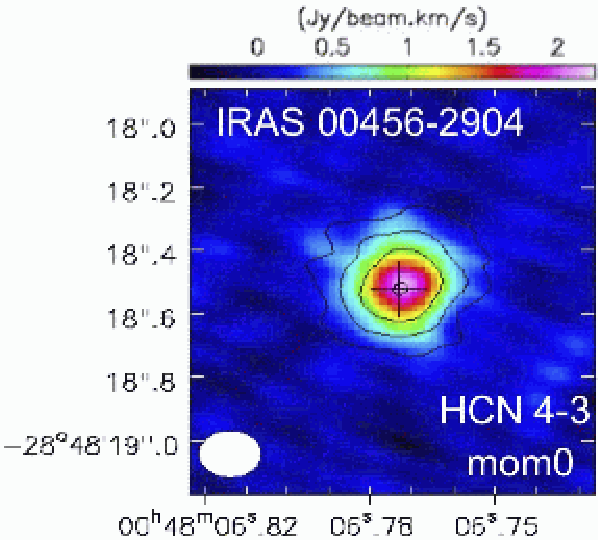} 
\includegraphics[angle=0,scale=.42]{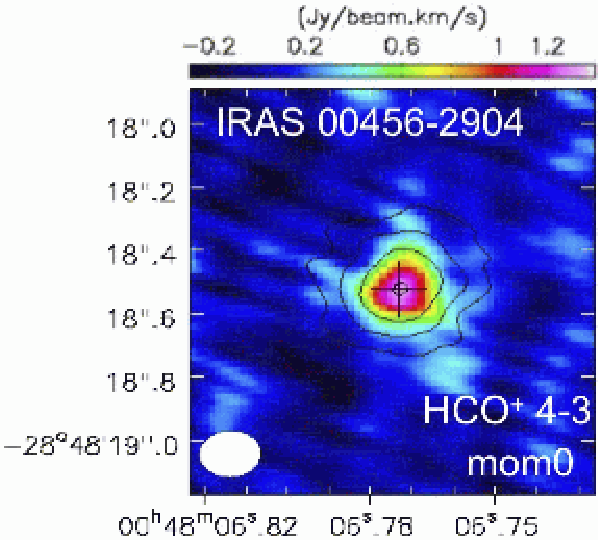} \\
\includegraphics[angle=0,scale=.42]{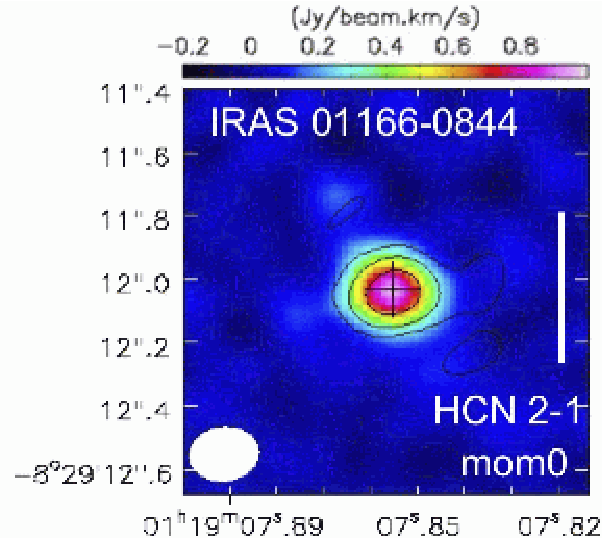} 
\includegraphics[angle=0,scale=.42]{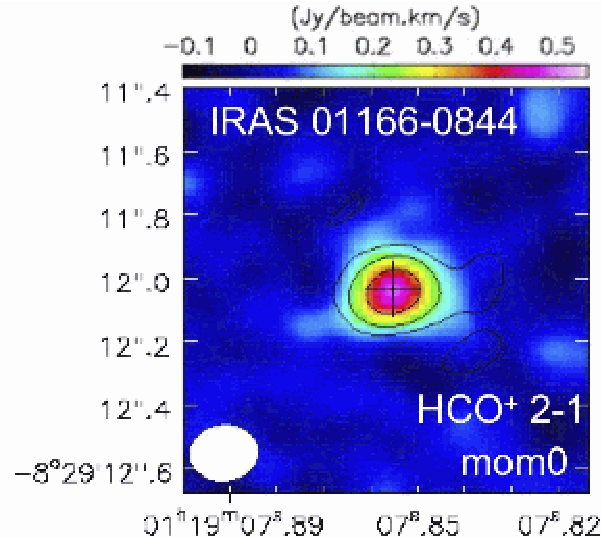} 
\includegraphics[angle=0,scale=.42]{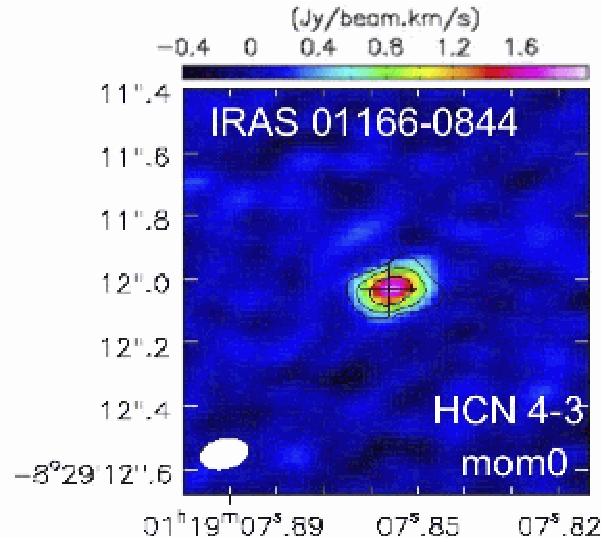} 
\includegraphics[angle=0,scale=.42]{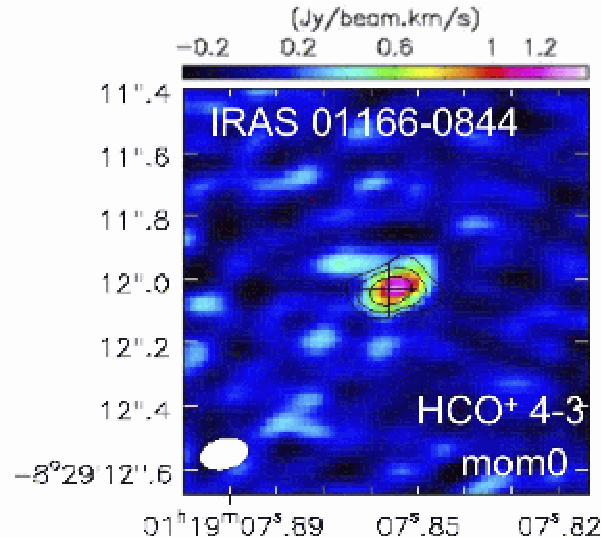} \\
\includegraphics[angle=0,scale=.42]{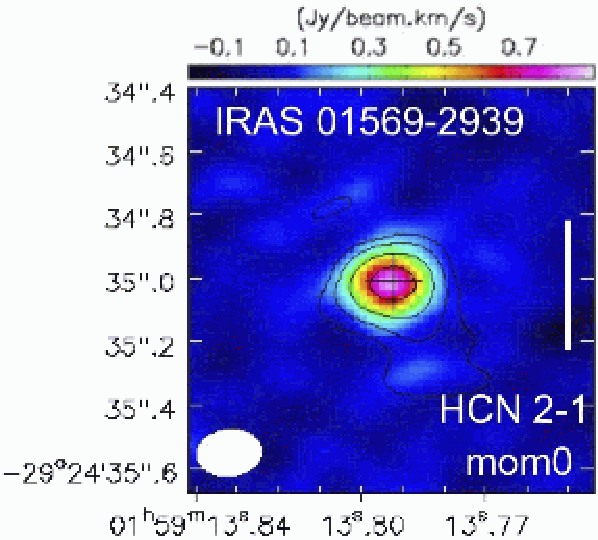} 
\includegraphics[angle=0,scale=.42]{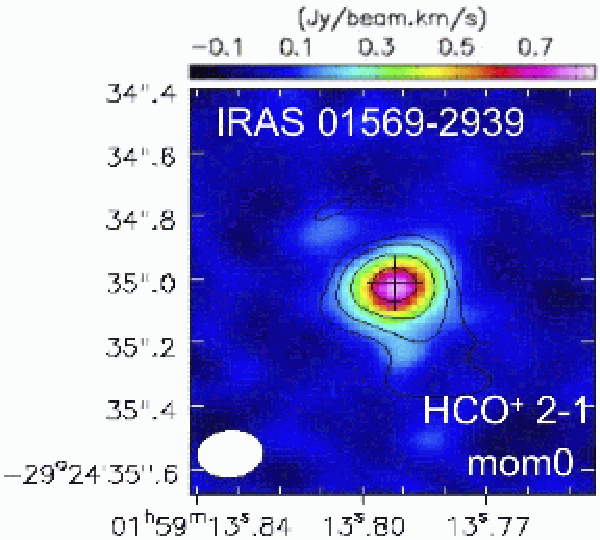} 
\includegraphics[angle=0,scale=.42]{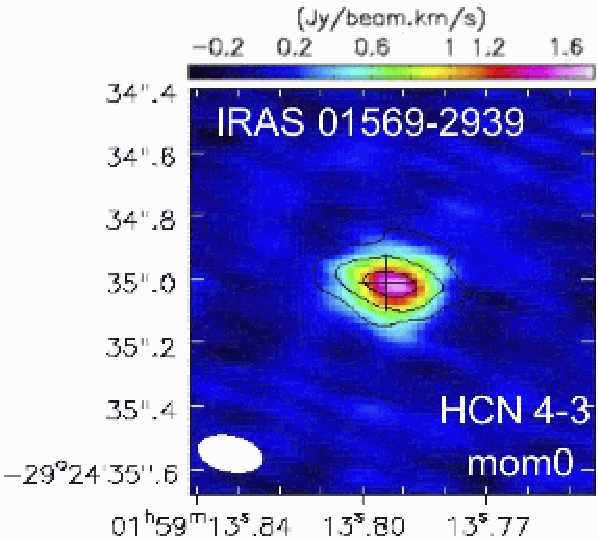} 
\includegraphics[angle=0,scale=.42]{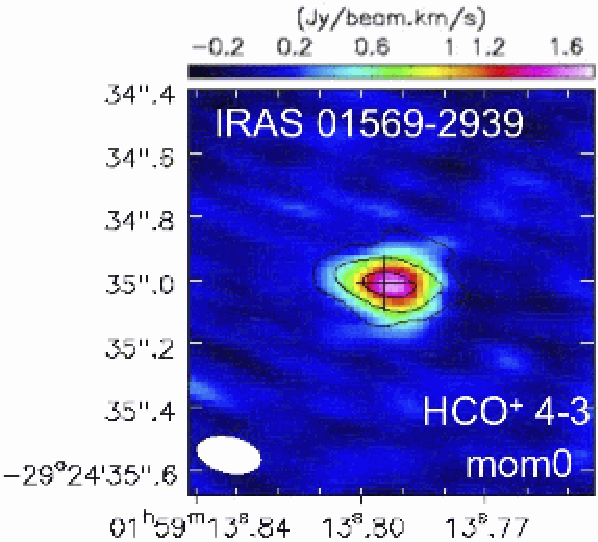} \\
\includegraphics[angle=0,scale=.42]{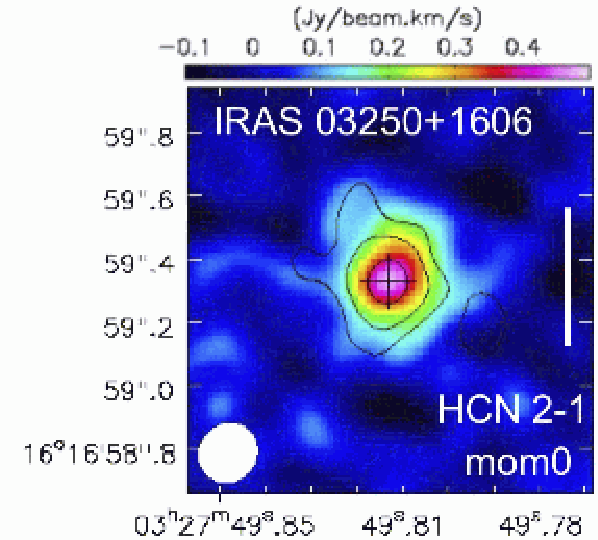} 
\includegraphics[angle=0,scale=.42]{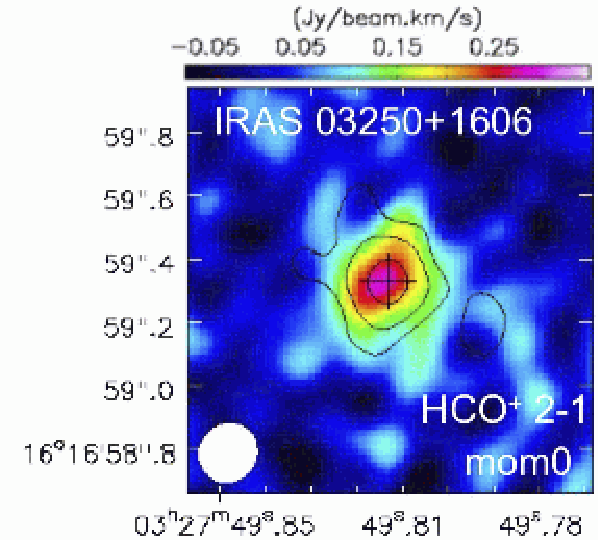} 
\includegraphics[angle=0,scale=.42]{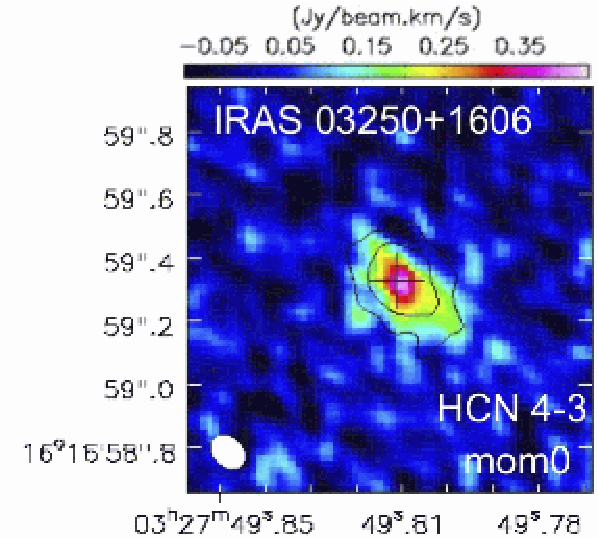} 
\includegraphics[angle=0,scale=.42]{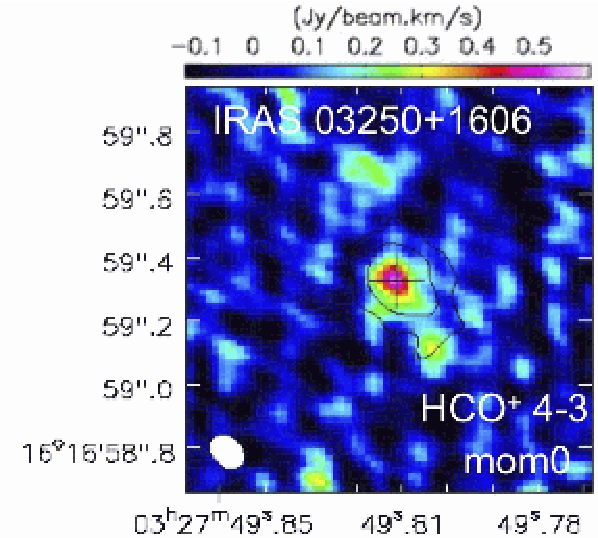} \\
\end{center}
\end{figure*}


\begin{deluxetable*}{lclrcrr}[!hbt]
\tabletypesize{\scriptsize}
\tablecaption{Continuum Emission Properties \label{tab:cont}}
\tablewidth{0pt}
\tablehead{
\colhead{Object} & \colhead{Data} & \colhead{Frequency} &
\colhead{Flux (Original beam)} & 
\colhead{Peak Coordinate} & \colhead{Flux (0.5 kpc)} & \colhead{Flux (1 kpc)} \\
\colhead{} & \colhead{} & \colhead{[GHz]} & \colhead{[mJy/beam] (kpc$\times$kpc)} & 
\colhead{(RA,DEC)ICRS} & \colhead{[mJy]} & \colhead{[mJy]} \\  
\colhead{(1)} & \colhead{(2)} & \colhead{(3)} & \colhead{(4)} 
& \colhead{(5)} & \colhead{(6)} & \colhead{(7)}   
}
\startdata 
IRAS 00091$-$0738 & J21 & 145.3--149.0, 157.1--160.8 (153) & 2.5 (98$\sigma$)
(0.40$\times$0.31) & (00$^{h}$11$^{m}$43.3$^{s}$,
$-$07$^{\circ}$22$'$07$''$) & 2.5 (83$\sigma$) & 2.7 (45$\sigma$) \\ 
(2.1 kpc/$''$) & J32 & 236.8--241.8 (239) & 5.5 (59$\sigma$) 
(0.39$\times$0.27) & (00$^{h}$11$^{m}$43.3$^{s}$,
$-$07$^{\circ}$22$'$07$''$) & 6.0 (44$\sigma$) & 6.9 (24$\sigma$) \\  
 & J43 & 304.4--308.1, 316.4--320.1 (312) & 10.7 (70$\sigma$)
(0.36$\times$0.22) & (00$^{h}$11$^{m}$43.3$^{s}$,
$-$07$^{\circ}$22$'$07$''$) & 12.2 (46$\sigma$) & 13.6 (25$\sigma$) \\ \hline  
IRAS 00188$-$0856 & J21 & 143.5--147.2, 155.6--159.3 (151) &
0.59 (38$\sigma$) (0.49$\times$0.38) &
(00$^{h}$21$^{m}$26.5$^{s}$, $-$08$^{\circ}$39$'$26$''$) & 0.64
(40$\sigma$) & 0.92 (32$\sigma$) \\
(2.3 kpc/$''$) & J32 & 234.7--239.7 (237) & 1.4 (35$\sigma$) 
(0.42$\times$0.29) & (00$^{h}$21$^{m}$26.5$^{s}$,
$-$08$^{\circ}$39$'$26$''$) & 1.8 (34$\sigma$) & 2.8 (22$\sigma$) \\ 
 & J43 & 301.6--305.3, 313.4--317.2 (309) & 3.1 (40$\sigma$)
(0.35$\times$0.23) & (00$^{h}$21$^{m}$26.5$^{s}$,
$-$08$^{\circ}$39$'$26$''$) & 4.9 (32$\sigma$) & 7.8 (21$\sigma$) \\ \hline 
IRAS 00456$-$2904 & J21 & 146.3--150.1, 158.2--162.0 (154) & 0.52
(35$\sigma$) (0.45$\times$0.33) &
(00$^{h}$48$^{m}$06.8$^{s}$, $-$28$^{\circ}$48$'$19$''$) & 0.63
(37$\sigma$) & 0.99 (28$\sigma$) \\ 
(2.0 kpc/$''$) & J32 & 238.5--243.4 (241) & 1.2 (25$\sigma$) 
(0.32$\times$0.24) & (00$^{h}$48$^{m}$06.8$^{s}$,
$-$28$^{\circ}$48$'$19$''$) & 2.0 (24$\sigma$) & 3.1 (16$\sigma$) \\  
 & J43 & 306.6--310.4, 318.7--322.4 (315) & 2.7 (33$\sigma$) 
(0.38$\times$0.27) & (00$^{h}$48$^{m}$06.8$^{s}$,
$-$28$^{\circ}$48$'$19$''$) & 4.0 (30$\sigma$) & 6.1 (19$\sigma$) \\ \hline  
IRAS 01166$-$0844 & J21 & 145.2--148.9, 157.2--160.9 (153) & 0.35 (24$\sigma$)
(0.45$\times$0.36) & (01$^{h}$19$^{m}$07.9$^{s}$,
$-$08$^{\circ}$29$'$12$''$) & 0.36 (25$\sigma$) & 0.49 (20$\sigma$) \\ 
(2.1 kpc/$''$) & J32 & 236.8--241.8 (239) & 0.68 (17$\sigma$)  
(0.25$\times$0.19) & (01$^{h}$19$^{m}$07.9$^{s}$,
$-$08$^{\circ}$29$'$12$''$) & 1.0 (15$\sigma$) & 1.5 (9.6$\sigma$) \\  
 & J43 & 304.6--308.4, 316.6--320.4 (313) & 1.7 (28$\sigma$)
(0.31$\times$0.19) & (01$^{h}$19$^{m}$07.9$^{s}$,
$-$08$^{\circ}$29$'$12$''$) & 2.0 (22$\sigma$) & 2.6 (14$\sigma$) \\ \hline 
IRAS 01569$-$2939 & J21 & 142.0--145.7, 154.0--157.7 (150) & 0.59 (43$\sigma$)
(0.51$\times$0.36) & (01$^{h}$59$^{m}$13.8$^{s}$,
$-$29$^{\circ}$24$'$35$''$) & 0.62 (43$\sigma$) & 0.80 (32$\sigma$) \\ 
(2.4 kpc/$''$) & J32 & 232.2--237.2 (235) & 0.70 (17$\sigma$) 
(0.27$\times$0.27) & (01$^{h}$59$^{m}$13.8$^{s}$,
$-$29$^{\circ}$24$'$35$''$) & 0.88 (16$\sigma$) & 1.3 (11$\sigma$) \\  
 & J43 & 298.4--302.1, 310.2--313.9 (306) & 1.1 (21$\sigma$)
(0.52$\times$0.21) & (01$^{h}$59$^{m}$13.8$^{s}$,
$-$29$^{\circ}$24$'$35$''$) & 1.5 (20$\sigma$) & 2.4 (17$\sigma$) \\ \hline 
IRAS 03250$+$1606 & J21 & 143.5--147.2, 155.6--159.2 (151) & 0.35 (20$\sigma$)
(0.43$\times$0.40) & (03$^{h}$27$^{m}$49.8$^{s}$,
$+$16$^{\circ}$16$'$59$''$) & 0.40 (23$\sigma$) & 0.62 (21$\sigma$) \\ 
(2.3 kpc/$''$) & J32 & 234.9--239.6 (237) & 0.41 (14$\sigma$)
(0.30$\times$0.23) & (03$^{h}$27$^{m}$49.8$^{s}$,
$+$16$^{\circ}$16$'$59$''$) & 0.81 (18$\sigma$) & 1.5 (15$\sigma$) \\  
 & J43 & 301.5--305.3, 313.4--317.1 (309) & 0.53 (15$\sigma$)
(0.25$\times$0.19) & (03$^{h}$27$^{m}$49.8$^{s}$,
$+$16$^{\circ}$16$'$59$''$) & 1.6 (22$\sigma$) & 3.0 (16$\sigma$) \\ \hline 
IRAS 10378$+$1108 & J21 & 142.4--146.2, 154.5--158.2 (150) & 0.27 (13$\sigma$)
(0.37$\times$0.34) & (10$^{h}$40$^{m}$29.2$^{s}$,
$+$10$^{\circ}$53$'$18$''$) & 0.33 (16$\sigma$) & 0.53 (15$\sigma$) \\ 
(2.4 kpc/$''$) & J32 & 233.0--238.0 (236) & 1.5 (33$\sigma$) 
(0.40$\times$0.35) & (10$^{h}$40$^{m}$29.2$^{s}$,
$+$10$^{\circ}$53$'$18$''$) & 1.7 (33$\sigma$) & 2.3 (21$\sigma$) \\ \hline 
IRAS 16090$-$0139 & J21 & 142.8--146.6, 154.9--158.6 (151) & 0.51 (23$\sigma$)
(0.45$\times$0.36) & (16$^{h}$11$^{m}$40.4$^{s}$,
$-$01$^{\circ}$47$'$06$''$) & 0.67 (29$\sigma$) & 1.2 (27$\sigma$) \\ 
(2.3 kpc/$''$) & J32 & 233.6--238.5 (236) & 1.4 (20$\sigma$)
(0.39$\times$0.35) & (16$^{h}$11$^{m}$40.4$^{s}$,
$-$01$^{\circ}$47$'$06$''$) & 1.9 (23$\sigma$) & 3.5 (17$\sigma$) \\  
 & J43 & 300.2--303.9, 312.0--315.8 (308) & 2.8 (33$\sigma$)
(0.42$\times$0.24) & (16$^{h}$11$^{m}$40.4$^{s}$,
$-$01$^{\circ}$47$'$06$''$) & 4.8 (32$\sigma$) & 7.8 (21$\sigma$) \\ \hline 
IRAS 22206$-$2715 & J21 & 143.0--146.7, 155.1--158.8 (151) & 0.55 (38$\sigma$)
(0.42$\times$0.34) & (22$^{h}$23$^{m}$28.9$^{s}$,
$-$27$^{\circ}$00$'$03$''$) & 0.58 (39$\sigma$) & 0.73 (27$\sigma$) \\  
(2.3 kpc/$''$) & J32 & 233.6--239.1 (236) & 1.3 (27$\sigma$)
(0.42$\times$0.31) & (22$^{h}$23$^{m}$28.9$^{s}$,
$-$27$^{\circ}$00$'$03$''$) & 1.4 (27$\sigma$) & 2.1 (18$\sigma$) \\ 
 & J43 & 300.6--304.3, 312.4--316.2 (308) & 2.8 (39$\sigma$)
(0.36$\times$0.23) & (22$^{h}$23$^{m}$28.9$^{s}$,
$-$27$^{\circ}$00$'$03$''$) & 3.4 (32$\sigma$) & 4.6 (20$\sigma$) \\ \hline 
IRAS 22491$-$1808  & J21 & 163.0--166.8, 175.0--178.6 (171) & 1.8
(69$\sigma$) (0.24$\times$0.18) &
(22$^{h}$51$^{m}$49.4$^{s}$, $-$17$^{\circ}$52$'$24$''$) & 2.2
(42$\sigma$) & 2.8 (24$\sigma$) \\ 
(1.5 kpc/$''$) & J32 & 245.6--251.1 (248) & 4.1 (44$\sigma$)
(0.34$\times$0.18) & 
(22$^{h}$51$^{m}$49.4$^{s}$, $-$17$^{\circ}$52$'$24$''$) & 4.9
(28$\sigma$) & 5.7 (16$\sigma$) \\ 
 & J43 & 316.2--319.7, 328.3--332.1 (324) & 5.4 (48$\sigma$)
(0.24$\times$0.15) & (22$^{h}$51$^{m}$49.4$^{s}$,
$-$17$^{\circ}$52$'$24$''$) & 8.7 (24$\sigma$) & 12.3 (13$\sigma$) \\ \hline 
IRAS 12112$+$0305 & J43 & 317.5--321.1, 329.7--333.5 (326) & 9.8 (73$\sigma$)
(0.16$\times$0.10) & (12$^{h}$13$^{m}$46.1$^{s}$,
$+$02$^{\circ}$48$'$42$''$) & 16.7 (26$\sigma$) & 20.0 (14$\sigma$) \\   
(1.4 kpc/$''$) & &  & & \\ \hline
NGC 1614 & J21 & 173.2--176.9, 185.4--189.1 (181) & --- \tablenotemark{a} 
(0.18$\times$0.12) & 
(04$^{h}$34$^{m}$00.0$^{s}$, $-$08$^{\circ}$34$'$45$''$) \tablenotemark{b} & 
6.0 (11$\sigma$) & 10.3 (7.5$\sigma$) \\
(0.32 kpc/$''$) & J32 & 260.8--265.5 (263) & --- \tablenotemark{a} 
(0.34$\times$0.19) & 
(04$^{h}$34$^{m}$00.0$^{s}$, $-$08$^{\circ}$34$'$45$''$) \tablenotemark{b} & 
7.3 (9.4$\sigma$) & 12.5 (6.9$\sigma$) \\
 & J43 & 336.1--338.1, 348.0--351.9 (344) & --- \tablenotemark{a} 
(0.44$\times$0.36) & 
(04$^{h}$34$^{m}$00.0$^{s}$, $-$08$^{\circ}$34$'$45$''$)
\tablenotemark{b} & 
14.9 (24$\sigma$) & 22.1 (16$\sigma$) \\ \hline
\enddata

\tablenotetext{a}{Multiple continuum peak positions are found in 
the original-beam-sized data \citep[e.g.,][]{ima16b,ima22}.}

\tablenotetext{b}{Continuum peak position in the 0.5 kpc beam-sized data.}

\tablecomments{Col.(1): Object name.
Physical scale in kpc arcsec$^{-1}$ is shown in parentheses for reference.
Col.(2): J21, J32, J43, respectively, mean continuum data
simultaneously taken during J=2--1, J=3--2, and J=4--3 observations of
HCN and HCO$^{+}$.
Col.(3): Frequency range in GHz used for continuum extraction is shown 
first.
Frequencies of obvious emission and absorption lines are removed.
The central frequency in GHz is shown in parenthesis.
Col.(4): Flux in mJy beam$^{-1}$ at the emission peak in the original beam.
Value at the highest flux pixel is extracted. 
The pixel scale is 0$\farcs$02 pixel$^{-1}$ for all ULIRGs' data 
(This paper; \citet{ima19}).  
For NGC 1614 J21, J32, and J43 data, the pixel scale is 
0$\farcs$05 pixel$^{-1}$, 0$\farcs$1 pixel$^{-1}$, and 0$\farcs$3
pixel$^{-1}$, respectively \citep{ima13a,ima16b,ima22}.  
Detection significance relative to the root mean square (rms)
noise (1$\sigma$) is shown in the first parentheses.  
Possible systematic uncertainties, coming from the absolute flux
calibration ambiguity in individual ALMA observation and 
choice of frequency range for the continuum level determination, 
are not included. 
Original beam size in kpc is shown in the second parentheses.
Col.(5): Coordinate of the continuum emission peak in ICRS in the
original-beam-sized map.
For NGC 1614, that in the 0.5 kpc beam-sized data is shown.
Cols.(6) and (7): Flux in mJy at the continuum emission peak in the
0.5 kpc and 1 kpc circular beam, respectively.   
Detection significance relative to the rms noise (1$\sigma$) is shown
in parentheses.  
}

\end{deluxetable*}


\begin{deluxetable*}{l|cc|cc|cc|c|c}[!hbt]
\tabletypesize{\scriptsize}
\tablecaption{Peak Flux of Molecular Emission Line in Integrated Intensity 
(Moment 0) Map with Original Beam Size \label{tab:mom0}} 
\tablewidth{0pt}
\tablehead{
\colhead{} & \multicolumn{8}{c}{Peak [Jy beam$^{-1}$ km s$^{-1}$]} \\ 
\colhead{Object} & \colhead{HCN J=2--1} & \colhead{HCO$^{+}$ J=2--1} &
\colhead{HCN J=3--2} & \colhead{HCO$^{+}$ J=3--2} & 
\colhead{HCN J=4--3} & \colhead{HCO$^{+}$ J=4--3} & \colhead{CS J=7--6} 
& \colhead{HC$_{3}$N J=18--17} \\
\colhead{(1)} & \colhead{(2)} & \colhead{(3)} & \colhead{(4)} & 
\colhead{(5)} & \colhead{(6)} & \colhead{(7)} & \colhead{(8)} &
\colhead{(9)}  
}
\startdata 
IRAS 00091$-$0738 & 1.7 (37$\sigma$) & 1.3 (18$\sigma$) & 2.8 (17$\sigma$) 
& 1.5 (15$\sigma$) & 3.3 (18$\sigma$) & 2.2 (7.4$\sigma$) & 1.6
(13$\sigma$) & 0.93 (24$\sigma$) \\
IRAS 00188$-$0856 & 1.2 (33$\sigma$) & 0.63 (20$\sigma$) & 1.6 (25$\sigma$) 
& 0.86 (15$\sigma$) & 2.0 (20$\sigma$) & 1.2 (10$\sigma$) & 0.42
(7.2$\sigma$) & 0.17 (7.1$\sigma$) \\ 
IRAS 00456$-$2904 & 0.98 (32$\sigma$) & 0.63 (22$\sigma$) & 1.5 (20$\sigma$) 
& 0.76 (14$\sigma$) & 2.2 (19$\sigma$) & 1.3 (9.4$\sigma$) & 0.31
(5.4$\sigma$) & 0.15 (5.1$\sigma$) \\
IRAS 01166$-$0844 & 0.86 (22$\sigma$) & 0.49 (14$\sigma$) & 1.1 (16$\sigma$) 
& 0.71 (10$\sigma$) & 1.9 (16$\sigma$) & 1.3 (9.1$\sigma$) & 0.72
(9.1$\sigma$) & 0.064 (3.1$\sigma$) \\
IRAS 01569$-$2939 & 0.85 (19$\sigma$) & 0.82 (18$\sigma$) & 1.1 (16$\sigma$) 
& 1.1 (14$\sigma$) & 1.7 (17$\sigma$) & 1.7 (15$\sigma$) & 0.55
(5.9$\sigma$) & $<$0.039 ($<$3$\sigma$) \\
IRAS 03250$+$1606 & 0.48 (11$\sigma$) & 0.29 (7.6$\sigma$) & 0.56 (11$\sigma$) 
& 0.40 (8.5$\sigma$) & 0.41 (8.0$\sigma$) & 0.53 (5.9$\sigma$) & 0.096
(3.4$\sigma$) & $<$0.066 ($<$3$\sigma$) \\ 
IRAS 10378$+$1108 & 0.58 (12$\sigma$) & 0.54 (10$\sigma$) & 1.9 (26$\sigma$) 
& 1.8 (24$\sigma$) & --- & --- & --- & $<$0.10 ($<$3$\sigma$) \\
IRAS 16090$-$0139 & 1.7 (26$\sigma$) & 1.3 (19$\sigma$) & 2.7 (26$\sigma$) 
& 2.1 (18$\sigma$) & 4.6 (25$\sigma$) & 3.3 (20$\sigma$) & 1.2
(12$\sigma$) & 0.31 (6.7$\sigma$) \\
IRAS 22206$-$2715 & 0.91 (25$\sigma$) & 0.54 (16$\sigma$) & 1.5 (18$\sigma$) 
& 1.2 (13$\sigma$) & 1.6 (17$\sigma$) & 1.2 (9.9$\sigma$) & 0.72
(8.1$\sigma$) & 0.13 (4.6$\sigma$) \\ 
IRAS 22491$-$1808 & 2.9 (37$\sigma$) & 1.8 (26$\sigma$) & 6.1 (31$\sigma$) 
& 4.6 (23$\sigma$) & 5.4 (9.5$\sigma$) & 4.0 (16$\sigma$) & 2.4
(16$\sigma$) & 0.50 (12$\sigma$) \tablenotemark{a} \\ \hline
IRAS 12112$+$0305 & --- & --- & --- & ---  & 3.9 (15$\sigma$) & 2.3
(8.3$\sigma$) & 1.2 (6.1$\sigma$) & --- \\  \hline
\enddata

\tablenotetext{a}{HC$_{3}$N J=21--20 emission line at 
$\nu_{\rm rest}$=191.040 GHz.}

\tablecomments{Col.(1): Object name. 
The LIRG NGC 1614 is not shown because there are multiple emission
peaks in the original-beam-sized moment 0 maps \citep[e.g.,][]{ima16b,ima22}. 
Cols.(2)--(9): Flux in Jy beam$^{-1}$ km s$^{-1}$ at the emission peak
in the moment 0 map with the original synthesized beam (Table
\ref{tab:beam}, column 2--4). 
Detection significance relative to the rms noise (1$\sigma$) in the 
moment 0 map is shown in parentheses. 
These original-beam-sized moment 0 maps are primarily used for 
the verification of significant molecular line detection at or very
close to the continuum emission peak position (tabulated in Table
\ref{tab:cont}, column 5). 
Col.(2): HCN J=2--1 (rest-frame frequency $\nu_{\rm rest}$=177.261 GHz).
Col.(3): HCO$^{+}$ J=2--1 ($\nu_{\rm rest}$=178.375 GHz).
Col.(4): HCN J=3--2 ($\nu_{\rm rest}$=265.886 GHz). 
Col.(5): HCO$^{+}$ J=3--2 ($\nu_{\rm rest}$=267.558 GHz). 
Col.(6): HCN J=4--3 ($\nu_{\rm rest}$=354.505 GHz).
Col.(7): HCO$^{+}$ J=4--3 ($\nu_{\rm rest}$=356.734 GHz).
The original beam size of the J=2--1, J=3--2, and J=4--3 line is 
virtually identical to that of the continuum J21, J32, and J43 data 
shown in column 2, 3, and 4 of Table \ref{tab:beam}, respectively.
Col.(8): CS J=7--6 ($\nu_{\rm rest}$=342.883 GHz).
Its original beam size is comparable to that of the continuum J43 data 
shown in Table \ref{tab:beam} (column 4).
Col.(9): HC$_{3}$N J=18--17 ($\nu_{\rm rest}$=163.753 GHz).
For IRAS 22491$-$1808, HC$_{3}$N J=21--20 ($\nu_{\rm rest}$=191.040
GHz) was covered, instead of HC$_{3}$N J=18--17.
The original beam size of the HC$_{3}$N lines is comparable to that of
the continuum J21 data shown in Table \ref{tab:beam} (column 2). 
}

\end{deluxetable*}

\begin{figure*}
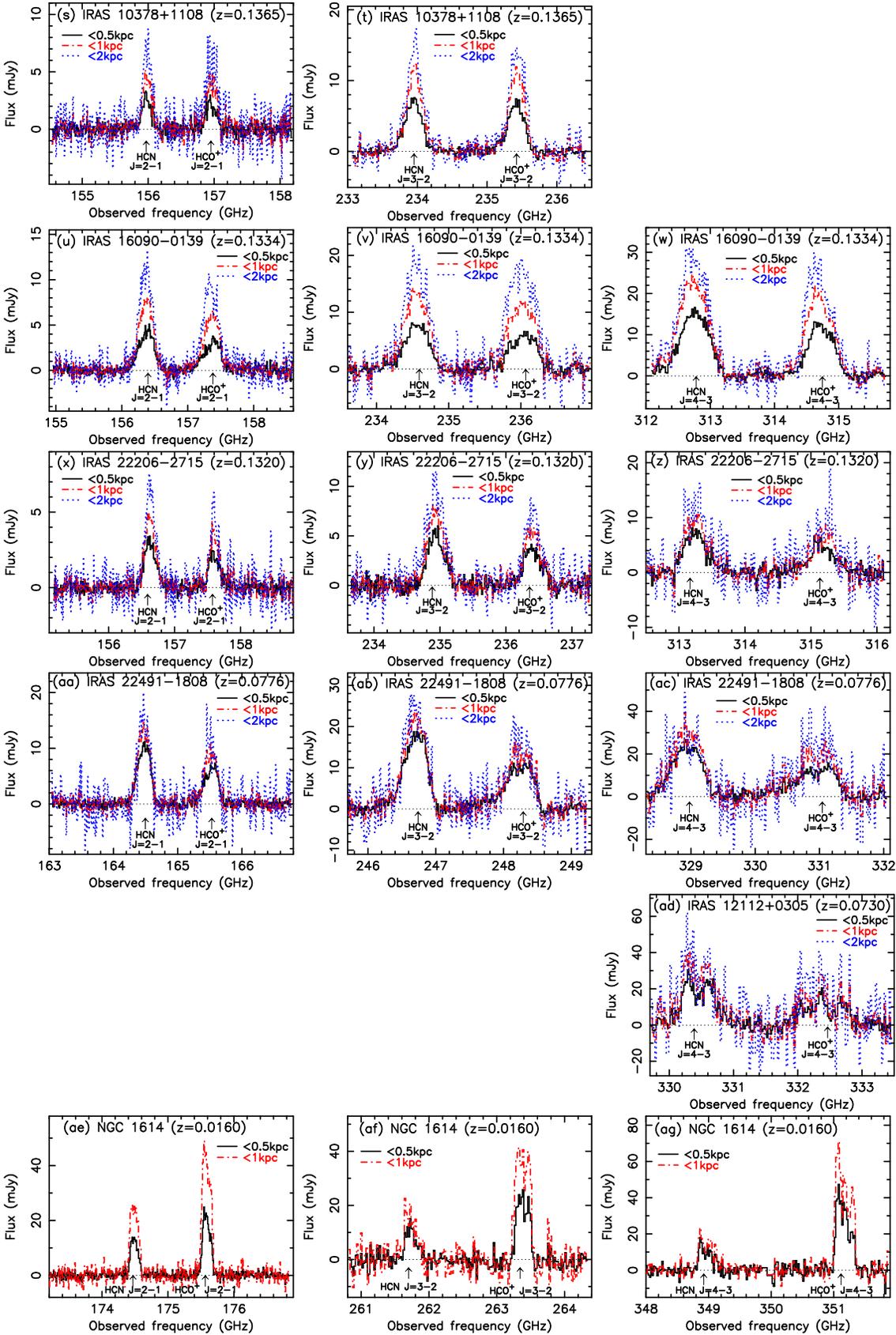

\includegraphics[angle=-90,scale=.22]{f2s.eps} 
\includegraphics[angle=-90,scale=.22]{f2t.eps} \\
\includegraphics[angle=-90,scale=.22]{f2u.eps} 
\includegraphics[angle=-90,scale=.22]{f2v.eps} 
\includegraphics[angle=-90,scale=.22]{f2w.eps} \\
\includegraphics[angle=-90,scale=.22]{f2x.eps} 
\includegraphics[angle=-90,scale=.22]{f2y.eps} 
\includegraphics[angle=-90,scale=.22]{f2z.eps} \\
\includegraphics[angle=-90,scale=.22]{f2aa.eps} 
\includegraphics[angle=-90,scale=.22]{f2ab.eps} 
\includegraphics[angle=-90,scale=.22]{f2ac.eps} \\
%
\hspace*{10.0cm}
\includegraphics[angle=-90,scale=.22]{f2ad.eps} \\
\includegraphics[angle=-90,scale=.22]{f2ae.eps} 
\includegraphics[angle=-90,scale=.22]{f2af.eps} 
\includegraphics[angle=-90,scale=.22]{f2ag.eps} 
\caption{
Spectra within central 0.5 kpc (black solid line), 
1 kpc (red dash-dotted line), and 2 kpc (blue dotted line) regions. 
The abscissa is observed frequency in GHz and the ordinate is flux
density in mJy. 
{\it (Left)}: J=2--1 of HCN and HCO$^{+}$.
{\it (Middle)}: J=3--2 of HCN and HCO$^{+}$.
{\it (Right)}: J=4--3 of HCN and HCO$^{+}$.
The expected frequency of HCN and HCO$^{+}$ J=2--1, J=3--2, and J=4--3
lines, at the adopted redshift of each (U)LIRG (column 2 of Table
\ref{tab:object}; also displayed at the top of each plot), 
is indicated with vertical arrow. 
The horizontal black thin dotted straight line indicates the zero flux level.
For the LIRG NGC 1614, $\lesssim$2 kpc spectra are not extracted 
because no meaningful information of molecular line emission 
at $\gtrsim$1 kpc is obtained from our ALMA data.  
\label{fig:SpectraA}
}
\end{figure*}

\begin{figure*}
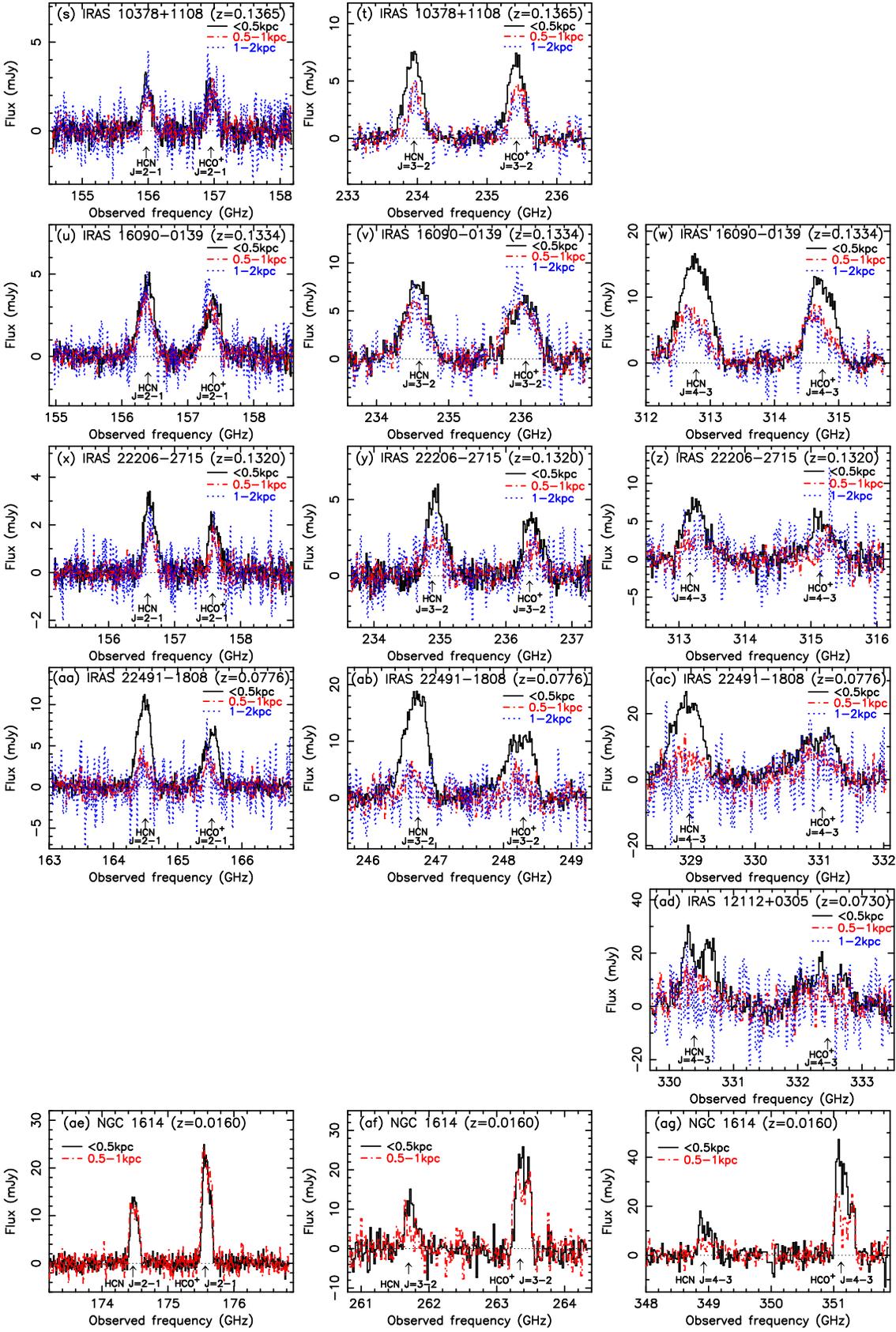

\includegraphics[angle=-90,scale=.22]{f3s.eps} 
\includegraphics[angle=-90,scale=.22]{f3t.eps} \\
\includegraphics[angle=-90,scale=.22]{f3u.eps} 
\includegraphics[angle=-90,scale=.22]{f3v.eps} 
\includegraphics[angle=-90,scale=.22]{f3w.eps} \\
\includegraphics[angle=-90,scale=.22]{f3x.eps} 
\includegraphics[angle=-90,scale=.22]{f3y.eps} 
\includegraphics[angle=-90,scale=.22]{f3z.eps} \\
\includegraphics[angle=-90,scale=.22]{f3aa.eps} 
\includegraphics[angle=-90,scale=.22]{f3ab.eps} 
\includegraphics[angle=-90,scale=.22]{f3ac.eps} \\
%
\hspace*{10.0cm}
\includegraphics[angle=-90,scale=.22]{f3ad.eps} \\
\includegraphics[angle=-90,scale=.22]{f3ae.eps} 
\includegraphics[angle=-90,scale=.22]{f3af.eps} 
\includegraphics[angle=-90,scale=.22]{f3ag.eps} 
\caption{
Spectra within central 0.5 kpc (black solid line), 0.5--1 kpc annular
(red dash-dotted line), and 1--2 kpc annular (blue dotted line)
regions, displayed in the same way as Figure \ref{fig:SpectraA}.
The 1--2 kpc spectra of the LIRG NGC 1614 are not shown 
for the same reason as explained in Figure \ref{fig:SpectraA}
caption. 
\label{fig:SpectraB}
}
\end{figure*}

To estimate nuclear dense molecular emission line 
(HCN and HCO$^{+}$) fluxes from the same physical scale, we modify
the original synthesized beam sizes (Table \ref{tab:beam}) to a $\sim$0.5
kpc diameter circle, using the CASA task ``imsmooth'' 
\citep{CASA22} to cleaned images, for all ULIRGs, and then extract
0.5 kpc beam-sized integrated flux spectra at the continuum
emission peak position.  
These spectra (called $\lesssim$0.5 kpc spectra) are shown in Figure
\ref{fig:SpectraA}.  
To investigate possible spatial variation of dense molecular 
line flux ratios, we also modify the original beam to 1 kpc and 2 kpc
diameter circles, and extract 1 kpc and 2 kpc beam-sized 
integrated flux spectra 
(called $\lesssim$1 kpc and $\lesssim$2 kpc spectra, respectively),
which are also shown in Figure \ref{fig:SpectraA}.
We also extract spectra of a 0.5--1 kpc (1--2 kpc) annular region, by
subtracting the 0.5 kpc (1 kpc) beam-sized spectrum from the 1 kpc (2
kpc) beam-sized spectrum.   
These spectra at the 0.5--1 kpc and 1--2 kpc annular regions 
(named 0.5--1 kpc and 1--2 kpc spectra, respectively) are shown in 
Figure \ref{fig:SpectraB}, by overplotting on the $\lesssim$0.5 kpc spectra. 
In making these new fixed-physical-scale spectra, we need to note
that in interferometric data, when we modify the originally very small
beam size to a large beam size, the resulting rms noise in units of
mJy beam$^{-1}$ increases.
The resulting large-beam-sized spectrum can become noisy with large
scatters.
In fact, the scatters of data points are generally larger for the 1--2
kpc and $\lesssim$2 kpc spectra than the 0.5--1 kpc and $\lesssim$0.5
kpc spectra (Figures \ref{fig:SpectraA} and \ref{fig:SpectraB}),
because we enlarge the originally small beam size ($\lesssim$0.5 kpc)
to 2 kpc.  
Thus, unless molecular emission line flux increases substantially, 
its detection significance decreases in the large-beam-sized spectra.
For the observed ULIRGs, HCN and HCO$^{+}$ emission line signals in 
the 1--2 kpc spectra are generally not large, even fainter than those
in the $\lesssim$0.5 kpc spectra (Figure \ref{fig:SpectraB}), despite
the fact that the signal-integrated area of the 1--2 kpc annular
region is a factor of 12 larger than that of the central 0.5 kpc
circular region. 
This is as expected because the bulk of dense molecular gas in nearby
ULIRGs' nuclei is usually concentrated into the central compact 
($\lesssim$1 kpc) regions \citep[e.g.,][]{ima18,ima22}.
Thus, we can obtain meaningful estimates of HCN and HCO$^{+}$ emission
line fluxes at the 1--2 kpc annular region only for a limited fraction
of the observed ULIRGs. 

\begin{figure*}[!hbt]
\includegraphics[angle=-90,scale=.22]{f2a.eps} 
\includegraphics[angle=-90,scale=.22]{f2b.eps} 
\includegraphics[angle=-90,scale=.22]{f2c.eps} \\
\includegraphics[angle=-90,scale=.22]{f2d.eps} 
\includegraphics[angle=-90,scale=.22]{f2e.eps} 
\includegraphics[angle=-90,scale=.22]{f2f.eps} \\
\includegraphics[angle=-90,scale=.22]{f2g.eps} 
\includegraphics[angle=-90,scale=.22]{f2h.eps} 
\includegraphics[angle=-90,scale=.22]{f2i.eps} \\
\includegraphics[angle=-90,scale=.22]{f2j.eps} 
\includegraphics[angle=-90,scale=.22]{f2k.eps} 
\includegraphics[angle=-90,scale=.22]{f2l.eps} \\
\includegraphics[angle=-90,scale=.22]{f2m.eps} 
\includegraphics[angle=-90,scale=.22]{f2n.eps} 
\includegraphics[angle=-90,scale=.22]{f2o.eps} \\
\includegraphics[angle=-90,scale=.22]{f2p.eps} 
\includegraphics[angle=-90,scale=.22]{f2q.eps} 
\includegraphics[angle=-90,scale=.22]{f2r.eps} \\
\end{figure*}

\begin{figure*}[!hbt]
\includegraphics[angle=-90,scale=.22]{f3a.eps} 
\includegraphics[angle=-90,scale=.22]{f3b.eps} 
\includegraphics[angle=-90,scale=.22]{f3c.eps} \\
\includegraphics[angle=-90,scale=.22]{f3d.eps} 
\includegraphics[angle=-90,scale=.22]{f3e.eps} 
\includegraphics[angle=-90,scale=.22]{f3f.eps} \\
\includegraphics[angle=-90,scale=.22]{f3g.eps} 
\includegraphics[angle=-90,scale=.22]{f3h.eps} 
\includegraphics[angle=-90,scale=.22]{f3i.eps} \\
\includegraphics[angle=-90,scale=.22]{f3j.eps} 
\includegraphics[angle=-90,scale=.22]{f3k.eps} 
\includegraphics[angle=-90,scale=.22]{f3l.eps} \\
\includegraphics[angle=-90,scale=.22]{f3m.eps} 
\includegraphics[angle=-90,scale=.22]{f3n.eps} 
\includegraphics[angle=-90,scale=.22]{f3o.eps} \\
\includegraphics[angle=-90,scale=.22]{f3p.eps} 
\includegraphics[angle=-90,scale=.22]{f3q.eps} 
\includegraphics[angle=-90,scale=.22]{f3r.eps} \\
\end{figure*}

Gaussian fits are applied to significantly detected HCN and
HCO$^{+}$ emission lines in the spectra at various regions.
Following \citet{ima23}, to simplify our flux estimates, we try
to apply a single Gaussian fit, as long as a line profile can be
approximated by a single emission component. 
We apply a double Gaussian fit only if an emission line displays 
a clear double-peaked profile with a deep central dip. 
Our final adopted best Gaussian fits are summarized in Appendix C.    
Table \ref{tab:flux} tabulates the derived Gaussian-fit
velocity-integrated emission line fluxes of HCN and HCO$^{+}$. 

\begin{figure*}[!hbt]
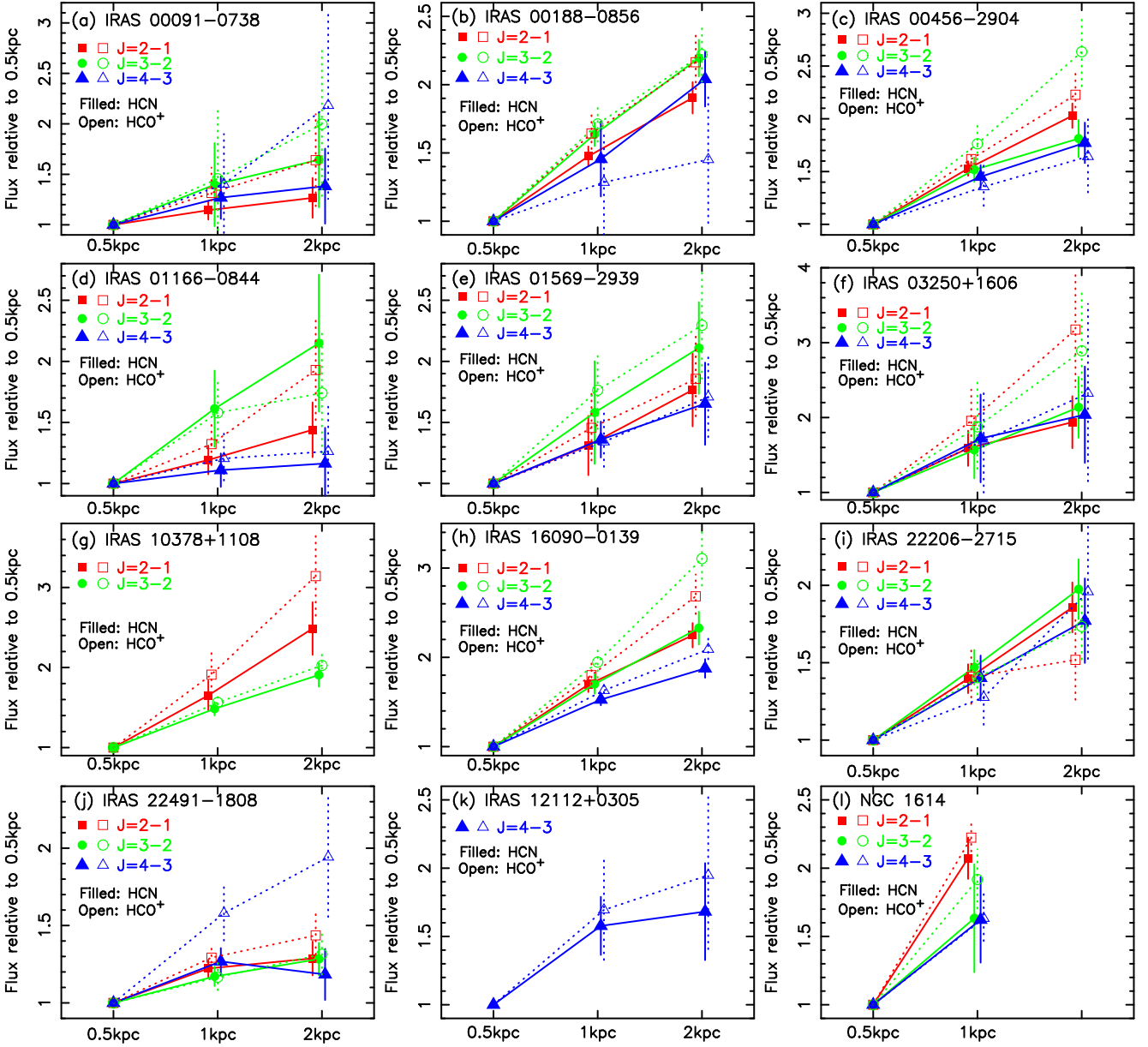

\includegraphics[angle=-90,scale=.26]{f4a.eps} 
\includegraphics[angle=-90,scale=.26]{f4b.eps} 
\includegraphics[angle=-90,scale=.26]{f4c.eps} \\ 
\includegraphics[angle=-90,scale=.26]{f4d.eps} 
\includegraphics[angle=-90,scale=.26]{f4e.eps} 
\includegraphics[angle=-90,scale=.26]{f4f.eps} \\
\includegraphics[angle=-90,scale=.26]{f4g.eps} 
\includegraphics[angle=-90,scale=.26]{f4h.eps} 
\includegraphics[angle=-90,scale=.26]{f4i.eps} \\
\includegraphics[angle=-90,scale=.26]{f4j.eps} 
\includegraphics[angle=-90,scale=.26]{f4k.eps} 
\includegraphics[angle=-90,scale=.26]{f4l.eps} 
\caption{
Curve of growth of HCN (filled symbol) and HCO$^{+}$ (open symbol)
emission line flux, based on our adopted Gaussian fit (Appendix C),
with increasing beam size from 0.5 kpc, through 1 kpc, to 2 kpc. 
Red square: J=2--1. Green circle: J=3--2. Blue triangle: J=4--3.
\label{fig:CoG}
}
\end{figure*}

Figure \ref{fig:CoG} shows the curve of growth of HCN and HCO$^{+}$
emission line fluxes at J=2--1, J=3--2, and J=4--3, with increasing
beam size from 0.5 kpc, through 1 kpc, to 2 kpc.
We see that in the majority of the observed ULIRGs, 
HCO$^{+}$ flux increases more significantly than HCN flux when
compared at the same J-transition, supporting a previous suggestion
that HCO$^{+}$ emission is spatially more extended than HCN emission
in nearby ULIRGs' nuclei \citep{ima19}.  
This is reasonable because the factor of $\sim$5 smaller critical 
density of HCO$^{+}$ than that of HCN at the same J-transition
\citep{shi15} can make HCO$^{+}$ excitation more efficient than HCN at
the outer nuclear (0.5--2 kpc) region where molecular gas density and
temperature are likely to be smaller than those at the innermost
($\lesssim$0.5 kpc) region.  
It is also found that the flux increase with increasing aperture size
tends to be smaller at J=4--3 than at J=2--1, which can also be
explained if gas density and temperature at the outer nuclear 
(0.5--2 kpc) region are not very high to sufficiently excite HCN and
HCO$^{+}$ to J=4. 

The CS J=7--6 ($\nu_{\rm rest}$ = 342.883 GHz) emission line was also
clearly detected in all ULIRGs during HCN and HCO$^{+}$ J=4--3 line 
observations. 
Appendix D summarizes the original-beam-sized moment 0 map, 0.5 kpc
beam-sized spectrum, and Gaussian fit in the spectrum, for the CS
J=7--6 line.  
  
HC$_{3}$N J=18--17 ($\nu_{\rm rest}$ = 163.753 GHz)
or J=21--20 ($\nu_{\rm rest}$ = 191.040 GHz) emission
line was also serendipitously detected during the HCN and HCO$^{+}$
J=2--1 line observations of all ULIRGs.  
Appendix E summarizes the original-beam-sized moment 0 map, 0.5 kpc
beam-sized spectrum, and Gaussian fit in the spectrum, for the
HC$_{3}$N lines.  
The peak fluxes of the CS J=7--6 and HC$_{3}$N emission lines in the
original-beam-sized moment 0 maps are also added in 
columns 8 and 9 of Table \ref{tab:mom0}, respectively. 

\begin{figure*}[!hbt]
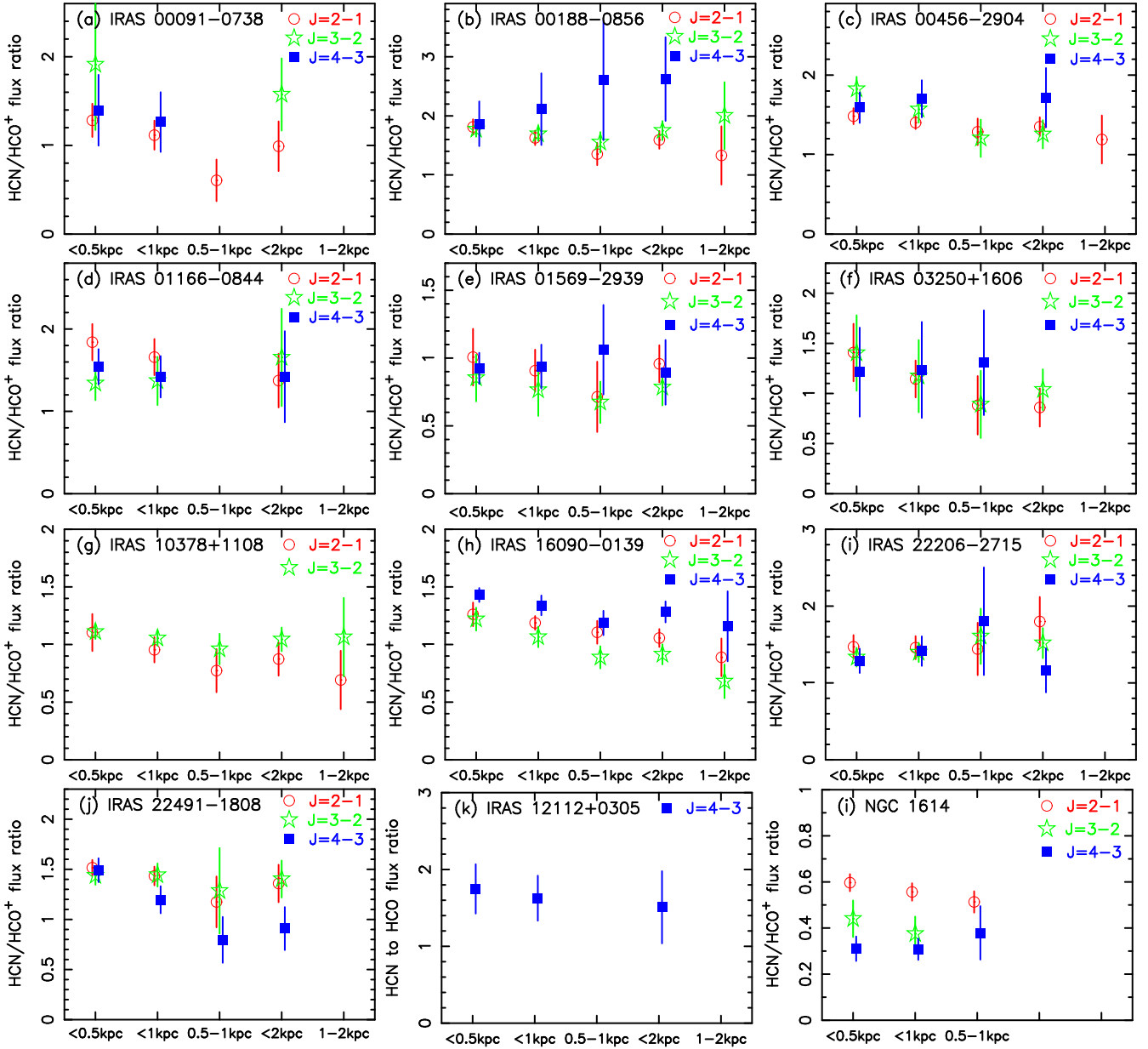

\includegraphics[angle=-90,scale=.26]{f5a.eps} 
\includegraphics[angle=-90,scale=.26]{f5b.eps} 
\includegraphics[angle=-90,scale=.26]{f5c.eps} \\ 
\includegraphics[angle=-90,scale=.26]{f5d.eps} 
\includegraphics[angle=-90,scale=.26]{f5e.eps} 
\includegraphics[angle=-90,scale=.26]{f5f.eps} \\
\includegraphics[angle=-90,scale=.26]{f5g.eps} 
\includegraphics[angle=-90,scale=.26]{f5h.eps} 
\includegraphics[angle=-90,scale=.26]{f5i.eps} \\
\includegraphics[angle=-90,scale=.26]{f5j.eps} 
\includegraphics[angle=-90,scale=.26]{f5k.eps} 
\includegraphics[angle=-90,scale=.26]{f5l.eps} 
\caption{
HCN-to-HCO$^{+}$ flux ratio measured in the $\lesssim$0.5 kpc, 
$\lesssim$1 kpc, 0.5--1 kpc, $\lesssim$2 kpc, and 1--2 kpc spectra.
Red open circle: J=2--1. Green open star: J=3--2. Blue filled square:
J=4--3. 
Only ratios with $\gtrsim$2.5$\sigma$ are plotted.
Note that the vertical axis range of the only one LIRG NGC 1614 is
much narrower than ULIRGs.   
\label{fig:HCNtoHCOratio}
}
\end{figure*}

\begin{figure*}[!hbt]
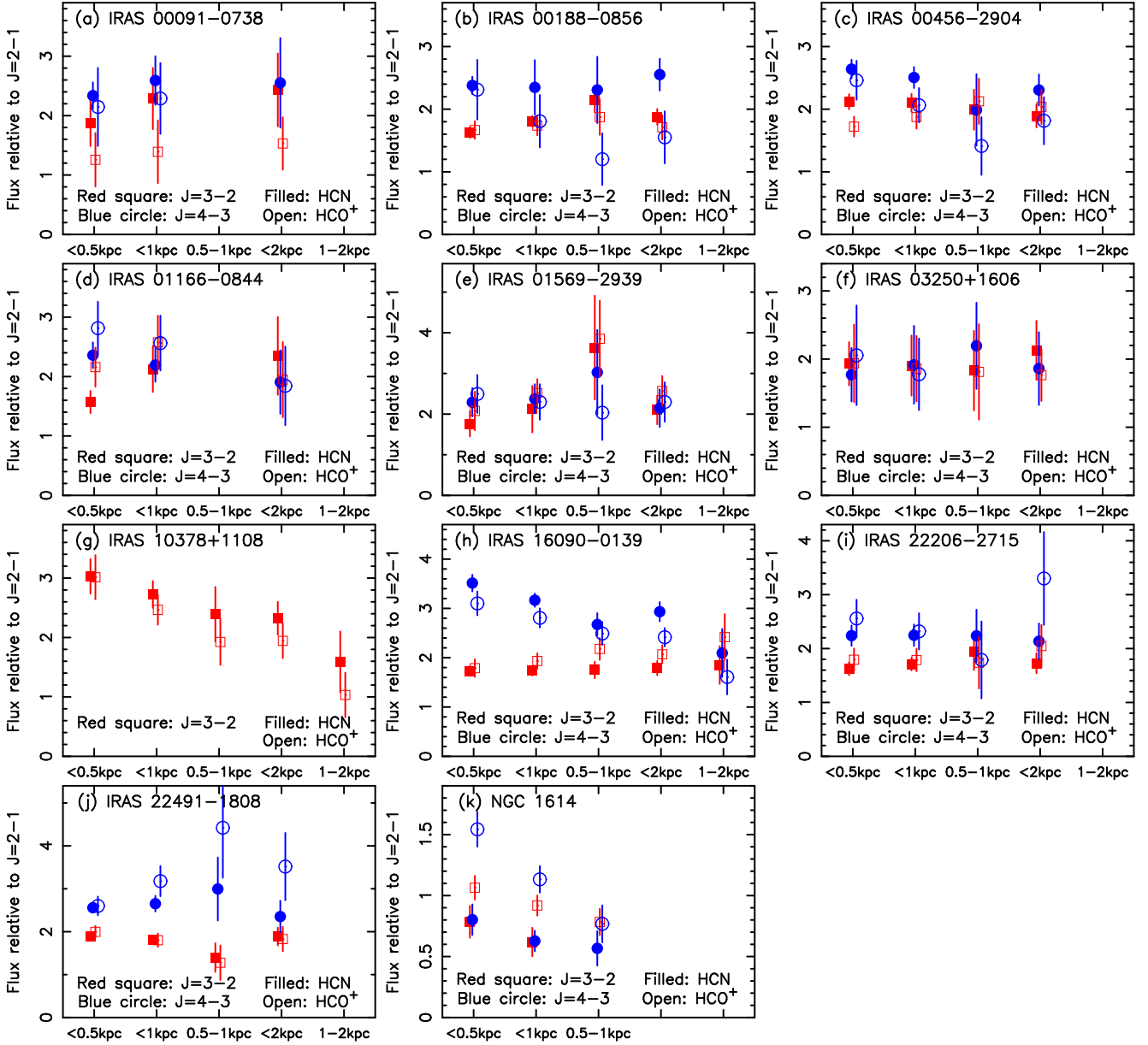

\includegraphics[angle=-90,scale=.26]{f6a.eps} 
\includegraphics[angle=-90,scale=.26]{f6b.eps} 
\includegraphics[angle=-90,scale=.26]{f6c.eps} \\ 
\includegraphics[angle=-90,scale=.26]{f6d.eps} 
\includegraphics[angle=-90,scale=.26]{f6e.eps} 
\includegraphics[angle=-90,scale=.26]{f6f.eps} \\
\includegraphics[angle=-90,scale=.26]{f6g.eps} 
\includegraphics[angle=-90,scale=.26]{f6h.eps} 
\includegraphics[angle=-90,scale=.26]{f6i.eps} \\
\includegraphics[angle=-90,scale=.26]{f6j.eps} 
\includegraphics[angle=-90,scale=.26]{f6k.eps} 
\caption{
J=4--3 (blue circle) or J=3--2 (red square) to J=2--1 flux
ratio of HCN (filled symbol) and HCO$^{+}$ (open symbol), measured in
the $\lesssim$0.5 kpc, $\lesssim$1 kpc, 0.5--1 kpc, $\lesssim$2 kpc,
and 1--2 kpc spectra.  
Only Gaussian fit statistical uncertainty is considered, and only
ratios with $\gtrsim$2.5$\sigma$ are plotted. 
The vertical axis range of the only one LIRG NGC 1614 is much narrower
than ULIRGs.  
\label{fig:Jratio}
}
\end{figure*}

We summarize in Appendix F (i) the observed HCN-to-HCO$^{+}$ flux
ratios at J=2--1, J=3--2, and J=4--3, and (ii) the observed high-J to
low-J flux ratios of HCN and HCO$^{+}$, based on the adopted Gaussian
fits (Appendix C), in the $\lesssim$0.5 kpc, $\lesssim$1 kpc,
$\lesssim$2 kpc, 0.5--1 kpc, and 1--2 kpc spectra.
These ratios are plotted in Figures \ref{fig:HCNtoHCOratio} and
\ref{fig:Jratio}, respectively, to visualize how the ratios vary 
in different regions.   

\clearpage

\startlongtable
\begin{deluxetable*}{ll|ccc|ccc}
\tabletypesize{\scriptsize}
\tablecaption{Gaussian-fit Velocity-integrated Flux of HCN and
HCO$^{+}$ Emission Lines \label{tab:flux}}   
\tablewidth{0pt}
\tablehead{
\colhead{Object} & \colhead{Region} & 
\multicolumn{6}{c}{Flux (Jy km s$^{-1}$)} \\
\colhead{} & \colhead{} & 
\multicolumn{3}{c}{HCN} & \multicolumn{3}{c}{HCO$^{+}$} \\
\colhead{} & \colhead{} & \colhead{J=2--1} & \colhead{J=3--2} & 
\colhead{J=4--3} & \colhead{J=2--1} & \colhead{J=3--2} &
\colhead{J=4--3} \\    
\colhead{(1)} & \colhead{(2)} & \colhead{(3)} & \colhead{(4)} &
\colhead{(5)} & \colhead{(6)} & \colhead{(7)} & \colhead{(8)} 
}
\startdata 
IRAS 00091$-$0738 & $\lesssim$0.5 kpc & 2.1$\pm$0.1 & 4.0$\pm$0.8 &
4.9$\pm$0.4 & 1.6$\pm$0.2 & 2.1$\pm$0.7 & 3.5$\pm$1.0 \\
 & $\lesssim$1 kpc & 2.4$\pm$0.2 & 5.5$\pm$1.2 & 6.3$\pm$0.9 &  2.2$\pm$0.3 &
3.0$\pm$1.1 \tablenotemark{a} & 5.0$\pm$1.1 \\
 & $\lesssim$2 kpc & 2.7$\pm$0.4 & 6.5$\pm$1.3 & 6.8$\pm$1.7 & 2.7$\pm$0.6 &
4.1$\pm$0.7 & 7.7$\pm$2.5 \\
 & 0.5--1 kpc & 0.30$\pm$0.07 & 1.4$\pm$0.4 & 1.7$\pm$1.4
\tablenotemark{a} & 0.49$\pm$0.16 & 0.94$\pm$0.27 & 0.53$\pm$0.34
\tablenotemark{a} \\ 
 & 1--2 kpc & 0.30$\pm$0.16 \tablenotemark{a} & 1.1$\pm$0.3 & --- & 
0.49$\pm$0.36 \tablenotemark{a} & 1.3$\pm$0.5 \tablenotemark{a} &  ---
\\ \hline 
IRAS 00188$-$0856 & $\lesssim$0.5 kpc & 1.5$\pm$0.1 & 2.4$\pm$0.1 &
3.6$\pm$0.2 & 0.83$\pm$0.05 & 1.4$\pm$0.1 & 1.9$\pm$0.4 \\
 & $\lesssim$1 kpc & 2.2$\pm$0.1 & 4.0$\pm$0.1 & 5.2$\pm$0.9 & 1.4$\pm$0.1 &
2.4$\pm$0.1 & 2.5$\pm$0.5 \\
 & $\lesssim$2 kpc & 2.9$\pm$0.1 & 5.4$\pm$0.3 & 7.3$\pm$0.6 & 1.8$\pm$0.1 &
3.1$\pm$0.2 & 2.8$\pm$0.7 \\
 & 0.5--1 kpc & 0.71$\pm$0.05 & 1.5$\pm$0.1 & 1.6$\pm$0.4 &
0.53$\pm$0.06 & 0.98$\pm$0.09 & 0.63$\pm$0.20 \\
 & 1--2 kpc & 0.63$\pm$0.09 & 1.4$\pm$0.2 & 1.9$\pm$0.6 &
0.48$\pm$0.16 \tablenotemark{a} & 0.69$\pm$0.17 & --- \\ \hline
IRAS 00456$-$2904 & $\lesssim$0.5 kpc & 1.3$\pm$0.1 & 2.9$\pm$0.1 &
3.6$\pm$0.2 & 0.91$\pm$0.05 & 1.6$\pm$0.1 & 2.2$\pm$0.2  \\
 & $\lesssim$1 kpc & 2.1$\pm$0.1 & 4.3$\pm$0.3 & 5.2$\pm$0.3 & 1.5$\pm$0.1 &
2.8$\pm$0.2 & 3.0$\pm$0.4 \\
 & $\lesssim$2 kpc & 2.7$\pm$0.1 & 5.2$\pm$0.4 & 6.3$\pm$0.6 & 2.0$\pm$0.1 &
4.1$\pm$0.5 & 3.7$\pm$0.7 \\
 & 0.5--1 kpc & 0.72$\pm$0.05 & 1.4$\pm$0.2 & 1.4$\pm$0.4 &
0.56$\pm$0.06 & 1.2$\pm$0.2 & 0.79$\pm$0.24 \\
 & 1--2 kpc & 0.67$\pm$0.10 & 0.88$\pm$0.29 & 1.2$\pm$0.5
\tablenotemark{a} & 0.56$\pm$0.12 & 1.3$\pm$0.4 & 0.88$\pm$0.41
\tablenotemark{a} \\ \hline 
IRAS 01166$-$0844 & $\lesssim$0.5 kpc & 1.2$\pm$0.1 & 1.8$\pm$0.2 &
2.8$\pm$0.2 & 0.64$\pm$0.07 & 1.4$\pm$0.2 & 1.8$\pm$0.2 \\
 & $\lesssim$1 kpc & 1.4$\pm$0.1 & 3.0$\pm$0.5 & 3.1$\pm$0.3 & 0.84$\pm$0.09
& 2.2$\pm$0.3 & 2.2$\pm$0.3 \\
 & $\lesssim$2 kpc & 1.7$\pm$0.2 & 4.0$\pm$0.9 & 3.2$\pm$0.8 & 1.2$\pm$0.2 &
2.4$\pm$0.6 & 2.3$\pm$0.7 \\
 & 0.5--1 kpc & 0.21$\pm$0.06 & 1.2$\pm$0.4 & 
0.35$\pm$0.51 \tablenotemark{a} & 0.20$\pm$0.07 \tablenotemark{a} & 
0.81$\pm$0.20 & 0.37$\pm$0.28 \tablenotemark{a} \\
 & 1--2 kpc & 0.21$\pm$0.14 \tablenotemark{a} & --- & --- &
0.45$\pm$0.40\tablenotemark{a} & --- & --- \\ \hline
IRAS 01569$-$2939 & $\lesssim$0.5 kpc & 1.0$\pm$0.1 & 1.8$\pm$0.2 &
2.3$\pm$0.2 & 1.0$\pm$0.2 & 2.1$\pm$0.3 & 2.5$\pm$0.3 \\
 & $\lesssim$1 kpc & 1.3$\pm$0.2 & 2.8$\pm$0.7 & 3.2$\pm$0.3 & 1.5$\pm$0.2 &
3.7$\pm$0.3 & 3.4$\pm$0.5 \\
 & $\lesssim$2 kpc & 1.8$\pm$0.2 & 3.8$\pm$0.5 & 3.8$\pm$0.7 & 1.9$\pm$0.2 &
4.8$\pm$0.5 & 4.3$\pm$0.8 \\
 & 0.5--1 kpc & 0.29$\pm$0.08 & 1.0$\pm$0.2 & 0.87$\pm$0.16 &
0.40$\pm$0.09 & 1.5$\pm$0.2 & 0.82$\pm$0.21 \\
 & 1--2 kpc & 0.43$\pm$0.22 \tablenotemark{a} & 0.84$\pm$0.34
\tablenotemark{a} & 0.67$\pm$0.26 \tablenotemark{a} & 0.40$\pm$0.13 &
1.1$\pm$0.35 & 0.94$\pm$0.54 \tablenotemark{a} \\ \hline
IRAS 03250$+$1606 & $\lesssim$0.5 kpc & 0.67$\pm$0.07 & 1.3$\pm$0.2 &
1.2$\pm$0.2 & 0.47$\pm$0.08 & 0.92$\pm$0.22 & 0.97$\pm$0.30 \\
 & $\lesssim$1 kpc & 1.1$\pm$0.1 & 2.0$\pm$0.4 & 2.0$\pm$0.6 & 0.92$\pm$0.12
& 1.7$\pm$0.4 & 1.6$\pm$0.4 \\
 & $\lesssim$2 kpc & 1.3$\pm$0.2 & 2.7$\pm$0.4 & 2.4$\pm$0.6 & 1.5$\pm$0.3 &
2.6$\pm$0.3 & 2.3$\pm$0.9 \tablenotemark{a} \\
 & 0.5--1 kpc & 0.39$\pm$0.08 & 0.71$\pm$0.17 & 0.86$\pm$0.17 &
0.44$\pm$0.12 & 0.80$\pm$0.23 & 0.66$\pm$0.23 \tablenotemark{a} \\
 & 1--2 kpc & 0.28$\pm$0.13 \tablenotemark{a} & 0.70$\pm$0.33
\tablenotemark{a} & --- &  0.61$\pm$0.25 \tablenotemark{a} &
0.85$\pm$0.21 & 0.56$\pm$0.36 \tablenotemark{a} \\ \hline 
IRAS 10378$+$1108 & $\lesssim$0.5 kpc & 0.90$\pm$0.08 & 2.7$\pm$0.1 & --- &
0.82$\pm$0.09 & 2.5$\pm$0.1 & --- \\
 & $\lesssim$1 kpc & 1.5$\pm$0.1 & 4.1$\pm$0.2 & --- & 1.6$\pm$0.1 &
3.8$\pm$0.2 & --- \\
 & $\lesssim$2 kpc & 2.2$\pm$0.2 & 5.2$\pm$0.4 & --- & 2.6$\pm$0.3 &
5.0$\pm$0.4 & --- \\
 & 0.5--1 kpc & 0.56$\pm$0.09 & 1.3$\pm$0.1 & --- & 0.72$\pm$0.13 &
1.4$\pm$0.1 & --- \\
 & 1--2 kpc & 0.73$\pm$0.18 & 1.2$\pm$0.2 & --- & 1.1$\pm$0.3 &
1.1$\pm$0.3 & --- \\ \hline
IRAS 16090$-$0139 & $\lesssim$0.5 kpc & 2.4$\pm$0.1 & 4.1$\pm$0.2 &
8.3$\pm$0.2 & 1.9$\pm$0.1 & 3.3$\pm$0.2 & 5.8$\pm$0.2 \\
 & $\lesssim$1 kpc & 4.0$\pm$0.1 & 6.9$\pm$0.3 & 12.6$\pm$0.4 & 3.4$\pm$0.1 &
6.5$\pm$0.4 & 9.4$\pm$0.5 \\
 & $\lesssim$2 kpc & 5.3$\pm$0.2 & 9.5$\pm$0.6 & 15.5$\pm$0.7 & 5.0$\pm$0.3 &
10.3$\pm$0.7 & 12.1$\pm$0.6 \\
 & 0.5--1 kpc & 1.6$\pm$0.1 & 2.8$\pm$0.2 & 4.3$\pm$0.3 & 1.5$\pm$0.1
& 3.2$\pm$0.2 & 3.6$\pm$0.2 \\
 & 1--2 kpc & 1.4$\pm$0.2 & 2.5$\pm$0.4 & 2.9$\pm$0.6 & 1.5$\pm$0.2 &
3.7$\pm$0.5 & 2.5$\pm$0.4 \\ \hline
IRAS 22206$-$2715 & $\lesssim$0.5 kpc & 1.3$\pm$0.1 & 2.1$\pm$0.1 &
2.8$\pm$0.2 & 0.86$\pm$0.08 & 1.5$\pm$0.1 & 2.2$\pm$0.2 \\
 & $\lesssim$1 kpc & 1.8$\pm$0.1 & 3.0$\pm$0.2 & 4.0$\pm$0.3 & 1.2$\pm$0.1 &
2.2$\pm$0.2 & 2.8$\pm$0.3 \\
 & $\lesssim$2 kpc & 2.4$\pm$0.2 & 4.1$\pm$0.3 & 5.0$\pm$0.7 & 1.3$\pm$0.2 &
2.7$\pm$0.3 & 4.3$\pm$0.9 \\
 & 0.5--1 kpc & 0.51$\pm$0.05 & 0.99$\pm$0.14 & 1.1$\pm$0.2 &
0.35$\pm$0.08 & 0.61$\pm$0.11 & 0.63$\pm$0.21 \\
 & 1--2 kpc & 0.58$\pm$0.13 & 1.2$\pm$0.4 & 1.1$\pm$0.6
\tablenotemark{a} & 0.23$\pm$0.11 \tablenotemark{a} &  
0.48$\pm$0.23 \tablenotemark{a} & 1.6$\pm$0.8 \tablenotemark{a} \\ \hline
IRAS 22491$-$1808 & $\lesssim$0.5 kpc & 4.8$\pm$0.1 & 9.2$\pm$0.3 &
12.4$\pm$0.4 & 3.2$\pm$0.1 & 6.4$\pm$0.3 & 8.3$\pm$0.6 \\
 & $\lesssim$1 kpc & 5.9$\pm$0.2 & 10.7$\pm$0.4 & 15.7$\pm$0.9 & 4.1$\pm$0.2
& 7.4$\pm$0.5 & 13.1$\pm$1.3 \\
 & $\lesssim$2 kpc & 6.2$\pm$0.5 & 11.8$\pm$0.9 & 14.7$\pm$2.0 & 4.6$\pm$0.5
& 8.4$\pm$0.9 & 16.1$\pm$3.1 \\
 & 0.5--1 kpc & 1.1$\pm$0.2 & 1.6$\pm$0.3 & 3.3$\pm$0.7 &
0.95$\pm$0.16 & 1.2$\pm$0.3 & 4.2$\pm$0.9 \\
 & 1--2 kpc & 0.50$\pm$0.33 \tablenotemark{a} & --- & --- &
0.56$\pm$0.22 \tablenotemark{a} & --- & --- \\ \hline
IRAS 12112$+$0305 & $\lesssim$0.5 kpc & --- & --- & 11.6$\pm$1.2 & --- & --- 
& 6.7$\pm$1.0 \\
 & $\lesssim$1 kpc & ---  & ---  & 18.4$\pm$1.6 &  --- & --- & 11.3$\pm$1.8 \\
 & $\lesssim$2 kpc & ---  & ---  & 19.6$\pm$3.6 & ---  & ---  & 13.0$\pm$3.3 \\
 & 0.5--1 kpc & ---  & ---  & 5.5$\pm$1.6 & ---  & ---  & 2.8$\pm$1.9
\tablenotemark{a} \\
 & 1--2 kpc & --- & --- & --- & --- & --- & --- \\ \hline
NGC 1614 & $\lesssim$0.5 kpc & 3.3$\pm$0.2 & 2.6$\pm$0.4 & 2.6$\pm$0.4 & 
5.5$\pm$0.2 & 5.9$\pm$0.5 & 8.5$\pm$0.7 \\ 
 & $\lesssim$1 kpc & 6.8$\pm$0.4  & 4.2$\pm$0.8  & 4.3$\pm$0.5 & 12.3$\pm$0.5
& 11.3$\pm$0.9 & 13.9$\pm$1.2 \\
 & 0.5--1 kpc & 3.5$\pm$0.3  & 1.6$\pm$0.7 \tablenotemark{a} & 
2.0$\pm$0.5 & 6.8$\pm$0.3 & 5.4$\pm$0.7 & 5.3$\pm$1.0 \\ \hline
\enddata

\tablenotetext{a}{Detection significance is $<$3$\sigma$.}

\tablecomments{Col.(1): Object name.
Col.(2): Region.
Central $\lesssim$0.5 kpc, $\lesssim$1 kpc, $\lesssim$2 kpc, 0.5--1
kpc annular, and 1--2 kpc annular regions.  
Cols.(3)--(8): Gaussian-fit velocity-integrated flux in units of 
Jy km s$^{-1}$ (Appendix C). 
Col.(3): HCN J=2--1.
Col.(4): HCN J=3--2.
Col.(5): HCN J=4--3.
Col.(6): HCO$^{+}$ J=2--1.
Col.(7): HCO$^{+}$ J=3--2.
Col.(8): HCO$^{+}$ J=4--3.
No value is shown when (a) no observations were conducted (J=4--3 of 
IRAS 10378$+$1108, and J=2--1 and J=3--2 of IRAS 12112$+$0305), or 
(b) there is no emission line signature at all, or 
(c) fitting uncertainty is too large to obtain meaningful information.
}

\end{deluxetable*}


\begin{figure*}[!hbt]
\begin{center}
\includegraphics[angle=0,scale=.4]{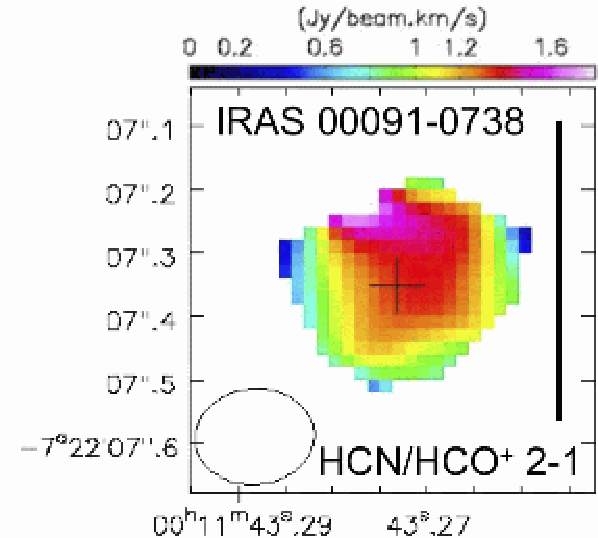} 
\includegraphics[angle=0,scale=.4]{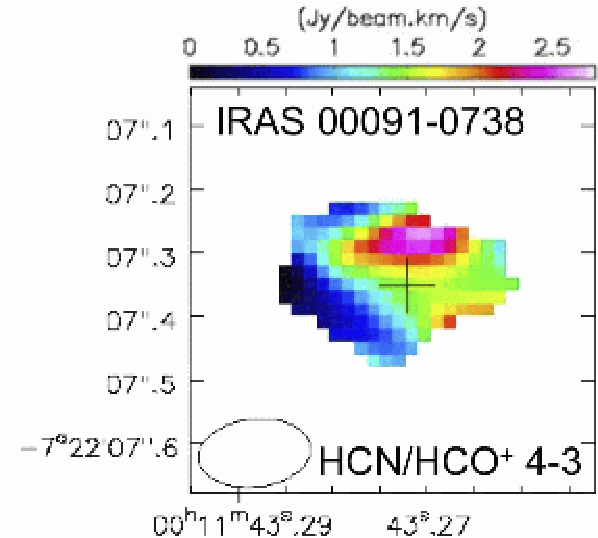}  
\includegraphics[angle=0,scale=.4]{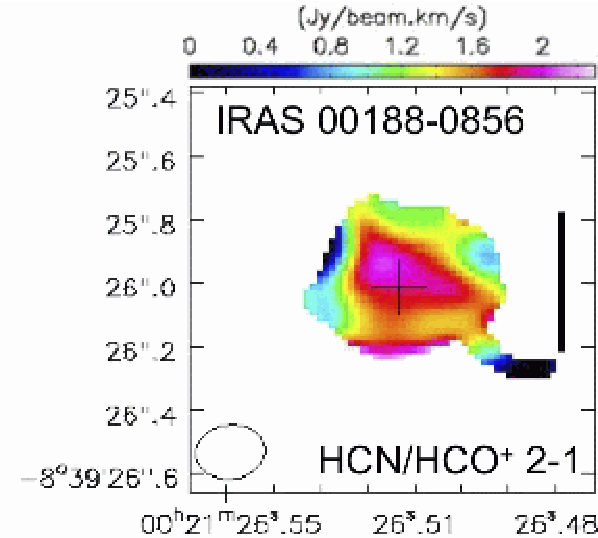} 
\includegraphics[angle=0,scale=.4]{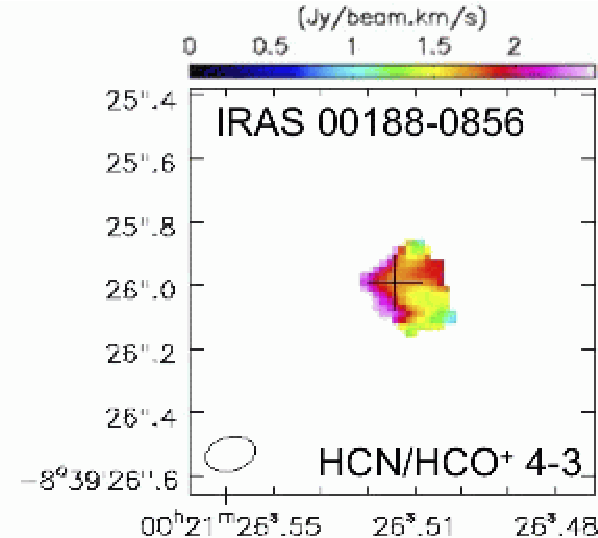} \\ 
\includegraphics[angle=0,scale=.4]{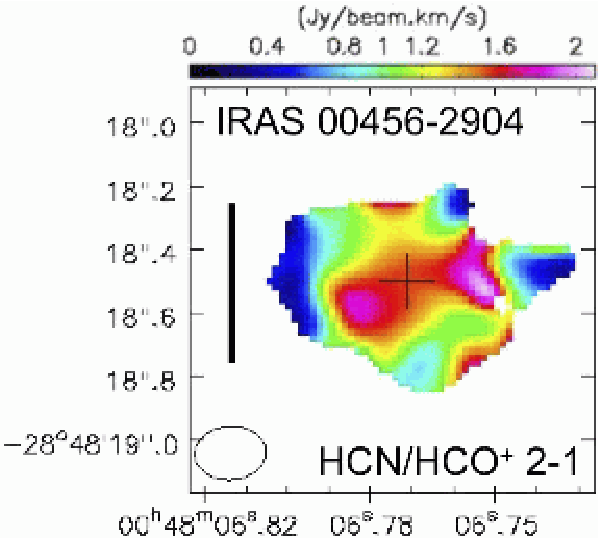} 
\includegraphics[angle=0,scale=.4]{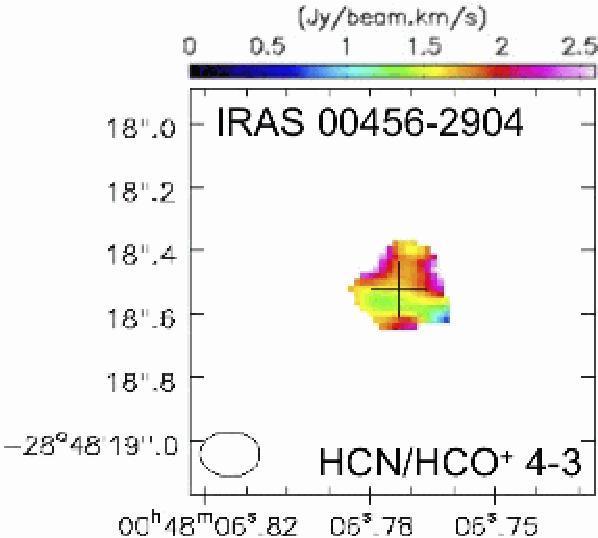}  
\includegraphics[angle=0,scale=.4]{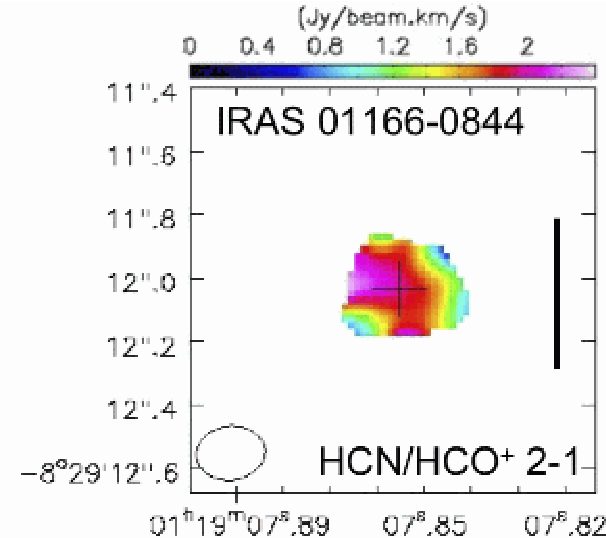} 
\includegraphics[angle=0,scale=.4]{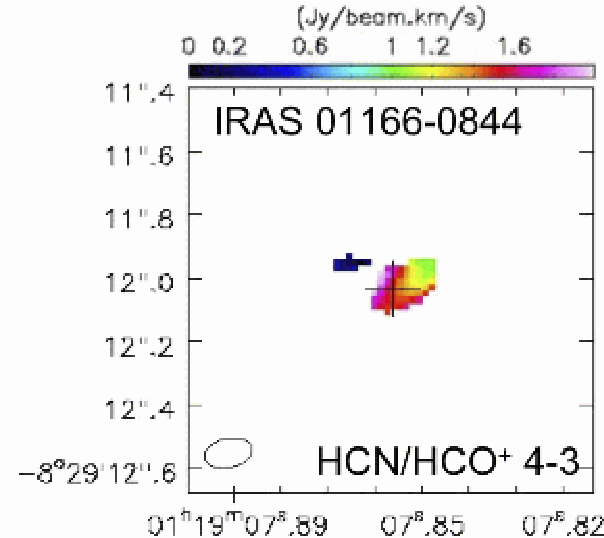} \\ 
\includegraphics[angle=0,scale=.4]{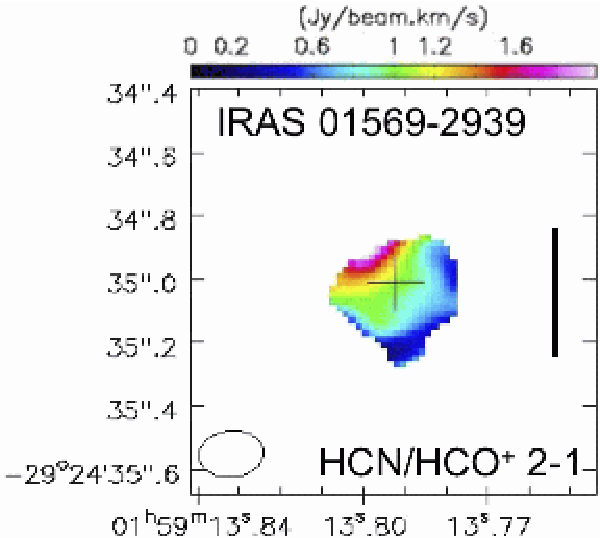} 
\includegraphics[angle=0,scale=.4]{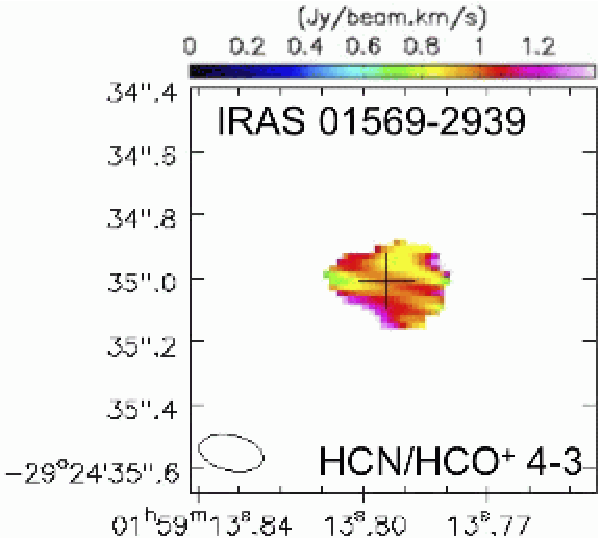}  
\includegraphics[angle=0,scale=.4]{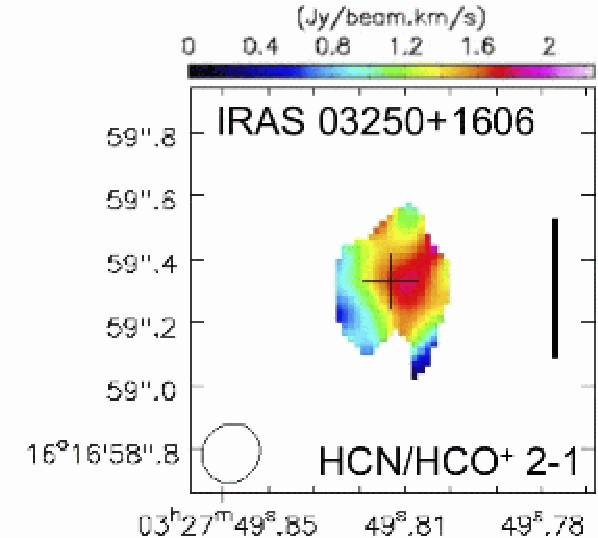} 
\includegraphics[angle=0,scale=.4]{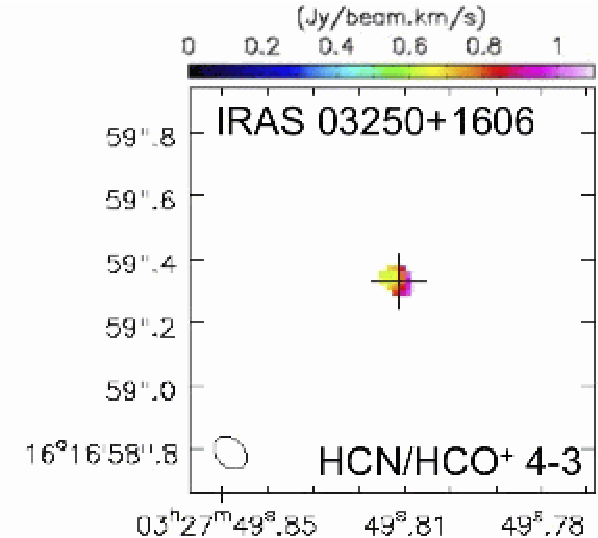} \\
\includegraphics[angle=0,scale=.4]{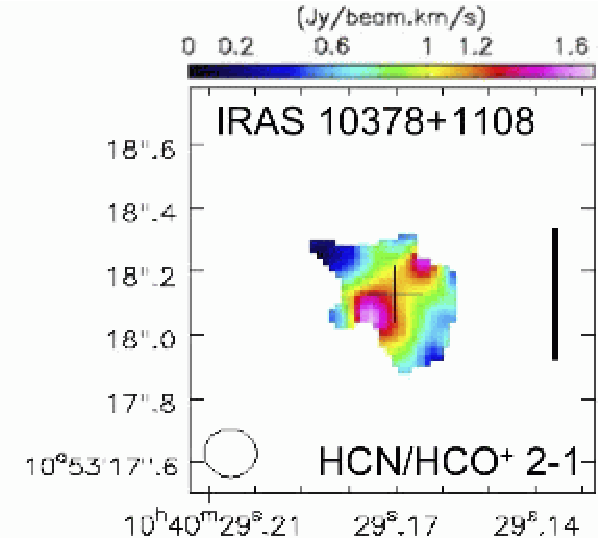} 
\includegraphics[angle=0,scale=.4]{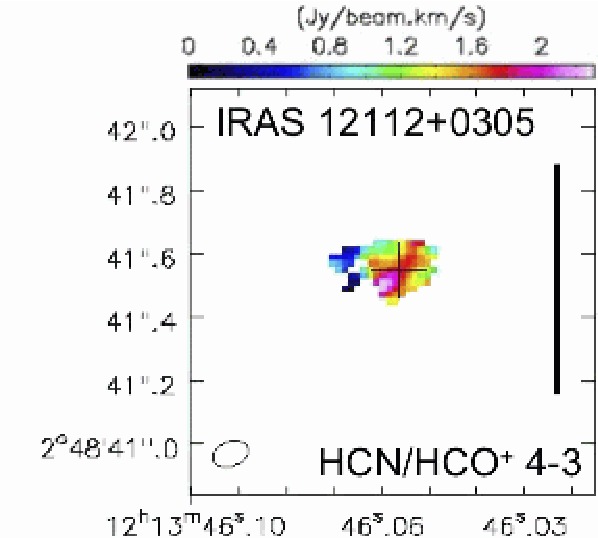}  
\includegraphics[angle=0,scale=.4]{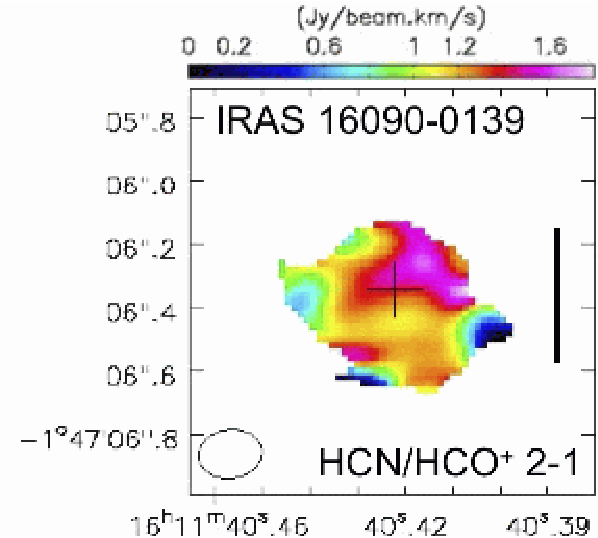} 
\includegraphics[angle=0,scale=.4]{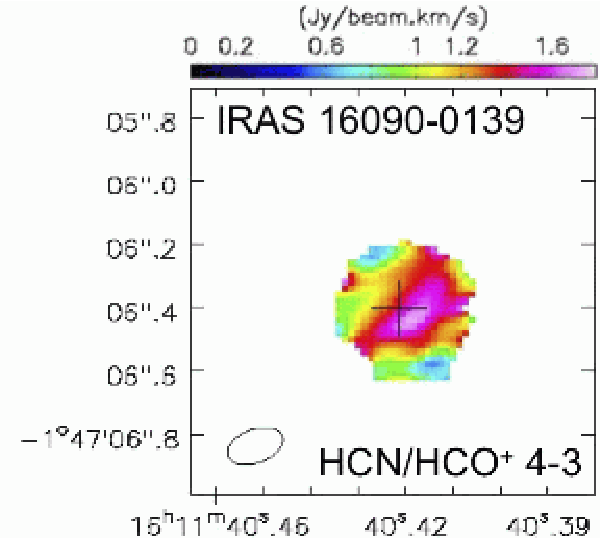} \\ 
\hspace*{-4.3cm}
\includegraphics[angle=0,scale=.4]{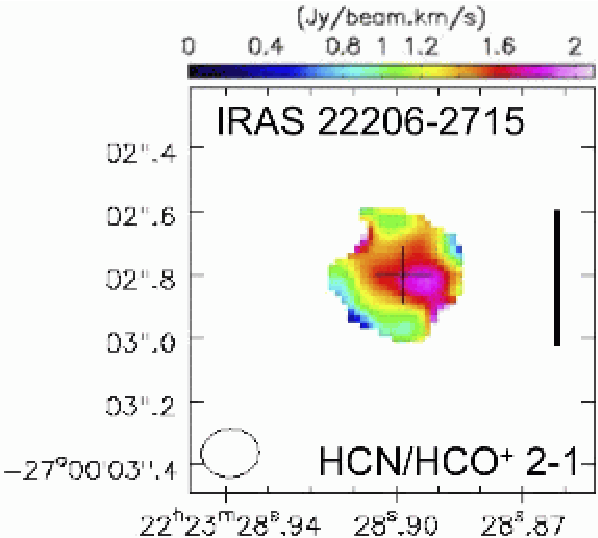} 
\includegraphics[angle=0,scale=.4]{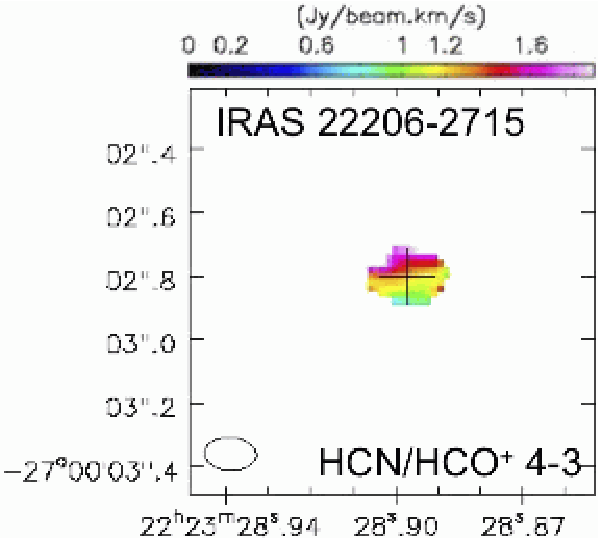}  
\includegraphics[angle=0,scale=.4]{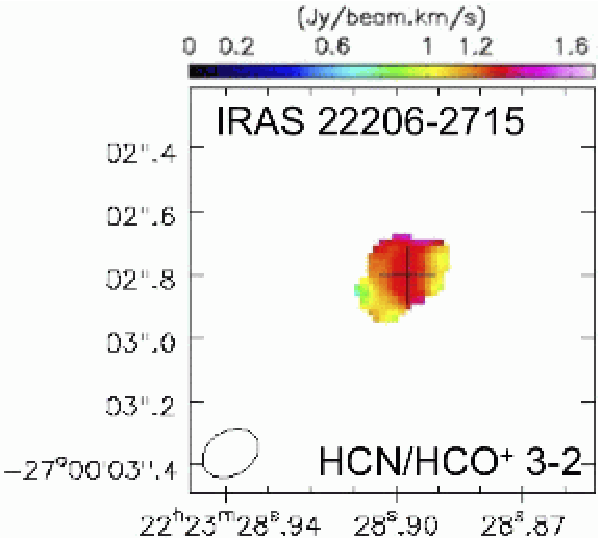} \\
\hspace*{-4.3cm}
\includegraphics[angle=0,scale=.4]{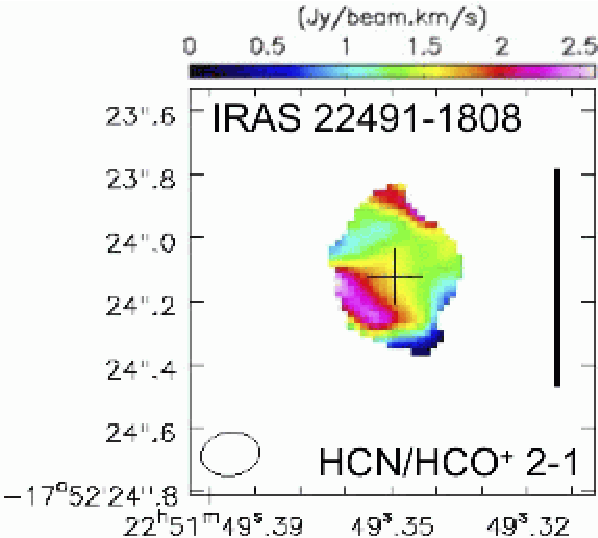} 
\includegraphics[angle=0,scale=.4]{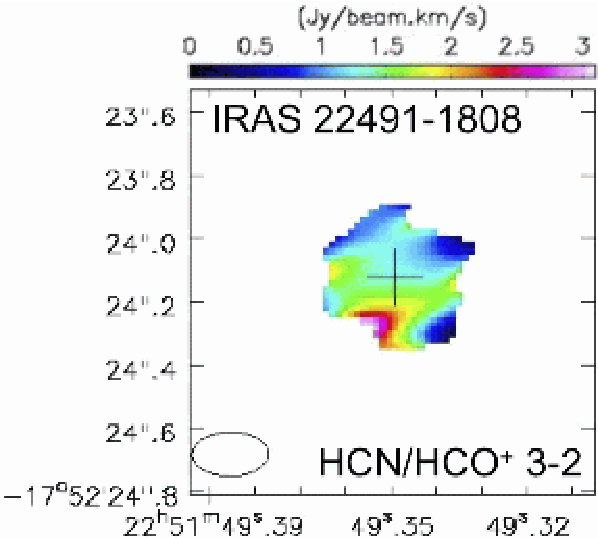} 
\includegraphics[angle=0,scale=.4]{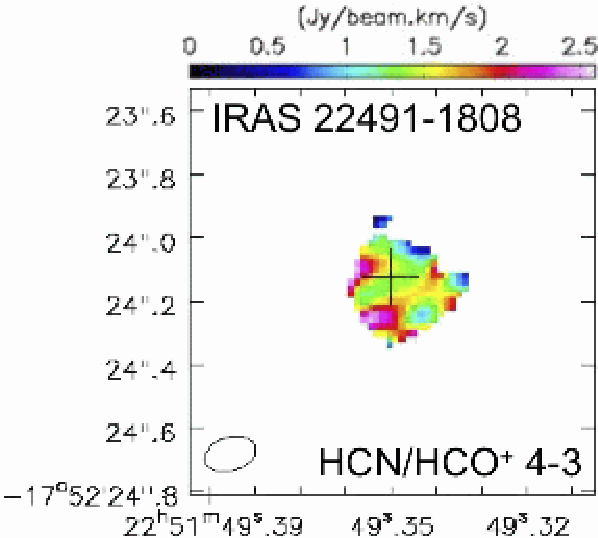}  
\end{center}
\caption{Original-beam-sized map of the observed
HCN-to-HCO$^{+}$ flux ratio calculated in units of Jy km s$^{-1}$. 
An appropriate cutoff ($>$2--3$\sigma$ depending on each ULIRG) is
applied for the HCO$^{+}$ flux (i.e., denominator), so that the 
resulting map is not dominated by noise. 
The plus mark denotes the continuum peak position which is regarded as
the nucleus of each ULIRG.
The length of the vertical black solid bar in the first image of each
object corresponds to 1 kpc. 
\label{fig:HCNdivHCOmap}
}
\end{figure*}

In each panel of Figure \ref{fig:HCNtoHCOratio}, 
in the left three ticks, the contribution from the innermost 
($\lesssim$0.5 kpc) molecular gas emission, relative to slightly extended
(0.5--1 kpc) emission, decreases from left to right. 
In the right two ticks, the contribution from the outermost nuclear (1--2
kpc) molecular gas emission, relative to inner ($\lesssim$1 kpc) emission,
increases from left to right. 
In many sources, we see a subtle trend that the observed
HCN-to-HCO$^{+}$ flux ratio at each J-transition slightly decreases
with decreasing contribution from the innermost ($\lesssim$0.5 kpc)
molecular emission 
(from left to right in the left three ticks) and with increasing
contribution from the outermost nuclear (1--2 kpc) emission (from left
to right in the right two ticks). 
This trend is most notably seen in IRAS 16090$-$0139 (Figure
\ref{fig:HCNtoHCOratio}h), because of small uncertainty in each data
point.  
It is thus suggested that the HCN-to-HCO$^{+}$ flux
ratio is higher inside and lower outside in a certain fraction of
nearby ULIRGs' nuclei.  
Figure \ref{fig:HCNdivHCOmap} displays the original-beam-sized maps of
the observed HCN-to-HCO$^{+}$ flux ratios at J=2--1, 
J=3--2, and J=4--3, created from newly
taken ALMA Cycle 7 data (Table \ref{tab:beam}).
The same maps at J=3--2 for some ULIRGs, created from ALMA Cycle
5 data (Table \ref{tab:beam}), are also found in \citet{ima19}. 
In some ULIRGs, the HCN-to-HCO$^{+}$ flux ratio is
confirmed to be higher at the very center (nuclear position) than
off-center regions (e.g., IRAS 16090$-$0139).   

We see a similar trend in some ULIRGs in Figure \ref{fig:Jratio} that
the observed high-J to low-J flux ratio slightly decreases with
decreasing (increasing) contribution from $\lesssim$0.5 kpc (1--2 kpc)
molecular emission, where the trend is most clearly seen in the J=3--2
to J=2--1 flux ratios of HCN and HCO$^{+}$ in IRAS 10378$+$1108 (Figure
\ref{fig:Jratio}g) and J=4--3 to J=2--1 flux ratios of HCN and
HCO$^{+}$ in IRAS 16090$-$0139 (Figure \ref{fig:Jratio}h).
These decreasing trends from the innermost to outermost nuclear
region, in the HCN-to-HCO$^{+}$ flux ratio and high-J to low-J flux
ratios of HCN and HCO$^{+}$, suggest that possible spatial variation
of dense molecular gas properties is discernible at 0.5 kpc
physical scales within the nuclear 2 kpc regions of some ULIRGs.   

\begin{figure}[!hbt]
\begin{center}
\includegraphics[angle=-90,scale=.36]{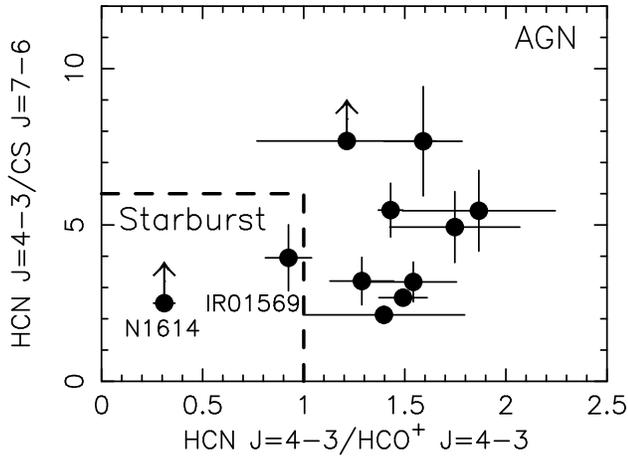} 
\end{center}
\caption{
HCN J=4--3 to HCO$^{+}$ J=4--3 flux ratio (abscissa)
and HCN J=4--3 to CS J=7--6 flux ratio (ordinate) derived from the 0.5 kpc
beam-sized spectra.
Flux ratio is calculated in units of Jy km s$^{-1}$.
\citet{izu16} proposed that starburst-dominated galaxies are
distributed in the lower-left region separated by the dashed straight
lines, while AGN-important galaxies are in the remaining upper-right
region. 
\label{fig:HCNHCOCS} 
}
\end{figure}

In Figure \ref{fig:HCNHCOCS}, we derive the HCN J=4--3 to HCO$^{+}$ J=4--3
and HCN J=4--3 to CS J=7--6 flux ratios measured in the 
$\lesssim$0.5 kpc spectra, to separate 
AGN-important and starburst-dominated sources, 
following the energy diagnostic diagram by \citet{izu16} where
these flux ratios are systematically higher in luminous AGNs than in
starbursts.   
The LIRG NGC 1614 is located in the region expected for
starburst-dominated galaxies, while ULIRGs (except IRAS 01569$-$2939)
are distributed in the region expected for AGN-important galaxies. 
The ULIRG IRAS 01569$-$2939 is located close to the borderline that
separates starburst-dominated and AGN-important galaxies. 
The energy diagnostic results in Figure \ref{fig:HCNHCOCS} thus
largely agree with the previously proposed infrared and
(sub)millimeter spectroscopic view that all ULIRGs are AGN important
and the LIRG NGC 1614 is starburst dominated (Table \ref{tab:object},
column 12 and footnote a).

\section{Discussion}

\subsection{Dense Molecular Gas Properties : Comparison with Non-LTE 
Model Calculations}  

We constrain nuclear molecular gas properties of the observed
(U)LIRGs at 0.5 kpc physical resolution, based on the three
J-transition line data (J=2--1, J=3--2, and J=4--3) of HCN and
HCO$^{+}$, by combining with non-LTE modeling.
The high-J to low-J flux ratios of HCN and HCO$^{+}$ can be used to
constrain the volume number density (n$_{\rm H_2}$) and kinetic temperature 
(T$_{\rm kin}$) of H$_{2}$ molecular gas, because high density and
temperature are needed to collisionally excite a significant fraction of
HCN and HCO$^{+}$ to J=4 or 3. 
The HCN-to-HCO$^{+}$ flux ratio at each J-transition contains information
of the HCN-to-HCO$^{+}$ abundance ratio, as was demonstrated by
\citet{ima23} for other nearby (U)LIRGs. 
The possible decrease of the HCN-to-HCO$^{+}$ flux ratio from low-J to
high-J can also contain H$_{2}$ gas density information 
\citep[e.g.,][]{ima23}; 
Because the critical density of HCN by H$_{2}$ collisional excitation
is a factor of $\sim$5 higher than that of HCO$^{+}$ at each
J-transition \citep{shi15}, the HCN-to-HCO$^{+}$ flux ratio can be
smaller at higher-J than at lower-J if H$_{2}$ gas density is not
sufficiently high.  

To derive molecular gas properties, we compare observed emission line
flux ratios with those calculated with the non-LTE radiative transfer 
code RADEX \citep{RADEX}, as we have done for other nearby
(U)LIRGs' nuclei using 1--2 kpc resolution data \citep{ima23}.  
Here, we do the same comparison for the newly observed nearby 
ULIRGs' nuclei (Table \ref{tab:object}). 
We also investigate the possible spatial variation of molecular gas
properties within the nuclear $\lesssim$2 kpc regions, using our
0.5 kpc resolution data.   
For all RADEX calculations, 
(1) gas geometry is assumed to be a one-zone uniform sphere,
(2) the cosmic microwave background with temperature of T$_{\rm bg}$ =
2.73 K is included, and   
(3) collisions with only H$_2$ are considered.
The molecular line width is commonly set to 500 km s$^{-1}$ as a
representative value, based on our Gaussian fits (Appendix C). 
Emission line flux ratios are calculated in units of Jy km s$^{-1}$
as listed in Tables \ref{tab:fluxratio} and \ref{tab:Jratio} 
in Appendix C. For convenience,
\texttt{pyradex}\footnote{\url{https://github.com/keflavich/pyradex}}, 
a Python wrapper for RADEX, is used.

In our model calculations, we adopt two approaches, following
\citet{ima23}. 
First, we constrain molecular gas density and temperature, by fixing
the HCN-to-HCO$^{+}$ abundance ratio and HCO$^{+}$ column density to
fiducial values ($\S$5.2--5.3), because the number of observational
constraints is limited.    
Next, we apply a Bayesian approach to constrain physical parameters,
by making all parameters free ($\S$5.4), with a caution that some
parameters may have systematic uncertainties, given so many free
parameters for the limited number of observational constrains.
We will then compare both results to confirm that our main arguments 
do not change and thus are robust.  

\subsection{HCN-to-HCO$^+$ Flux Ratios}

\begin{figure*}[!hbt]
\begin{center}
\includegraphics[angle=0,scale=.31]{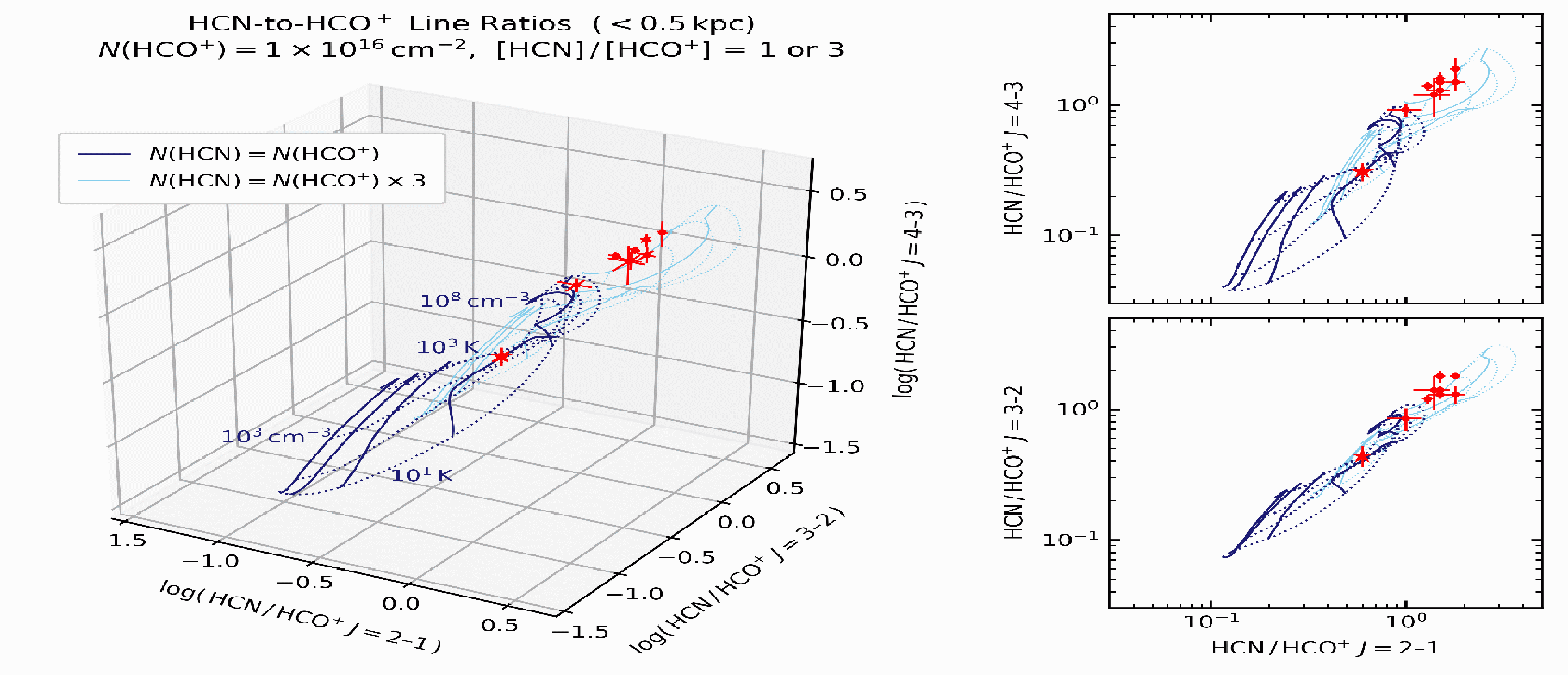} 
\end{center}
\caption{
Comparison of the HCN-to-HCO$^+$ flux ratios measured in the
0.5 kpc beam-sized spectra and those calculated with RADEX.
{\it (Left panel)}: 3D plot of the flux ratios at J=2--1, J=3--2, and
J=4--3. 
The red points represent the observed values.
The RADEX-calculated flux ratios are computed for 
log n$_{\rm  H_2}$/cm$^{-3}$ = 3--8 in steps of 0.1 (density) and 
log (T$_{\rm kin}$/K) = 1--3 in steps of 0.05 (temperature).
The ratios are shown as a blue mesh consisting of iso-density curves
at log n$_{\rm H_2}$ = 3, 4, $\cdots$, 8 (solid), and iso-temperature
curves at log T$_{\rm kin}$ = 1.0, 1.5, $\cdots$, 3.0 (dotted). 
The molecular line width and HCO$^+$ column density are fixed at 
$\Delta$v = 500 km s$^{-1}$ and N$_{\rm HCO+}$ = 1 $\times$ 10$^{16}$
cm$^{-2}$, respectively.  
The results of the HCN-to-HCO$^{+}$ abundance ratio of [HCN]/[HCO$^+$]
= 1 and 3 are drawn in dark and light colors, respectively. 
{\it (Right panels)}: Projections of the 3D plot along the J=3--2
{\it (upper)} and J=4--3 {\it (lower)} directions. 
\label{fig:RadexComp}
}
\end{figure*}

Figure \ref{fig:RadexComp} compares the HCN-to-HCO$^+$ flux ratios at
J=2--1, J=3--2, and J=4--3 measured in the 0.5 kpc beam-sized spectra, 
with predicted flux ratios by RADEX; 
RADEX calculations are made at densities 10$^{3-8}$ cm$^{-3}$
and temperatures 10$^{1-3}$ K.
The HCO$^+$ column density is fixed at N$_{\rm HCO+}$ = 1 $\times$
10$^{16}$ cm$^{-2}$, 
based on the assumption that the observed (U)LIRGs suffer from
modestly Compton thick (N$_{\rm H}$ $\sim$ a few $\times$ 10$^{24}$
cm$^{-2}$) absorption and the HCO$^{+}$-to-H$_{2}$ abundance ratio is
$\sim$10$^{-8}$ \citep[e.g.,][]{mar06,sai18}.
The HCN-to-HCO$^{+}$ abundance ratio is tested for two cases, 
[HCN]/[HCO$^+$] = 1 and 3.

Except for the LIRG NGC 1614 (the most bottom-left red filled star in
Figure \ref{fig:RadexComp}), the observed HCN-to-HCO$^{+}$ flux ratios
of all ULIRGs can be better explained by the HCN-to-HCO$^{+}$
abundance ratio of [HCN]/[HCO$^+$] = 3 rather than 1, under the above
assumed gas density and temperature ranges.  
We also try different HCN-to-HCO$^{+}$ abundance ratio 
([HCN]/[HCO$^+$] = 7) and HCO$^{+}$ column density 
(N$_{\rm HCO+}$ = 1 $\times$ 10$^{15}$ cm$^{-2}$ and 3 $\times$
10$^{16}$ cm$^{-2}$), but our conclusion that the observed
HCN-to-HCO$^{+}$ flux ratios of ULIRGs are better reproduced with an
enhanced ($>$1) HCN-to-HCO$^{+}$ abundance ratio, remains
unchanged (Appendix G), as previously confirmed for other nearby
ULIRGs, calculated with different line widths \citep{ima23}.   
We conservatively adopt [HCN]/[HCO$^+$] = 3 as a fiducial value.

\subsection{High-J to Low-J Flux Ratios}

We fit the observed high-J to low-J flux ratios of HCN and HCO$^+$
with RADEX to estimate molecular gas density and temperature. 
The method is the same as that employed by \citet{ima23}.
The least-squares fitting for log n$_{\rm H_2}$ (density) and log
T$_{\rm kin}$ (temperature) is performed with the conventional
Levenberg-Marquardt algorithm using the Python package \texttt{lmfit} 
\citep{lmfit}. 
Confidence intervals for the parameters are examined
by grid computing $\Delta\chi^2\equiv\chi^2-\chi^2_{\rm best}$
with log n$_{\rm H_2}$ ranging from 2 to 6 and log T$_{\rm kin}$
from 1 to 3.

As described by \citet{ima23} and in $\S$3, the high-J to low-J
flux ratios of HCN and HCO$^{+}$ can be affected by possible absolute flux
calibration uncertainty of individual ALMA observations, because
J=2--1, J=3--2, and J=4--3 data were taken at different times.
This systematic uncertainty needs to be taken into account when we
compare the observed and RADEX-calculated flux ratios. 
As our second calculations, we allow scaling of absolute flux within
maximum 5$\%$ for J=2--1 and 10$\%$ for J=3--2 and J=4--3 ($\S$3).
We divide the fitting into two stages:
in the first stage, gas density, temperature, and scaling of each
emission line flux within the calibration uncertainty are left free, and
the residuals are minimized using the L-BFGS-B method; 
in the second stage, the scaling factors are fixed to the obtained values.
We then derive gas density and temperature using the
Levenberg-Marquardt algorithm with the Python package \texttt{lmfit}
\citep{lmfit}, in the same way as above.

\begin{figure*}[!hbt]
\begin{center}
\includegraphics[angle=0,scale=.23]{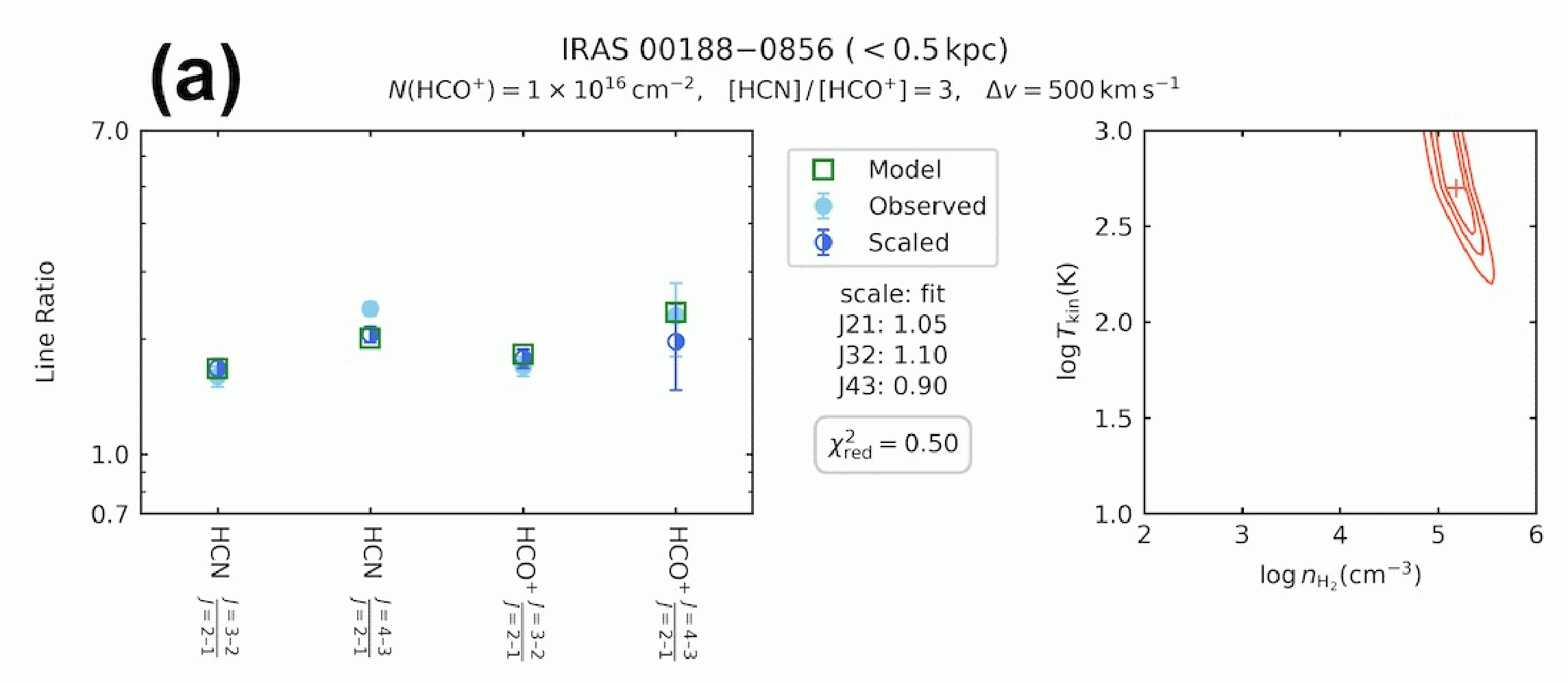} 
\includegraphics[angle=0,scale=.23]{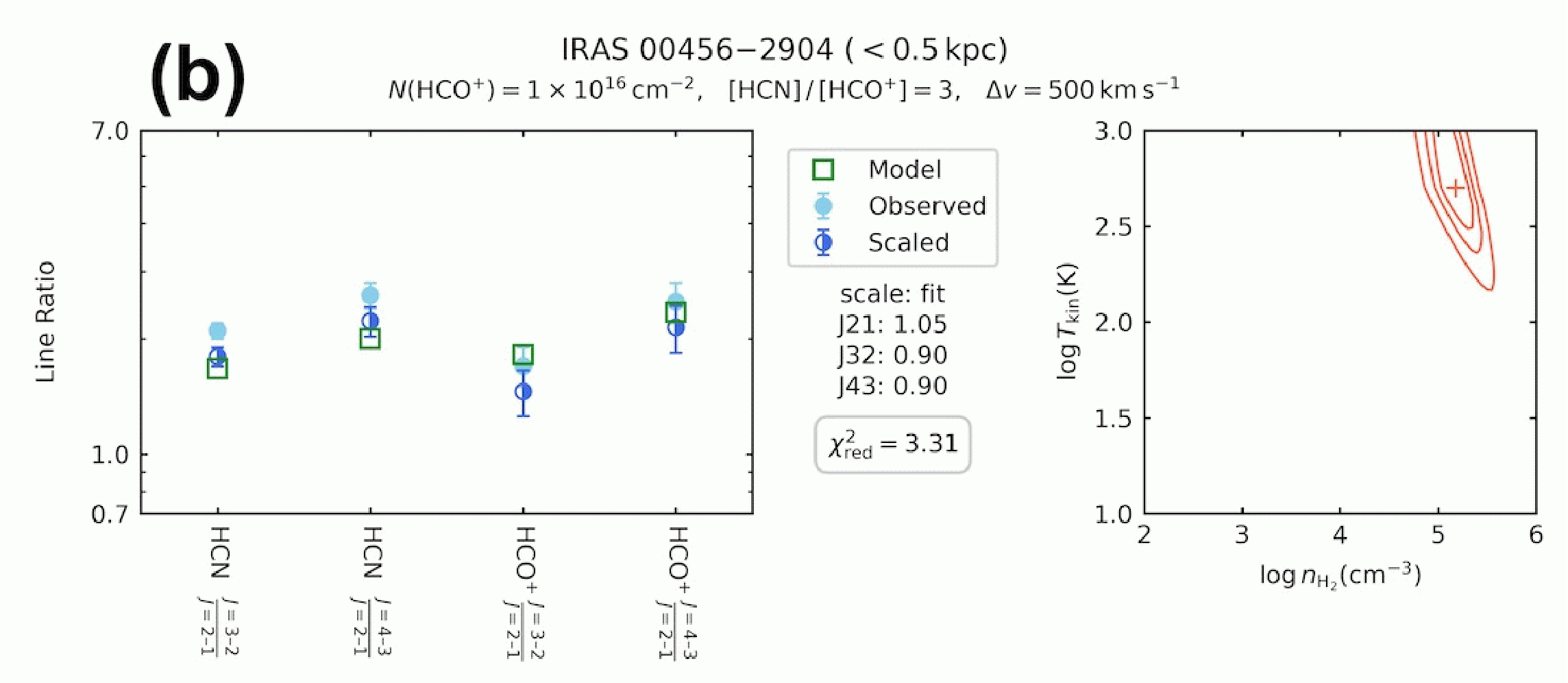} \\
\includegraphics[angle=0,scale=.23]{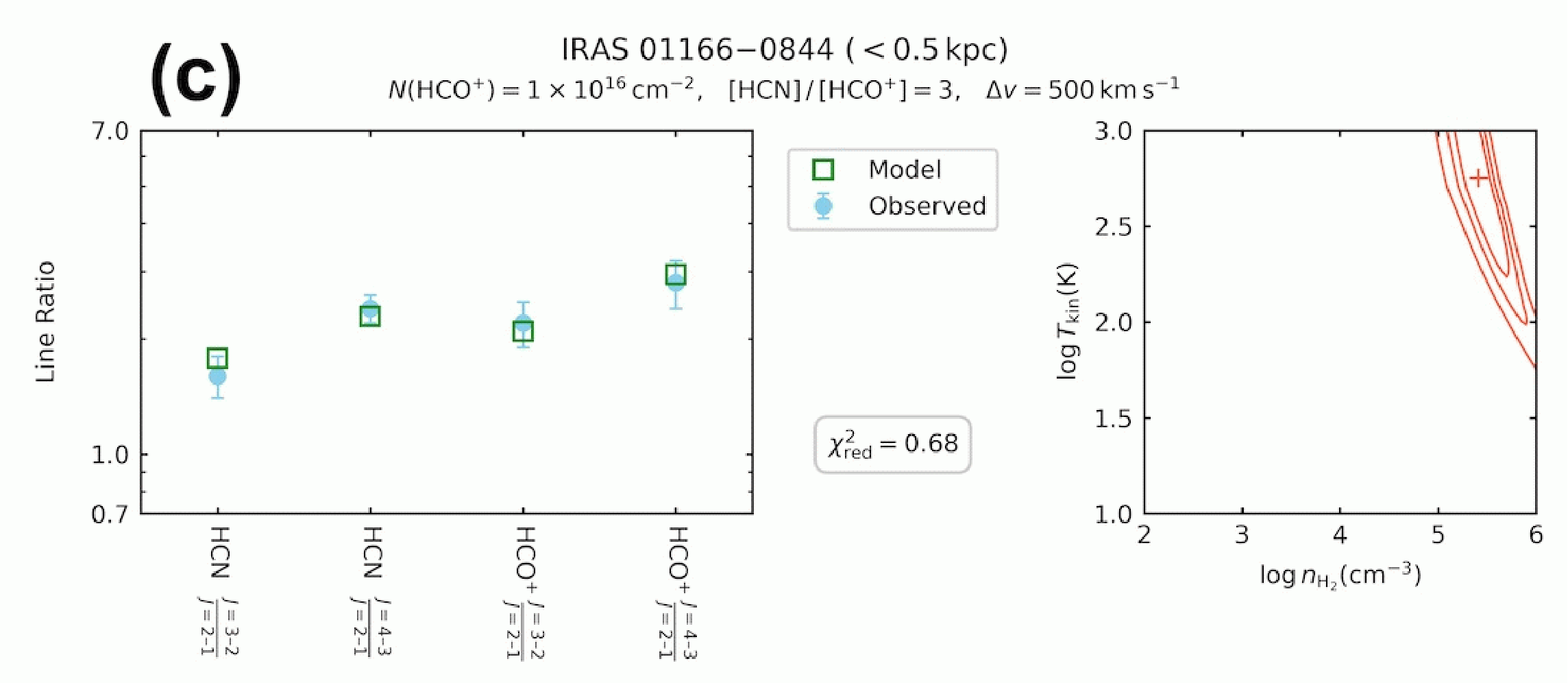} 
\includegraphics[angle=0,scale=.23]{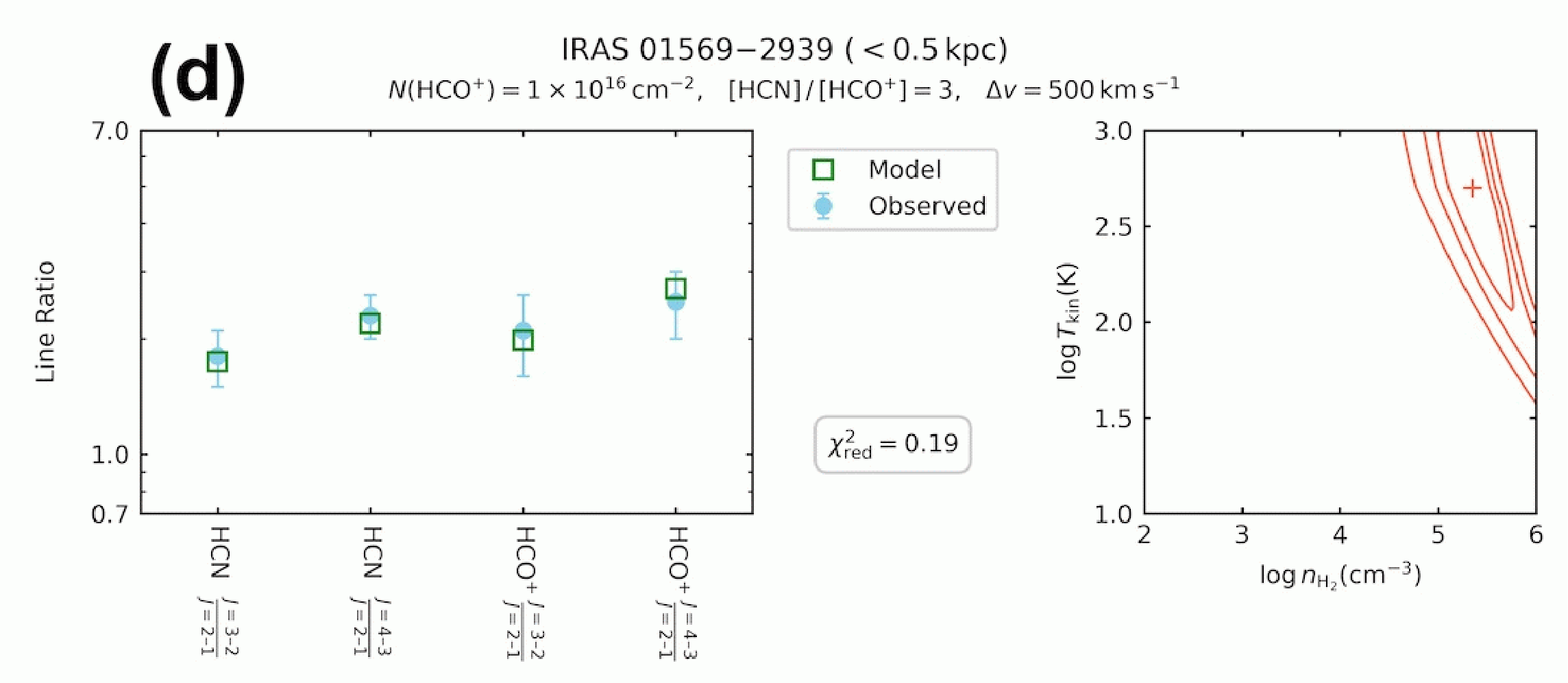} \\
\includegraphics[angle=0,scale=.23]{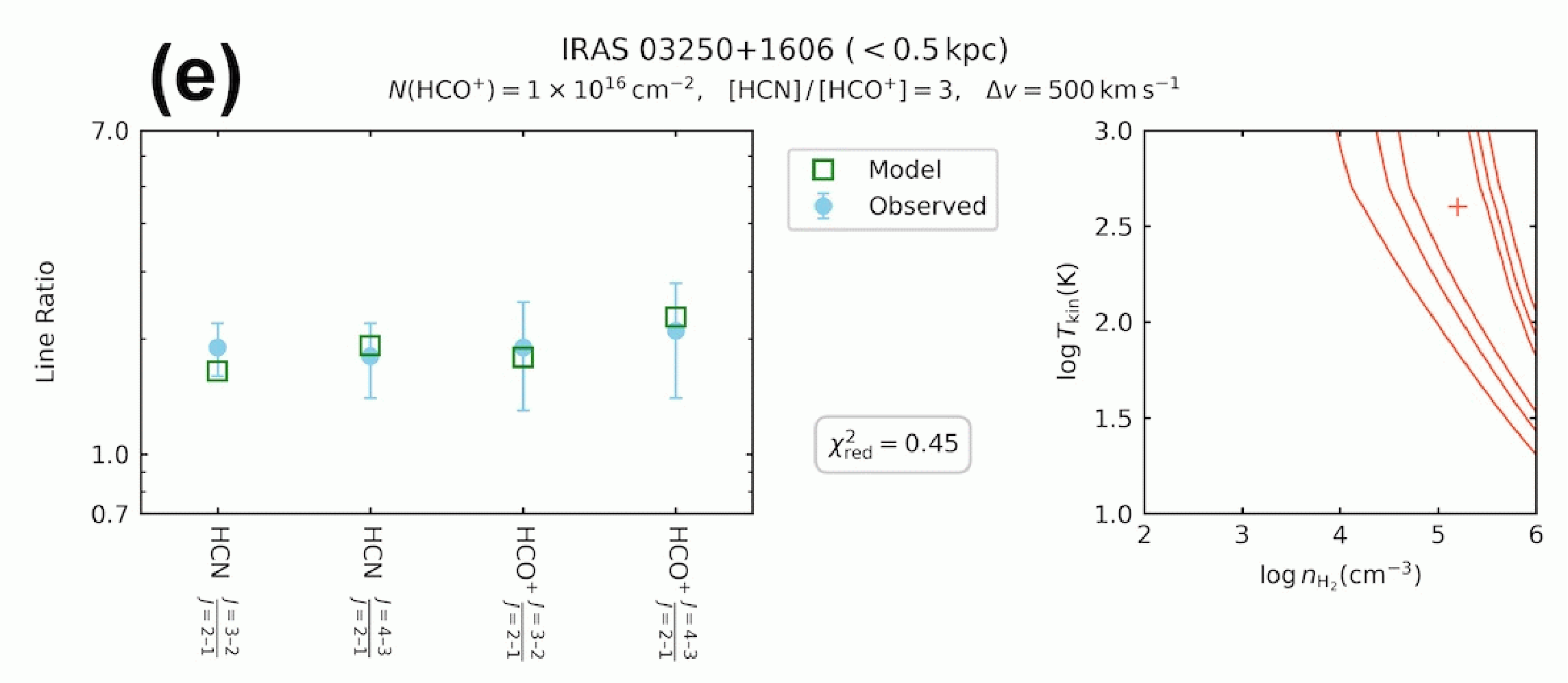} 
\includegraphics[angle=0,scale=.23]{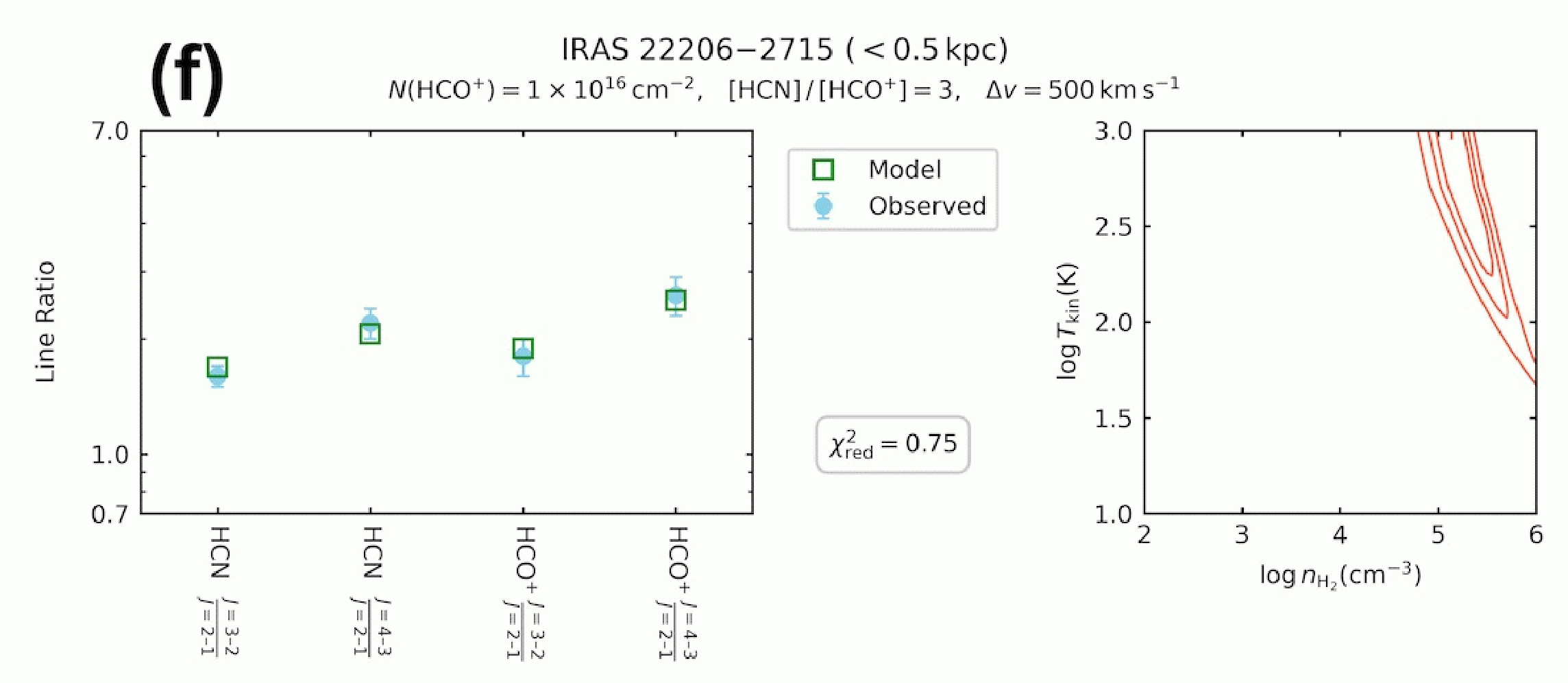} \\
\includegraphics[angle=0,scale=.23]{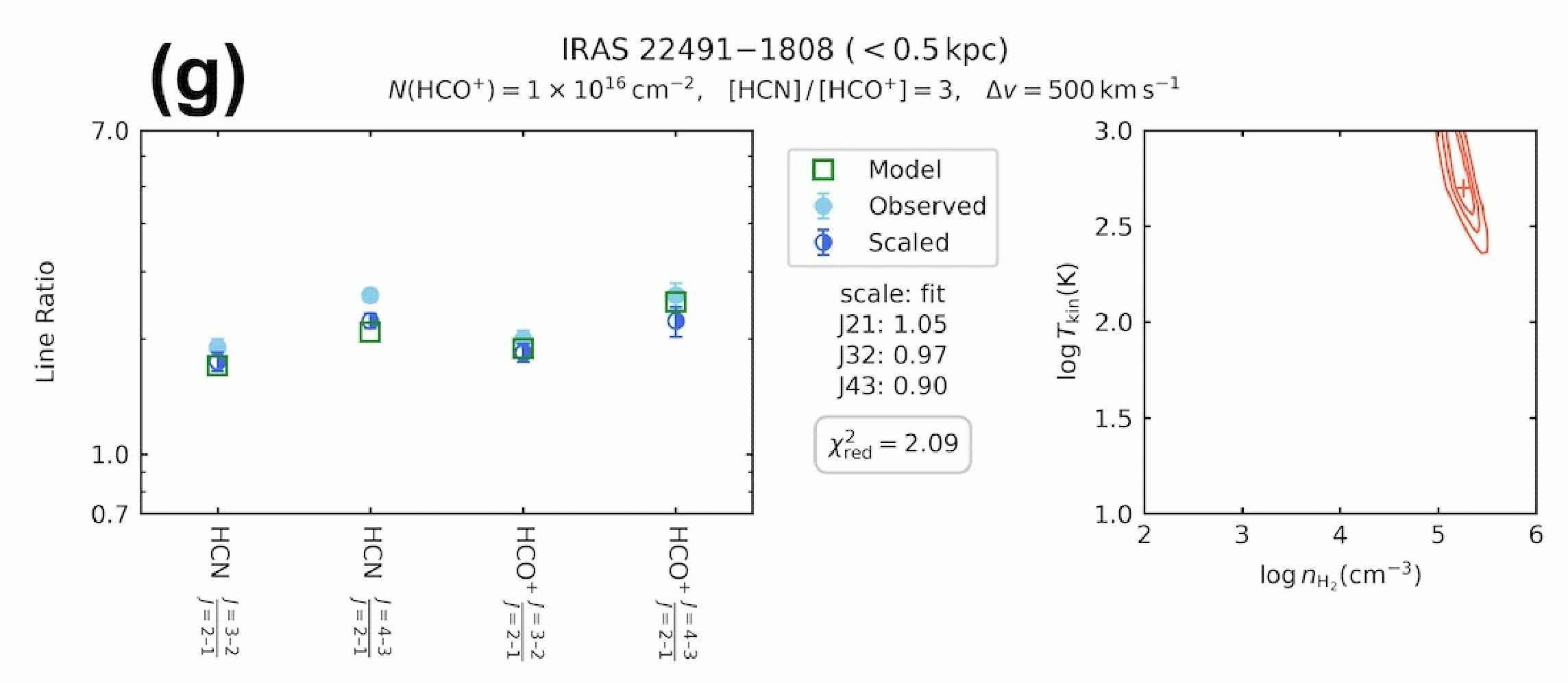} 
\includegraphics[angle=0,scale=.23]{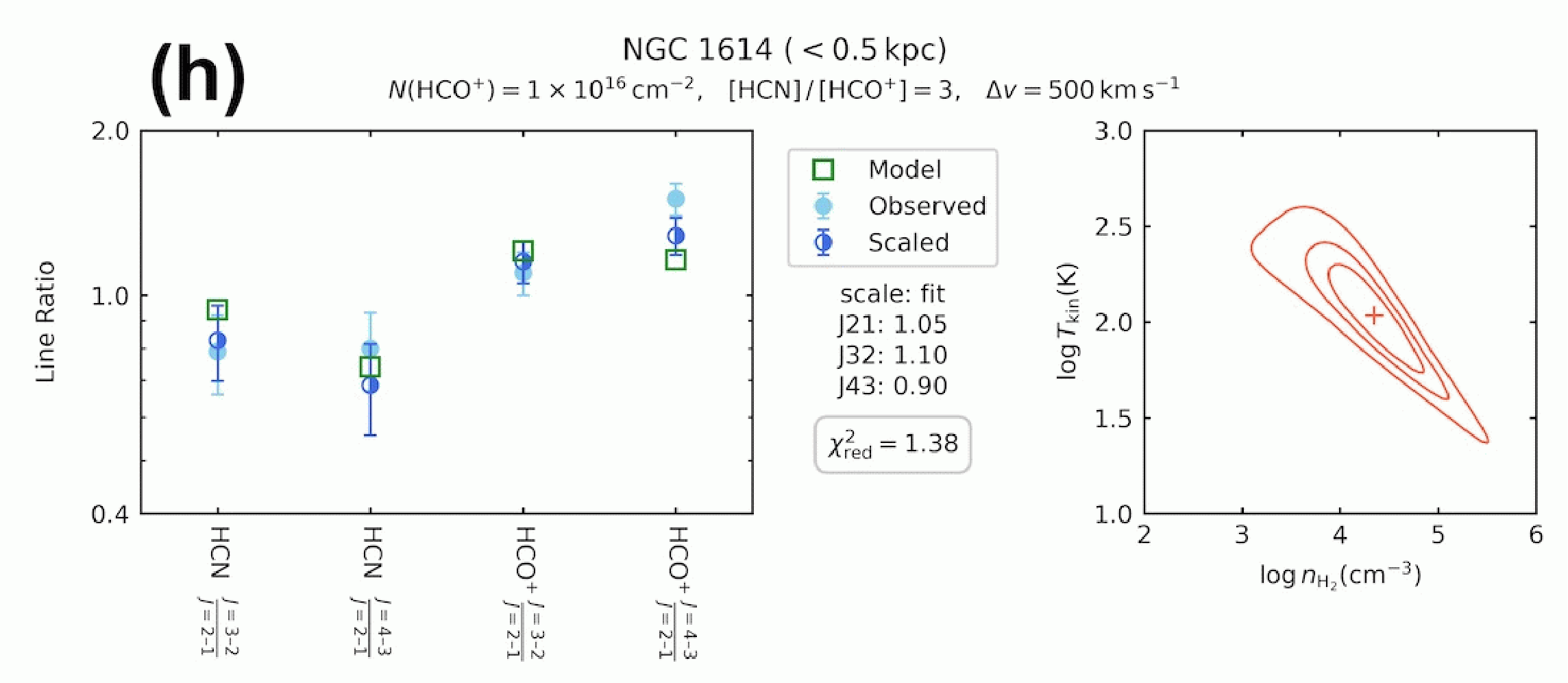} \\
\end{center}
\caption{
Fitting results for the high-J to low-J flux ratios of HCN and HCO$^+$
for the central $\lesssim$0.5 kpc regions of (U)LIRGs.
{\it (Left panel)}: Comparison of the observed and RADEX-calculated
emission line flux ratios.   
The light blue filled circles show the observed flux ratios listed in
Table \ref{tab:Jratio}.
The green open squares indicate the flux ratios of the best-fit model,
whose gas density and temperature are denoted as the plus sign in the
right panel.
Other fixed parameters (column density, abundance ratio, and line
width) are noted at the top of the figure. 
We adopt the second fitting results (flux scaling adjustment allowed; 
see $\S$5.3) 
for IRAS 00188$-$0856, IRAS 00456$-$2904, IRAS 22491$-$1808, and NGC
1614, for which the observed flux ratios after scaling adjustment are
displayed in the dark blue half-filled circles. 
The scaling factors for each observation are listed below the legend.
The reduced $\chi^2$ value of the best-fit model is also noted.
{\it (Right panel)}: Confidence contours for gas density and temperature
quoted at 68\%, 90\%, and 99\% levels ($\Delta\chi^2=2.28, 4.61, 9.21$).
Result of IRAS 16090$-$0139 is displayed separately in Figure
\ref{fig:IR16090FitLM}a.  
\label{fig:FitLM}
}
\end{figure*}

\begin{figure*}[!hbt]
\begin{center}
\includegraphics[angle=0,scale=.24]{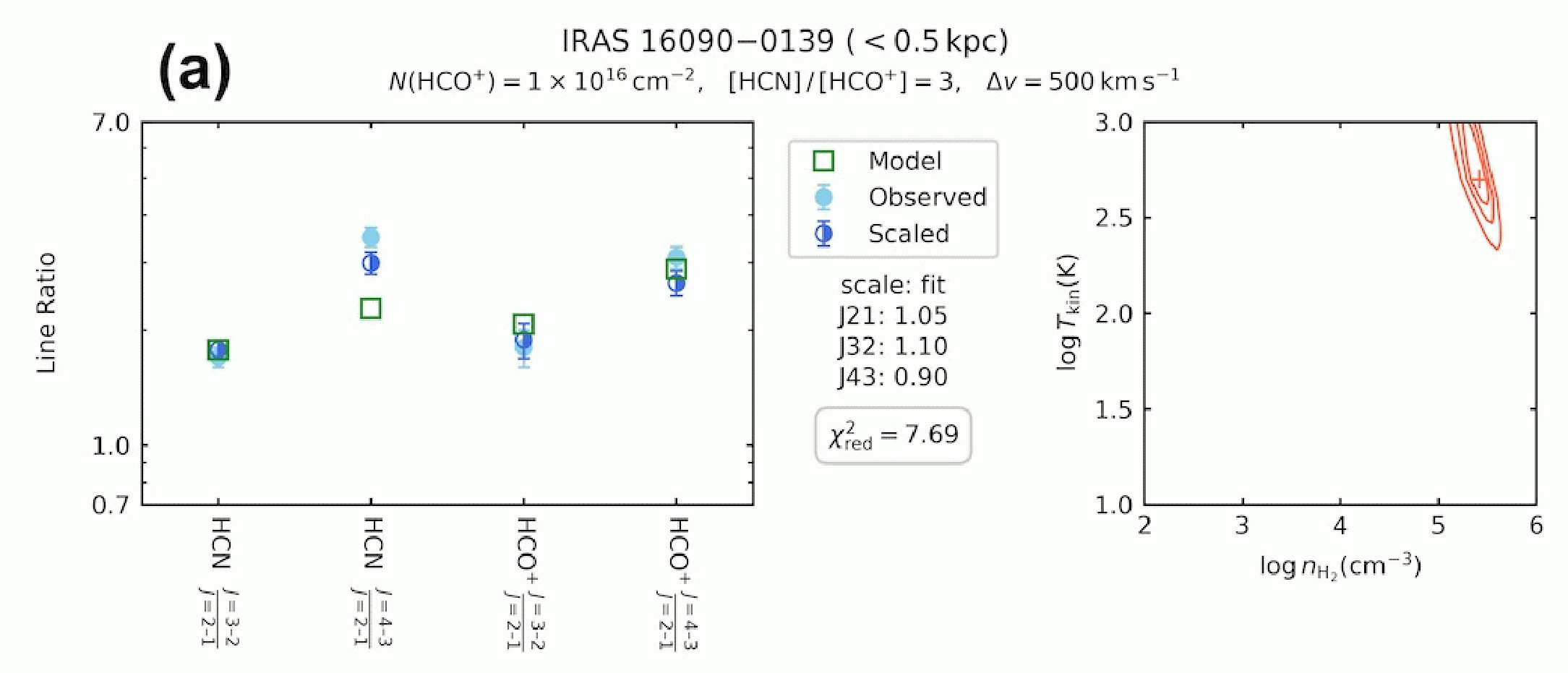} 
\includegraphics[angle=0,scale=.24]{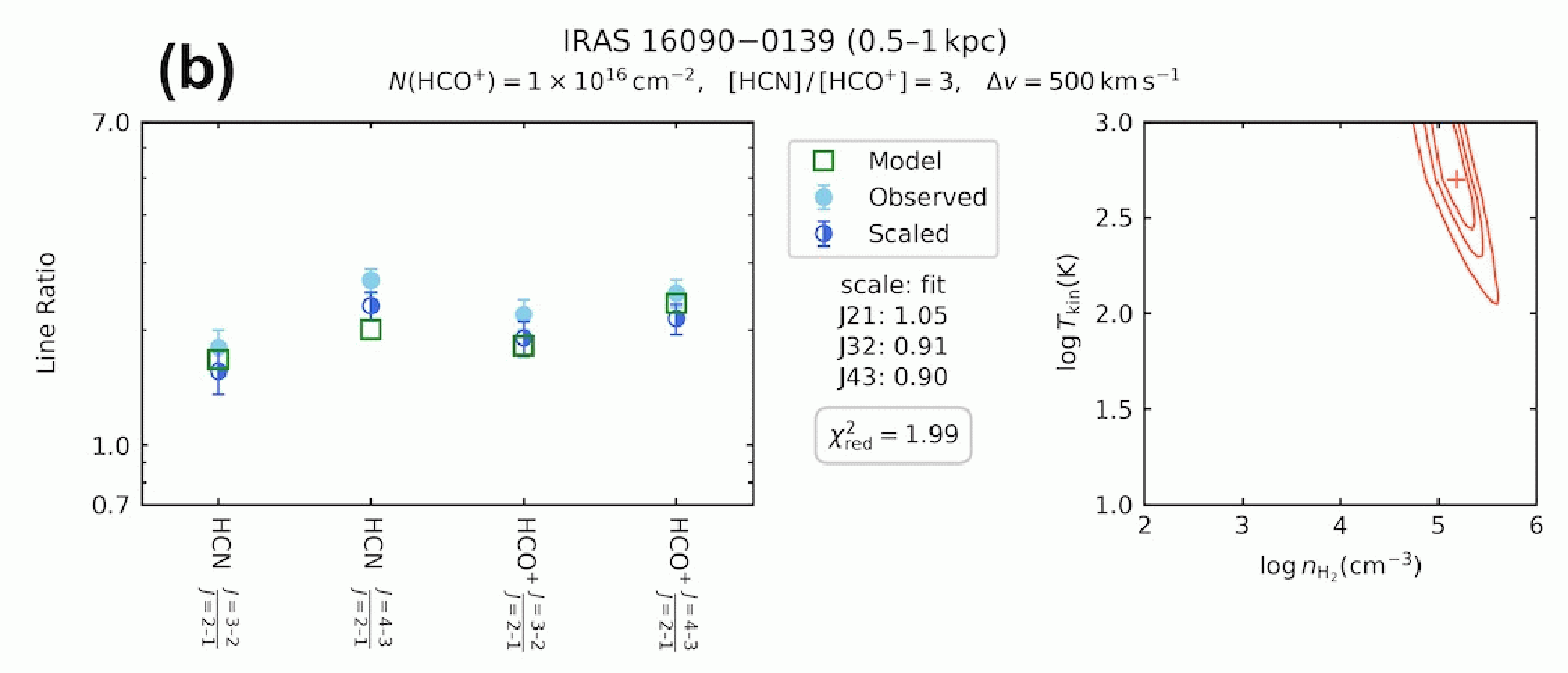} \\ 
\includegraphics[angle=0,scale=.24]{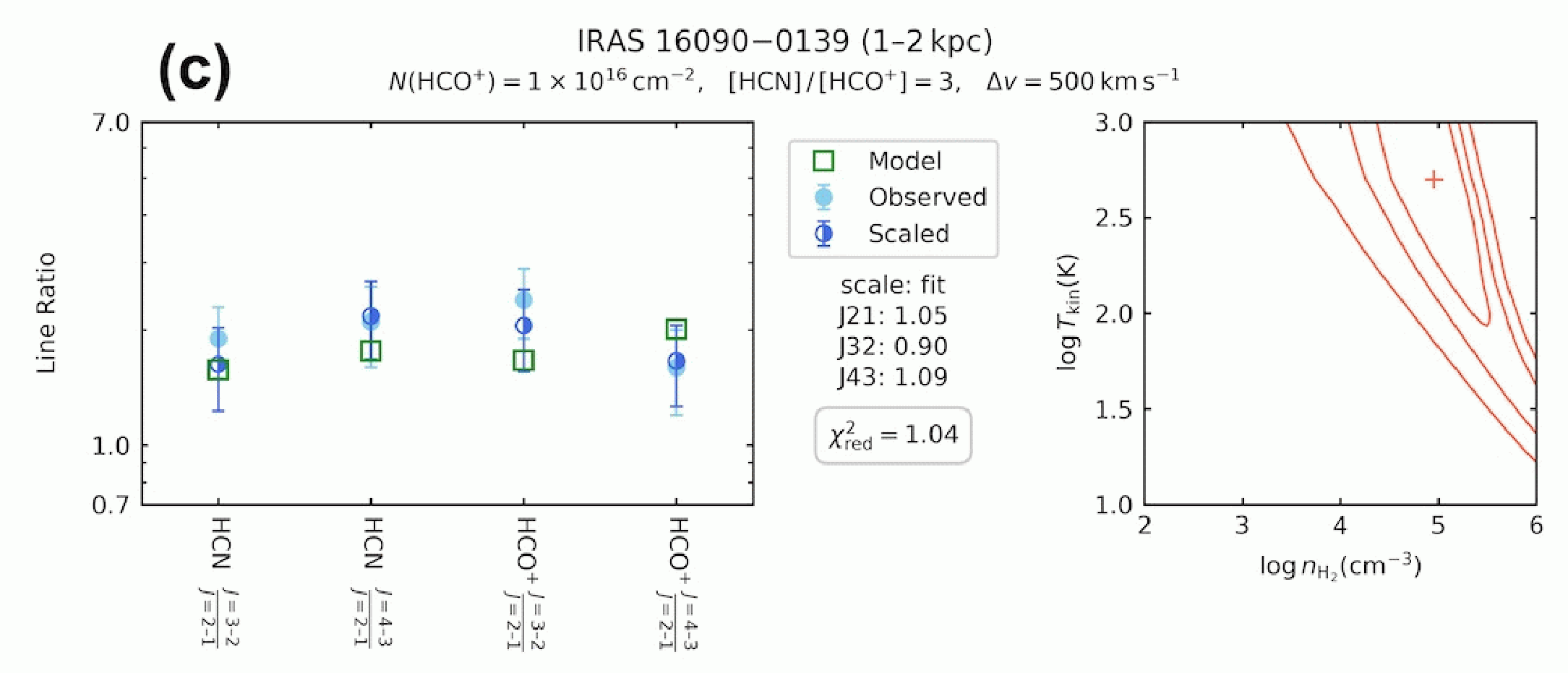} 
\includegraphics[angle=0,scale=.24]{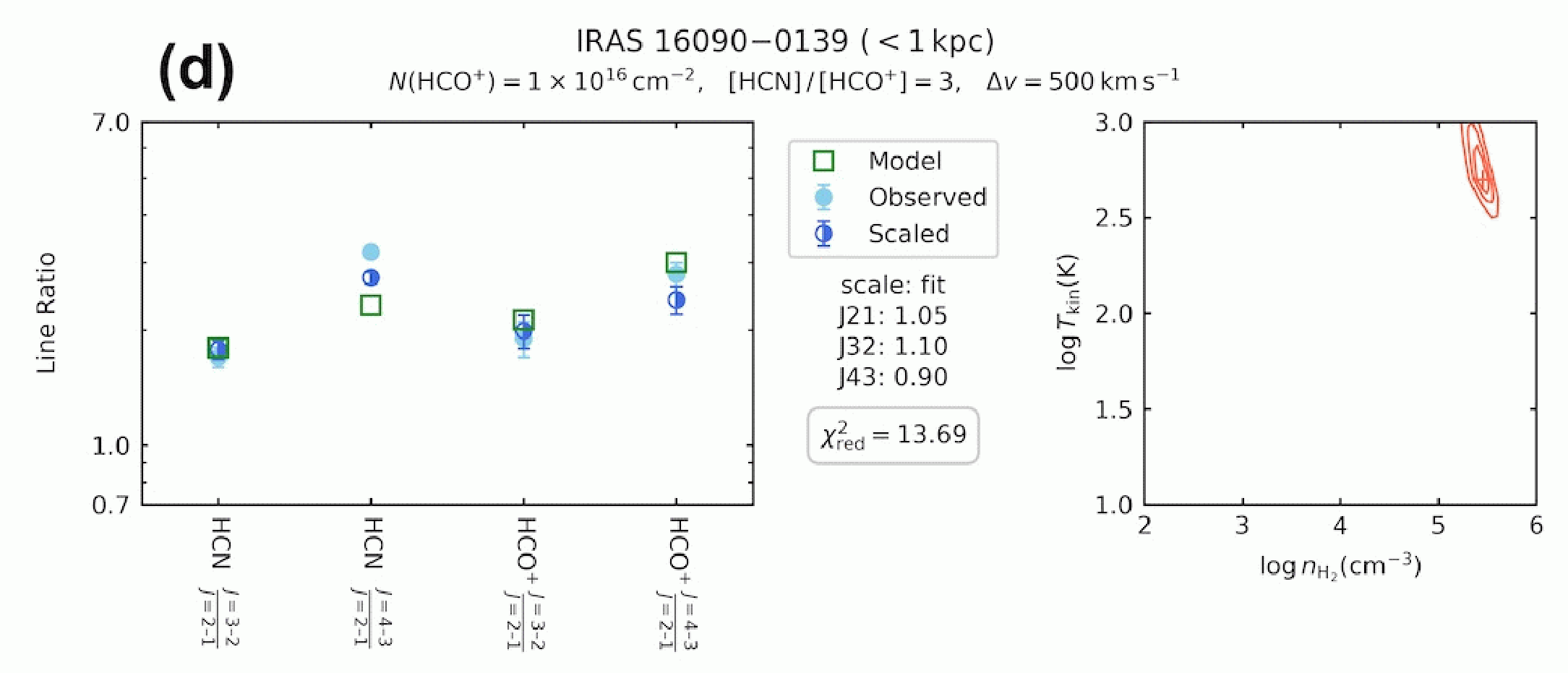} \\ 
\hspace*{-9cm}
\includegraphics[angle=0,scale=.24]{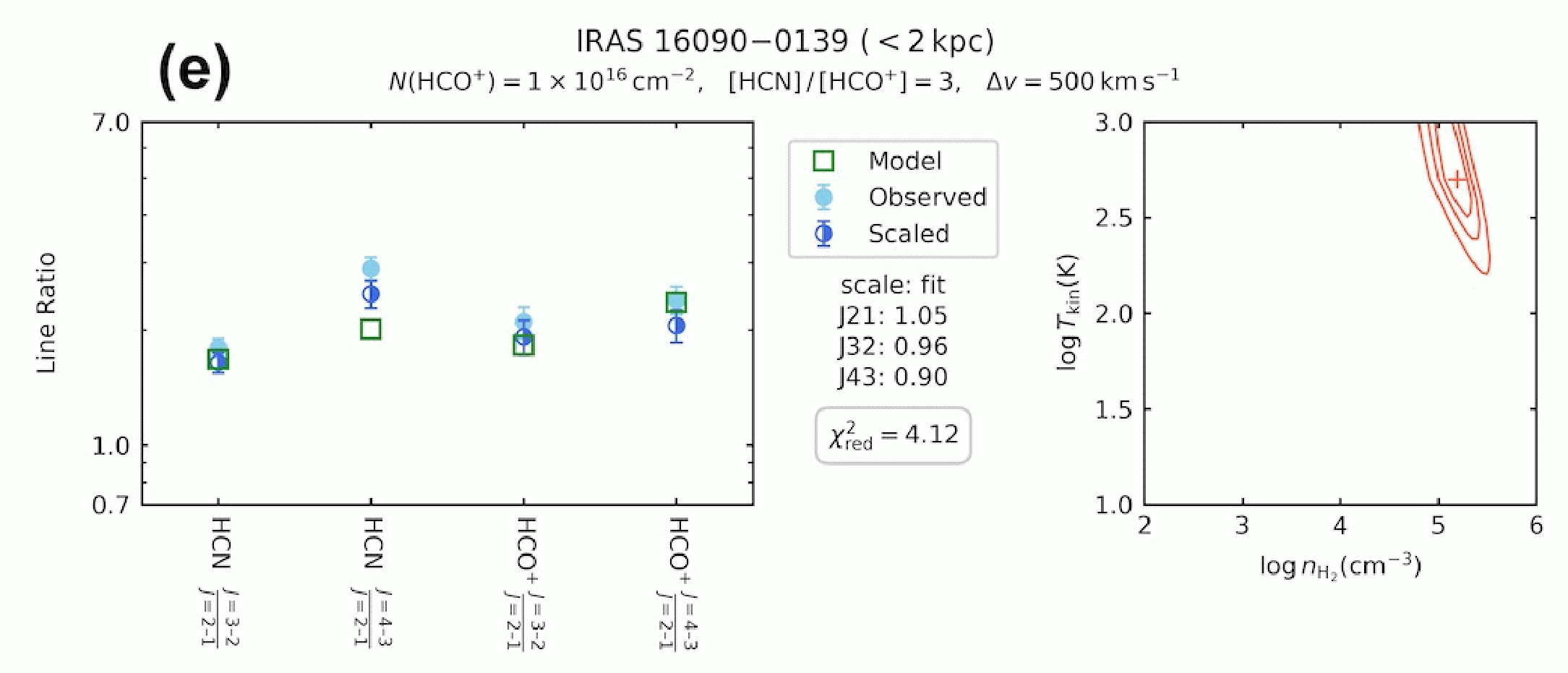} 
\end{center}
\caption{
Fitting results for the high-J to low-J flux ratios of HCN and HCO$^+$
for various regions of IRAS 16090$-$0139.
{\it (a)}: central $\lesssim$0.5 kpc region.
{\it (b)}: 0.5--1 kpc annular region.
{\it (c)}: 1--2 kpc annular region.
{\it (d)}: central $\lesssim$1 kpc region.
{\it (e)}: central $\lesssim$2 kpc region.
We adopt the second fitting result (flux scaling adjustment allowed) 
for all regions.
The content of each panel and symbols are the same as Figure
\ref{fig:FitLM}.  
\label{fig:IR16090FitLM}
}
\end{figure*}

We derive molecular gas physical parameters, by 
(1) using the observed high-J to low-J flux ratios as they are (no
flux scaling), and 
(2) allowing flux scale adjustment for individual J=2--1, J=3--2, and
J=4--3 data within the above allowable range (5--10$\%$).
The derived gas density and temperature are generally comparable between
the first and second methods, but the reduced $\chi^{2}$ 
value is usually smaller in the second fitting result (scaling
on) than the first one (scaling off).
We adopt the first one as much as possible, but
refer to the second one only if the first one cannot determine the
best fit value or provides a very large reduced $\chi^{2}$ value.  
Our adopted final results for the central $\lesssim$0.5 kpc region are
presented in Figures \ref{fig:FitLM} and \ref{fig:IR16090FitLM}a, and
are summarized in Table \ref{tab:bestfit}.
The same results for the central $\lesssim$1 kpc and $\lesssim$2 kpc
regions are presented in Appendix H, to be compared with those
derived for different nearby ULIRGs with comparable 1--2 kpc
resolutions by \citet{ima23}.  
We exclude IRAS 10378$+$1108 and IRAS 12112$+$0305 because not all the
three J-transition line data are available.
We exclude also IRAS 00091$-$0738 because negative signals 
below the continuum levels, clearly detected at the HCO$^{+}$
central absorption dips in Figures \ref{fig:SpectraA} and 
\ref{fig:SpectraB}, suggest self-absorption (see also Appendix C,
Table \ref{tab:Gaussfit}, footnote a).
Namely, molecular gas consists of more than one component
(i.e., emission and absorption components), which complicates
comparison between the observed data and one-zone RADEX model
calculations.   


\begin{deluxetable}{lccccc}[!hbt]
\tablecaption{Summary of the Best Fit Values \label{tab:bestfit}}
\tablewidth{0pt}
\tablehead{\colhead{Object} & \colhead{Region} & \colhead{Scaling} & 
\colhead{log n$_{\rm H_2}$} & \colhead{log T$_{\rm kin}$} & 
\colhead{Reduced} \\
\colhead{} & \colhead{} & \colhead{} & \colhead{[cm$^{-3}$]} & 
\colhead{[K]} & \colhead{$\chi^{2}$} \\ 
\colhead{(1)} & \colhead{(2)} & \colhead{(3)} & \colhead{(4)} &
\colhead{(5)} & \colhead{(6)}  
} 
\startdata
IRAS 00188$-$0856 & $<$0.5 kpc & on & 5.2$^{+0.1}_{-0.2}$ & 2.7$^{+0.6}_{-0.1}$
& 0.50 \\ 
IRAS 00456$-$2904 & $<$0.5 kpc & on & 5.2$^{+0.1}_{-0.2}$ & 2.7$^{+0.3}_{-0.1}$
& 3.3 \\ 
IRAS 01166$-$0844 & $<$0.5 kpc & off & 5.4$^{+0.2}_{-0.2}$ & 2.8$^{+\infty}_{-0.4}$
& 0.68 \\ 
IRAS 01569$-$2939 & $<$0.5 kpc & off & 5.4$^{+0.3}_{-0.3}$ & 2.7$^{+\infty}_{-0.4}$
& 0.19 \\ 
IRAS 03250$+$1606 & $<$0.5 kpc & off & 5.2$^{+0.8}_{-0.4}$ & 2.6$^{+\infty}_{-0.9}$
& 0.45 \\ 
IRAS 22206$-$2715 & $<$0.5 kpc & off & 5.1$^{+0.3}_{-0.1}$ &
3.0$^{+\infty}_{-0.6}$ \tablenotemark{a} & 0.75 \\ 
IRAS 22491$-$1808 & $<$0.5 kpc & on & 5.3$^{+0.1}_{-0.1}$ & 2.7$^{+0.2}_{-0.1}$
& 2.1 \\  
NGC 1614 & $<$0.5 kpc & on & 4.3$^{+0.3}_{-0.3}$ & 2.0$^{+0.2}_{-0.2}$ & 1.4 \\ 
IRAS 16090$-$0139 & $<$0.5 kpc & on & 5.4$^{+0.1}_{-0.1}$ & 2.7$^{+0.2}_{-0.1}$
& 7.7 \\ 
 & 0.5--1 kpc & on & 5.2$^{+0.1}_{-0.2}$ & 2.7$^{+0.5}_{-0.1}$ & 2.0 \\ 
 & 1--2 kpc & on & 5.0$^{+0.3}_{-0.4}$ & 2.7$^{+\infty}_{-0.4}$ & 1.0 \\ 
 & $<$1 kpc & on & 5.5$^{+0.1}_{-0.1}$ & 2.7$^{+0.1}_{-0.1}$ & 13.7 \\ 
 & $<$2 kpc & on & 5.2$^{+0.1}_{-0.1}$ & 2.7$^{+0.3}_{-0.1}$ & 4.1 \\ 
\enddata

\tablenotetext{a}{The best fit value is our adopted upper
bound of T$_{\rm kin}$ = 1000 K.}

\tablecomments{
Col.(1): Object name. 
Col.(2): Region.
Col.(3): Scaling on or off.
Col.(4): Decimal logarithm of H$_{2}$ gas density in units of cm$^{-3}$. 
Col.(5): Decimal logarithm of gas kinetic temperature in units of K. 
Col.(6): Reduced $\chi^{2}$ value.
The HCO$^{+}$ column density, HCN-to-HCO$^{+}$ abundance ratio, and
molecular line width are fixed at N$_{\rm HCO^+}$ = 1 $\times$
10$^{16}$ cm$^{-2}$, [HCN]/[HCO$^+$] = 3, and $\Delta$v = 500 km
s$^{-1}$, respectively. 
}

\end{deluxetable}

We clearly see in Figure \ref{fig:FitLM} and Table \ref{tab:bestfit}
that molecular gas in the central $\lesssim$0.5 kpc regions of all the
observed ULIRGs is very dense ($\gtrsim$10$^{5}$ cm$^{-3}$)
and warm ($\gtrsim$10$^{2.5}$ K or $\gtrsim$300 K).
Molecular gas density and temperature are estimated to be very
high also in the $\lesssim$1 kpc and $\lesssim$2 kpc region data of
ULIRGs (Appendix H).  
On the other hand, the starburst-dominated LIRG NGC 1614 contains 
less dense ($\lesssim$10$^{4.3}$ cm$^{-3}$) and cooler ($\sim$10$^{2}$ K) 
molecular gas both at the central $\lesssim$0.5 kpc and $\lesssim$1
kpc regions (Appendix H).    
This is as expected because the high-J to low-J flux ratios of HCN and
HCO$^{+}$ in NGC 1614 are distinctly smaller than those of ULIRGs 
at the $\lesssim$0.5 kpc, $\lesssim$1 kpc, and 0.5--1 kpc 
regions (Figure \ref{fig:Jratio}). 
Systematic difference of nuclear gas density and temperature between
nearby ULIRGs and LIRGs has previously been seen also at 1--2 kpc
resolution for a different ULIRG sample \citep{ima23}. 
Nearby ULIRGs are usually energetically dominated by compact 
($\lesssim$1 kpc) nuclear regions
\citep[e.g.,][]{soi00,dia10,ima11,per21}, while in nearby LIRGs,  
compact nuclear regions are energetically less dominant, relative to
spatially extended ($\gtrsim$a few kpc) star-formation activity
\citep{soi01}. 
It is also found that nearby ULIRGs show luminous AGN signatures
more frequently than nearby LIRGs do \citep[e.g.,][]{vei09,nar10,ima10b}.
A natural scenario for the derived denser and warmer molecular gas 
at the innermost ($\lesssim$0.5 kpc) regions of nearby ULIRGs is that 
(1) a larger amount of nuclear concentrated molecular gas can be
a fuel to a central SMBH, and (2) the resulting enhanced AGN activity
can make the innermost ($\lesssim$0.5 kpc) molecular gas warmer than 
starburst-dominated LIRGs 
\footnote{
\citet{ima23} found a trend of denser and warmer nuclear molecular gas
in AGN-important sources than in starburst-dominated ones in the ULIRG
population (L$_{\rm IR}$ $\gtrsim$ 10$^{12}$L$_{\odot}$), but
\citet{kri08} did not find any such trend between AGNs and starbursts
at lower infrared luminosities.
}.
It is very likely that the warm molecular gas that we detect in the 
$\lesssim$0.5 kpc spectra of nearby ULIRGs largely comes from the
innermost molecular gas surrounding the central luminous AGNs, as
probed by infrared 4--5 $\mu$m ro-vibrational CO absorption study
\citep{bab18}. 

Figure \ref{fig:HCNtoHCOratio}h shows that for IRAS 16090$-$0139,  
the statistical uncertainty of the HCN-to-HCO$^{+}$ flux ratios 
is very small and thus the clearest decreasing trend of the
ratios from left to right is seen among the observed (U)LIRGs. 
For IRAS 16090$-$0139, the decreasing trend is also recognizable in 
the J=4--3 to J=2--1 flux ratios of HCN and HCO$^{+}$ (Figure
\ref{fig:Jratio}h). 
We investigate how the derived gas density and temperature spatially
change in IRAS 16090$-$0139.
Figure \ref{fig:IR16090FitLM} and Table \ref{tab:bestfit} show the
results. 
We see some sign that the derived best-fit gas density tends to
decrease from the innermost $\lesssim$0.5 kpc region, through 0.5--1 kpc
annular region, to 1--2 kpc annular region 
(Figures \ref{fig:IR16090FitLM}a--c and Table \ref{tab:bestfit}). 
The derived gas density also tends to decrease by increasing the beam
size from 0.5 kpc to 2 kpc (Figures \ref{fig:IR16090FitLM}a,e and 
Table \ref{tab:bestfit}). 
The detection of the decreasing gas density trend from the innermost
($\lesssim$0.5 kpc) to outer nuclear (0.5--2 kpc) region in 
IRAS 16090$-$0139 suggests that it may be feasible to investigate
the spatial variation of molecular gas physical parameters within
nuclear $\sim$2 kpc regions in more detail at least for some nearby
(U)LIRGs with significant molecular line detection. 

In principle, possible spatial variation of molecular gas physical
parameters can be seen more clearly from $\lesssim$0.5 kpc to 0.5--1
kpc and 1--2 kpc annular regions, than that from $\lesssim$0.5 kpc to
$\lesssim$1 kpc and $\lesssim$ 2 kpc circular regions, because each
region is separated more clearly in the former.  
If nuclear ($\lesssim$2 kpc) molecular gas emission is dominated by 
the innermost $\lesssim$0.5 kpc region, possible spatial variation of 
the gas physical parameters can be diluted in the latter comparison. 
In Figures \ref{fig:HCNtoHCOratio} and \ref{fig:Jratio}, (i) the
HCN-to-HCO$^{+}$ flux ratios at J=2--1, J=3--2, and J=4--3, and
(ii) high-J to low-J flux ratios of HCN and HCO$^{+}$ for both J=4--3 to
J=2--1 and J=3--2 to J=2--1, are all derived with sufficiently high
S/N ratios, in both the 0.5--1 kpc and 1--2 kpc annular regions,  
only for IRAS 16090$-$0139. 
There are two reasons for this.
First, dense molecular line emission at 0.5--1 kpc and 1--2 kpc
annular regions is generally significantly fainter than that at the
innermost ($\lesssim$0.5 kpc) region (Figure \ref{fig:SpectraB}),
despite larger signal-integrated areas in the former by a
factor of 3 and 12, respectively.
Second, scatters of spectral data points are inevitably large 
particularly in the 1--2 kpc spectra, because 
(1) enlarging originally small-beam-sized data to large beams,
increases rms noise in units of mJy beam$^{-1}$ ($\S$4) and 
(2) subtraction of two spectra results in a further noise increase by
a factor of $\sqrt{2}$.  
Thus, for other ULIRGs than IRAS 16090$-$0139, we primarily
investigate the possible spatial variation of molecular gas physical
parameters from the $\lesssim$0.5 kpc region to the 0.5--1 kpc annular,
and $\lesssim$1 kpc and $\lesssim$2 kpc circular regions, with the
above caveat that the possible spatial variation may be
diluted in the comparison using the latter circular regions.   


\subsection{Bayesian Analysis of Both Types of Ratios}

For selected regions of some (U)LIRGs where molecular emission lines
are significantly detected with high S/N ratios,
we fit both the HCN-to-HCO$^+$ flux ratios and HCO$^+$ high-J to low-J
flux ratios simultaneously with RADEX to derive the gas physical
parameters in detail, without fixing the HCO$^{+}$ column density and
HCN-to-HCO$^{+}$ abundance, by using a Bayesian approach. 
Although the number of available independent emission line flux ratios
(HCN-to-HCO$^+$ flux ratio at J=2--1, J=3--2, and J=4--3, HCO$^+$ J=3--2
to J=2--1, and HCO$^{+}$ J=4--3 to J=2--1) is fewer than the total
number of parameters, including the absolute flux scaling factors, the
Bayesian technique is able to sample the posterior probability
distribution naturally including the indeterminacy of the solution. 
 
We use a Markov Chain Monte Carlo (MCMC) sampler
implemented in the \texttt{emcee} package \citep{emcee}
to explore the parameter space.
Flat priors having upper and lower bounds listed in Table
\ref{tab:bounds_2} are employed. 
The chain is run with 100 walkers initialized around a first guess
obtained by the L-BFGS-B solver ($\S$5.3) with the column density and
abundance ratio unfixed. 
By defining $\tau$ as the longest autocorrelation time of the parameters,
the chain is continued up to 100$\tau$ steps,
with the first 5$\tau$ steps discarded as ``burn-in'',
and finally thinned out by 0.5$\tau$ steps to leave independent samples.
The total number of sampling of the posterior probability distribution
is thus $100\times(100-5)/0.5=19,\!000$.

\begin{deluxetable}{ccc}[!hbt]
\tabletypesize{\scriptsize}
\tablecaption{Bounds of the flat priors\label{tab:bounds_2}}
\tablewidth{0pt}
\tablehead{\colhead{parameter} & \colhead{lower} & \colhead{upper}}
\startdata
log(n$_{\rm H_2}$/cm$^{-3}$)    &  2  &  6 \\
log(T$_{\rm kin}$/K)           &  1  &  3 \\
log(N$_{\rm HCO^+}$/cm$^{-2}$)  & 14  & 17  \\
$\mathrm{[HCN]/[HCO^+]}$ &  0.1  & 10    \\
J21 scaling & 0.95 & 1.05 \\
J32 scaling & 0.9  & 1.1  \\
J43 scaling & 0.9  & 1.1  \\ \hline
\enddata
\end{deluxetable}

\begin{figure*}
\begin{center}
\includegraphics[angle=0,scale=.86]{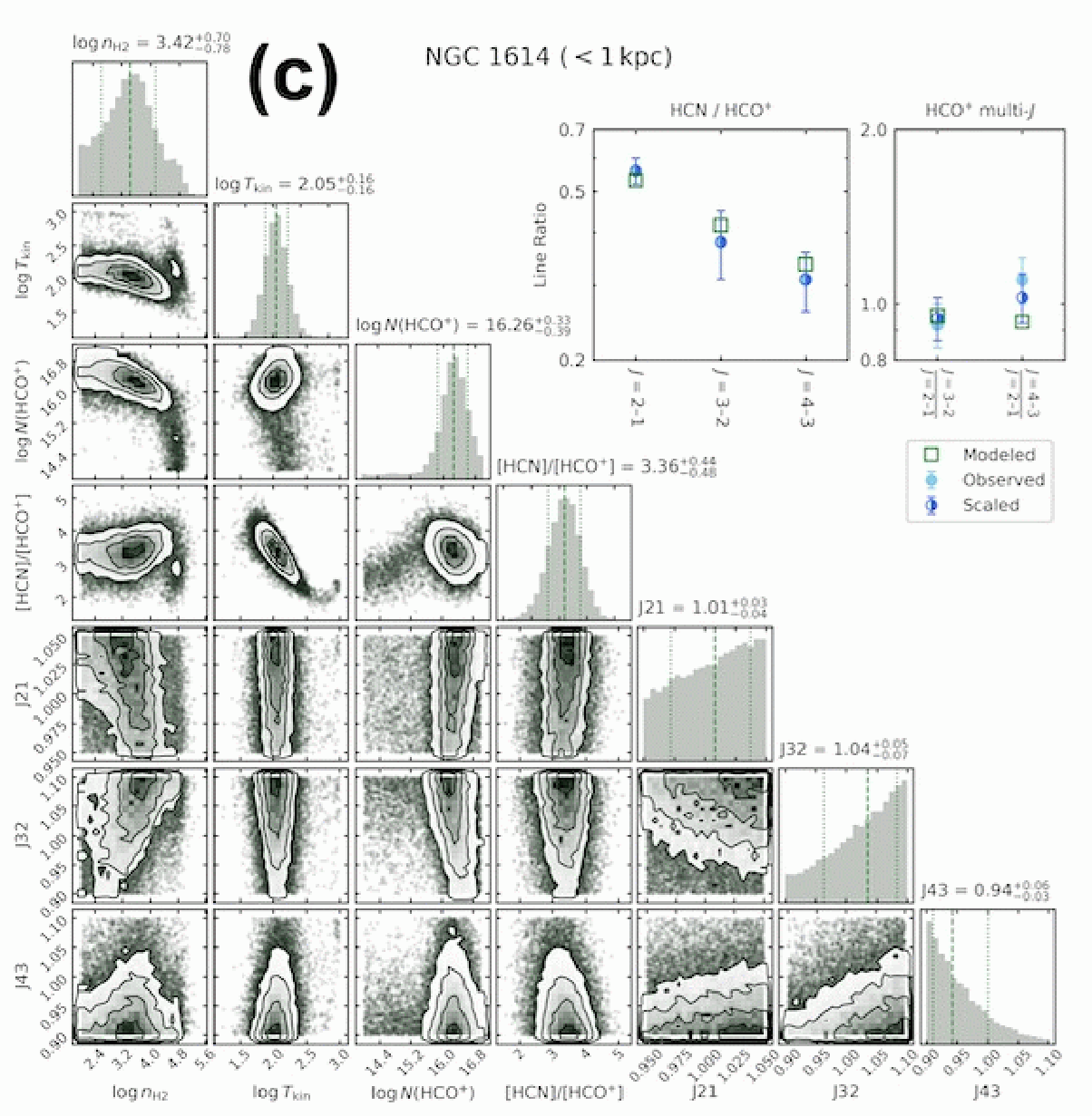} 
\end{center}
\caption{
Example result of the MCMC parameter estimation
using the HCN-to-HCO$^+$ and HCO$^+$ multiple J-transition line flux
ratios. 
{\it (a)}: Central $\lesssim$0.5 kpc region of IRAS 16090$-$0139. 
{\it (b)}: Central $\lesssim$0.5 kpc region of IRAS 22491$-$1808. 
{\it (c)}: Central $\lesssim$1 kpc region of NGC 1614. 
The corner plot on the left shows 1D and 2D projections of the
posterior distribution for any parameters and parameter pairs,
respectively. 
The median and 68\% credible bounds of each parameter are given in
the label above each column 
and are denoted by the dashed and dotted lines, respectively, in the
1D posterior. 
The panel on the upper right shows a comparison of the observed and
modeled emission line flux ratios. 
The model flux ratios (green open squares) are calculated at the medians of 
log n$_{\rm H_2}$, log T$_{\rm kin}$, log N$_{\rm HCO^+}$, and [HCN]/[HCO$^+$]. 
The observed flux ratios (light blue filled circles) are quoted from
Tables \ref{tab:fluxratio} and \ref{tab:Jratio} and then scaled 
with the medians of the J21, J32, and J43 scaling factors (dark blue
half-filled circles).
\label{fig:MCMC}
}
\end{figure*}

We apply this MCMC analysis to the data of selected regions of (U)LIRGs.
IRAS 10378$+$1108, IRAS 12112$+$0305, and IRAS 00091$-$0738 are
excluded for the same reasons as before ($\S$5.3).
Figure \ref{fig:MCMC} shows example results for IRAS 16090$-$0139
($\lesssim$0.5 kpc region), IRAS 22491$-$1808 ($\lesssim$0.5 kpc
region), and NGC 1614 ($\lesssim$1 kpc region).  
As previously derived in $\S$5.3, the presence of dense
($\gtrsim$10$^{5}$ cm$^{-3}$) and warm 
($\gtrsim$10$^{2.5}$ K or $\gtrsim$300 K) molecular gas at the
nuclear $\lesssim$0.5 kpc regions of the two ULIRGs is confirmed with
this new MCMC analysis as well.
For the starburst-dominated LIRG NGC 1614 ($\lesssim$1 kpc physical
scale), this new MCMC analysis derives even less dense ($\lesssim$10$^{4}$
cm$^{-3}$) and similar temperature ($\sim$10$^{2}$ K) molecular
gas, when compared to the previous estimate using the
Levenberg-Marquardt method, with the fixed fiducial HCO$^{+}$ column
density and HCN-to-HCO$^{+}$ abundance ratio (Table
\ref{tab:bestfit2}). 

\begin{figure*}[!hbt]
\begin{center}
\includegraphics[angle=0,scale=.86]{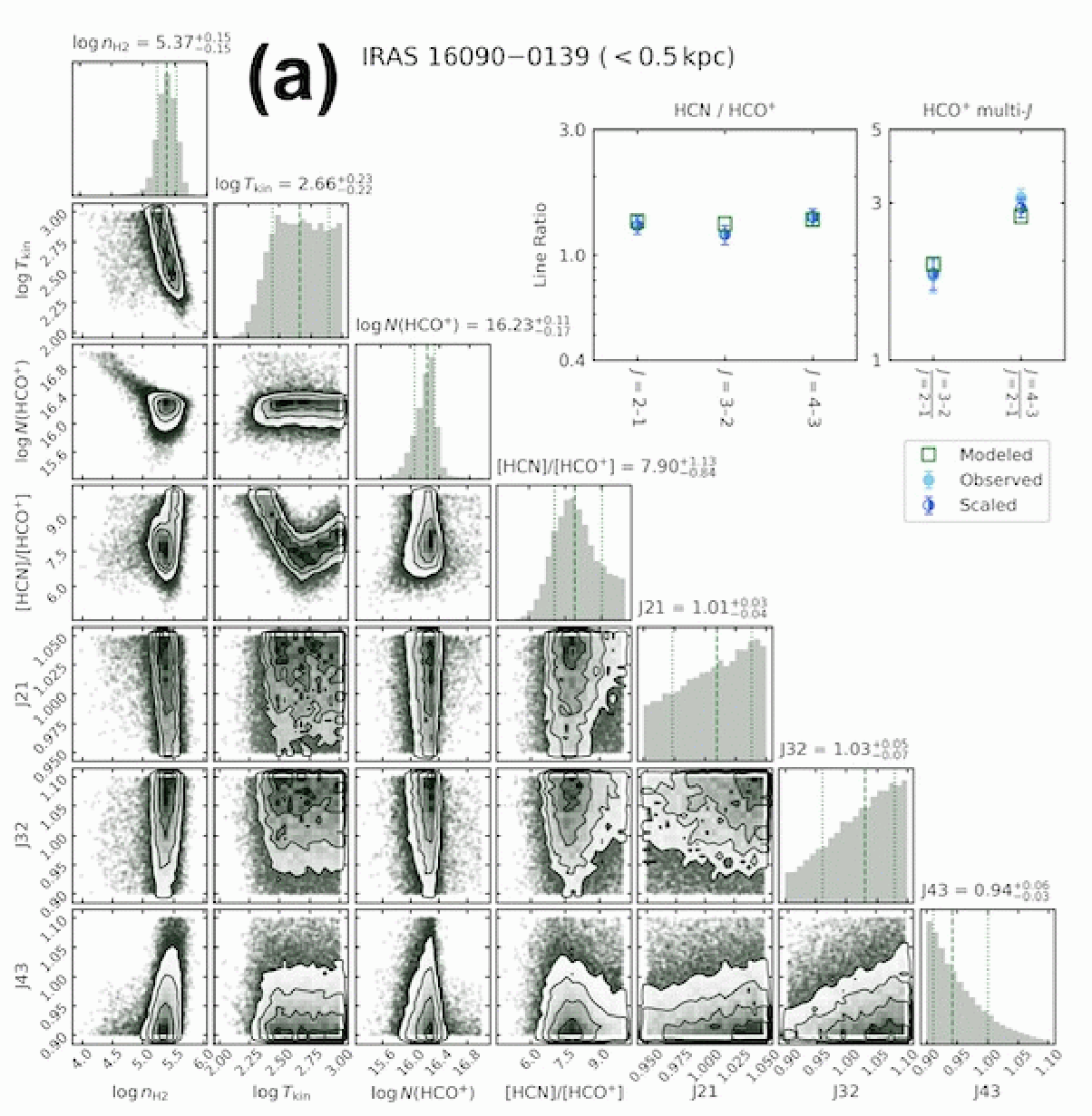} \\
\end{center}
\end{figure*}


\begin{figure*}
\begin{center}
\includegraphics[angle=0,scale=.86]{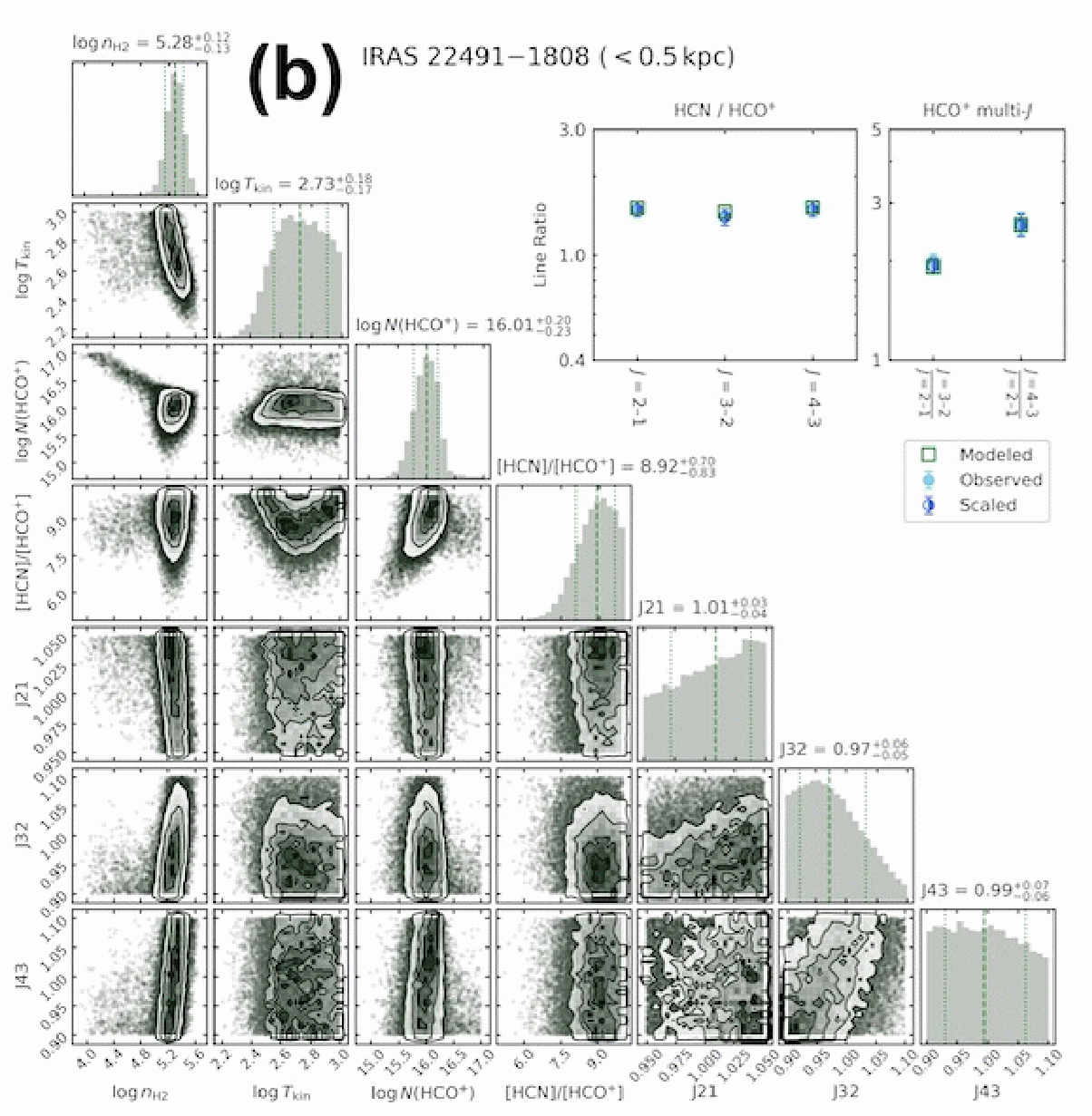} 
\end{center}
\end{figure*}


\begin{figure*}[!hbt]
\begin{center}
\includegraphics[angle=0,scale=.23]{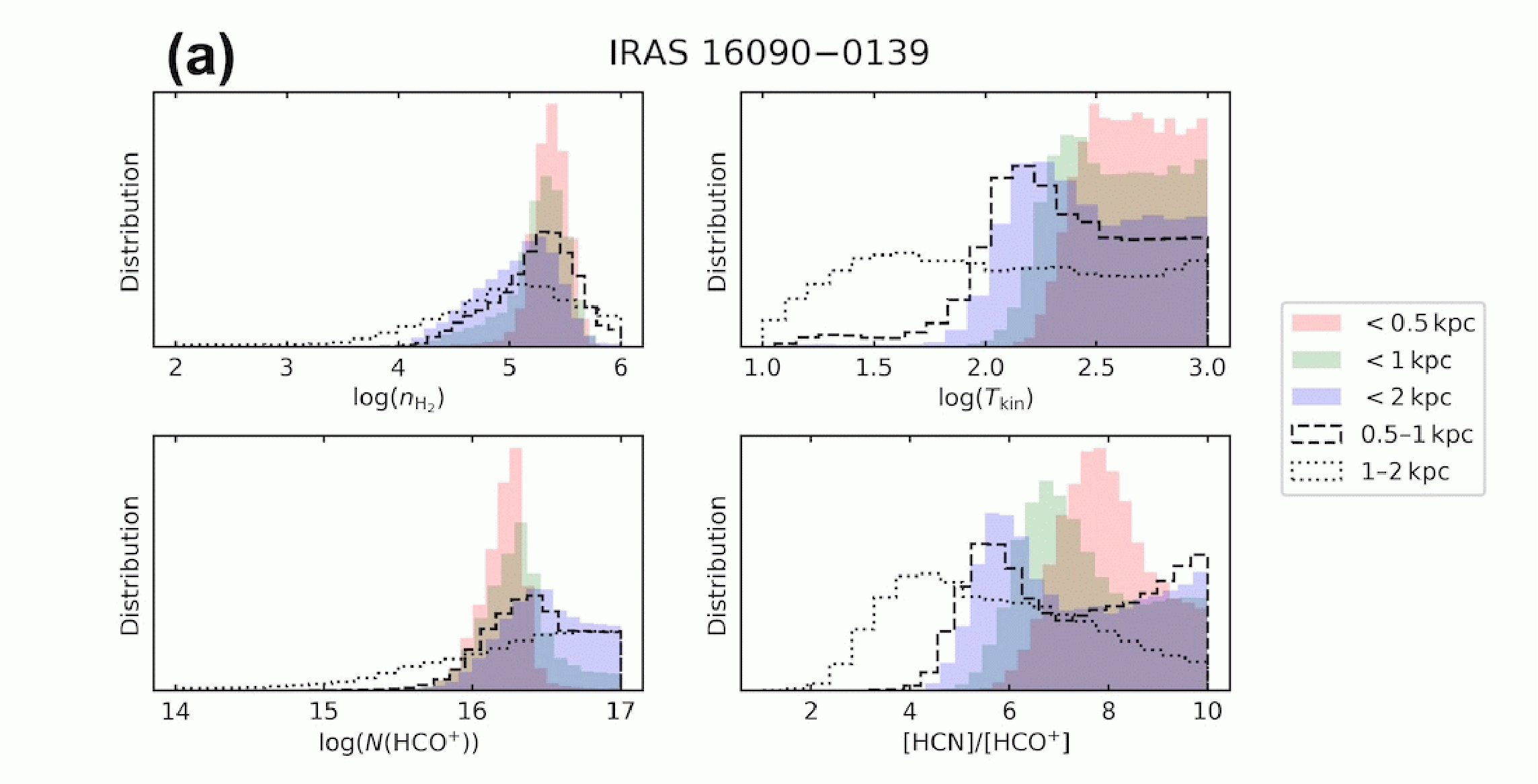} 
\includegraphics[angle=0,scale=.23]{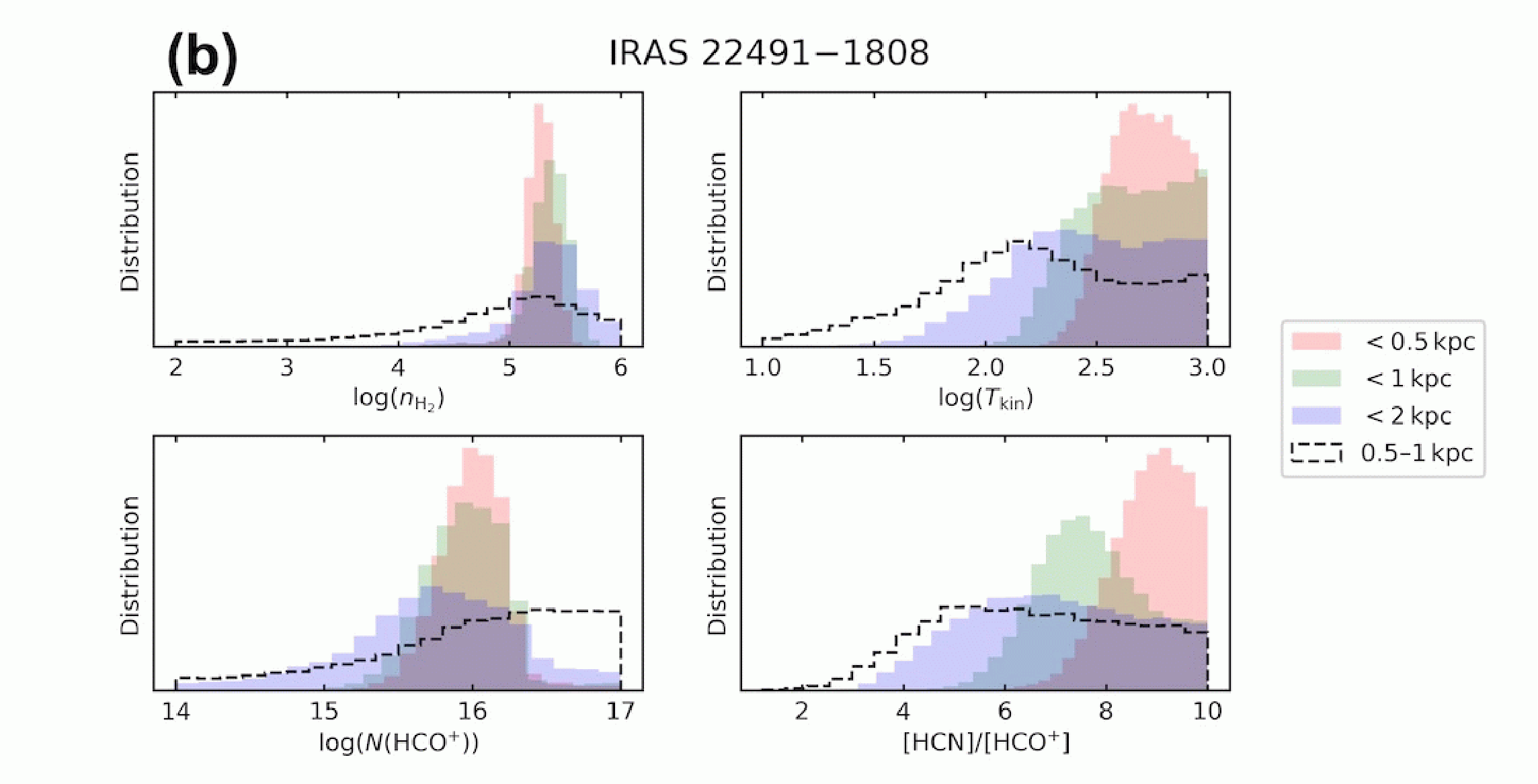} \\
\hspace*{-9cm}
\includegraphics[angle=0,scale=.23]{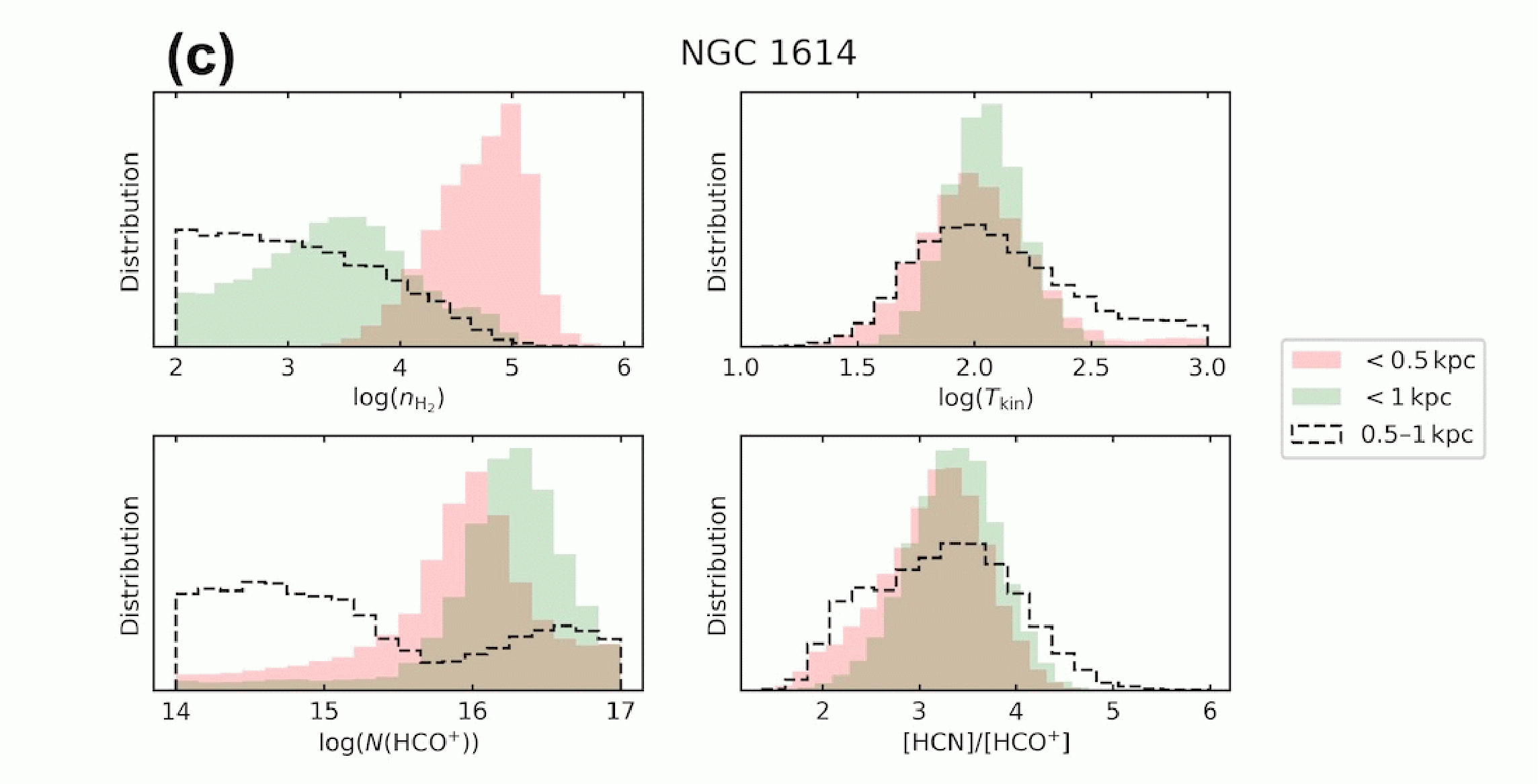} 
\end{center}
\caption{
Comparison of the posteriors of gas parameters obtained for 
different regions in {\it (a)}: IRAS 16090$-$0139, 
{\it (b)}: IRAS 22491$-$1808, and {\it (c)}: NGC 1614.
{\it (Upper left)}: H$_{2}$ gas density in cm$^{-3}$.
{\it (Upper right)}: H$_{2}$ gas kinetic temperature in K. 
{\it (Lower left)}: HCO$^{+}$ column density in cm$^{-2}$.
{\it (Lower right)}: HCN-to-HCO$^{+}$ abundance ratio.
The red, green, blue filled histograms, and dashed and dotted line
histograms correspond to the results for the central $\lesssim$0.5
kpc, $\lesssim$1 kpc, $\lesssim$2 kpc, 0.5--1 kpc annular, and 1--2
kpc annular regions, respectively.
Note that the horizontal axis range of the HCN-to-HCO$^{+}$ abundance
ratio {\it (Lower right)} is much narrower for NGC 1614 than for the
other two ULIRGs.
\label{fig:MCMCcompa}
}
\end{figure*}

We then compare the posteriors of the gas parameters obtained in
different regions of the same (U)LIRG, to illustrate how molecular gas
physical parameters spatially change. 
Figure \ref{fig:MCMCcompa} displays the comparison for three (U)LIRGs
as the representatives to discuss the possible spatial variation of
some physical parameters.
In Figure \ref{fig:MCMCcompa}a and \ref{fig:MCMCcompa}b, 
for the two ULIRGs IRAS 16090$-$0139 and IRAS 22491$-$1808, 
while molecular gas density is estimated to be very high
($\gtrsim$10$^{5}$ cm$^{-3}$) at all the $\lesssim$0.5 kpc,
$\lesssim$1 kpc, $\lesssim$2 kpc, 0.5--1 kpc, and 1--2 kpc regions,  
a trend of systematically higher temperature and HCN-to-HCO$^{+}$
abundance ratio at the innermost ($\lesssim$0.5 kpc) regions than the outer
nuclear regions (0.5--2 kpc), is seen. 
Both sources are diagnosed to contain luminous buried AGNs 
(Table \ref{tab:object} and Figure \ref{fig:HCNHCOCS}).
It is possible that the luminous AGNs create high gas temperature at
the innermost part.
It is also reported that an HCN-to-HCO$^{+}$ abundance ratio can be
enhanced in dense molecular gas in the vicinity of, and affected by, a
luminous AGN    
\citep[e.g.,][]{ala15,sai18,tak19,nak18,kam20,ima20,but22,nak23}.
The trend seen in these two ULIRGs can be caused by a luminous AGN.

In Figure \ref{fig:MCMCcompa}c, the starburst-dominated LIRG NGC 1614
shows 
(1) much smaller gas temperature and HCN-to-HCO$^{+}$ abundance ratio
than the other two AGN-hosting ULIRGs with the same physical
apertures, and 
(2) no discernible spatial variation of the gas temperature and
HCN-to-HCO$^{+}$ abundance ratio at 0.5 kpc physical scales
within the central $\lesssim$1 kpc region. 
These results can naturally be explained by our view that NGC 1614 is
energetically dominated by $\sim$1 kpc wide starburst 
activity, without significant contribution from a central compact
luminous AGN (Table \ref{tab:object}). 

\begin{figure*}[!hbt]
\begin{center}
\includegraphics[angle=0,scale=.23]{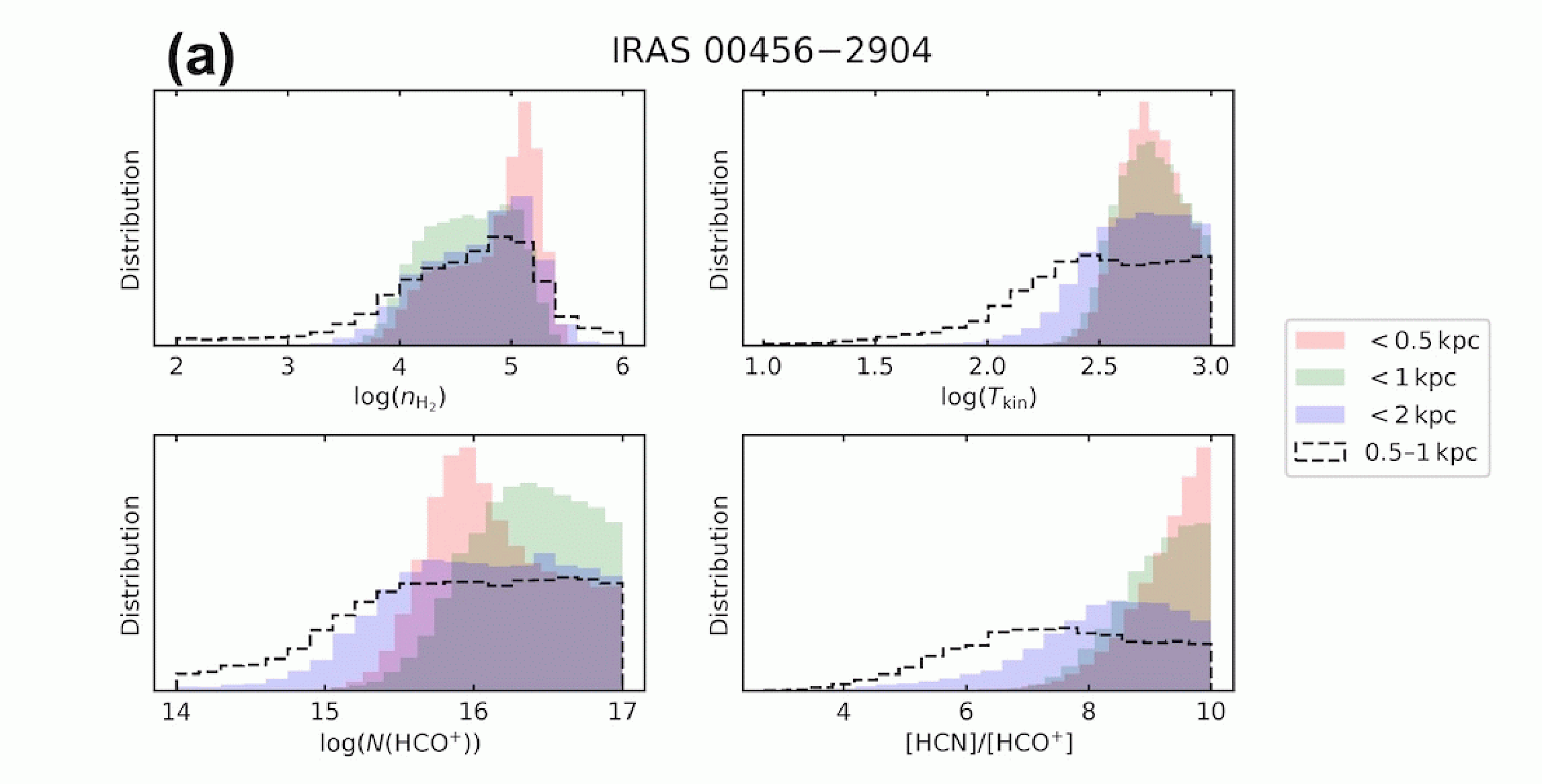} 
\includegraphics[angle=0,scale=.23]{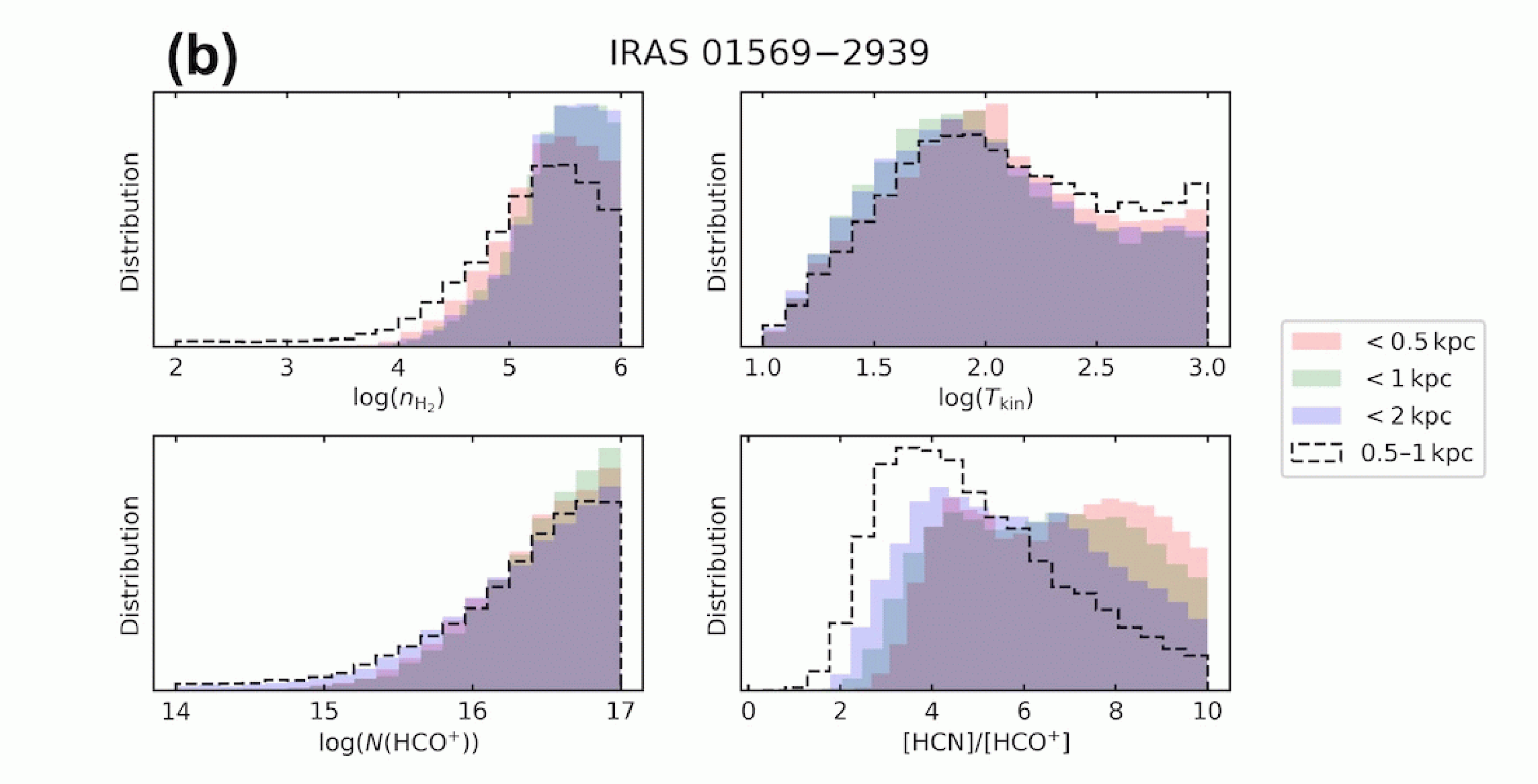} \\
\includegraphics[angle=0,scale=.23]{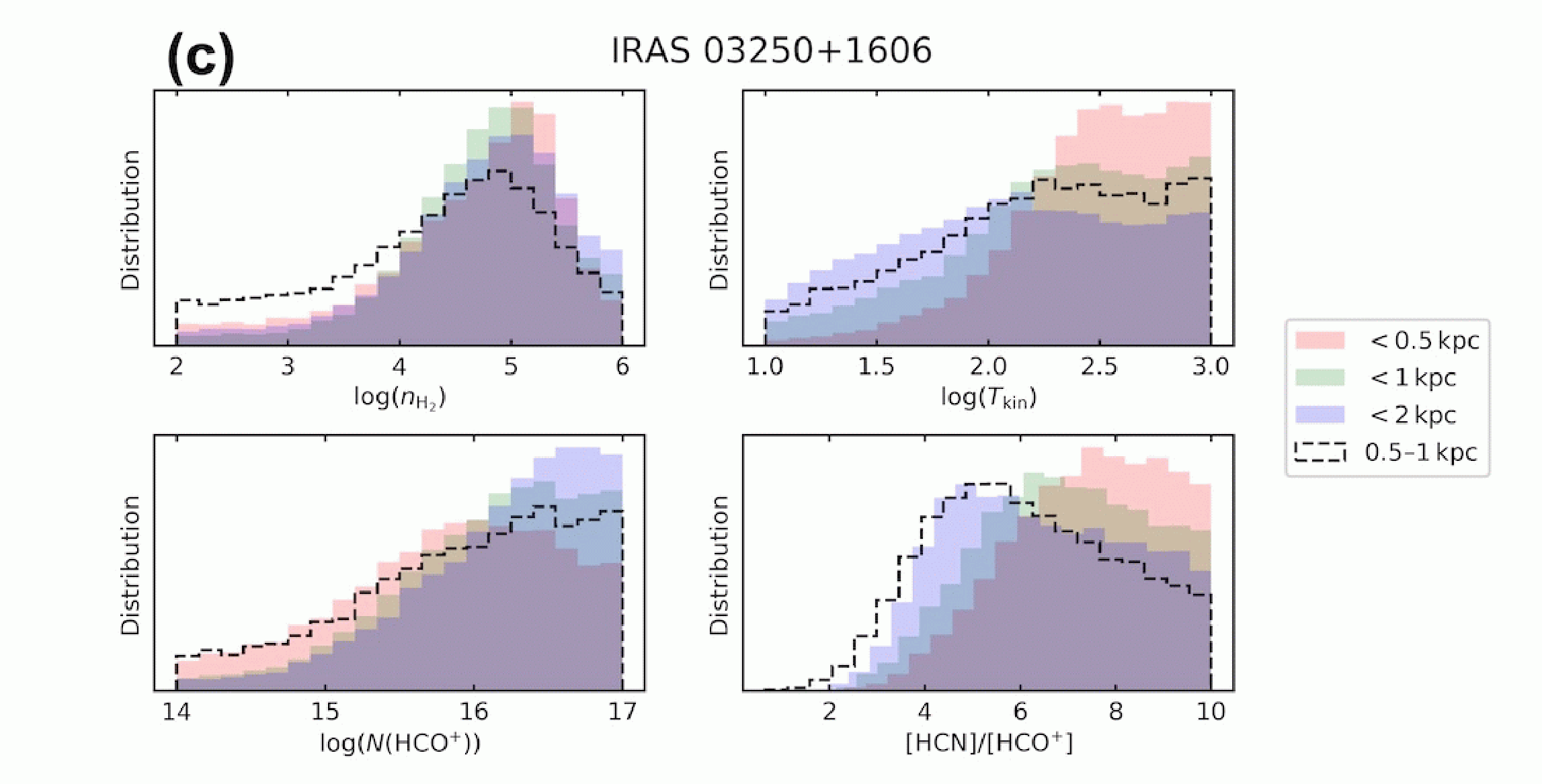} 
\includegraphics[angle=0,scale=.23]{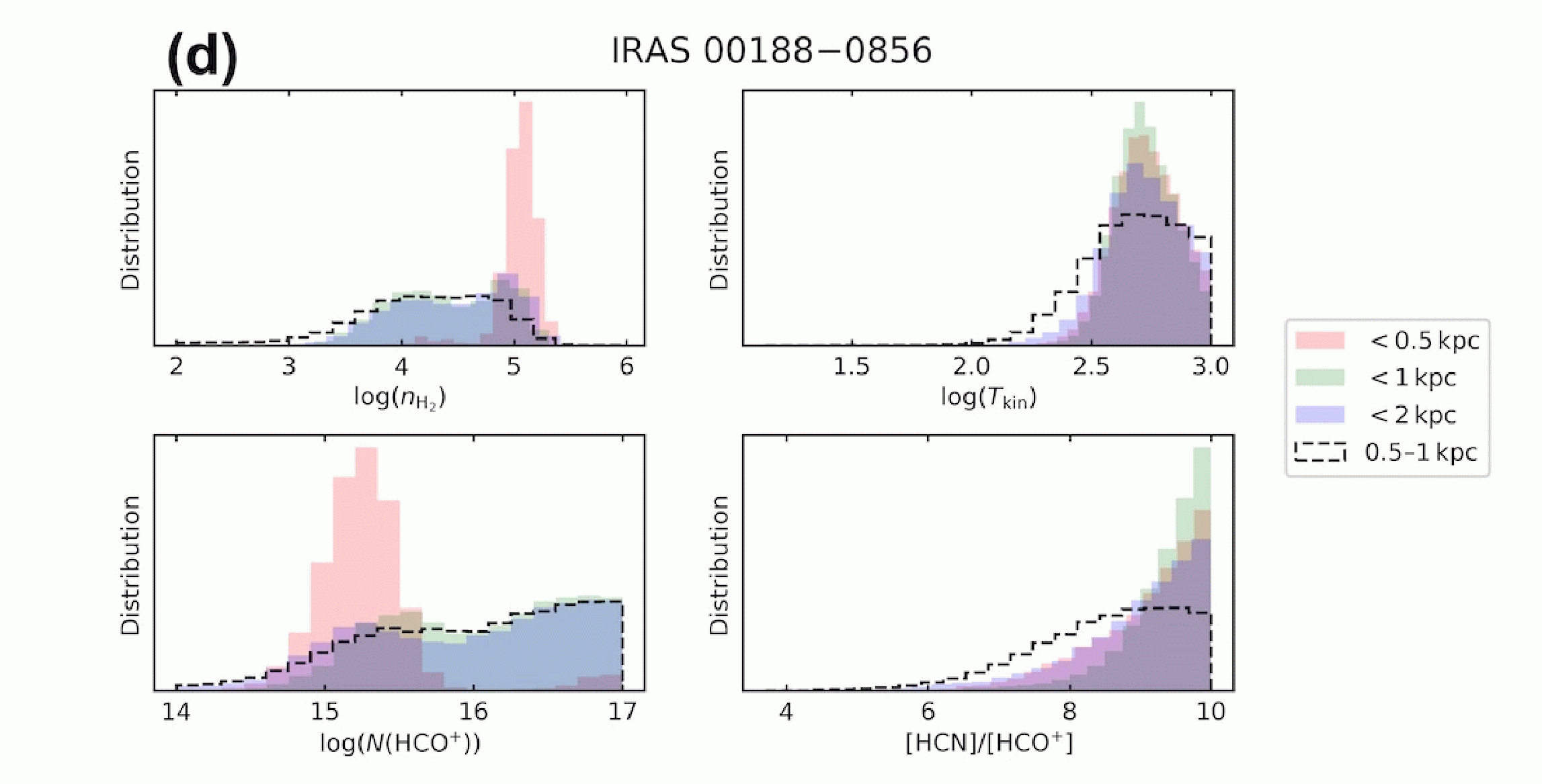} \\
\includegraphics[angle=0,scale=.23]{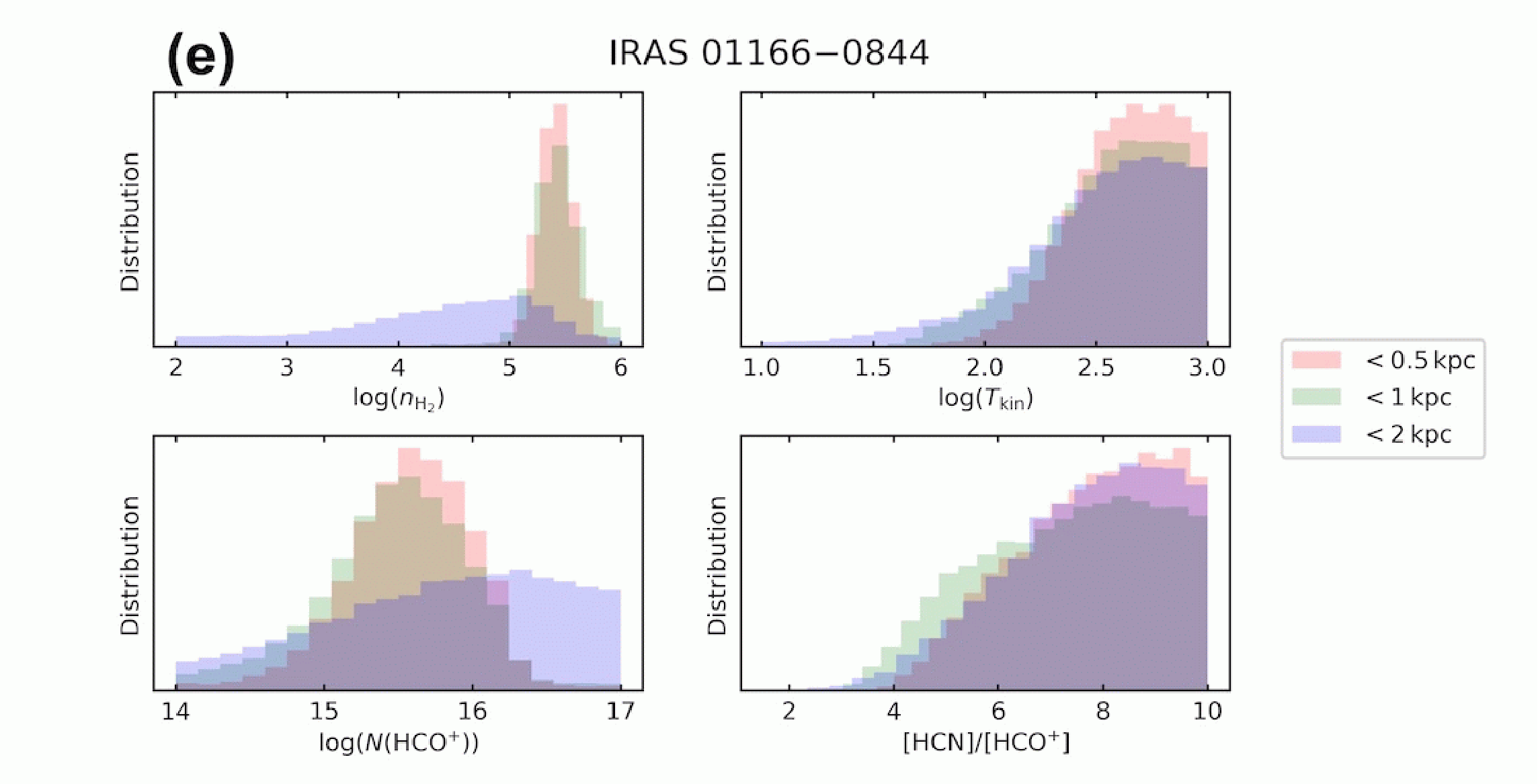} 
\includegraphics[angle=0,scale=.23]{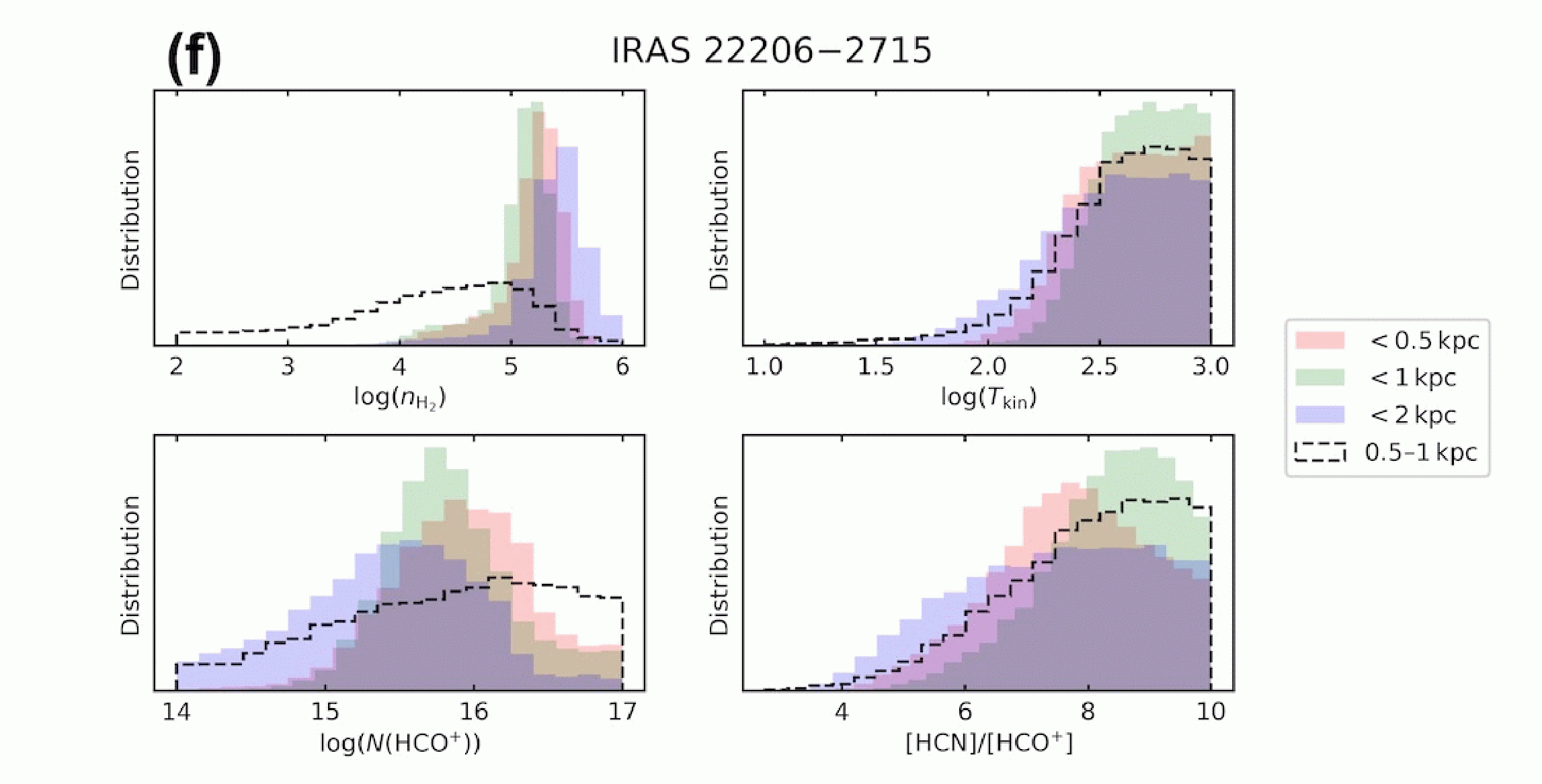} \\
\end{center}
\caption{
Same as Figure \ref{fig:MCMCcompa}, but for 
{\it (a)}: IRAS 00456$-$2904, {\it (b)}: IRAS 01569$-$2939,
{\it (c)}: IRAS 03250$+$1606, {\it (d)}: IRAS 00188$-$0856, 
{\it (e)}: IRAS 01166$-$0844,  and {\it (f)}: IRAS 22206$-$2715.
A subtle sign of increasing HCN-to-HCO$^{+}$ abundance ratio and/or
increasing temperature toward the innermost $\lesssim$0.5 kpc region is seen 
in {\it (a)--(c)}, but not in {\it (d)--(f)}. 
\label{fig:MCMCcompb}
}
\end{figure*}

Figure \ref{fig:MCMCcompb} shows the MCMC results for the remaining ULIRGs.
The presence of very dense ($\gtrsim$10$^{5}$ cm$^{-3}$) and warm
($\gtrsim$300 K) molecular gas is confirmed with this MCMC method in 
all the regions of all these ULIRGs.
For IRAS 00456$-$2904, IRAS 01569$-$2939, and IRAS 03250$+$1606
(Figure \ref{fig:MCMCcompb}a--c), there might be a very subtle sign of
higher HCN-to-HCO$^{+}$ abundance ratio and/or higher gas temperature
at the innermost ($\lesssim$0.5 kpc) region than outer nuclear (0.5--2
kpc) region, but the trend is much weaker than the previously
discussed IRAS 16090$-$0139 and IRAS 22491$-$1808 (Figures
\ref{fig:MCMCcompa}a and \ref{fig:MCMCcompa}b).    
For the remaining ULIRGs, IRAS 00188$-$0856, IRAS 01166$-$0844, and
IRAS 22206$-$2715, we see no such trend at all.
The absence of such trend can be real, but we note that for ULIRGs
for which we have to compare physical parameters among overlapped regions
($\lesssim$0.5 kpc, $\lesssim$1 kpc, and $\lesssim$2 kpc), rather than
non-overlapped annular regions ($\lesssim$0.5 kpc, 0.5--1 kpc, and
1--2 kpc), possible spatial variation of gas physical parameters can
be diluted if emission is dominated by the innermost $\lesssim$0.5 kpc 
region ($\S$5.3).   

\section{Summary} 

We presented the results of our ALMA $\lesssim$0.5 kpc-resolution, 
three rotational transition line (J=2--1, J=3--2, and J=4--3)
observations of HCN and HCO$^{+}$ for 11 ULIRGs with luminous buried
AGN signatures, and one starburst-dominated LIRG NGC 1614. 
We extracted spectra at the central 0.5 kpc, 1 kpc, and 2 kpc 
regions, as well as 0.5--1 kpc and 1--2 kpc annular regions, to 
(1) derive (i) the HCN-to-HCO$^{+}$ flux ratios at J=2--1, J=3--2, and
J=4--3, and (ii) high-J to low-J (J=4--3 to J=2--1 and J=3--2 to J=2--1)
flux ratios of HCN and HCO$^{+}$, in individual regions, and 
(2) investigate the possible spatial variations of these ratios among
the different regions.  
We ran RADEX non-LTE model calculations to constrain molecular gas
properties for nine (U)LIRGs after excluding three ULIRGs for which 
(a) not all the J=2--1, J=3--2, and J=4--3 data are available (two
sources), and 
(b) one zone model cannot be applied (one source).
We (1) used the Levenberg-Marquardt method by fixing the HCO$^{+}$
column density and HCN-to-HCO$^{+}$ abundance ratio at fiducial values
and (2) applied a Bayesian approach by making all parameters free.  
We found the following main results.

\begin{enumerate}

\item
HCN and HCO$^{+}$ emission at J=2--1, J=3--2, and J=4--3 were clearly
detected in the 0.5 kpc, 1 kpc, and 2 kpc beam-sized spectra of the
majority of the observed (U)LIRGs, suggesting the abundant presence of
dense and warm molecular gas at the nuclear regions.

\item 
We quantitatively found that molecular gas at ULIRGs' innermost 
($\lesssim$0.5 kpc) and whole nuclear ($\lesssim$1--2 kpc) regions
is very dense ($\gtrsim$10$^{5}$ cm$^{-3}$) and warm ($\gtrsim$300 K),
and that it is also modestly dense (10$^{4-4.5}$ cm$^{-3}$) and warm
($\sim$100 K) in one starburst-dominated LIRG's nucleus ($\lesssim$1
kpc).    

\item 
We saw a signature that the HCN-to-HCO$^{+}$ flux ratios at J=2--1,
J=3--2, and J=4--3, and high-J to low-J flux ratios of HCN and
HCO$^{+}$, decrease from the innermost ($\lesssim$0.5 kpc) to outer
nuclear (0.5--2 kpc) region for some fraction of the observed ULIRGs.

\item 
For the above ULIRGs showing the signature, we conducted RADEX non-LTE 
model calculations by freeing all parameters, based on a Bayesian approach,
and detected an increasing trend of the HCN-to-HCO$^{+}$ abundance  
ratio and gas kinetic temperature from the outer nuclear (0.5--2 kpc)
to the innermost ($\lesssim$0.5 kpc) regions in two ULIRGs with luminous AGN 
signatures (IRAS 16090$-$0139 and IRAS 22491$-$1808) significantly 
and in additional three ULIRGs (IRAS 00456$-$2904, IRAS 01569$-$2939,
and IRAS 03250$+$1606) marginally. 
We interpreted that the trend could naturally be explained by luminous
AGN effects to the innermost molecular gas.

\item Our Bayesian approach also demonstrated that the LIRG NGC 1614
displayed (a) much lower gas temperature and HCN-to-HCO$^{+}$
abundance ratio than the observed other nearby ULIRGs' nuclei,
and 
(b) no discernible spatial change in these two parameters 
at 0.5 kpc physical scales within the central $\lesssim$1 kpc 
region.  
This can naturally be explained by a scenario that NGC 1614 is
energetically dominated by $\sim$1 kpc wide starburst activity.

\end{enumerate}

We demonstrated that ALMA multiple molecular, multiple rotational
transition line observations, with a combination of non-LTE modeling,
are a very unique tool to constrain the spatial variations of physical
and chemical properties of molecular gas within nearby (U)LIRGs'
nuclei ($\lesssim$2 kpc), thanks to achievable high spatial resolution
($\lesssim$0.5 kpc) and high sensitivity.  

\begin{acknowledgments}
We thank the anonymous referee for his/her valuable
comments, which helped improve the clarity of this manuscript.
This paper made use of the following ALMA data:
ADS/JAO.ALMA\#2015.1.00027.S and \#2019.0.00027.S.
ALMA is a partnership of ESO (representing its member states), NSF (USA) 
and NINS (Japan), together with NRC (Canada), NSC and ASIAA
(Taiwan), and KASI (Republic of Korea), in cooperation with the Republic
of Chile. The Joint ALMA Observatory is operated by ESO, AUI/NRAO, and
NAOJ. 
M.I., K.N., and T.I. are supported by JP21K03632, JP19K03937, and
JP20K14531, respectively.  
S.B. is supported by JP19J00892 and JP21H04496.
Data analysis was in part carried out on the open use data analysis
computer system at the Astronomy Data Center, ADC, of the National
Astronomical Observatory of Japan. 
This research has made use of NASA's Astrophysics Data System and the
NASA/IPAC Extragalactic Database (NED) which is operated by the Jet
Propulsion Laboratory, California Institute of Technology, under
contract with the National Aeronautics and Space Administration. 
\end{acknowledgments}

%

\vspace{5mm}
\facilities{ALMA}
\software{CASA \citep{CASA22},
RADEX \citep{RADEX},
pyradex (\url{https://github.com/keflavich/pyradex}),
IPython \citep{IPython},
Jupyter Notebook \citep{JupyterNotebook},
NumPy \citep{NumPy},
SciPy \citep{SciPy},
Pandas \citep{Pandas},
Matplotlib \citep{Matplotlib},
Astropy \citep{Astropy_v5},
lmfit \citep{lmfit},
emcee \citep{emcee},
corner \citep{corner}}






\clearpage

\appendix

\section{Continuum Spectral Energy Distribution}

For four ULIRGs, photometric data at 250 $\mu$m, 350 $\mu$m, and 
500 $\mu$m, taken with the Herschel Space Observatory, are available
\citep{cle18}.  
Figure \ref{fig:contSED} overplots our ALMA 1 kpc beam-sized
continuum flux measurements on the Herschel data as well IRAS 60
$\mu$m and 100 $\mu$m data.
Both Herschel and IRAS data were obtained with much larger
aperture sizes ($\gtrsim$5$''$ or $\gtrsim$6 kpc at $z \gtrsim$ 0.07)
than our ALMA measurements. 
Infrared 60--500 $\mu$m emission is usually dominated by dust thermal
radiation.  
Our ALMA photometric measurements roughly agree with extrapolation
from shorter wavelength IRAS and Herschel data, suggesting that our
ALMA 1 kpc beam-sized data cover the bulk of dust thermal radiation
from these ULIRGs, as expected from compact ($\lesssim$1 kpc) nature
of energetically dominant regions in nearby ULIRGs in general
\citep[e.g.,][]{soi00,dia10,ima11,per21}.   

\begin{figure}[!hbt]
\begin{center}
\includegraphics[angle=-90,scale=.19]{fA1a.eps} 
\includegraphics[angle=-90,scale=.19]{fA1b.eps} 
\includegraphics[angle=-90,scale=.19]{fA1c.eps} 
\includegraphics[angle=-90,scale=.19]{fA1d.eps} 
\caption{
Continuum spectral energy distribution from 60 $\mu$m to 
$\sim$2000 $\mu$m for four ULIRGs.
Herschel 250 $\mu$m, 350 $\mu$m, and  500 $\mu$m \citep{cle18}, and  
IRAS 60 $\mu$m and 100 $\mu$m photometric measurements 
(Table \ref{tab:object}) are displayed as black filled circles. 
Our ALMA 1 kpc beam-sized continuum flux measurements at 850--2000 $\mu$m 
(Table \ref{tab:cont}, column 7) are plotted as open stars.
The best-fit graybody curve derived by \citet{cle18}, using only
infrared data at $\lesssim$500 $\mu$m, is overplotted as thick curved
line after flux normalization at 250 $\mu$m. 
Object name and redshift are shown in each panel.
\label{fig:contSED}
}
\end{center}
\end{figure}


\section{Intensity-weighted Mean Velocity (Moment 1) Map}

\begin{figure}
\begin{center}
\includegraphics[angle=0,scale=.4]{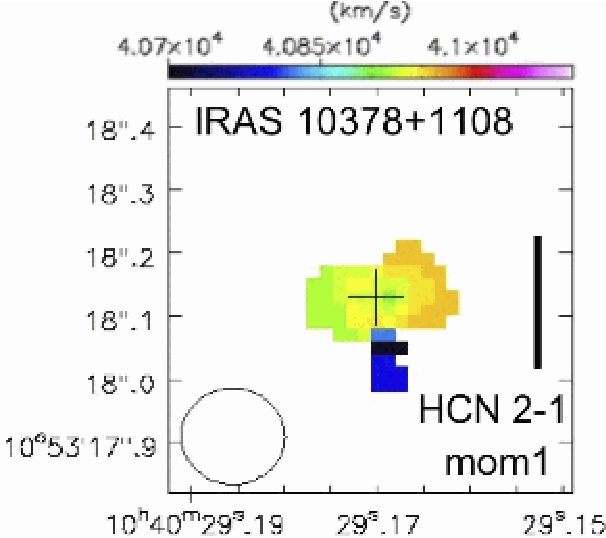} 
\includegraphics[angle=0,scale=.4]{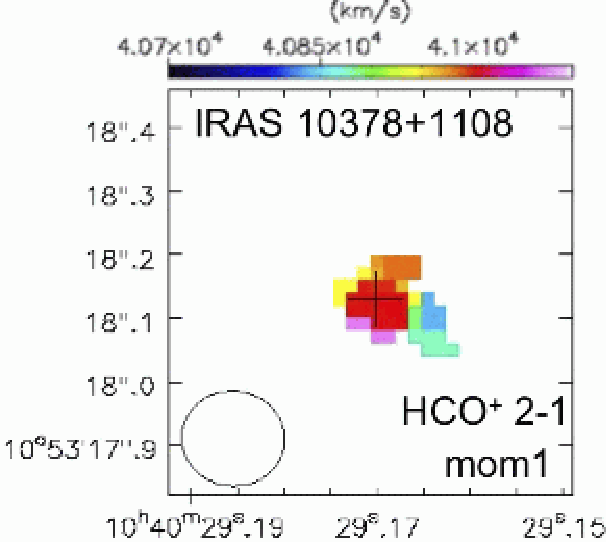} 
\includegraphics[angle=0,scale=.4]{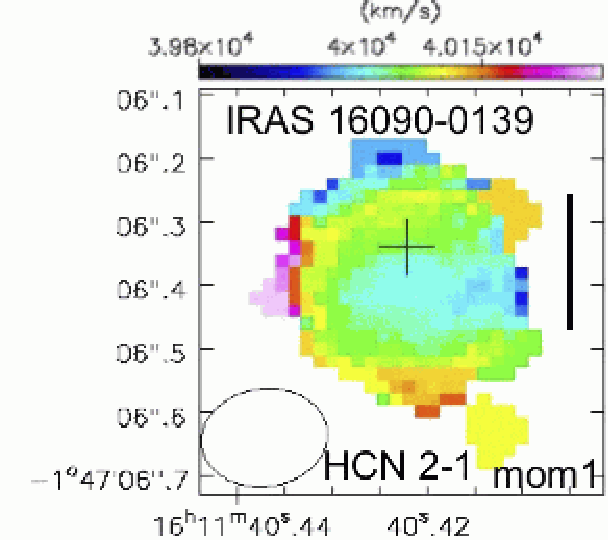} 
\includegraphics[angle=0,scale=.4]{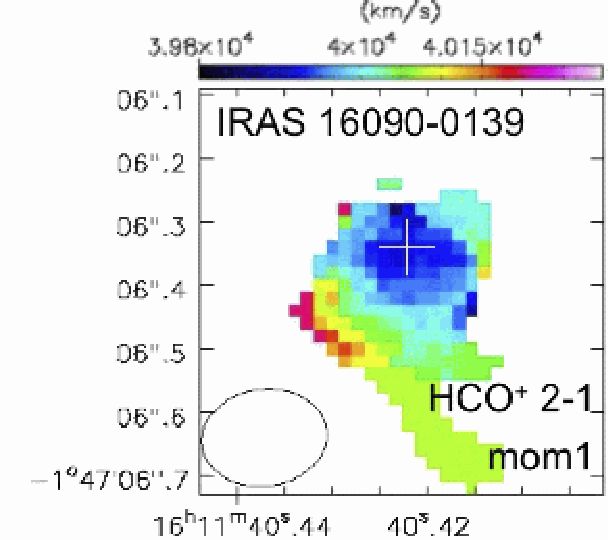} \\
\includegraphics[angle=0,scale=.4]{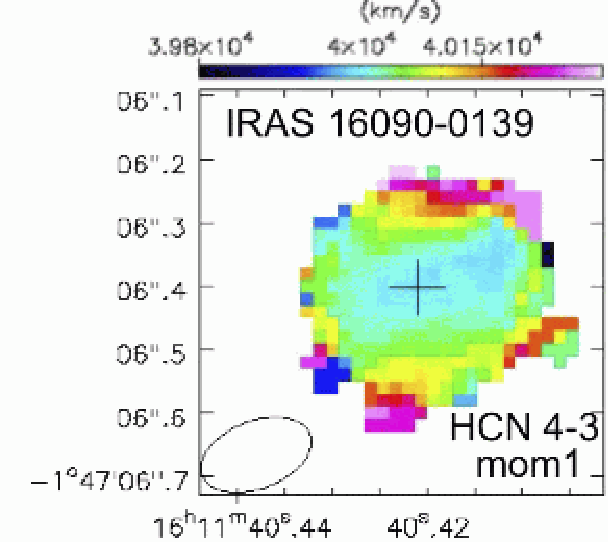} 
\includegraphics[angle=0,scale=.4]{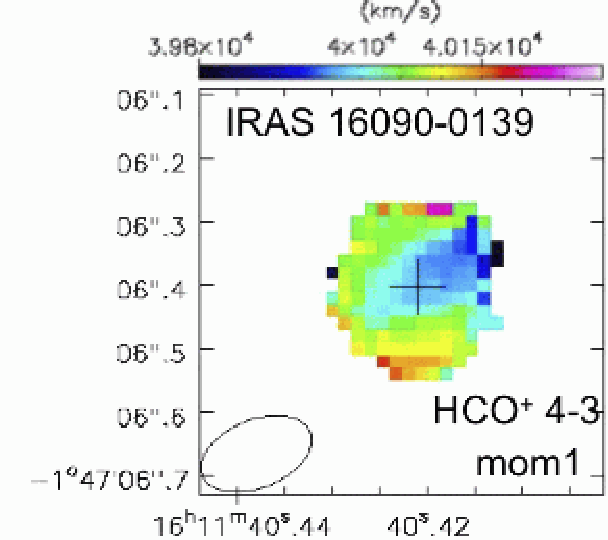}  
\includegraphics[angle=0,scale=.4]{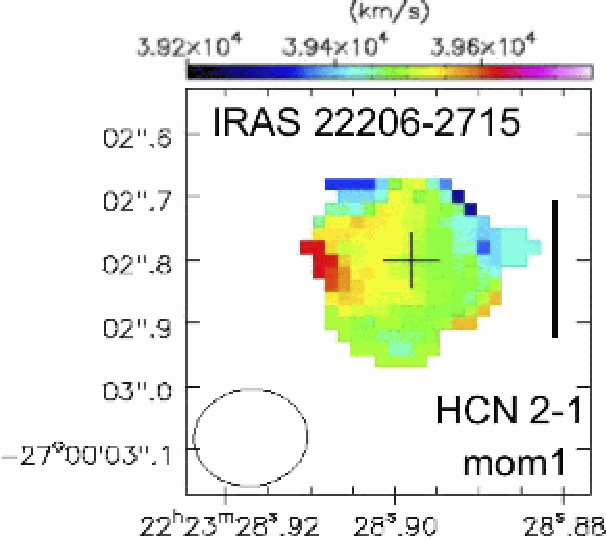} 
\includegraphics[angle=0,scale=.4]{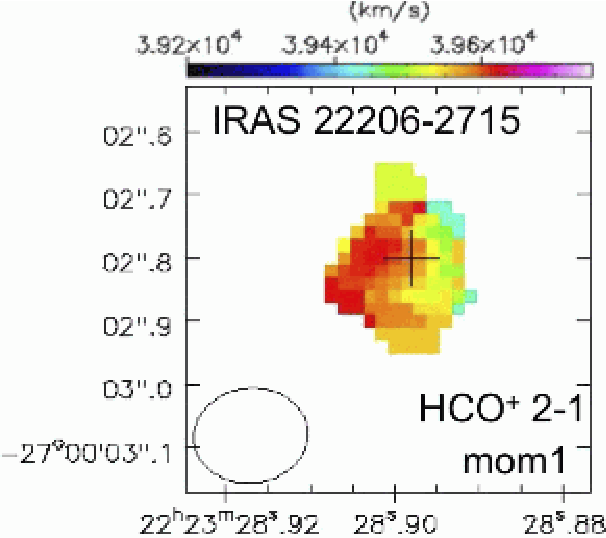} \\
\includegraphics[angle=0,scale=.4]{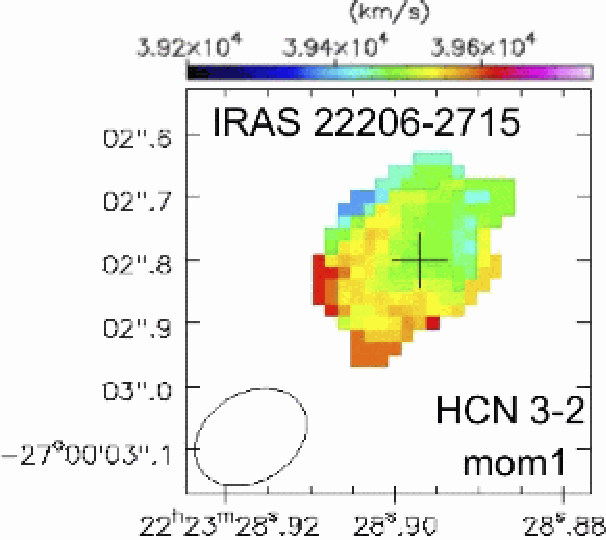} 
\includegraphics[angle=0,scale=.4]{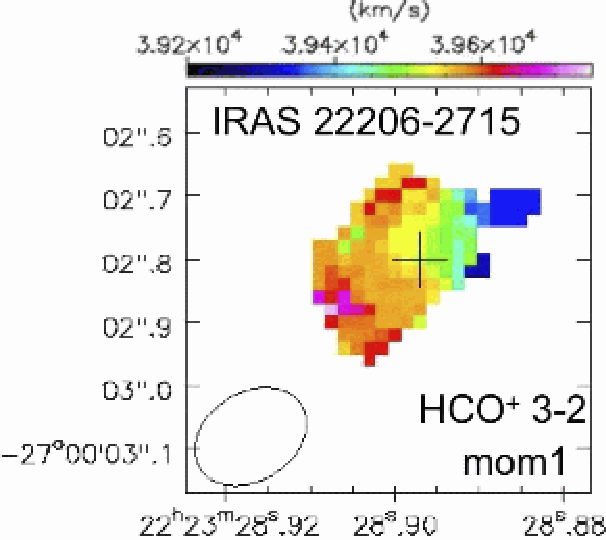} 
\includegraphics[angle=0,scale=.4]{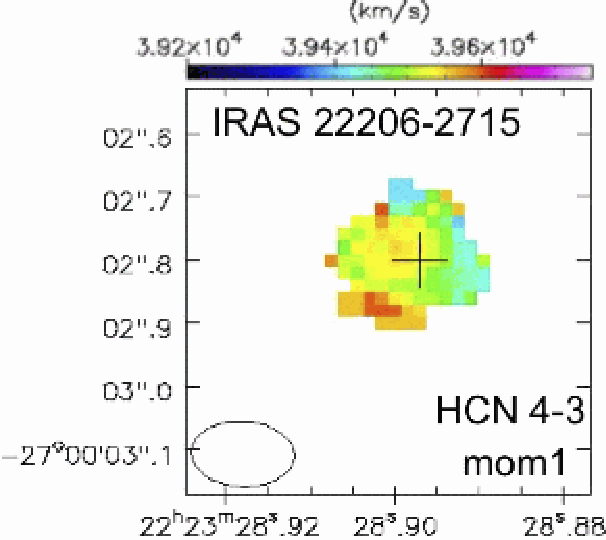} 
\includegraphics[angle=0,scale=.4]{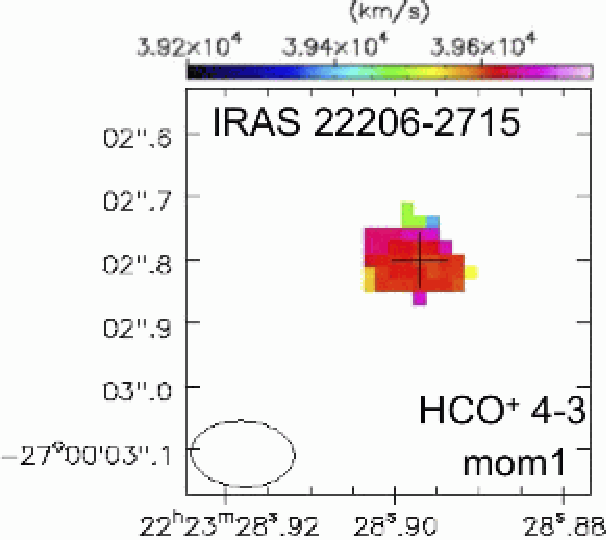} \\
\includegraphics[angle=0,scale=.4]{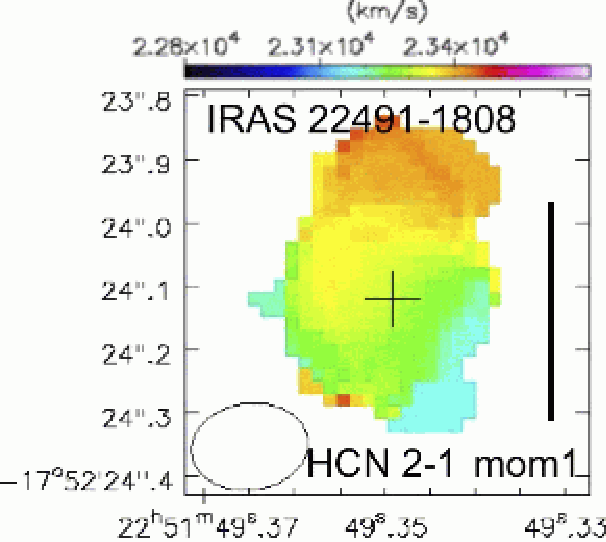} 
\includegraphics[angle=0,scale=.4]{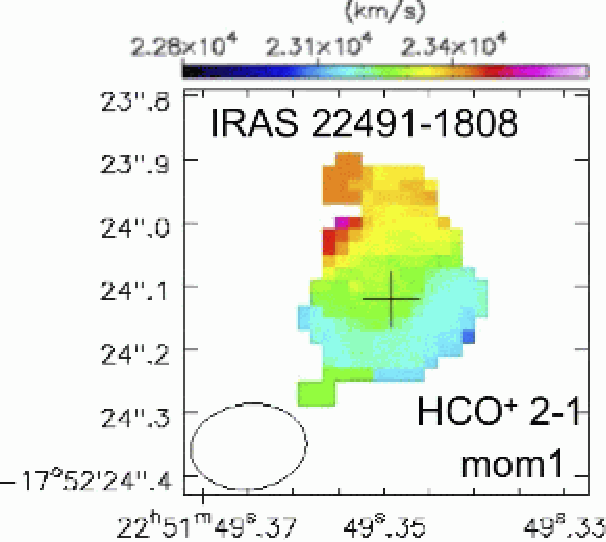}  
\includegraphics[angle=0,scale=.4]{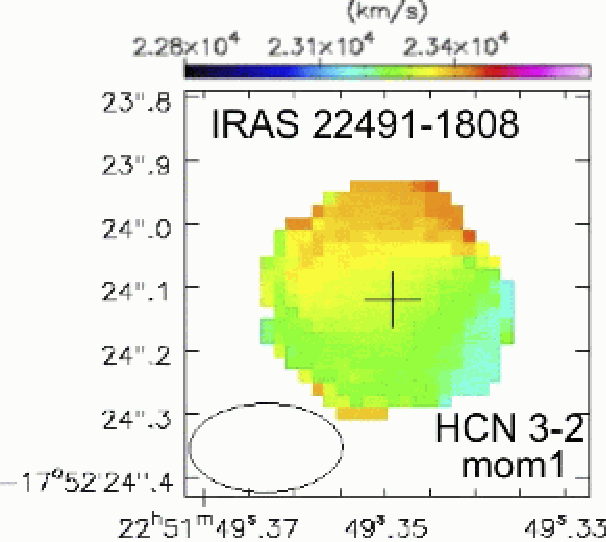} 
\includegraphics[angle=0,scale=.4]{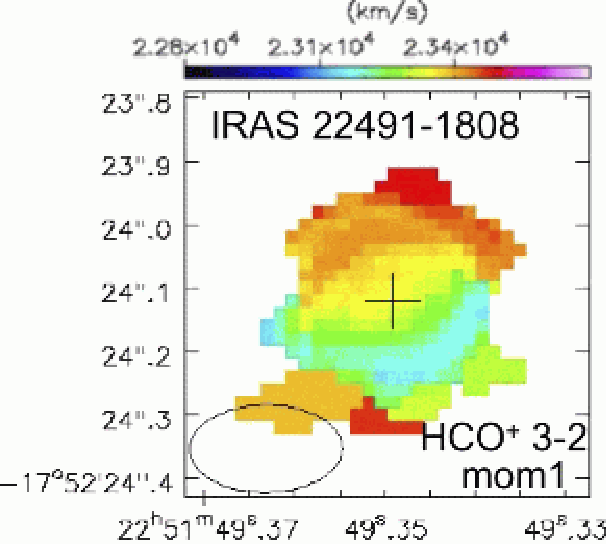} \\
\includegraphics[angle=0,scale=.4]{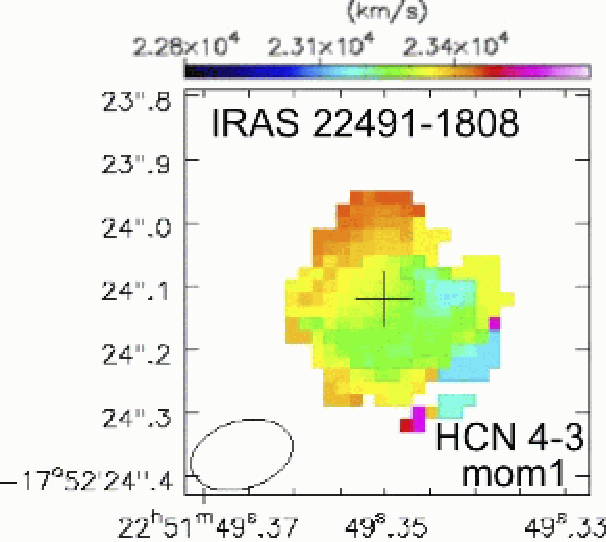} 
\includegraphics[angle=0,scale=.4]{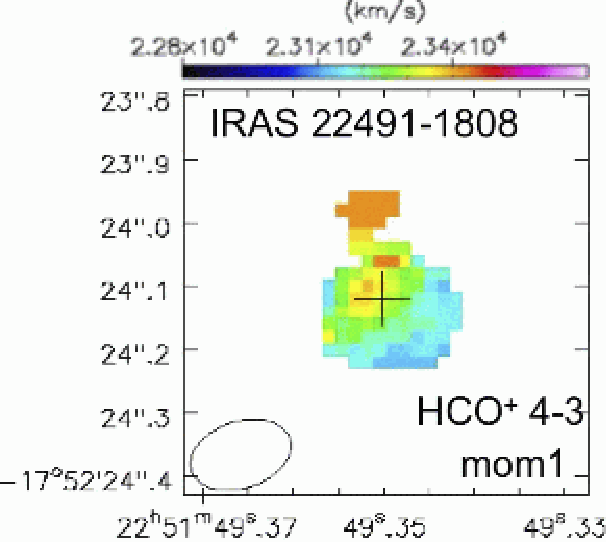} 
\includegraphics[angle=0,scale=.4]{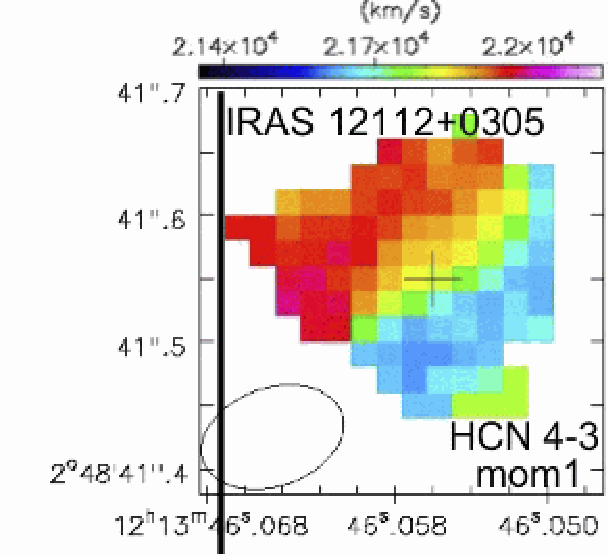} 
\includegraphics[angle=0,scale=.4]{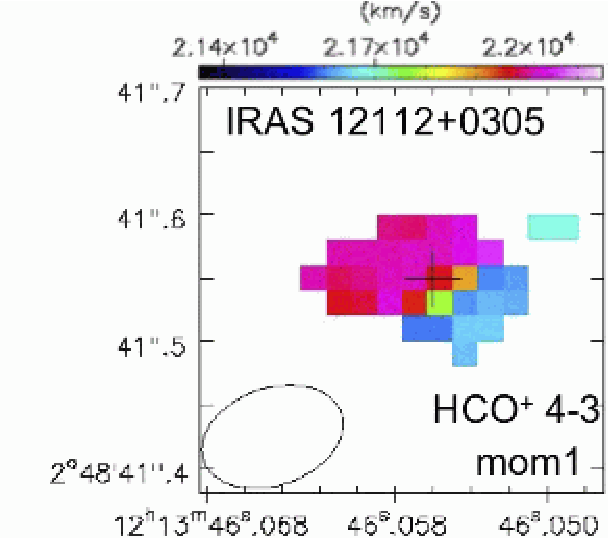} \\ 
\end{center}
\caption{Intensity-weighted mean velocity (moment 1) map of HCN and 
HCO$^{+}$ lines created from the original-beam-sized data
(Table \ref{tab:beam}, columns 2--4) taken in ALMA Cycle 7. 
The map of HCO$^{+}$ J=4--3 line for IRAS 03250$+$1606 is not shown
because of insufficient S/N ratios. 
The peak position of the simultaneously taken continuum emission is
shown as a cross. 
For each object, the field of view and velocity display range are set
as the same for all lines. 
The length of the vertical black solid bar in the first image of each
object corresponds to 0.5 kpc. 
Beam size for each moment 1 map is shown as an open circle 
in the lower-left region.
\label{fig:mom1}
}
\end{figure}

Figure \ref{fig:mom1} displays the original-beam-sized 
($\lesssim$0.5 kpc; Table \ref{tab:beam}, column 2--4),
intensity-weighted mean velocity (moment 1) maps of HCN and 
HCO$^{+}$ for ULIRGs observed in ALMA Cycle 7, to show
dynamical properties of dense molecular line emission. 
The same maps of other HCN and HCO$^{+}$ J-transition lines for some
(U)LIRGs, observed in ALMA Cycle 5 or earlier, have been presented in
previous publications \citep{ima13a,ima19,ima22} and thus are not shown
here.   
  

\begin{figure}[!hbt]
\begin{center}
\includegraphics[angle=0,scale=.4]{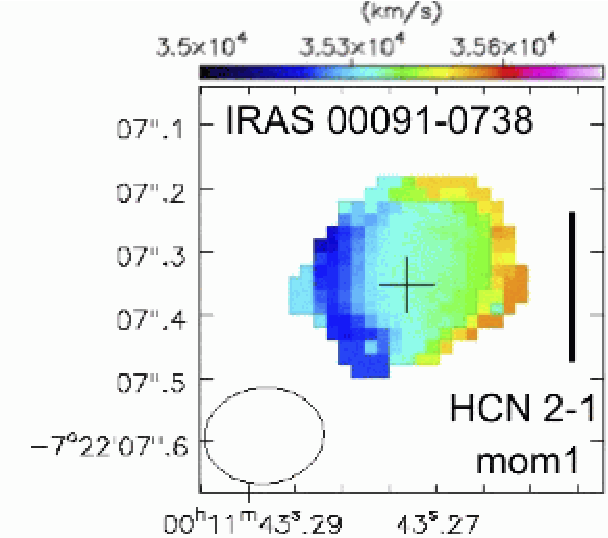} 
\includegraphics[angle=0,scale=.4]{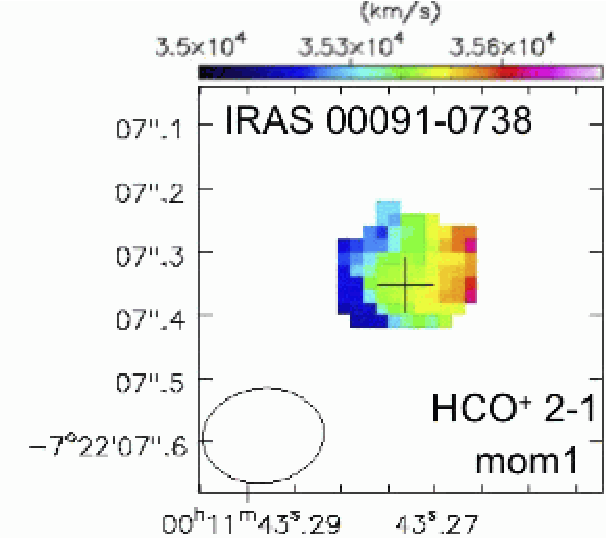} 
\includegraphics[angle=0,scale=.4]{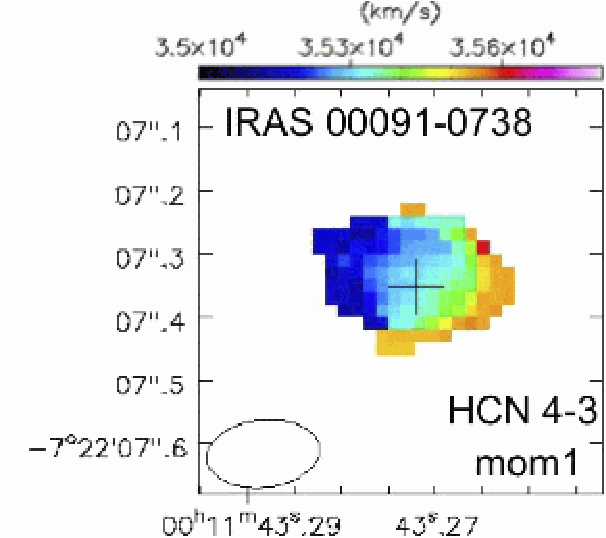} 
\includegraphics[angle=0,scale=.4]{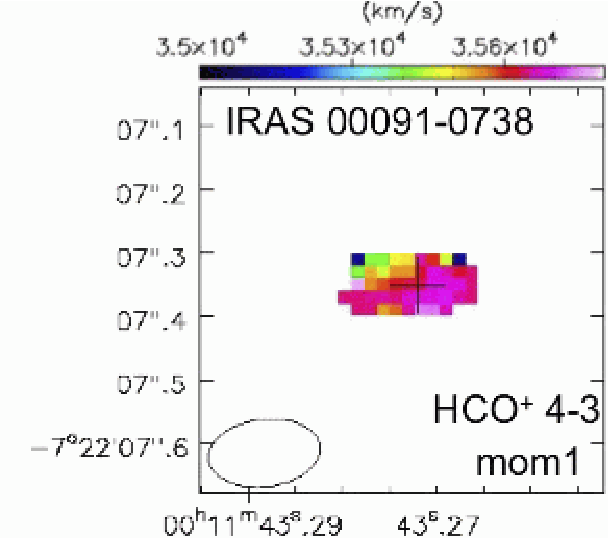} \\ 
\includegraphics[angle=0,scale=.4]{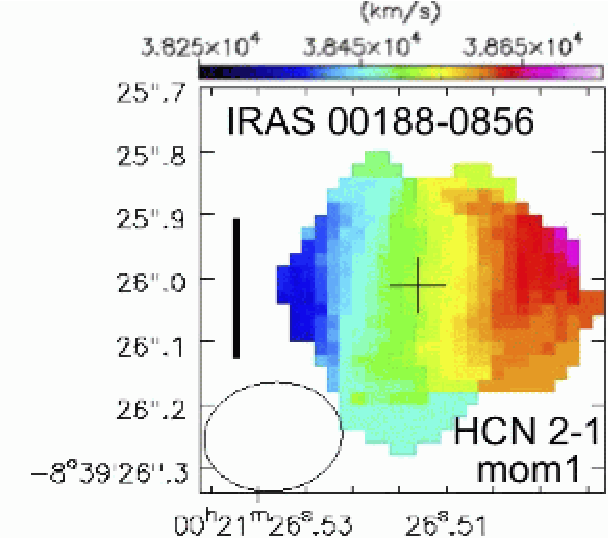} 
\includegraphics[angle=0,scale=.4]{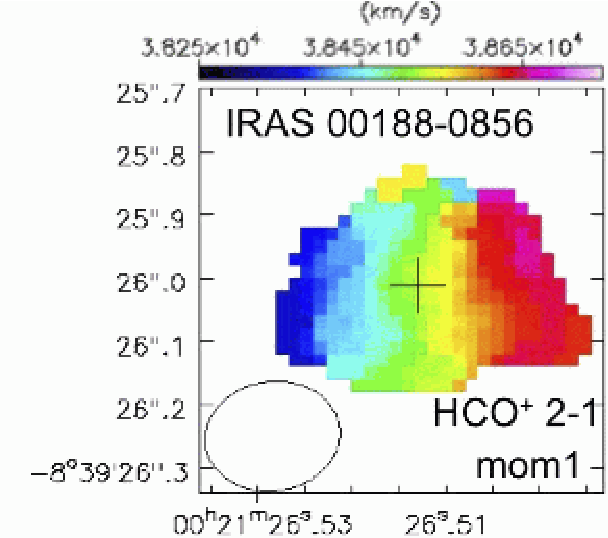} 
\includegraphics[angle=0,scale=.4]{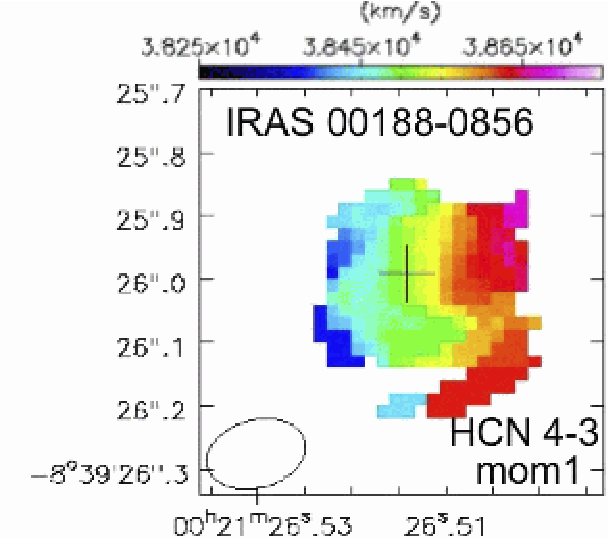} 
\includegraphics[angle=0,scale=.4]{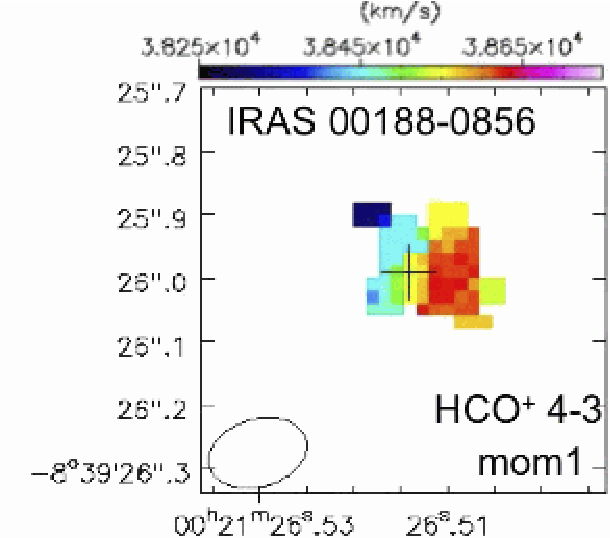} \\ 
\includegraphics[angle=0,scale=.4]{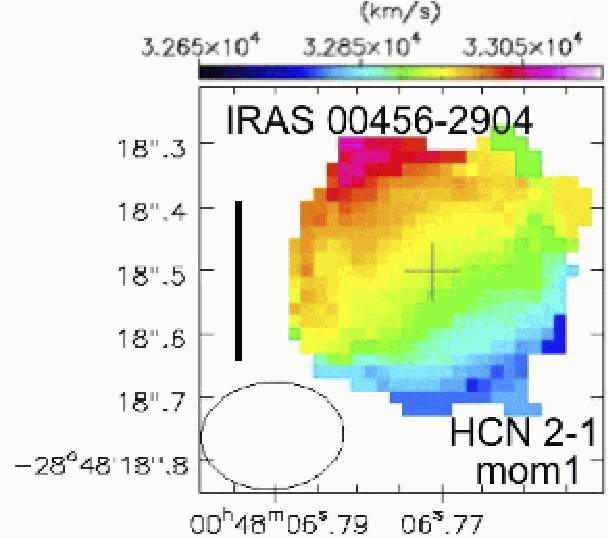} 
\includegraphics[angle=0,scale=.4]{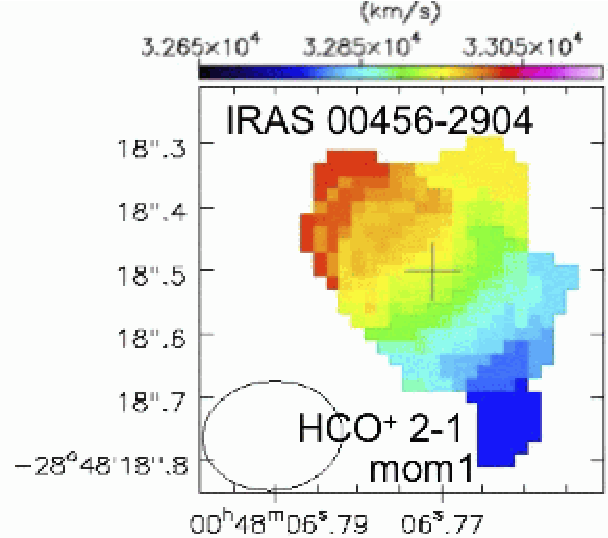} 
\includegraphics[angle=0,scale=.4]{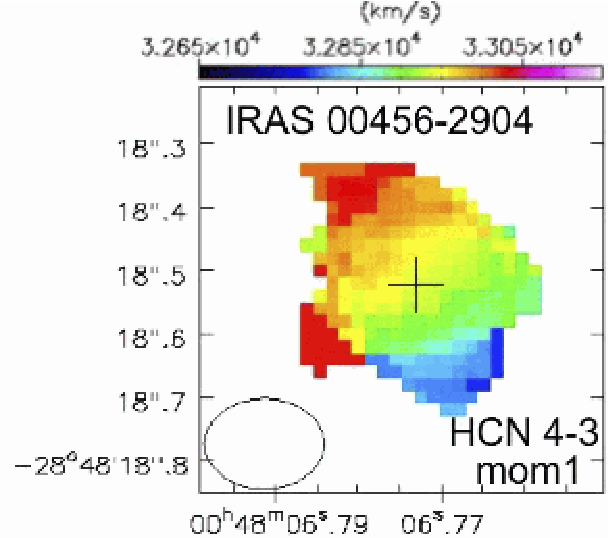} 
\includegraphics[angle=0,scale=.4]{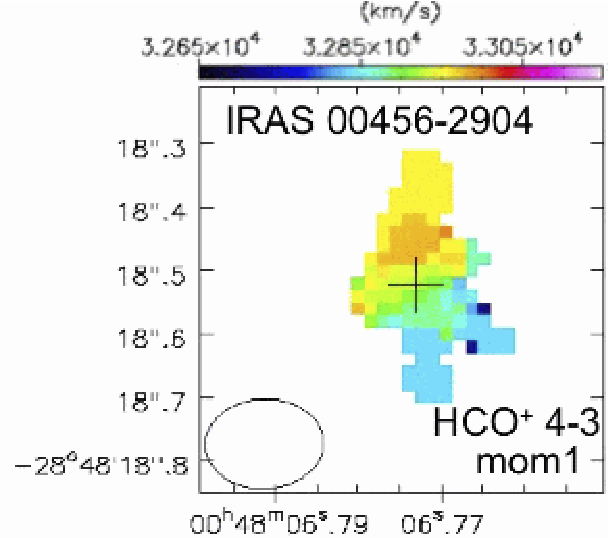} \\ 
\includegraphics[angle=0,scale=.4]{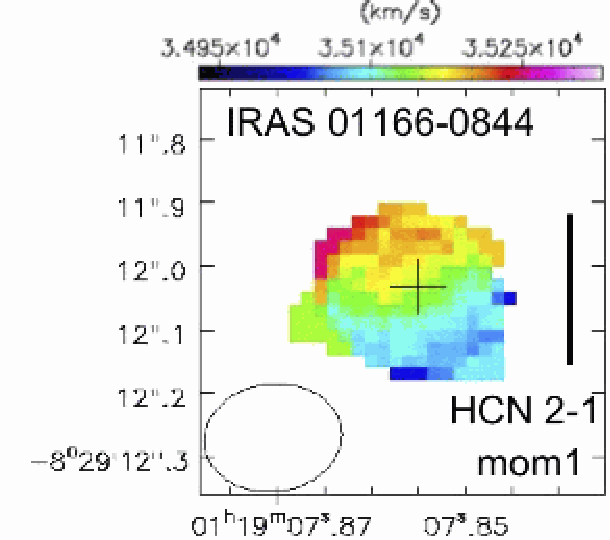} 
\includegraphics[angle=0,scale=.4]{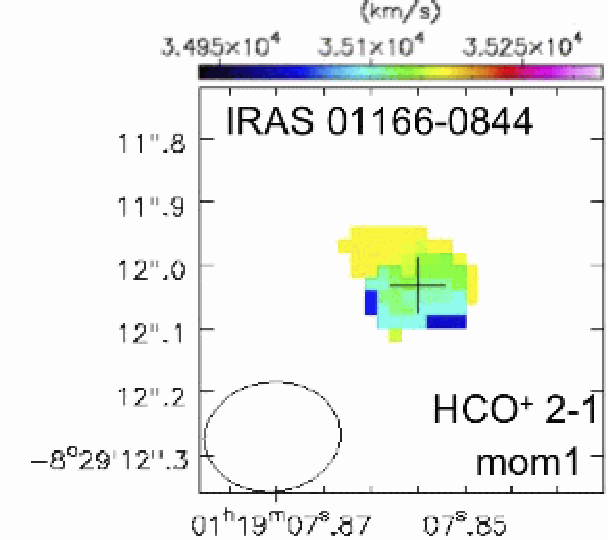} 
\includegraphics[angle=0,scale=.4]{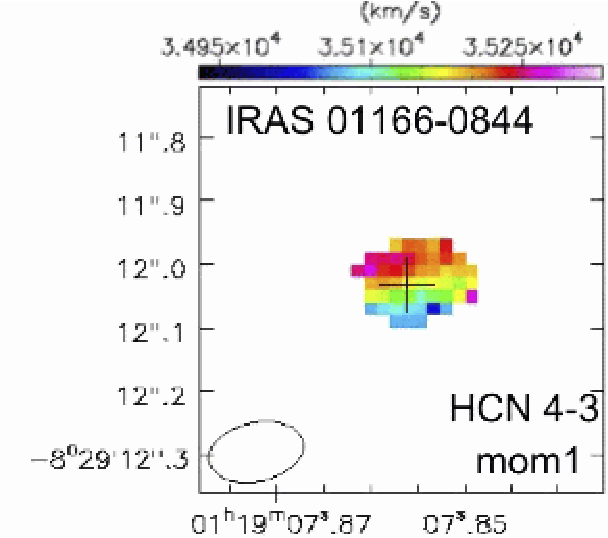} 
\includegraphics[angle=0,scale=.4]{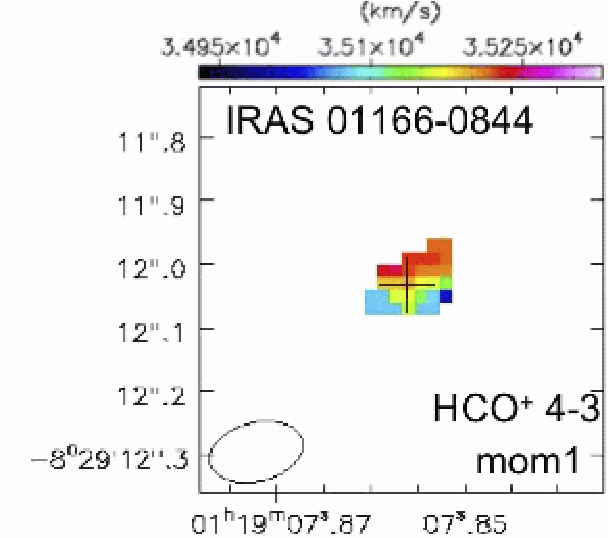} \\ 
\includegraphics[angle=0,scale=.4]{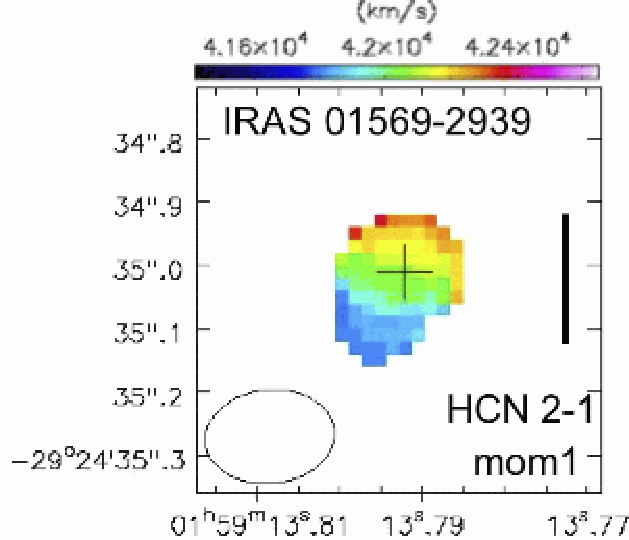} 
\includegraphics[angle=0,scale=.4]{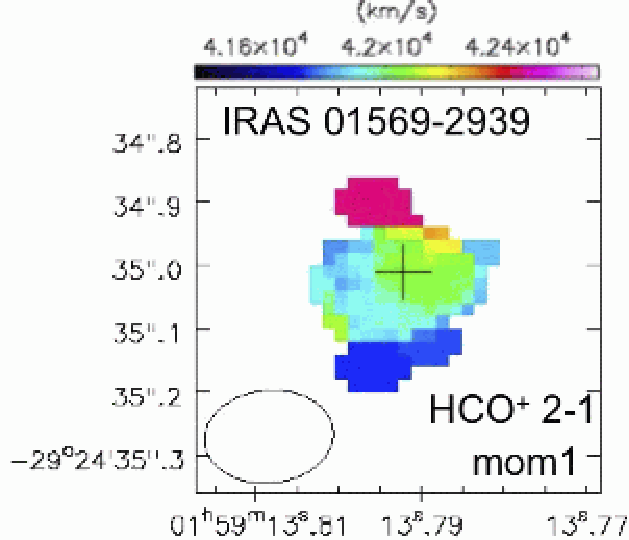} 
\includegraphics[angle=0,scale=.4]{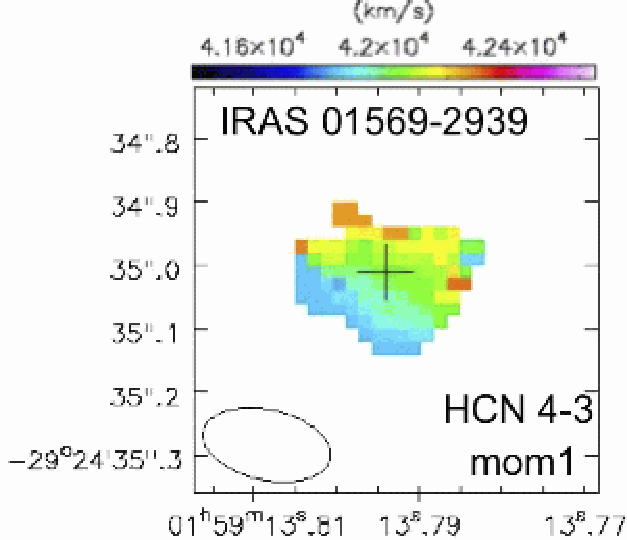} 
\includegraphics[angle=0,scale=.4]{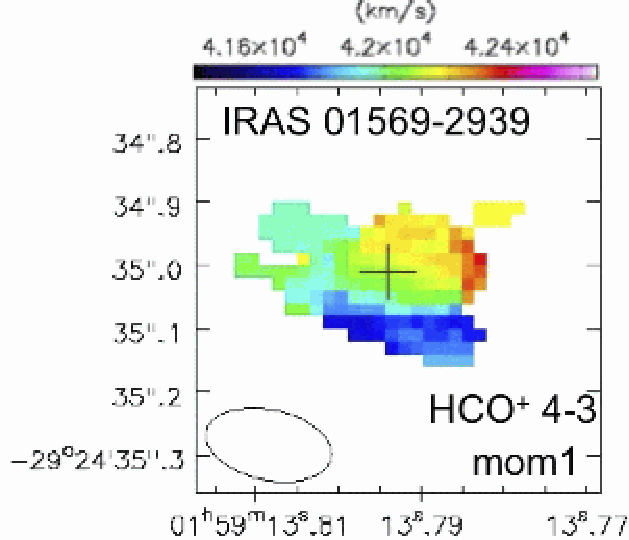}
\\ 
\hspace*{-4.7cm}
\includegraphics[angle=0,scale=.4]{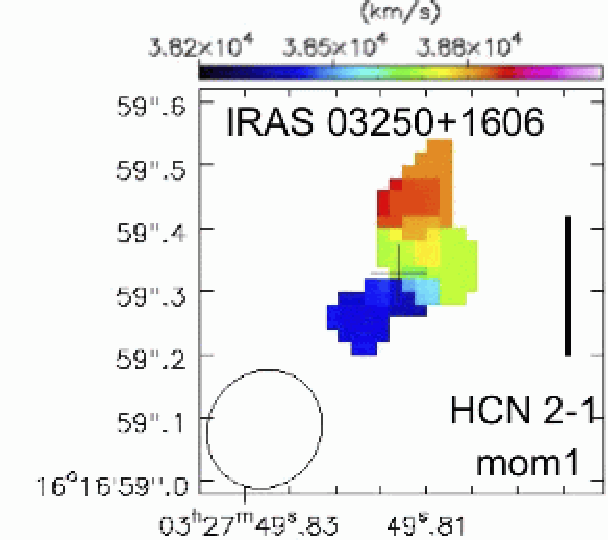} 
\includegraphics[angle=0,scale=.4]{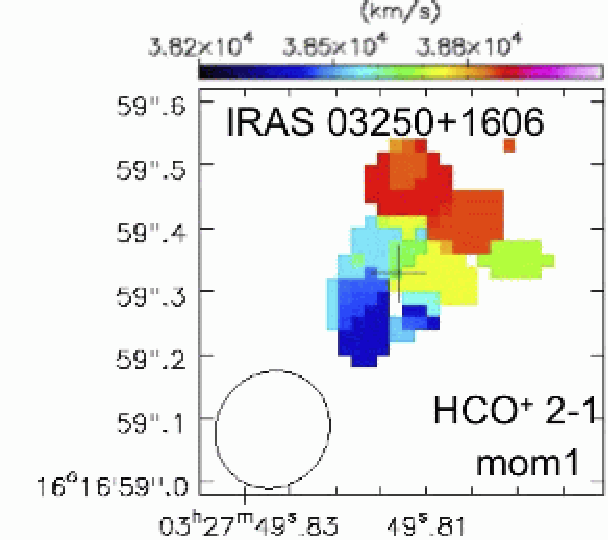} 
\includegraphics[angle=0,scale=.4]{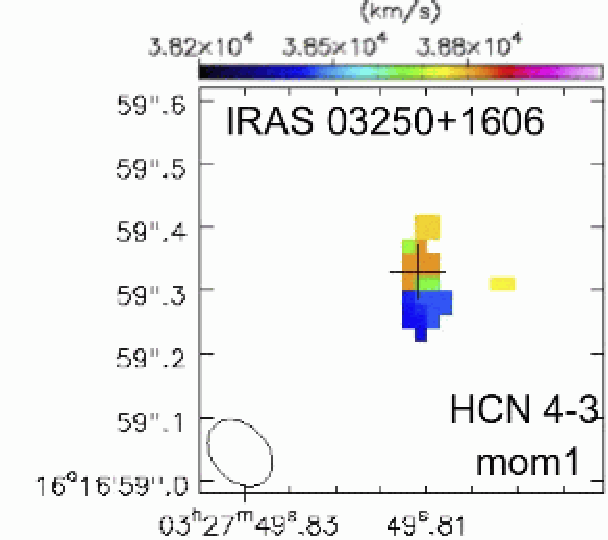} \\
\end{center}
\end{figure}


The moment 1 maps in Figure \ref{fig:mom1} are created
by integrating channels that show significant line signals, relative
to continuum flux level.
For IRAS 00091$-$0738 and IRAS 22206$-$2715, we see that the
HCO$^{+}$ J=4--3 emission line displays distinctly strong redshifted
components, compared to other emission lines. 
We investigate this origin.

\begin{figure}[!hbt]
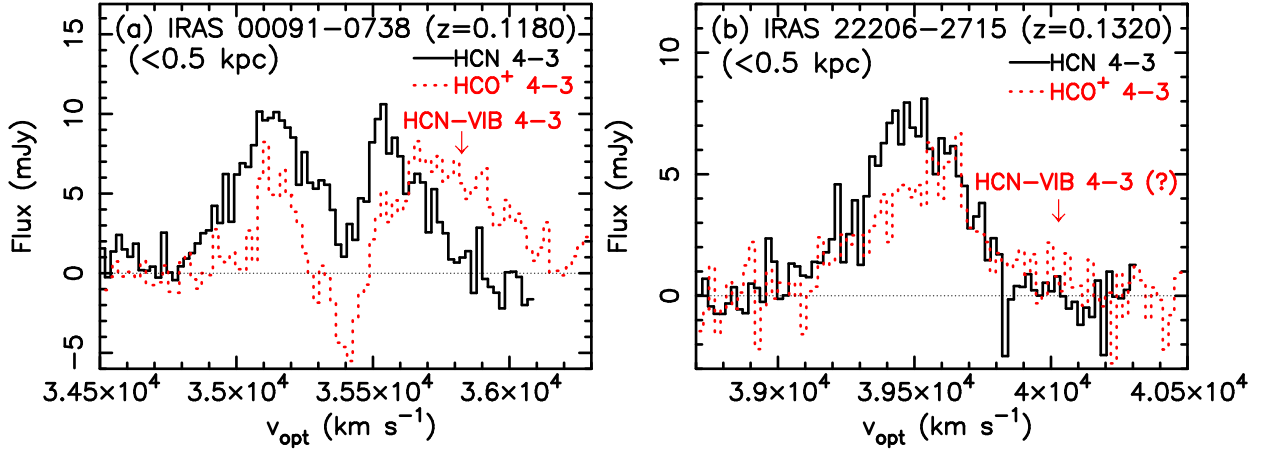

\begin{center}
\includegraphics[angle=-90,scale=.35]{fB2a.eps} 
\includegraphics[angle=-90,scale=.35]{fB2b.eps} 
\end{center}
\caption{
Comparison of velocity profile of HCN J=4--3 (black solid) and 
HCO$^{+}$ J=4--3 (red dotted) lines for {\it (a)} IRAS
00091$-$0738 and {\it (b)} IRAS 22206$-$2715 extracted from the 0.5
kpc beam-sized spectra.
The abscissa is optical local standard of rest (LSR) velocity in km
s$^{-1}$ and the ordinate is flux density in mJy. 
The expected velocity of the HCN-VIB J=4--3 line at the adopted
redshift (Table \ref{tab:object}, column 2) is indicated with a red
downward arrow. 
The horizontal black dotted straight line indicates the zero flux level.
\label{fig:HCNVIB43}
}
\end{figure}

It is well known that for sources with non-small molecular line
widths ($\gtrsim$300 km s$^{-1}$ in full width at half maximum
[FWHM]), it is often difficult to clearly separate HCO$^{+}$ and
vibrationally excited HCN v$_{2}$=1, l=1f (HCN-VIB) emission lines because
the latter line is only $\sim$400 km s$^{-1}$ redshifted in velocity
at the same J-transition \citep[e.g.,][]{aal15b,ima16b,ima18,fal19,fal21}.     
The HCN-VIB emission line flux in units of Jy km s$^{-1}$ can be
higher at higher J-transition, in the case of thermal excitation.  
Figure \ref{fig:HCNVIB43} compares the velocity profile of HCN J=4--3
and HCO$^{+}$ J=4--3 lines of IRAS 00091$-$0738 and 
IRAS 22206$-$2715.
IRAS 00091$-$0738 displays significant flux excess at the expected
frequency of HCN-VIB J=4--3 at the redshifted side of HCO$^{+}$ J=4--3
(Figure \ref{fig:HCNVIB43}a), suggesting that the HCN-VIB J=4--3
emission line is significantly detected and its contamination is a
cause of the apparently strong redshifted HCO$^{+}$ J=4--3 emission
component in Figure \ref{fig:mom1}. 
For IRAS 22206$-$2715, however, the signature of the HCN-VIB J=4--3
emission line is not clear, but the blue emission component is weaker
for HCO$^{+}$ J=4--3 than for HCN J=4--3 (Figure \ref{fig:HCNVIB43}b).
The apparently strong redshifted emission component seen in the
HCO$^{+}$ J=4--3 moment 1 map (Figure \ref{fig:mom1}) could be explained
by the intrinsic velocity difference, possibly caused by different
spatial distribution; HCO$^{+}$ emission is usually spatially more
extended than HCN emission at the same J-transition in nearby ULIRGs'
nuclei (\citet{ima19}; $\S$4 and Figure \ref{fig:CoG} of this
paper).  

\section{Gaussian fit}

\begin{figure*}
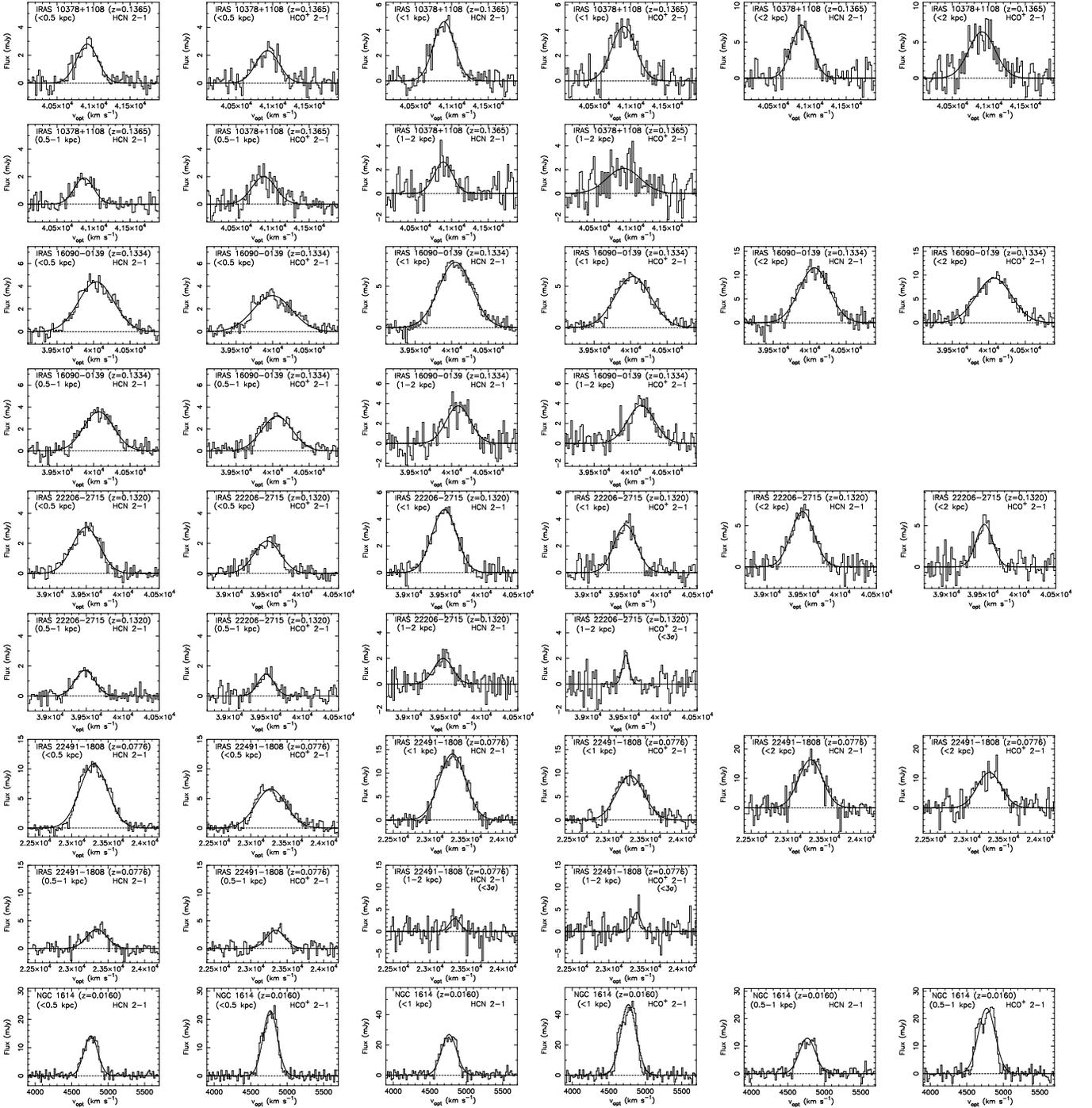

\includegraphics[angle=-90,scale=.118]{fC1bi.eps} \hspace{0.15cm}
\includegraphics[angle=-90,scale=.118]{fC1bj.eps} \hspace{0.15cm}
\includegraphics[angle=-90,scale=.118]{fC1bk.eps} \hspace{0.15cm}
\includegraphics[angle=-90,scale=.118]{fC1bl.eps} \hspace{0.15cm}
\includegraphics[angle=-90,scale=.118]{fC1bm.eps} \hspace{0.15cm}
\includegraphics[angle=-90,scale=.118]{fC1bn.eps} \\
\includegraphics[angle=-90,scale=.118]{fC1bo.eps} \hspace{0.15cm}
\includegraphics[angle=-90,scale=.118]{fC1bp.eps} \hspace{0.15cm}
\includegraphics[angle=-90,scale=.118]{fC1bq.eps} \hspace{0.15cm}
\includegraphics[angle=-90,scale=.118]{fC1br.eps} \\
\includegraphics[angle=-90,scale=.118]{fC1bs.eps} \hspace{0.15cm}
\includegraphics[angle=-90,scale=.118]{fC1bt.eps} \hspace{0.15cm}
\includegraphics[angle=-90,scale=.118]{fC1bu.eps} \hspace{0.15cm}
\includegraphics[angle=-90,scale=.118]{fC1bv.eps} \hspace{0.15cm}
\includegraphics[angle=-90,scale=.118]{fC1bw.eps} \hspace{0.15cm}
\includegraphics[angle=-90,scale=.118]{fC1bx.eps} \\
\includegraphics[angle=-90,scale=.118]{fC1by.eps} \hspace{0.15cm}
\includegraphics[angle=-90,scale=.118]{fC1bz.eps} \hspace{0.15cm}
\includegraphics[angle=-90,scale=.118]{fC1ca.eps} \hspace{0.15cm}
\includegraphics[angle=-90,scale=.118]{fC1cb.eps} \\
\includegraphics[angle=-90,scale=.118]{fC1cc.eps} 
\includegraphics[angle=-90,scale=.118]{fC1cd.eps} 
\includegraphics[angle=-90,scale=.118]{fC1ce.eps} 
\includegraphics[angle=-90,scale=.118]{fC1cf.eps} 
\includegraphics[angle=-90,scale=.118]{fC1cg.eps} 
\includegraphics[angle=-90,scale=.118]{fC1ch.eps} \\
\includegraphics[angle=-90,scale=.118]{fC1ci.eps} 
\includegraphics[angle=-90,scale=.118]{fC1cj.eps} 
\includegraphics[angle=-90,scale=.118]{fC1ck.eps} 
\includegraphics[angle=-90,scale=.118]{fC1cl.eps} \\
\includegraphics[angle=-90,scale=.118]{fC1cm.eps} \hspace{0.15cm}
\includegraphics[angle=-90,scale=.118]{fC1cn.eps} \hspace{0.15cm}
\includegraphics[angle=-90,scale=.118]{fC1co.eps} \hspace{0.15cm}
\includegraphics[angle=-90,scale=.118]{fC1cp.eps} \hspace{0.15cm}
\includegraphics[angle=-90,scale=.118]{fC1cq.eps} \hspace{0.15cm}
\includegraphics[angle=-90,scale=.118]{fC1cr.eps} \\
\includegraphics[angle=-90,scale=.118]{fC1cs.eps} \hspace{0.15cm}
\includegraphics[angle=-90,scale=.118]{fC1ct.eps} \hspace{0.15cm}
\includegraphics[angle=-90,scale=.118]{fC1cu.eps} \hspace{0.15cm}
\includegraphics[angle=-90,scale=.118]{fC1cv.eps} \\
\includegraphics[angle=-90,scale=.118]{fC1cw.eps} \hspace{0.15cm}
\includegraphics[angle=-90,scale=.118]{fC1cx.eps} \hspace{0.15cm}
\includegraphics[angle=-90,scale=.118]{fC1cy.eps} \hspace{0.15cm}
\includegraphics[angle=-90,scale=.118]{fC1cz.eps} \hspace{0.15cm}
\includegraphics[angle=-90,scale=.118]{fC1da.eps} \hspace{0.15cm}
\includegraphics[angle=-90,scale=.118]{fC1db.eps} \\
\caption{
Adopted Gaussian fits (solid curved lines) of the HCN and HCO$^{+}$ 
J=2--1 emission lines extracted in the 0.5 kpc (left two panels), 
1 kpc (middle two panels), and 2 kpc (right two panels) beam-sized
spectra in the first row for each ULIRG.
In the second row, those extracted from the spectra of 0.5--1 kpc (left
two panels) and 1--2 kpc (middle two panels) annular regions are
shown.
For the LIRG NGC 1614, only those from the spectra of the central
$\lesssim$0.5 kpc (left two panels), $\lesssim$1 kpc (middle two
panels), and 0.5--1 kpc annular (right two panels) regions are
displayed. 
The abscissa is optical LSR velocity in km s$^{-1}$ and the ordinate 
is flux density in mJy.
The horizontal thin dotted straight line indicates the zero flux 
level.
\label{fig:GaussFit21}
}
\end{figure*}

\begin{figure*}
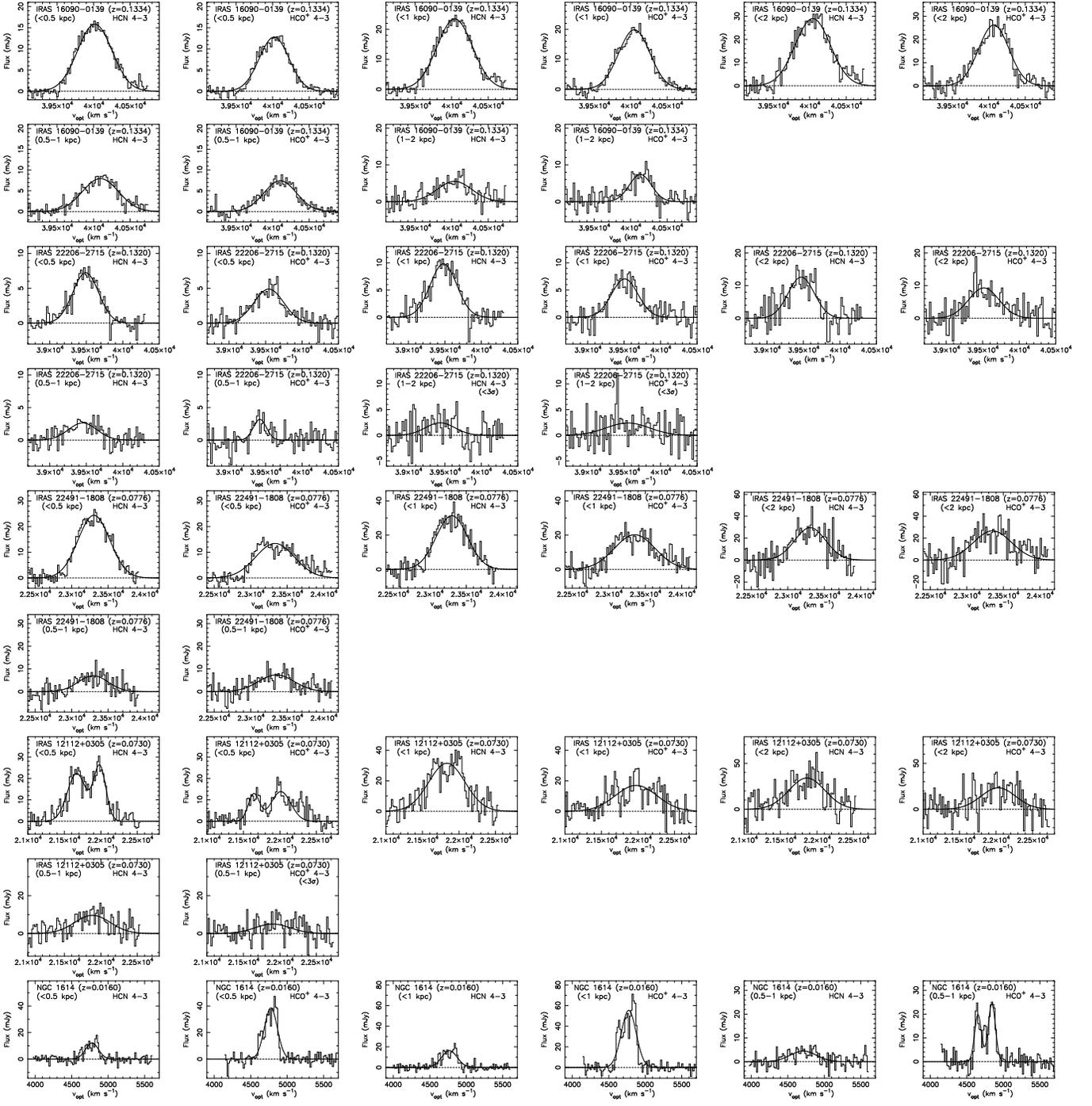

\includegraphics[angle=-90,scale=.118]{fC3bc.eps} \hspace{0.15cm}
\includegraphics[angle=-90,scale=.118]{fC3bd.eps} \hspace{0.15cm}
\includegraphics[angle=-90,scale=.118]{fC3be.eps} \hspace{0.15cm}
\includegraphics[angle=-90,scale=.118]{fC3bf.eps} \hspace{0.15cm}
\includegraphics[angle=-90,scale=.118]{fC3bg.eps} \hspace{0.15cm}
\includegraphics[angle=-90,scale=.118]{fC3bh.eps} \\
\includegraphics[angle=-90,scale=.118]{fC3bi.eps} \hspace{0.15cm}
\includegraphics[angle=-90,scale=.118]{fC3bj.eps} \hspace{0.15cm}
\includegraphics[angle=-90,scale=.118]{fC3bk.eps} \hspace{0.15cm}
\includegraphics[angle=-90,scale=.118]{fC3bl.eps} \\
\includegraphics[angle=-90,scale=.118]{fC3bm.eps} 
\includegraphics[angle=-90,scale=.118]{fC3bn.eps} 
\includegraphics[angle=-90,scale=.118]{fC3bo.eps} 
\includegraphics[angle=-90,scale=.118]{fC3bp.eps} 
\includegraphics[angle=-90,scale=.118]{fC3bq.eps} 
\includegraphics[angle=-90,scale=.118]{fC3br.eps} \\
\includegraphics[angle=-90,scale=.118]{fC3bs.eps} 
\includegraphics[angle=-90,scale=.118]{fC3bt.eps} 
\includegraphics[angle=-90,scale=.118]{fC3bu.eps} 
\includegraphics[angle=-90,scale=.118]{fC3bv.eps} \\
\includegraphics[angle=-90,scale=.118]{fC3bw.eps} \hspace{0.15cm}
\includegraphics[angle=-90,scale=.118]{fC3bx.eps} \hspace{0.15cm}
\includegraphics[angle=-90,scale=.118]{fC3by.eps} \hspace{0.15cm}
\includegraphics[angle=-90,scale=.118]{fC3bz.eps} \hspace{0.15cm}
\includegraphics[angle=-90,scale=.118]{fC3ca.eps} \hspace{0.15cm}
\includegraphics[angle=-90,scale=.118]{fC3cb.eps} \\
\includegraphics[angle=-90,scale=.118]{fC3cc.eps} \hspace{0.15cm}
\includegraphics[angle=-90,scale=.118]{fC3cd.eps} \\
\includegraphics[angle=-90,scale=.118]{fC3ce.eps} \hspace{0.15cm}
\includegraphics[angle=-90,scale=.118]{fC3cf.eps} \hspace{0.15cm}
\includegraphics[angle=-90,scale=.118]{fC3cg.eps} \hspace{0.15cm}
\includegraphics[angle=-90,scale=.118]{fC3ch.eps} \hspace{0.15cm}
\includegraphics[angle=-90,scale=.118]{fC3ci.eps} \hspace{0.15cm}
\includegraphics[angle=-90,scale=.118]{fC3cj.eps} \\
\includegraphics[angle=-90,scale=.118]{fC3ck.eps} \hspace{0.15cm}
\includegraphics[angle=-90,scale=.118]{fC3cl.eps} \\
\includegraphics[angle=-90,scale=.118]{fC3cm.eps} \hspace{0.15cm}
\includegraphics[angle=-90,scale=.118]{fC3cn.eps} \hspace{0.15cm}
\includegraphics[angle=-90,scale=.118]{fC3co.eps} \hspace{0.15cm}
\includegraphics[angle=-90,scale=.118]{fC3cp.eps} \hspace{0.15cm}
\includegraphics[angle=-90,scale=.118]{fC3cq.eps} \hspace{0.15cm}
\includegraphics[angle=-90,scale=.118]{fC3cr.eps} \\
\caption{Same as Figure \ref{fig:GaussFit21}, but for the J=4--3 lines 
of HCN and HCO$^{+}$. 
No Gaussian fit is applied when there is no emission line signature at
all, particularly in the 1--2 kpc annular region spectra.
\label{fig:GaussFit43}
}
\end{figure*}

Table \ref{tab:Gaussfit} summarizes the best Gaussian fit of
significantly detected HCN and HCO$^{+}$ emission lines at J=2--1,
J=3--2, and J=4--3 in the $\lesssim$0.5 kpc, $\lesssim$1 kpc, $\lesssim$2
kpc, 0.5--1 kpc, and 1--2 kpc spectra.  
For an emission line with only a subtle central dip and/or with
possible multiple peaks, but with large data scatters, we adopt a
single Gaussian fitting result, after confirming that single and
double Gaussian fits provide fluxes which agree well within $\sim$10\%.
The adopted best Gaussian fits are overplotted to the observed
HCN and HCO$^{+}$ emission lines in Figure \ref{fig:GaussFit21}
(for J=2--1), Figure \ref{fig:GaussFit32} (for J=3--2), and Figure
\ref{fig:GaussFit43} (for J=4--3).   
These overplots can be used to visually inspect the goodness of the 
Gaussian fits and check the reliability of our discussion of dense
molecular emission line flux ratios.


\startlongtable
\begin{deluxetable}{lllc|cccc}
\tabletypesize{\scriptsize}
\tablecaption{Gaussian Fit of Emission Lines \label{tab:Gaussfit}} 
\tablewidth{0pt}
\tablehead{
\colhead{Object} & \colhead{Region} & \colhead{Molecule} & \colhead{Line} &  
\multicolumn{4}{c}{Gaussian fit} \\  
\colhead{} & \colhead{} & \colhead{} & \colhead{} & \colhead{Velocity}
& \colhead{Peak} & \colhead{FWHM} & \colhead{Flux} \\ 
\colhead{} & \colhead{} & \colhead{} & \colhead{} & \colhead{[km s$^{-1}$]} 
& \colhead{[mJy]} & \colhead{[km s$^{-1}$]} &
\colhead{[Jy km s$^{-1}$]} \\  
\colhead{(1)} & \colhead{(2)} & \colhead{(3)} & \colhead{(4)} & 
\colhead{(5)} & \colhead{(6)} & \colhead{(7)} & \colhead{(8)} 
}
\startdata 
IRAS 00091$-$0738 & $\lesssim$0.5 kpc & HCN & J=2--1 &
35179$\pm$4,35560$\pm$6 & 4.7$\pm$0.1, 3.6$\pm$0.1 & 260$\pm$11, 278$\pm$16 
& 2.1$\pm$0.1 \\
& & & J=3--2 & 35364$\pm$12, 35392$\pm$8 & 7.5$\pm$1.0, $-$6.0$\pm$0.9
\tablenotemark{a} &
699$\pm$49, 180$\pm$33 & 4.0$\pm$0.8 \\
& & & J=4--3 & 35132$\pm$10, 35580$\pm$12 & 9.6$\pm$0.6, 8.3$\pm$0.7 &
332$\pm$26, 238$\pm$30 & 4.9$\pm$0.4 \\ 
 & & HCO$^{+}$ & J=2--1 & 35441$\pm$13, 35405$\pm$5 & 3.3$\pm$0.3,
$-$3.9$\pm$0.3 \tablenotemark{a} & 721$\pm$35, 161$\pm$18 & 1.6$\pm$0.2 \\ 
& & & J=3--2 & 35489$\pm$18, 35414$\pm$7 & 5.6$\pm$0.9, $-$7.2$\pm$0.9
\tablenotemark{a} &
680$\pm$44, 230$\pm$23 & 2.1$\pm$0.7 \\ 
& & & J=4--3 & 35533$\pm$25, 35409$\pm$9 & 8.5$\pm$1.1,
$-$11.6$\pm$1.1 \tablenotemark{a} &
766$\pm$52, 241$\pm$24 & 3.5$\pm$1.0 \\  
 & $\lesssim$1 kpc & HCN & J=2--1 & 35183$\pm$9, 35547$\pm$11 &  5.0$\pm$0.2,
4.1$\pm$0.2 & 263$\pm$19, 301$\pm$28 & 2.4$\pm$0.2 \\
 &          &     & J=3--2 & 35372$\pm$15, 35397$\pm$9 & 9.6$\pm$1.5,
$-$7.5$\pm$1.2 \tablenotemark{a} & 718$\pm$58, 148$\pm$46 & 5.5$\pm$1.2 \\
 &          &     & J=4--3 & 35129$\pm$18, 35569$\pm$19 &
11.7$\pm$1.0, 9.5$\pm$1.5 & 344$\pm$53, 271$\pm$51 & 6.3$\pm$0.9 \\  
 &          & HCO$^{+}$ & J=2--1 & 35437$\pm$16, 35404$\pm$7 &
3.8$\pm$0.3, $-$3.8$\pm$0.4 \tablenotemark{a} & 749$\pm$45, 140$\pm$21
& 2.2$\pm$0.3 \\ 
 &          &     & J=3--2 & 35491$\pm$21, 35415$\pm$9 & 7.5$\pm$1.4,
$-$8.6$\pm$1.3 \tablenotemark{a} & 692$\pm$53, 233$\pm$32 &
3.0$\pm$1.1 ($<$3$\sigma$) \\ 
 &          &     & J=4--3 & 35573$\pm$43, 35408$\pm$8 & 8.1$\pm$0.9,
$-$12.6$\pm$1.5 \tablenotemark{a} & 892$\pm$96, 158$\pm$30 & 5.0$\pm$1.1 \\  
 & $\lesssim$2 kpc & HCN & J=2--1 & 35207$\pm$21, 35546$\pm$29 & 5.6$\pm$0.5,
4.5$\pm$0.5 &  273$\pm$38, 285$\pm$68 & 2.7$\pm$0.4 \\
 &          &     & J=3--2 & 35373$\pm$18, 35396$\pm$9 & 12.0$\pm$1.7,
$-$10.5$\pm$1.6 \tablenotemark{a} & 676$\pm$54, 122$\pm$40 & 6.5$\pm$1.3 \\
 &          &     & J=4--3 & 35143$\pm$26, 35568$\pm$24 &
12.4$\pm$1.6, 11.8$\pm$2.6 & 375$\pm$60, 212$\pm$123 & 6.8$\pm$1.7 \\  
 &          & HCO$^{+}$ & J=2--1 & 35422$\pm$31, 35405$\pm$12 &
4.6$\pm$0.7, $-$4.1$\pm$0.9 \tablenotemark{a} & 730$\pm$87, 130$\pm$38
& 2.7$\pm$0.6 \\ 
 &          &     & J=3--2 & 35173$\pm$21, 35747$\pm$24 & 6.0$\pm$1.1,
6.9$\pm$0.8 & 218$\pm$47, 441$\pm$70 & 4.1$\pm$0.7 \\
 &          &     & J=4--3 & 35633$\pm$105, 35406$\pm$10 & 8.8$\pm$1.2,
$-$17.3$\pm$3.0 \tablenotemark{a} & 1145$\pm$246, 115$\pm$31 & 7.7$\pm$2.5 \\  
 & 0.5--1 kpc & HCN & J=2--1 & 35397$\pm$29 & 0.75$\pm$0.11 & 417$\pm$73
& 0.30$\pm$0.07 \\ 
& & & J=3--2 & 35189$\pm$29, 35623$\pm$130 & 2.3$\pm$0.7, 1.7$\pm$0.2
& 301$\pm$62, 477$\pm$204 & 1.4$\pm$0.4 \\ 
& & & J=4--3 & 35221$\pm$192 & 1.8$\pm$0.4 & 990$\pm$782 & 1.7$\pm$1.4
($<$3$\sigma$) \\
 & & HCO$^{+}$ & J=2--1 & 35402$\pm$60 & 0.74$\pm$0.12 & 698$\pm$191 &
0.49$\pm$0.16  \\
& & & J=3--2 & 35153$\pm$45, 35733$\pm$51 & 1.2$\pm$0.4, 1.3$\pm$0.2 &
286$\pm$127, 488$\pm$136 & 0.94$\pm$0.27 \\ 
& & & J=4--3 & 35220$\pm$50 & 2.2$\pm$0.9 & 257$\pm$117 &
0.53$\pm$0.34 ($<$3$\sigma$) \\ 
 & 1--2 kpc & HCN & J=2--1 & 35370$\pm$53 & 0.95$\pm$0.29 &
326$\pm$143 & 0.30$\pm$0.16 ($<$3$\sigma$) \\
& & & J=3--2 & 35379$\pm$66 & 1.8$\pm$0.3 & 642$\pm$152 & 1.1$\pm$0.3 \\
& & & J=4--3 & --- & --- & --- & --- \\
& & HCO$^{+}$ & J=2--1 & 35340$\pm$103 & 1.0$\pm$0.3 & 494$\pm$340 &
0.49$\pm$0.36 ($<$3$\sigma$) \\ 
& & & J=3--2 & 35613$\pm$113 & 1.5$\pm$0.4 & 905$\pm$265 & 1.3$\pm$0.5
($<$3$\sigma$) \\
& & & J=4--3 & --- & --- & --- & --- \\ \hline
IRAS 00188$-$0856 & $\lesssim$0.5 kpc & HCN & J=2--1 & 38533$\pm$4 &
4.7$\pm$0.1 & 338$\pm$9 & 1.5$\pm$0.1 \\
& & & J=3--2 & 38540$\pm$4 & 7.7$\pm$0.2 & 336$\pm$10 & 2.4$\pm$0.1 \\
& & & J=4--3 & 38525$\pm$5 & 11.1$\pm$0.3 & 339$\pm$12 & 3.6$\pm$0.2 \\ 
 & & HCO$^{+}$ & J=2--1 & 38523$\pm$7 & 2.9$\pm$0.1 & 303$\pm$14 &
0.83$\pm$0.05 \\ 
& & & J=3--2 & 38530$\pm$7 & 4.5$\pm$0.2 & 328$\pm$14 & 1.4$\pm$0.1 \\
& & & J=4--3 & 38454$\pm$25, 38633$\pm$14 & 5.8$\pm$0.5, 5.3$\pm$1.4 &
225$\pm$51, 136$\pm$27 & 1.9$\pm$0.4 \\
 & $\lesssim$1 kpc & HCN & J=2--1 & 38532$\pm$4 & 6.8$\pm$0.1 & 343$\pm$8 &
2.2$\pm$0.1 \\
 &          &     & J=3--2 & 38544$\pm$4 & 12.7$\pm$0.3 & 334$\pm$9 &
4.0$\pm$0.1 \\
 &          &     & J=4--3 & 38418$\pm$20, 38611$\pm$23 &
11.9$\pm$2.6, 14.4$\pm$2.5 & 195$\pm$40, 220$\pm$32 & 5.2$\pm$0.9 \\  
 &          & HCO$^{+}$ & J=2--1 & 38519$\pm$7 & 4.5$\pm$0.2 &
317$\pm$16 & 1.4$\pm$0.1 \\
 &          &     & J=3--2 & 38528$\pm$7 & 7.6$\pm$0.3 & 329$\pm$14 &
2.4$\pm$0.1 \\
 &          &     & J=4--3 & 38449$\pm$27, 38628$\pm$15 & 8.0$\pm$0.9,
9.3$\pm$1.9 & 179$\pm$56, 126$\pm$22 & 2.5$\pm$0.5 \\  
 & $\lesssim$2 kpc & HCN & J=2--1 & 38532$\pm$5 & 8.7$\pm$0.3 & 348$\pm$13 &
2.9$\pm$0.1 \\
 &          &     & J=3--2 & 38544$\pm$5 & 16.9$\pm$0.5 & 337$\pm$12 &
5.4$\pm$0.3 \\
 &          &     & J=4--3 & 38532$\pm$10 & 20.8$\pm$1.2 & 371$\pm$25
& 7.3$\pm$0.6 \\  
 &          & HCO$^{+}$ & J=2--1 & 38519$\pm$10 & 5.6$\pm$0.3 &
339$\pm$23 & 1.8$\pm$0.1 \\
 &          &     & J=3--2 & 38517$\pm$9 & 10.0$\pm$0.5 & 325$\pm$18 &
3.1$\pm$0.2 \\
 &          &     & J=4--3 & 38519$\pm$24 & 9.2$\pm$1.4 & 319$\pm$65 & 
2.8$\pm$0.7  \\  
 & 0.5--1 kpc & HCN & J=2--1 & 38530$\pm$9 & 2.1$\pm$0.1 & 353$\pm$18 
& 0.71$\pm$0.05 \\
& & & J=3--2 & 38548$\pm$7 & 5.0$\pm$0.2 & 325$\pm$16 & 1.5$\pm$0.1 \\
& & & J=4--3 & 38442$\pm$27, 38633$\pm$41 & 5.0$\pm$1.2, 5.3$\pm$1.1 &
166$\pm$34, 172$\pm$37 & 1.6$\pm$0.4 \\  
 & & HCO$^{+}$ & J=2--1 & 38510$\pm$13 & 1.7$\pm$0.1 & 329$\pm$27 &
0.53$\pm$0.06 \\ 
& & & J=3--2 & 38524$\pm$10 & 3.2$\pm$0.2 & 331$\pm$24 & 0.98$\pm$0.09 \\
& & & J=4--3 & 38414$\pm$12, 38608$\pm$18 & 2.9$\pm$1.0, 3.0$\pm$0.6 &
54$\pm$40, 170$\pm$48 & 0.63$\pm$0.20 \\ 
 & 1--2 kpc & HCN & J=2--1 & 38532$\pm$20 & 1.8$\pm$0.2 & 365$\pm$33 &
0.63$\pm$0.09 \\
& & & J=3--2 & 38549$\pm$14 & 4.2$\pm$0.3 & 351$\pm$39 & 1.4$\pm$0.2 \\
& & & J=4--3 & 38531$\pm$45 & 4.4$\pm$0.9 & 463$\pm$109 & 1.9$\pm$0.6 \\
& & HCO$^{+}$ & J=2--1 & 38525$\pm$54 & 1.0$\pm$0.2 & 483$\pm$143 &
0.48$\pm$0.16 ($<$3$\sigma$) \\ 
& & & J=3--2 & 38484$\pm$22 & 2.5$\pm$0.4 & 293$\pm$56 & 0.69$\pm$0.17 \\
& & & J=4--3 & --- & --- & --- & --- \\ \hline
IRAS 00456$-$2904 & $\lesssim$0.5 kpc & HCN & J=2--1 & 32938$\pm$3 &
5.2$\pm$0.1 & 270$\pm$6 & 1.3$\pm$0.1 \\
& & & J=3--2 & 32937$\pm$5 & 9.9$\pm$0.3 & 301$\pm$10 & 2.9$\pm$0.1 \\
& & & J=4--3 & 32934$\pm$4 & 12.6$\pm$0.4 & 294$\pm$11 & 3.6$\pm$0.2 \\
 & & HCO$^{+}$ & J=2--1 & 32934$\pm$4 & 3.6$\pm$0.1 & 262$\pm$12 &
0.91$\pm$0.05 \\ 
& & & J=3--2 & 32936$\pm$6 & 6.0$\pm$0.3 & 272$\pm$14 & 1.6$\pm$0.1 \\
& & & J=4--3 & 32924$\pm$9 & 8.1$\pm$0.6 & 289$\pm$25 & 2.2$\pm$0.2 \\ 
 & $\lesssim$1 kpc & HCN & J=2--1 & 32934$\pm$3 & 7.6$\pm$0.2 & 281$\pm$7 &
2.1$\pm$0.1 \\
 &          &     & J=3--2 & 32936$\pm$6 & 14.7$\pm$0.6 & 307$\pm$13 &
4.3$\pm$0.3 \\
 &          &     & J=4--3 & 32932$\pm$6 & 17.4$\pm$0.6 & 309$\pm$14 &
5.2$\pm$0.3 \\  
 &          & HCO$^{+}$ & J=2--1 & 32925$\pm$4 & 5.8$\pm$0.2 &
263$\pm$9 & 1.5$\pm$0.1 \\
 &          &     & J=3--2 & 32931$\pm$8 & 10.0$\pm$0.6 & 286$\pm$18 &
2.8$\pm$0.2 \\
 &          &     & J=4--3 & 32940$\pm$11 & 11.2$\pm$1.0 & 283$\pm$26
& 3.0$\pm$0.4 \\  
 & $\lesssim$2 kpc & HCN & J=2--1 & 32933$\pm$4 & 10.3$\pm$0.3 & 278$\pm$10 &
2.7$\pm$0.1 \\
 &          &     & J=3--2 & 32933$\pm$9 & 18.5$\pm$1.1 & 291$\pm$18 &
5.2$\pm$0.4 \\
 &          &     & J=4--3 & 32926$\pm$10 & 21.0$\pm$1.3 & 314$\pm$24
& 6.3$\pm$0.6 \\  
 &          & HCO$^{+}$ & J=2--1 & 32922$\pm$6 & 8.1$\pm$0.4 &
261$\pm$14 & 2.0$\pm$0.1 \\
 &          &     & J=3--2 & 32923$\pm$12 & 14.5$\pm$1.1 & 296$\pm$24 &
4.1$\pm$0.5 \\
 &          &     & J=4--3 & 32943$\pm$14 & 17.2$\pm$2.0 & 222$\pm$35
& 3.7$\pm$0.7 \\  
 & 0.5--1 kpc & HCN & J=2--1 & 32926$\pm$7 & 2.4$\pm$0.1 & 305$\pm$17  
& 0.72$\pm$0.05 \\
& & & J=3--2 & 32936$\pm$13 & 4.8$\pm$0.4 & 309$\pm$36 & 1.4$\pm$0.2 \\
& & & J=4--3 & 32853$\pm$13, 33035$\pm$11 & 5.4$\pm$0.7, 5.2$\pm$0.9 &
166$\pm$29, 111$\pm$67 & 1.4$\pm$0.4 \\
 & & HCO$^{+}$ & J=2--1 & 32910$\pm$9 & 2.2$\pm$0.1 & 260$\pm$22 &
0.56$\pm$0.06  \\
& & & J=3--2 & 32923$\pm$14 & 4.1$\pm$0.4 & 299$\pm$28 & 1.2$\pm$0.2 \\
& & & J=4--3 & 32949$\pm$171, 33025$\pm$34 & 2.5$\pm$1.0, 4.4$\pm$1.9
& 243 (fix) \tablenotemark{b}, 46 (fix) \tablenotemark{b} & 0.79$\pm$0.24 \\ 
 & 1--2 kpc & HCN & J=2--1 & 32929$\pm$13 & 2.6$\pm$0.3 & 269$\pm$29 &
0.67$\pm$0.10  \\
& & & J=3--2 & 32920$\pm$27 & 3.8$\pm$0.8 & 239$\pm$63 & 0.88$\pm$0.29 \\
& & & J=4--3 & 32893$\pm$51 & 3.5$\pm$0.9 & 351$\pm$115 & 1.2$\pm$0.5
($<$3$\sigma$) \\ 
& & HCO$^{+}$ & J=2--1 & 32917$\pm$18 & 2.2$\pm$0.3 & 267$\pm$38 &
0.56$\pm$0.12 \\
& & & J=3--2 & 32910$\pm$24 & 4.7$\pm$0.7 & 300$\pm$65 & 1.3$\pm$0.4 \\
& & & J=4--3 & 32946$\pm$26 & 5.8$\pm$1.6 & 159$\pm$59 & 0.88$\pm$0.41
($<$3$\sigma$) \\ \hline
IRAS 01166$-$0844 & $\lesssim$0.5 kpc & HCN & J=2--1 & 35139$\pm$8 &
2.6$\pm$0.1 & 473$\pm$21 & 1.2$\pm$0.1 \\
& & & J=3--2 & 35151$\pm$15 & 3.9$\pm$0.2 & 498$\pm$42 & 1.8$\pm$0.2 \\
& & & J=4--3 & 35161$\pm$10 & 6.3$\pm$0.3 & 464$\pm$25 & 2.8$\pm$0.2 \\
 & & HCO$^{+}$ & J=2--1 & 35143$\pm$12 & 1.8$\pm$0.1 & 371$\pm$29 &
0.64$\pm$0.07 \\ 
& & & J=3--2 & 35144$\pm$15 & 3.5$\pm$0.3 & 408$\pm$34 & 1.4$\pm$0.2 \\
& & & J=4--3 & 35170$\pm$16 & 4.3$\pm$0.3 & 434$\pm$40 & 1.8$\pm$0.2 \\ 
 & $\lesssim$1 kpc & HCN & J=2--1 & 35134$\pm$9 & 3.4$\pm$0.2 & 433$\pm$27 &
1.4$\pm$0.1 \\
 &          &     & J=3--2 & 35155$\pm$24 & 5.3$\pm$0.5 & 586$\pm$80 &
3.0$\pm$0.5 \\
 &          &     & J=4--3 & 35151$\pm$14 & 7.1$\pm$0.5 & 454$\pm$35 &
3.1$\pm$0.3 \\  
 &          & HCO$^{+}$ & J=2--1 & 35129$\pm$12 & 2.4$\pm$0.2 &
363$\pm$29 & 0.84$\pm$0.09 \\
 &          &     & J=3--2 & 35127$\pm$19 & 5.7$\pm$0.5 & 398$\pm$41 &
2.2$\pm$0.3 \\
 &          &     & J=4--3 & 35162$\pm$18 & 5.9$\pm$0.5 & 387$\pm$44 &
2.2$\pm$0.3 \\  
 & $\lesssim$2 kpc & HCN & J=2--1 & 35135$\pm$16 & 4.2$\pm$0.4 & 422$\pm$47 &
1.7$\pm$0.2 \\
 &          &     & J=3--2 & 35145$\pm$48 & 6.1$\pm$0.8 & 686$\pm$137
& 4.0$\pm$0.9 \\
 &          &     & J=4--3 & 35139$\pm$35 & 7.1$\pm$1.0 & 475$\pm$93 &
3.2$\pm$0.8 \\  
 &          & HCO$^{+}$ & J=2--1 & 35109$\pm$22 & 2.9$\pm$0.3 &
441$\pm$65 & 1.2$\pm$0.2 \\
 &          &     & J=3--2 & 35137$\pm$34 & 6.7$\pm$1.2 & 375$\pm$75 &
2.4$\pm$0.6 \\
 &          &     & J=4--3 & 35177$\pm$22 & 9.7$\pm$1.7 & 245$\pm$62 &
2.3$\pm$0.7 \\  
 & 0.5--1 kpc & HCN & J=2--1 & 35105$\pm$15 & 1.0$\pm$0.2 & 218$\pm$46 
& 0.21$\pm$0.06 \\
& & & J=3--2 & 35154$\pm$93 & 1.5$\pm$0.3 & 821$\pm$225 & 1.2$\pm$0.4 \\
& & & J=4--3 & 35077$\pm$231 & 0.77$\pm$0.31 & 475$\pm$674 &
0.35$\pm$0.51 ($<$3$\sigma$) \\
 & & HCO$^{+}$ & J=2--1 & 35096$\pm$28 & 0.67$\pm$0.16 & 314$\pm$89 &
0.20$\pm$0.07 ($<$3$\sigma$) \\
& & & J=3--2 & 35101$\pm$32 & 2.2$\pm$0.4 & 385$\pm$72 & 0.81$\pm$0.20 \\
& & & J=4--3 & 35159$\pm$61 & 1.4$\pm$0.9 & 272$\pm$119 &
0.37$\pm$0.28 ($<$3$\sigma$) \\ 
 & 1--2 kpc & HCN & J=2--1 & 35093$\pm$23 & 1.3$\pm$0.6 & 178$\pm$81 &
0.21$\pm$0.14 ($<$3$\sigma$) \\
& & & J=3--2 & --- & ---  & ---  & --- \\
& & & J=4--3 & --- & ---  & ---  & --- \\
& & HCO$^{+}$ & J=2--1 & 35040$\pm$209 & 0.60$\pm$0.18 & 780$\pm$670 &
0.45$\pm$0.40 ($<$3$\sigma$) \\
& & & J=3--2 & --- & --- & --- & --- \\
& & & J=4--3 & --- & --- & --- & --- \\ \hline
IRAS 01569$-$2939 & $\lesssim$0.5 kpc & HCN & J=2--1 & 41838$\pm$21, 42280$\pm$16 &
1.4$\pm$0.1, 1.5$\pm$0.2 & 434$\pm$78, 308$\pm$43 & 1.0$\pm$0.1 \\
& & & J=3--2 & 41854$\pm$22, 42281$\pm$19 & 2.8$\pm$0.2, 2.9$\pm$0.2 &
378$\pm$53, 297$\pm$47 & 1.8$\pm$0.2 \\ 
& & & J=4--3 & 41863$\pm$14, 42285$\pm$12 & 3.7$\pm$0.2, 3.6$\pm$0.2 &
400$\pm$32, 276$\pm$26 & 2.3$\pm$0.2 \\ 
 & & HCO$^{+}$ & J=2--1 & 41823$\pm$27, 42265$\pm$30 & 1.6$\pm$0.1, 1.5$\pm$0.1 &
340$\pm$77, 363$\pm$64 & 1.0$\pm$0.2 \\
& & & J=3--2 & 41799$\pm$40, 42243$\pm$54 & 2.6$\pm$0.5, 2.5$\pm$0.3 &
388$\pm$65, 491$\pm$93 & 2.1$\pm$0.3 \\ 
& & & J=4--3 & 41806$\pm$23, 42262$\pm$22 & 3.3$\pm$0.3, 3.8$\pm$0.2 &
321$\pm$48, 435$\pm$50 & 2.5$\pm$0.3 \\  
 & $\lesssim$1 kpc & HCN & J=2--1 & 41830$\pm$29, 42284$\pm$29 & 1.9$\pm$0.1,
1.7$\pm$0.2 & 412$\pm$67, 373$\pm$54 & 1.3$\pm$0.2 \\
 &          &     & J=3--2 & 41844$\pm$29, 42291$\pm$30 & 4.4$\pm$0.4,
3.9$\pm$0.5 & 408$\pm$144, 313$\pm$70 & 2.8$\pm$0.7 \\
 &          &     & J=4--3 & 41865$\pm$16, 42293$\pm$13 & 5.2$\pm$0.2,
4.6$\pm$0.3 & 418$\pm$38, 261$\pm$30 & 3.2$\pm$0.3 \\  
 &          & HCO$^{+}$ & J=2--1 & 41813$\pm$18, 42064$\pm$44 &
1.4$\pm$0.2, 1.8$\pm$0.1 & 142$\pm$94, 743$\pm$45 & 1.5$\pm$0.2 \\
 &          &     & J=3--2 & 42004$\pm$18 & 5.4$\pm$0.3 & 727$\pm$41 &
3.7$\pm$0.3 \\
 &          &     & J=4--3 & 41804$\pm$24, 42244$\pm$46 & 5.1$\pm$0.6,
4.5$\pm$0.3 & 302$\pm$55, 454$\pm$91 & 3.4$\pm$0.5 \\  
 & $\lesssim$2 kpc & HCN & J=2--1 & 42032$\pm$26 & 2.4$\pm$0.2 & 798$\pm$63 &
1.8$\pm$0.2 \\
 &          &     & J=3--2 & 41991$\pm$33 & 5.7$\pm$0.5 & 707$\pm$70 &
3.8$\pm$0.5 \\
 &          &     & J=4--3 & 41873$\pm$35, 42294$\pm$40 & 6.5$\pm$0.5,
5.0$\pm$0.8 & 427$\pm$87, 262$\pm$89 & 3.8$\pm$0.7 \\  
 &          & HCO$^{+}$ & J=2--1 & 41985$\pm$24 & 2.8$\pm$0.2 &
703$\pm$52 & 1.9$\pm$0.2 \\
 &          &     & J=3--2 & 42000$\pm$24 & 7.1$\pm$0.5 & 726$\pm$56 &
4.8$\pm$0.5 \\
 &          &     & J=4--3 & 41764$\pm$12, 42144$\pm$66 & 6.1$\pm$1.0,
5.4$\pm$0.4 & 179$\pm$75, 646$\pm$120 & 4.3$\pm$0.8 \\  
 & 0.5--1 kpc & HCN & J=2--1 & 41952$\pm$57 & 0.47$\pm$0.09 & 656$\pm$145 
& 0.29$\pm$0.08 \\
& & & J=3--2 & 41954$\pm$48 & 1.5$\pm$0.2 & 753$\pm$114 & 1.0$\pm$0.2 \\
& & & J=4--3 & 41984$\pm$44 & 1.3$\pm$0.2 & 709$\pm$96 & 0.87$\pm$0.16 \\
 & & HCO$^{+}$ & J=2--1 & 41922$\pm$44 & 0.67$\pm$0.10 & 646$\pm$101 & 0.40$\pm$0.09 \\
& & & J=3--2 & 41979$\pm$26 & 2.5$\pm$0.2 & 673$\pm$62 & 1.5$\pm$0.2 \\
& & & J=4--3 & 41913$\pm$43 & 1.6$\pm$0.3 & 535$\pm$100 & 0.82$\pm$0.21 \\ 
 & 1--2 kpc & HCN & J=2--1 & 42089$\pm$164 & 0.48$\pm$0.11 &
955$\pm$443 & 0.43$\pm$0.22 ($<$3$\sigma$) \\
& & & J=3--2 & 41940$\pm$78 & 1.6$\pm$0.4 & 576$\pm$190 &
0.84$\pm$0.34 ($<$3$\sigma$) \\
& & & J=4--3 & 41946$\pm$66 & 1.3$\pm$0.3 & 543$\pm$167 &
0.67$\pm$0.26 ($<$3$\sigma$) \\
& & HCO$^{+}$ & J=2--1 & 41956$\pm$67 & 0.70$\pm$0.14 & 620$\pm$156 & 0.40$\pm$0.13 \\
& & & J=3--2 & 41987$\pm$70 & 1.6$\pm$0.3 & 724$\pm$186 & 1.1$\pm$0.35  \\
& & & J=4--3 & 41897$\pm$89 & 1.3$\pm$0.5 & 768$\pm$302 &
0.94$\pm$0.54 ($<$3$\sigma$)  \\ \hline
IRAS 03250$+$1606 & $\lesssim$0.5 kpc & HCN & J=2--1 & 38607$\pm$21 & 1.2$\pm$0.1 & 582$\pm$49 & 0.67$\pm$0.07 \\
& & & J=3--2 & 38407$\pm$10, 38700$\pm$30 & 2.6$\pm$0.4, 2.2$\pm$0.2 &
130$\pm$34, 453$\pm$50 & 1.3$\pm$0.2 \\ 
& & & J=4--3 & 38443$\pm$20, 38780$\pm$38 & 2.7$\pm$0.4, 2.2$\pm$0.2 &
199$\pm$40, 317$\pm$84 & 1.2$\pm$0.2 \\ 
 & & HCO$^{+}$ & J=2--1 & 38607$\pm$32 & 0.94$\pm$0.10 & 533$\pm$71 & 0.47$\pm$0.08 \\
& & & J=3--2 & 38449$\pm$23, 38751$\pm$43 & 2.1$\pm$0.3, 1.8$\pm$0.2 &
221$\pm$76, 279$\pm$77 & 0.92$\pm$0.22 \\ 
& & & J=4--3 & 38505$\pm$35, 38844$\pm$62 & 2.2$\pm$0.3, 1.8$\pm$0.3 &
287$\pm$78, 212$\pm$132 & 0.97$\pm$0.30 \\  
 & $\lesssim$1 kpc & HCN & J=2--1 & 38615$\pm$20 & 1.9$\pm$0.1 & 583$\pm$43 &
1.1$\pm$0.1 \\
 &          &     & J=3--2 & 38455$\pm$25, 38800$\pm$52 & 4.6$\pm$0.4,
3.5$\pm$0.4 & 233$\pm$55, 298$\pm$95 & 2.0$\pm$0.4 \\
 &          &     & J=4--3 & 38445$\pm$32, 38793$\pm$62 & 4.1$\pm$0.6,
3.7$\pm$0.4 & 240$\pm$80, 309$\pm$126 & 2.0$\pm$0.6 \\  
 &          & HCO$^{+}$ & J=2--1 & 38614$\pm$26 & 1.7$\pm$0.1 &
577$\pm$53 & 0.92$\pm$0.12 \\
 &          &     & J=3--2 & 38447$\pm$23, 38769$\pm$34 & 3.5$\pm$0.5,
3.7$\pm$0.3 & 210$\pm$83, 295$\pm$69 & 1.7$\pm$0.4 \\
 &          &     & J=4--3 & 38463$\pm$36, 38780$\pm$23 & 3.7$\pm$0.7,
3.9$\pm$0.5 & 148$\pm$94, 309$\pm$62 & 1.6$\pm$0.4 \\  
 & $\lesssim$2 kpc & HCN & J=2--1 & 38580$\pm$24 & 2.7$\pm$0.2 & 512$\pm$55 &
1.3$\pm$0.2 \\
 &          &     & J=3--2 & 38601$\pm$27 & 5.2$\pm$0.6 & 555$\pm$55 &
2.7$\pm$0.4 \\
 &          &     & J=4--3 & 38423$\pm$30, 38796$\pm$39 & 4.9$\pm$0.9,
4.9$\pm$0.8 & 205$\pm$81, 316$\pm$81 & 2.4$\pm$0.6 \\  
 &          & HCO$^{+}$ & J=2--1 & 38638$\pm$38 & 2.4$\pm$0.3 &
659$\pm$85 & 1.5$\pm$0.3 \\
 &          &     & J=3--2 & 38665$\pm$23 & 5.5$\pm$0.5 & 509$\pm$51 &
2.6$\pm$0.3 \\
 &          &     & J=4--3 & 38452$\pm$121, 38761$\pm$34 &
4.5$\pm$1.6, 5.8$\pm$1.0 & 137$\pm$176, 305$\pm$72 & 2.3$\pm$0.9 ($<$3$\sigma$) \\  
 & 0.5--1 kpc & HCN & J=2--1 & 38626$\pm$38 & 0.71$\pm$0.09 & 581$\pm$93 & 0.39$\pm$0.08 \\
& & & J=3--2 & 38481$\pm$24, 38834$\pm$29 & 1.8$\pm$0.2, 1.5$\pm$0.3 &
240$\pm$65, 223$\pm$78 & 0.71$\pm$0.17 \\ 
& & & J=4--3 & 38634$\pm$40 & 1.6$\pm$0.2 & 575$\pm$94 & 0.86$\pm$0.17 \\
 & & HCO$^{+}$ & J=2--1 & 38629$\pm$49 & 0.78$\pm$0.11 & 605$\pm$130 & 0.44$\pm$0.12 \\
& & & J=3--2 & 38456$\pm$42, 38796$\pm$49 & 1.4$\pm$0.4, 1.9$\pm$0.2 &
217$\pm$133, 283$\pm$56 & 0.80$\pm$0.23 \\ 
& & & J=4--3 & 38757$\pm$34 & 1.9$\pm$0.5 & 366$\pm$90 & 0.66$\pm$0.23
($<$3$\sigma$) \\ 
 & 1--2 kpc & HCN & J=2--1 & 38529$\pm$58 & 0.79$\pm$0.22 &
372$\pm$136 & 0.28$\pm$0.13 ($<$3$\sigma$) \\
& & & J=3--2 & 38592$\pm$68 & 1.3$\pm$0.6 & 555$\pm$112 &
0.70$\pm$0.33 ($<$3$\sigma$) \\
& & & J=4--3 & --- & --- & --- & --- \\
& & HCO$^{+}$ & J=2--1 & 38676$\pm$133 & 0.76$\pm$0.22 & 854$\pm$253 &
0.61$\pm$0.25 ($<$3$\sigma$) \\
& & & J=3--2 & 38715$\pm$34 & 2.1$\pm$0.3 & 440$\pm$89 & 0.85$\pm$0.21 \\
& & & J=4--3 & 38725$\pm$63 & 1.9$\pm$0.8 & 306$\pm$149 &
0.56$\pm$0.36 ($<$3$\sigma$) \\ \hline
IRAS 10378$+$1108 & $\lesssim$0.5 kpc & HCN & J=2--1 & 40913$\pm$9 & 2.8$\pm$0.2 & 342$\pm$24 
& 0.90$\pm$0.08 \\
& & & J=3--2 & 40934$\pm$4 & 7.2$\pm$0.2 & 402$\pm$11 & 2.7$\pm$0.1 \\
 & & HCO$^{+}$ & J=2--1 & 40930$\pm$13 & 2.3$\pm$0.2 & 373$\pm$34 & 0.82$\pm$0.09 \\
& & & J=3--2 & 40934$\pm$5 & 6.9$\pm$0.2 & 381$\pm$12 & 2.5$\pm$0.1 \\ 
 & $\lesssim$1 kpc & HCN & J=2--1 & 40896$\pm$8 & 4.7$\pm$0.2 & 341$\pm$18 &
1.5$\pm$0.1 \\
 &          &     & J=3--2 & 40915$\pm$5 & 11.3$\pm$0.3 & 383$\pm$13 &
4.1$\pm$0.2 \\
 &          & HCO$^{+}$ & J=2--1 & 40906$\pm$12 & 4.3$\pm$0.3 &
388$\pm$27 & 1.6$\pm$0.1 \\
 &          &     & J=3--2 & 40916$\pm$5 & 11.3$\pm$0.3 & 362$\pm$12 &
3.8$\pm$0.2 \\
 & $\lesssim$2 kpc & HCN & J=2--1 & 40892$\pm$10 & 7.3$\pm$0.5 & 327$\pm$24 &
2.2$\pm$0.2 \\
 &          &     & J=3--2 & 40913$\pm$7 & 14.3$\pm$0.7 & 388$\pm$20 &
5.2$\pm$0.4 \\
 &          & HCO$^{+}$ & J=2--1 & 40907$\pm$16 & 6.4$\pm$0.5 &
427$\pm$46 & 2.6$\pm$0.3 \\
 &          &     & J=3--2 & 40906$\pm$9 & 14.5$\pm$0.6 & 366$\pm$21 &
5.0$\pm$0.4 \\
 & 0.5--1 kpc & HCN & J=2--1 & 40869$\pm$15 & 1.9$\pm$0.2 & 312$\pm$44 
& 0.56$\pm$0.09 \\
& & & J=3--2 & 40886$\pm$9 & 4.2$\pm$0.3 & 339$\pm$24 & 1.3$\pm$0.1 \\
 & & HCO$^{+}$ & J=2--1 & 40879$\pm$19 & 2.0$\pm$0.2 & 385$\pm$47 & 0.72$\pm$0.13 \\
& & & J=3--2 & 40889$\pm$10 & 4.5$\pm$0.3 & 328$\pm$28 & 1.4$\pm$0.1 \\ 
 & 1--2 kpc & HCN & J=2--1 & 40886$\pm$21 & 2.7$\pm$0.4 & 294$\pm$58 & 0.73$\pm$0.18 \\
& & & J=3--2 & 40903$\pm$23 & 3.1$\pm$0.4 & 403$\pm$63 & 1.2$\pm$0.2 \\
& & HCO$^{+}$ & J=2--1 & 40903$\pm$43 & 2.1$\pm$0.3 & 535$\pm$117 & 1.1$\pm$0.3 \\
& & & J=3--2 & 40872$\pm$27 & 3.3$\pm$0.5 & 357$\pm$66 & 1.1$\pm$0.3 \\ \hline
IRAS 16090$-$0139 & $\lesssim$0.5 kpc & HCN & J=2--1 & 40036$\pm$7 & 4.4$\pm$0.1 & 570$\pm$17 
& 2.4$\pm$0.1 \\
& & & J=3--2 & 40041$\pm$7 & 8.0$\pm$0.2 & 538$\pm$18 & 4.1$\pm$0.2 \\
& & & J=4--3 & 40017$\pm$5 & 15.8$\pm$0.3 & 556$\pm$12 & 8.3$\pm$0.2 \\
 & & HCO$^{+}$ & J=2--1 & 39998$\pm$14 & 3.2$\pm$0.1 & 624$\pm$36 & 1.9$\pm$0.1 \\
& & & J=3--2 & 40035$\pm$13 & 6.2$\pm$0.3 & 570$\pm$31 & 3.3$\pm$0.2 \\
& & & J=4--3 & 40021$\pm$5 & 12.8$\pm$0.3 & 480$\pm$11 & 5.8$\pm$0.2 \\ 
 & $\lesssim$1 kpc & HCN & J=2--1 & 40052$\pm$5 & 7.9$\pm$0.2 & 537$\pm$12 &
4.0$\pm$0.1 \\
 &          &     & J=3--2 & 40054$\pm$9 & 13.6$\pm$0.4 & 540$\pm$20 &
6.9$\pm$0.3 \\
 &          &     & J=4--3 & 40043$\pm$5 & 23.5$\pm$0.5 & 572$\pm$13 &
12.6$\pm$0.4 \\  
 &          & HCO$^{+}$ & J=2--1 & 40030$\pm$7 & 6.2$\pm$0.2 &
579$\pm$18 & 3.4$\pm$0.1 \\
 &          &     & J=3--2 & 40066$\pm$13 & 12.0$\pm$0.5 & 576$\pm$31 &
6.5$\pm$0.4 \\
 &          &     & J=4--3 & 40050$\pm$5 & 19.9$\pm$0.4 & 505$\pm$27 &
9.4$\pm$0.5 \\  
 & $\lesssim$2 kpc & HCN & J=2--1 & 40064$\pm$7 & 11.5$\pm$0.3 & 487$\pm$18 &
5.3$\pm$0.2 \\
 &          &     & J=3--2 & 40057$\pm$12 & 18.6$\pm$0.8 & 541$\pm$28 &
9.5$\pm$0.6 \\
 &          &     & J=4--3 & 40042$\pm$9 & 29.0$\pm$0.9 & 570$\pm$21 &
15.5$\pm$0.7 \\  
 &          & HCO$^{+}$ & J=2--1 & 40072$\pm$10 & 9.4$\pm$0.4 &
569$\pm$24 & 5.0$\pm$0.3 \\
 &          &     & J=3--2 & 40087$\pm$13 & 18.5$\pm$0.9 & 596$\pm$30 &
10.3$\pm$0.7 \\
 &          &     & J=4--3 & 40069$\pm$8 & 26.2$\pm$0.9 & 490$\pm$20 &
12.1$\pm$0.6 \\  
 & 0.5--1 kpc & HCN & J=2--1 & 40070$\pm$9 & 3.6$\pm$0.1 & 483$\pm$21 & 1.6$\pm$0.1 \\
& & & J=3--2 & 40072$\pm$13 & 5.6$\pm$0.3 & 535$\pm$32 & 2.8$\pm$0.2 \\
& & & J=4--3 & 40094$\pm$11 & 7.9$\pm$0.3 & 580$\pm$29 & 4.3$\pm$0.3 \\
 & & HCO$^{+}$ & J=2--1 & 40061$\pm$11 & 3.2$\pm$0.1 & 495$\pm$27 & 1.5$\pm$0.1 \\
& & & J=3--2 & 40094$\pm$15 & 5.9$\pm$0.3 & 581$\pm$32 & 3.2$\pm$0.2 \\
& & & J=4--3 & 40104$\pm$10 & 7.4$\pm$0.3 & 523$\pm$25 & 3.6$\pm$0.2 \\ 
 & 1--2 kpc & HCN & J=2--1 & 40091$\pm$15 & 3.8$\pm$0.3 & 389$\pm$34 & 1.4$\pm$0.2 \\
& & & J=3--2 & 40065$\pm$28 & 5.0$\pm$0.5 & 544$\pm$72 & 2.5$\pm$0.4 \\
& & & J=4--3 & 40039$\pm$29 & 5.5$\pm$0.7 & 556$\pm$89 & 2.9$\pm$0.6 \\
& & HCO$^{+}$ & J=2--1 & 40143$\pm$16 & 3.8$\pm$0.3 & 432$\pm$47 & 1.5$\pm$0.2 \\
& & & J=3--2 & 40129$\pm$25 & 6.8$\pm$0.6 & 587$\pm$59 & 3.7$\pm$0.5 \\
& & & J=4--3 & 40133$\pm$20 & 7.4$\pm$0.7 & 358$\pm$48 & 2.5$\pm$0.4 \\ \hline
IRAS 22206$-$2715 & $\lesssim$0.5 kpc & HCN & J=2--1 & 39496$\pm$7 & 3.1$\pm$0.1 & 439$\pm$16 
& 1.3$\pm$0.1 \\
& & & J=3--2 & 39495$\pm$8 & 5.2$\pm$0.2 & 420$\pm$18 & 2.1$\pm$0.1 \\
& & & J=4--3 & 39492$\pm$9 & 7.2$\pm$0.3 & 417$\pm$22 & 2.8$\pm$0.2 \\
 & & HCO$^{+}$ & J=2--1 & 39524$\pm$12 & 2.2$\pm$0.1 & 423$\pm$29 & 0.86$\pm$0.08 \\
& & & J=3--2 & 39517$\pm$10 & 3.7$\pm$0.2 & 438$\pm$24 & 1.5$\pm$0.1 \\
& & & J=4--3 & 39545$\pm$14 & 4.9$\pm$0.3 & 479$\pm$37 & 2.2$\pm$0.2 \\ 
 & $\lesssim$1 kpc & HCN & J=2--1 & 39493$\pm$6 & 4.7$\pm$0.1 & 398$\pm$14 &
1.8$\pm$0.1 \\
 &          &     & J=3--2 & 39500$\pm$7 & 7.4$\pm$0.2 & 432$\pm$17 &
3.0$\pm$0.2 \\
 &          &     & J=4--3 & 39481$\pm$10 & 9.8$\pm$0.5 & 430$\pm$24 &
4.0$\pm$0.3 \\  
 &          & HCO$^{+}$ & J=2--1 & 39517$\pm$10 & 3.6$\pm$0.2 &
361$\pm$24 & 1.2$\pm$0.1 \\
 &          &     & J=3--2 & 39513$\pm$10 & 5.6$\pm$0.3 & 412$\pm$23 &
2.2$\pm$0.2 \\
 &          &     & J=4--3 & 39508$\pm$15 & 7.0$\pm$0.5 & 424$\pm$36 &
2.8$\pm$0.3 \\  
 & $\lesssim$2 kpc & HCN & J=2--1 & 39489$\pm$9 & 6.7$\pm$0.3 & 373$\pm$21 &
2.4$\pm$0.2 \\
 &          &     & J=3--2 & 39513$\pm$11 & 9.3$\pm$0.5 & 463$\pm$29 &
4.1$\pm$0.3 \\
 &          &     & J=4--3 & 39470$\pm$20 & 12.6$\pm$1.2 & 423$\pm$42
& 5.0$\pm$0.7 \\  
 &          & HCO$^{+}$ & J=2--1 & 39523$\pm$12 & 5.1$\pm$0.5 &
270$\pm$34 & 1.3$\pm$0.2 \\
 &          &     & J=3--2 & 39508$\pm$13 & 7.4$\pm$0.5 & 383$\pm$30 &
2.7$\pm$0.3 \\
 &          &     & J=4--3 & 39515$\pm$34 & 9.3$\pm$1.1 & 493$\pm$83 &
4.3$\pm$0.9 \\  
 & 0.5--1 kpc & HCN & J=2--1 & 39482$\pm$10 & 1.7$\pm$0.1 & 317$\pm$25 
& 0.51$\pm$0.05 \\
& & & J=3--2 & 39518$\pm$20 & 2.2$\pm$0.2 & 474$\pm$53 & 0.99$\pm$0.14 \\
& & & J=4--3 & 39450$\pm$31 & 2.6$\pm$0.3 & 460$\pm$65 & 1.1$\pm$0.2 \\
 & & HCO$^{+}$ & J=2--1 & 39506$\pm$16 & 1.5$\pm$0.2 & 259$\pm$43 & 0.35$\pm$0.08 \\
& & & J=3--2 & 39506$\pm$18 & 1.9$\pm$0.2 & 349$\pm$46 & 0.61$\pm$0.11 \\
& & & J=4--3 & 39422$\pm$23 & 3.2$\pm$0.7 & 212$\pm$54 & 0.63$\pm$0.21 \\ 
 & 1--2 kpc & HCN & J=2--1 & 39477$\pm$20 & 2.0$\pm$0.3 & 306$\pm$50 &
0.58$\pm$0.13 \\ 
& & & J=3--2 & 39601$\pm$74 & 1.9$\pm$0.3 & 669$\pm$196 &
1.2$\pm$0.4 \\
& & & J=4--3 & 39435$\pm$84 & 2.4$\pm$1.1 & 474$\pm$140 & 1.1$\pm$0.6 ($<$3$\sigma$) \\
& & HCO$^{+}$ & J=2--1 & 39523$\pm$12 & 2.3$\pm$0.6 & 109$\pm$42 &
0.23$\pm$0.11 ($<$3$\sigma$) \\
& & & J=3--2 & 39499$\pm$38 & 1.8$\pm$0.5 & 277$\pm$113 &
0.48$\pm$0.23 ($<$3$\sigma$) \\
& & & J=4--3 & 39550$\pm$137 & 2.4$\pm$0.7 & 695$\pm$295 & 1.6$\pm$0.8
($<$3$\sigma$) \\ \hline
IRAS 22491$-$1808 & $\lesssim$0.5 kpc & HCN & J=2--1 & 23303$\pm$4 & 10.8$\pm$0.2 & 452$\pm$9 
& 4.8$\pm$0.1 \\
& & & J=3--2 & 23310$\pm$5 & 18.6$\pm$0.4 & 499$\pm$13 & 9.2$\pm$0.3 \\
& & & J=4--3 & 23305$\pm$6 & 24.4$\pm$0.6 & 514$\pm$14 & 12.4$\pm$0.4 \\
 & & HCO$^{+}$ & J=2--1 & 23276$\pm$8 & 6.6$\pm$0.2 & 490$\pm$17 & 3.2$\pm$0.1 \\
& & & J=3--2 & 23334$\pm$12 & 10.9$\pm$0.4 & 592$\pm$24 & 6.4$\pm$0.3 \\
& & & J=4--3 & 23333$\pm$16 & 13.4$\pm$0.6 & 626$\pm$34 & 8.3$\pm$0.6 \\ 
 & $\lesssim$1 kpc & HCN & J=2--1 & 23311$\pm$5 & 14.1$\pm$0.4 & 425$\pm$12 &
5.9$\pm$0.2 \\
 &          &     & J=3--2 & 23323$\pm$6 & 22.9$\pm$0.6 & 474$\pm$14 &
10.7$\pm$0.4 \\
 &          &     & J=4--3 & 23303$\pm$10 & 31.4$\pm$1.2 & 505$\pm$22 &
15.7$\pm$0.9 \\  
 &          & HCO$^{+}$ & J=2--1 & 23295$\pm$8 & 9.4$\pm$0.3 &
442$\pm$18 & 4.1$\pm$0.2 \\
 &          &     & J=3--2 & 23341$\pm$12 & 14.3$\pm$0.6 & 525$\pm$28 &
7.4$\pm$0.5 \\
 &          &     & J=4--3 & 23356$\pm$22 & 20.2$\pm$1.2 & 656$\pm$51
& 13.1$\pm$1.3 \\  
 & $\lesssim$2 kpc & HCN & J=2--1 & 23313$\pm$10 & 16.4$\pm$0.9 & 385$\pm$23 &
6.2$\pm$0.5 \\
 &          &     & J=3--2 & 23326$\pm$11 & 25.6$\pm$1.2 & 464$\pm$26
& 11.8$\pm$0.9 \\
 &          &     & J=4--3 & 23306$\pm$23 & 29.6$\pm$2.6 & 502$\pm$52 &
14.7$\pm$2.0 \\  
 &          & HCO$^{+}$ & J=2--1 & 23308$\pm$14 & 12.0$\pm$0.9 &
388$\pm$32 & 4.6$\pm$0.5 \\
 &          &     & J=3--2 & 23354$\pm$18 & 17.5$\pm$1.2 & 485$\pm$40
& 8.4$\pm$0.9 \\
 &          &     & J=4--3 & 23352$\pm$38 & 26.7$\pm$2.8 & 612$\pm$99 &
16.1$\pm$3.1 \\  
 & 0.5--1 kpc & HCN & J=2--1 & 23345$\pm$15 & 3.5$\pm$0.3 & 325$\pm$33 
& 1.1$\pm$0.2 \\
& & & J=3--2 & 23371$\pm$18 & 4.8$\pm$0.6 & 325$\pm$50 & 1.6$\pm$0.3 \\
& & & J=4--3 & 23296$\pm$32 & 7.0$\pm$0.9 & 482$\pm$74 & 3.3$\pm$0.7 \\
 & & HCO$^{+}$ & J=2--1 & 23343$\pm$16 & 3.3$\pm$0.3 & 292$\pm$38 & 0.95$\pm$0.16 \\
& & & J=3--2 & 23369$\pm$24 & 3.8$\pm$0.7 & 326$\pm$65 & 1.2$\pm$0.3 \\
& & & J=4--3 & 23351$\pm$37 & 7.4$\pm$0.9 & 575$\pm$98 & 4.2$\pm$0.9 \\
 & 1--2 kpc & HCN & J=2--1 & 23344$\pm$38 & 2.9$\pm$1.0 & 174$\pm$94 &
0.50$\pm$0.33 ($<$3$\sigma$) \\
& & & J=3--2 & --- & --- & --- & --- \\
& & & J=4--3 & --- & --- & --- & --- \\ 
& & HCO$^{+}$ & J=2--1 & 23381$\pm$24 & 4.3$\pm$1.1 & 131$\pm$39 &
0.56$\pm$0.22 ($<$3$\sigma$) \\
& & & J=3--2 & --- & --- & --- & --- \\
& & & J=4--3 & --- & --- & --- & --- \\ \hline
IRAS 12112$+$0305 & $\lesssim$0.5 kpc & HCN & J=4--3 & 21668$\pm$15, 21985$\pm$9 &
22.3$\pm$1.6, 25.3$\pm$1.9 & 301$\pm$38, 199$\pm$23 & 11.6$\pm$1.2 \\
 & & HCO$^{+}$ & J=4--3 & 21643$\pm$13, 22013$\pm$17 & 12.6$\pm$1.5,
13.8$\pm$1.2 & 181$\pm$33, 321$\pm$58 & 6.7$\pm$1.0 \\ 
 & $\lesssim$1 kpc & HCN & J=4--3 & 21835$\pm$17 & 31.5$\pm$1.9 & 587$\pm$38
& 18.4$\pm$1.6 \\
 &          & HCO$^{+}$ & J=4--3 & 21975$\pm$35 & 17.0$\pm$1.7 &
670$\pm$80 & 11.3$\pm$1.8 \\
 & $\lesssim$2 kpc & HCN & J=4--3 & 21861$\pm$32 & 34.0$\pm$3.5 & 580$\pm$89
& 19.6$\pm$3.6 \\
 &          & HCO$^{+}$ & J=4--3 & 22031$\pm$46 & 23.3$\pm$3.7 &
561$\pm$110 & 13.0$\pm$3.3 \\
 & 0.5--1 kpc & HCN & J=4--3 & 21885$\pm$40 & 9.6$\pm$1.3 & 574$\pm$143 
& 5.5$\pm$1.6 \\
 & & HCO$^{+}$ & J=4--3 & 21900 (fix) & 5.0$\pm$1.3 & 568$\pm$357 &
2.8$\pm$1.9 ($<$3$\sigma$) \\
 & 1--2 kpc & HCN & J=4--3 & --- & --- & --- & --- \\
& & HCO$^{+}$ & J=4--3 & --- & --- & --- & --- \\ \hline
NGC 1614 & $\lesssim$0.5 kpc & HCN & J=2--1 & 4763$\pm$4 & 14.2$\pm$0.5 & 222$\pm$8 
& 3.3$\pm$0.2 \\
& & & J=3--2 & 4762$\pm$11 & 11.9$\pm$1.3 & 209$\pm$25 & 2.6$\pm$0.4 \\
& & & J=4--3 & 4779$\pm$10 & 12.2$\pm$1.2 & 208$\pm$23 & 2.6$\pm$0.4 \\
 & & HCO$^{+}$ & J=2--1 & 4767$\pm$3 & 23.1$\pm$0.5 & 229$\pm$7 & 5.5$\pm$0.2 \\
& & & J=3--2 & 4753$\pm$7 & 24.1$\pm$1.5 & 233$\pm$13 & 5.9$\pm$0.5 \\
& & & J=4--3 & 4776$\pm$6 & 38.7$\pm$2.2 & 211$\pm$13 & 8.5$\pm$0.7 \\ 
 & $\lesssim$1 kpc & HCN & J=2--1 & 4765$\pm$4 & 26.9$\pm$0.9 & 243$\pm$10 &
6.8$\pm$0.4 \\
 &          &     & J=3--2 & 4776$\pm$14 & 16.5$\pm$1.9 & 245$\pm$34 &
4.2$\pm$0.8 \\
 &          &     & J=4--3 & 4769$\pm$10 & 16.3$\pm$1.3 & 252$\pm$24 &
4.3$\pm$0.5 \\  
 &          & HCO$^{+}$ & J=2--1 & 4769$\pm$4 & 46.7$\pm$1.2 &
252$\pm$8 & 12.3$\pm$0.5 \\
 &          &     & J=3--2 & 4757$\pm$8 & 42.1$\pm$2.5 & 256$\pm$13 &
11.3$\pm$0.9 \\
 &          &     & J=4--3 & 4773$\pm$7 & 55.4$\pm$3.0 & 240$\pm$16 &
13.9$\pm$1.2 \\  
 & 0.5--1 kpc & HCN & J=2--1 & 4767$\pm$6 & 12.9$\pm$0.6 & 259$\pm$16 
& 3.5$\pm$0.3 \\
& & & J=3--2 & 4835$\pm$41 & 5.4$\pm$1.4 & 272$\pm$107 & 1.6$\pm$0.7
($<$3$\sigma$) \\ 
& & & J=4--3 & 4712$\pm$34 & 4.5$\pm$0.7 & 425$\pm$73 & 2.0$\pm$0.5 \\
 & & HCO$^{+}$ & J=2--1 & 4772$\pm$4 & 23.8$\pm$0.7 & 274$\pm$9 & 6.8$\pm$0.3 \\
& & & J=3--2 & 4661$\pm$10, 4840$\pm$12 & 18.6$\pm$2.1, 19.6$\pm$1.9 &
119$\pm$19, 148$\pm$22 & 5.4$\pm$0.7 \\
& & & J=4--3 & 4653$\pm$13, 4842$\pm$9 & 19.9$\pm$2.1, 24.4$\pm$2.6 &
109$\pm$26, 117$\pm$30 & 5.3$\pm$1.0 \\ \hline
\enddata

\tablenotetext{a}{
We adopt one broad Gaussian emission and one narrow Gaussian 
absorption components, because negative signals below the continuum 
level at the HCO$^{+}$ central dip, observed in IRAS 00091$-$0738
(Figures \ref{fig:SpectraA} and \ref{fig:SpectraB}), 
cannot be reproduced by two Gaussian emission components.
This flux estimate agrees within $\sim$10\% with that based on 
two Gaussian emission components. 
}

\tablenotetext{b}{Fixed to the best fit value.}

\tablecomments{
Col.(1): Object name.
Col.(2): Region.
Col.(3): Molecule.
Col.(4): J-transition.
Cols.(5)--(8): Gaussian fit of emission line.
``---'' means that no Gaussian fit is applied, because of no emission
line signature in a spectrum.
Col.(5): Optical LSR velocity (v$_{\rm opt}$) of emission line peak in
km s$^{-1}$.  
Col.(6): Peak flux in mJy. 
Col.(7): Observed full width at half maximum (FWHM) in km s$^{-1}$.
Col.(8): Gaussian-fit velocity-integrated flux in Jy km s$^{-1}$. 
When fitting results of two Gaussian components are adopted, fluxes of
the two components are added. 
Only Gaussian fitting error (statistical uncertainty) is considered.
}

\end{deluxetable}

\begin{figure*}
\includegraphics[angle=-90,scale=.118]{fC1a.eps} \hspace{0.15cm}
\includegraphics[angle=-90,scale=.118]{fC1b.eps} \hspace{0.15cm}
\includegraphics[angle=-90,scale=.118]{fC1c.eps} \hspace{0.15cm}
\includegraphics[angle=-90,scale=.118]{fC1d.eps} \hspace{0.15cm}
\includegraphics[angle=-90,scale=.118]{fC1e.eps} \hspace{0.15cm}
\includegraphics[angle=-90,scale=.118]{fC1f.eps} \hspace{0.15cm} \\
\includegraphics[angle=-90,scale=.118]{fC1g.eps} \hspace{0.15cm}
\includegraphics[angle=-90,scale=.118]{fC1h.eps} \hspace{0.15cm}
\includegraphics[angle=-90,scale=.118]{fC1i.eps} \hspace{0.15cm}
\includegraphics[angle=-90,scale=.118]{fC1j.eps} \\
\includegraphics[angle=-90,scale=.118]{fC1k.eps} \hspace{0.15cm}
\includegraphics[angle=-90,scale=.118]{fC1l.eps} \hspace{0.15cm}
\includegraphics[angle=-90,scale=.118]{fC1m.eps} \hspace{0.15cm}
\includegraphics[angle=-90,scale=.118]{fC1n.eps} \hspace{0.15cm}
\includegraphics[angle=-90,scale=.118]{fC1o.eps} \hspace{0.15cm}
\includegraphics[angle=-90,scale=.118]{fC1p.eps} \\
\includegraphics[angle=-90,scale=.118]{fC1q.eps} \hspace{0.15cm}
\includegraphics[angle=-90,scale=.118]{fC1r.eps} \hspace{0.15cm}
\includegraphics[angle=-90,scale=.118]{fC1s.eps} \hspace{0.15cm}
\includegraphics[angle=-90,scale=.118]{fC1t.eps} \\
\includegraphics[angle=-90,scale=.118]{fC1u.eps} \hspace{0.15cm}
\includegraphics[angle=-90,scale=.118]{fC1v.eps} \hspace{0.15cm}
\includegraphics[angle=-90,scale=.118]{fC1w.eps} \hspace{0.15cm}
\includegraphics[angle=-90,scale=.118]{fC1x.eps} \hspace{0.15cm}
\includegraphics[angle=-90,scale=.118]{fC1y.eps} \hspace{0.15cm}
\includegraphics[angle=-90,scale=.118]{fC1z.eps} \\
\includegraphics[angle=-90,scale=.118]{fC1aa.eps} \hspace{0.15cm}
\includegraphics[angle=-90,scale=.118]{fC1ab.eps} \hspace{0.15cm}
\includegraphics[angle=-90,scale=.118]{fC1ac.eps} \hspace{0.15cm}
\includegraphics[angle=-90,scale=.118]{fC1ad.eps} \\
\includegraphics[angle=-90,scale=.118]{fC1ae.eps} \hspace{0.05cm}
\includegraphics[angle=-90,scale=.118]{fC1af.eps} \hspace{0.05cm}
\includegraphics[angle=-90,scale=.118]{fC1ag.eps} \hspace{0.05cm}
\includegraphics[angle=-90,scale=.118]{fC1ah.eps} \hspace{0.05cm}
\includegraphics[angle=-90,scale=.118]{fC1ai.eps} \hspace{0.05cm}
\includegraphics[angle=-90,scale=.118]{fC1aj.eps} \\
\includegraphics[angle=-90,scale=.118]{fC1ak.eps} \hspace{0.05cm}
\includegraphics[angle=-90,scale=.118]{fC1al.eps} \hspace{0.05cm}
\includegraphics[angle=-90,scale=.118]{fC1am.eps} \hspace{0.05cm}
\includegraphics[angle=-90,scale=.118]{fC1an.eps} \\
\includegraphics[angle=-90,scale=.118]{fC1ao.eps} 
\includegraphics[angle=-90,scale=.118]{fC1ap.eps} \hspace{0.01cm}
\includegraphics[angle=-90,scale=.118]{fC1aq.eps} 
\includegraphics[angle=-90,scale=.118]{fC1ar.eps} \hspace{0.01cm}
\includegraphics[angle=-90,scale=.118]{fC1as.eps} 
\includegraphics[angle=-90,scale=.118]{fC1at.eps} \\
\includegraphics[angle=-90,scale=.118]{fC1au.eps} 
\includegraphics[angle=-90,scale=.118]{fC1av.eps} \hspace{0.01cm}
\includegraphics[angle=-90,scale=.118]{fC1aw.eps} 
\includegraphics[angle=-90,scale=.118]{fC1ax.eps} \\
\includegraphics[angle=-90,scale=.118]{fC1ay.eps} 
\includegraphics[angle=-90,scale=.118]{fC1az.eps} 
\includegraphics[angle=-90,scale=.118]{fC1ba.eps} 
\includegraphics[angle=-90,scale=.118]{fC1bb.eps} 
\includegraphics[angle=-90,scale=.118]{fC1bc.eps} 
\includegraphics[angle=-90,scale=.118]{fC1bd.eps} \\
\includegraphics[angle=-90,scale=.118]{fC1be.eps} 
\includegraphics[angle=-90,scale=.118]{fC1bf.eps} 
\includegraphics[angle=-90,scale=.118]{fC1bg.eps} 
\includegraphics[angle=-90,scale=.118]{fC1bh.eps} \\
\end{figure*}


\begin{figure*}
\includegraphics[angle=-90,scale=.118]{fC2a.eps} \hspace{0.15cm}
\includegraphics[angle=-90,scale=.118]{fC2b.eps} \hspace{0.15cm}
\includegraphics[angle=-90,scale=.118]{fC2c.eps} \hspace{0.15cm}
\includegraphics[angle=-90,scale=.118]{fC2d.eps} \hspace{0.15cm}
\includegraphics[angle=-90,scale=.118]{fC2e.eps} \hspace{0.15cm}
\includegraphics[angle=-90,scale=.118]{fC2f.eps} \\
\includegraphics[angle=-90,scale=.118]{fC2g.eps} \hspace{0.15cm}
\includegraphics[angle=-90,scale=.118]{fC2h.eps} \hspace{0.15cm}
\includegraphics[angle=-90,scale=.118]{fC2i.eps} \hspace{0.15cm}
\includegraphics[angle=-90,scale=.118]{fC2j.eps} \\
\includegraphics[angle=-90,scale=.118]{fC2k.eps} \hspace{0.15cm}
\includegraphics[angle=-90,scale=.118]{fC2l.eps} \hspace{0.15cm}
\includegraphics[angle=-90,scale=.118]{fC2m.eps} \hspace{0.15cm}
\includegraphics[angle=-90,scale=.118]{fC2n.eps} \hspace{0.15cm}
\includegraphics[angle=-90,scale=.118]{fC2o.eps} \hspace{0.15cm}
\includegraphics[angle=-90,scale=.118]{fC2p.eps} \\
\includegraphics[angle=-90,scale=.118]{fC2q.eps} \hspace{0.15cm}
\includegraphics[angle=-90,scale=.118]{fC2r.eps} \hspace{0.15cm}
\includegraphics[angle=-90,scale=.118]{fC2s.eps} \hspace{0.15cm}
\includegraphics[angle=-90,scale=.118]{fC2t.eps} \\
\includegraphics[angle=-90,scale=.118]{fC2u.eps} \hspace{0.15cm}
\includegraphics[angle=-90,scale=.118]{fC2v.eps} \hspace{0.15cm}
\includegraphics[angle=-90,scale=.118]{fC2w.eps} \hspace{0.15cm}
\includegraphics[angle=-90,scale=.118]{fC2x.eps} \hspace{0.15cm}
\includegraphics[angle=-90,scale=.118]{fC2y.eps} \hspace{0.15cm}
\includegraphics[angle=-90,scale=.118]{fC2z.eps} \\
\includegraphics[angle=-90,scale=.118]{fC2aa.eps} \hspace{0.15cm}
\includegraphics[angle=-90,scale=.118]{fC2ab.eps} \hspace{0.15cm}
\includegraphics[angle=-90,scale=.118]{fC2ac.eps} \hspace{0.15cm}
\includegraphics[angle=-90,scale=.118]{fC2ad.eps} \\
\includegraphics[angle=-90,scale=.118]{fC2ae.eps} \hspace{0.05cm}
\includegraphics[angle=-90,scale=.118]{fC2af.eps} \hspace{0.05cm}
\includegraphics[angle=-90,scale=.118]{fC2ag.eps} \hspace{0.05cm}
\includegraphics[angle=-90,scale=.118]{fC2ah.eps} \hspace{0.05cm}
\includegraphics[angle=-90,scale=.118]{fC2ai.eps} \hspace{0.05cm}
\includegraphics[angle=-90,scale=.118]{fC2aj.eps} \\
\includegraphics[angle=-90,scale=.118]{fC2ak.eps} \hspace{0.05cm}
\includegraphics[angle=-90,scale=.118]{fC2al.eps} \\
\includegraphics[angle=-90,scale=.118]{fC2am.eps} \hspace{0.01cm}
\includegraphics[angle=-90,scale=.118]{fC2an.eps} 
\includegraphics[angle=-90,scale=.118]{fC2ao.eps} \hspace{0.01cm}
\includegraphics[angle=-90,scale=.118]{fC2ap.eps} 
\includegraphics[angle=-90,scale=.118]{fC2aq.eps} \hspace{0.01cm}
\includegraphics[angle=-90,scale=.118]{fC2ar.eps} \\
\includegraphics[angle=-90,scale=.118]{fC2as.eps} \hspace{0.01cm}
\includegraphics[angle=-90,scale=.118]{fC2at.eps} 
\includegraphics[angle=-90,scale=.118]{fC2au.eps} \hspace{0.01cm}
\includegraphics[angle=-90,scale=.118]{fC2av.eps} \\
\includegraphics[angle=-90,scale=.118]{fC2aw.eps} 
\includegraphics[angle=-90,scale=.118]{fC2ax.eps} 
\includegraphics[angle=-90,scale=.118]{fC2ay.eps} 
\includegraphics[angle=-90,scale=.118]{fC2az.eps} 
\includegraphics[angle=-90,scale=.118]{fC2ba.eps} 
\includegraphics[angle=-90,scale=.118]{fC2bb.eps} \\
\includegraphics[angle=-90,scale=.118]{fC2bc.eps} 
\includegraphics[angle=-90,scale=.118]{fC2bd.eps} 
\includegraphics[angle=-90,scale=.118]{fC2be.eps} 
\includegraphics[angle=-90,scale=.118]{fC2bf.eps} \\
\end{figure*}


\begin{figure*}
\includegraphics[angle=-90,scale=.118]{fC2bg.eps} \hspace{0.15cm}
\includegraphics[angle=-90,scale=.118]{fC2bh.eps} \hspace{0.15cm}
\includegraphics[angle=-90,scale=.118]{fC2bi.eps} \hspace{0.15cm}
\includegraphics[angle=-90,scale=.118]{fC2bj.eps} \hspace{0.15cm}
\includegraphics[angle=-90,scale=.118]{fC2bk.eps} \hspace{0.15cm}
\includegraphics[angle=-90,scale=.118]{fC2bl.eps} \\
\includegraphics[angle=-90,scale=.118]{fC2bm.eps} \hspace{0.15cm}
\includegraphics[angle=-90,scale=.118]{fC2bn.eps} \hspace{0.15cm}
\includegraphics[angle=-90,scale=.118]{fC2bo.eps} \hspace{0.15cm}
\includegraphics[angle=-90,scale=.118]{fC2bp.eps} \\
\includegraphics[angle=-90,scale=.118]{fC2bq.eps} \hspace{0.15cm}
\includegraphics[angle=-90,scale=.118]{fC2br.eps} \hspace{0.15cm}
\includegraphics[angle=-90,scale=.118]{fC2bs.eps} \hspace{0.15cm}
\includegraphics[angle=-90,scale=.118]{fC2bt.eps} \hspace{0.15cm}
\includegraphics[angle=-90,scale=.118]{fC2bu.eps} \hspace{0.15cm}
\includegraphics[angle=-90,scale=.118]{fC2bv.eps} \\
\includegraphics[angle=-90,scale=.118]{fC2bw.eps} \hspace{0.15cm}
\includegraphics[angle=-90,scale=.118]{fC2bx.eps} \hspace{0.15cm}
\includegraphics[angle=-90,scale=.118]{fC2by.eps} \hspace{0.15cm}
\includegraphics[angle=-90,scale=.118]{fC2bz.eps} \\
\includegraphics[angle=-90,scale=.118]{fC2ca.eps} 
\includegraphics[angle=-90,scale=.118]{fC2cb.eps} 
\includegraphics[angle=-90,scale=.118]{fC2cc.eps} 
\includegraphics[angle=-90,scale=.118]{fC2cd.eps} 
\includegraphics[angle=-90,scale=.118]{fC2ce.eps} 
\includegraphics[angle=-90,scale=.118]{fC2cf.eps} \\
\includegraphics[angle=-90,scale=.118]{fC2cg.eps} 
\includegraphics[angle=-90,scale=.118]{fC2ch.eps} 
\includegraphics[angle=-90,scale=.118]{fC2ci.eps} 
\includegraphics[angle=-90,scale=.118]{fC2cj.eps} \\
\includegraphics[angle=-90,scale=.118]{fC2ck.eps} \hspace{0.15cm}
\includegraphics[angle=-90,scale=.118]{fC2cl.eps} \hspace{0.15cm}
\includegraphics[angle=-90,scale=.118]{fC2cm.eps} \hspace{0.15cm}
\includegraphics[angle=-90,scale=.118]{fC2cn.eps} \hspace{0.15cm}
\includegraphics[angle=-90,scale=.118]{fC2co.eps} \hspace{0.15cm}
\includegraphics[angle=-90,scale=.118]{fC2cp.eps} \hspace{0.15cm} \\
\includegraphics[angle=-90,scale=.118]{fC2cq.eps} \hspace{0.15cm}
\includegraphics[angle=-90,scale=.118]{fC2cr.eps} \\
\includegraphics[angle=-90,scale=.118]{fC2cs.eps} \hspace{0.15cm}
\includegraphics[angle=-90,scale=.118]{fC2ct.eps} \hspace{0.15cm}
\includegraphics[angle=-90,scale=.118]{fC2cu.eps} \hspace{0.15cm}
\includegraphics[angle=-90,scale=.118]{fC2cv.eps} \hspace{0.15cm}
\includegraphics[angle=-90,scale=.118]{fC2cw.eps} \hspace{0.15cm}
\includegraphics[angle=-90,scale=.118]{fC2cx.eps} \\
\caption{Same as Figure \ref{fig:GaussFit21}, but for the J=3--2 lines 
of HCN and HCO$^{+}$. 
No Gaussian fit is applied when there is no emission line signature at
all, particularly in the 1--2 kpc annular region spectra.
\label{fig:GaussFit32}
}
\end{figure*}

\begin{figure*}
\includegraphics[angle=-90,scale=.118]{fC3a.eps} \hspace{0.15cm}
\includegraphics[angle=-90,scale=.118]{fC3b.eps} \hspace{0.15cm}
\includegraphics[angle=-90,scale=.118]{fC3c.eps} \hspace{0.15cm}
\includegraphics[angle=-90,scale=.118]{fC3d.eps} \hspace{0.15cm}
\includegraphics[angle=-90,scale=.118]{fC3e.eps} \hspace{0.15cm}
\includegraphics[angle=-90,scale=.118]{fC3f.eps} \hspace{0.15cm} \\
\includegraphics[angle=-90,scale=.118]{fC3g.eps} \hspace{0.15cm}
\includegraphics[angle=-90,scale=.118]{fC3h.eps} \\
\includegraphics[angle=-90,scale=.118]{fC3i.eps} \hspace{0.15cm}
\includegraphics[angle=-90,scale=.118]{fC3j.eps} \hspace{0.15cm}
\includegraphics[angle=-90,scale=.118]{fC3k.eps} \hspace{0.15cm}
\includegraphics[angle=-90,scale=.118]{fC3l.eps} \hspace{0.15cm}
\includegraphics[angle=-90,scale=.118]{fC3m.eps} \hspace{0.15cm}
\includegraphics[angle=-90,scale=.118]{fC3n.eps} \\
\includegraphics[angle=-90,scale=.118]{fC3o.eps} \hspace{0.15cm}
\includegraphics[angle=-90,scale=.118]{fC3p.eps} \hspace{0.15cm}
\includegraphics[angle=-90,scale=.118]{fC3q.eps} \\
\includegraphics[angle=-90,scale=.118]{fC3r.eps} \hspace{0.15cm}
\includegraphics[angle=-90,scale=.118]{fC3s.eps} \hspace{0.15cm}
\includegraphics[angle=-90,scale=.118]{fC3t.eps} \hspace{0.15cm}
\includegraphics[angle=-90,scale=.118]{fC3u.eps} \hspace{0.15cm}
\includegraphics[angle=-90,scale=.118]{fC3v.eps} \hspace{0.15cm}
\includegraphics[angle=-90,scale=.118]{fC3w.eps} \\
\includegraphics[angle=-90,scale=.118]{fC3x.eps} \hspace{0.15cm}
\includegraphics[angle=-90,scale=.118]{fC3y.eps} \hspace{0.15cm}
\includegraphics[angle=-90,scale=.118]{fC3z.eps} \hspace{0.15cm}
\includegraphics[angle=-90,scale=.118]{fC3aa.eps} \\
\includegraphics[angle=-90,scale=.118]{fC3ab.eps} \hspace{0.05cm}
\includegraphics[angle=-90,scale=.118]{fC3ac.eps} \hspace{0.05cm}
\includegraphics[angle=-90,scale=.118]{fC3ad.eps} \hspace{0.05cm}
\includegraphics[angle=-90,scale=.118]{fC3ae.eps} 
\includegraphics[angle=-90,scale=.118]{fC3af.eps} \hspace{0.05cm}
\includegraphics[angle=-90,scale=.118]{fC3ag.eps} \\
\includegraphics[angle=-90,scale=.118]{fC3ah.eps} \hspace{0.05cm}
\includegraphics[angle=-90,scale=.118]{fC3ai.eps} \\
\includegraphics[angle=-90,scale=.118]{fC3aj.eps} \hspace{0.01cm}
\includegraphics[angle=-90,scale=.118]{fC3ak.eps} 
\includegraphics[angle=-90,scale=.118]{fC3al.eps}  
\includegraphics[angle=-90,scale=.118]{fC3am.eps} \hspace{0.01cm}
\includegraphics[angle=-90,scale=.118]{fC3an.eps} 
\includegraphics[angle=-90,scale=.118]{fC3ao.eps} \\
\includegraphics[angle=-90,scale=.118]{fC3ap.eps}  
\includegraphics[angle=-90,scale=.118]{fC3aq.eps} \hspace{0.01cm}
\includegraphics[angle=-90,scale=.118]{fC3ar.eps} 
\includegraphics[angle=-90,scale=.118]{fC3as.eps} \\
\includegraphics[angle=-90,scale=.118]{fC3at.eps} 
\includegraphics[angle=-90,scale=.118]{fC3au.eps} \hspace{0.01cm}
\includegraphics[angle=-90,scale=.118]{fC3av.eps} 
\includegraphics[angle=-90,scale=.118]{fC3aw.eps} \hspace{0.01cm}
\includegraphics[angle=-90,scale=.118]{fC3ax.eps} 
\includegraphics[angle=-90,scale=.118]{fC3ay.eps} \\
\includegraphics[angle=-90,scale=.118]{fC3az.eps} 
\includegraphics[angle=-90,scale=.118]{fC3ba.eps} \hspace{2.9cm}
\includegraphics[angle=-90,scale=.118]{fC3bb.eps} \\
\end{figure*}


\clearpage

\section{CS J=7--6 line}

\begin{figure*}[!hbt]
\begin{center}
\includegraphics[angle=0,scale=.34]{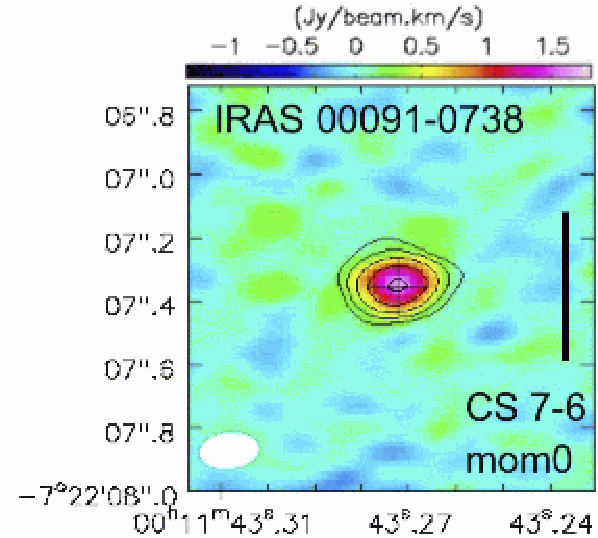} 
\includegraphics[angle=0,scale=.34]{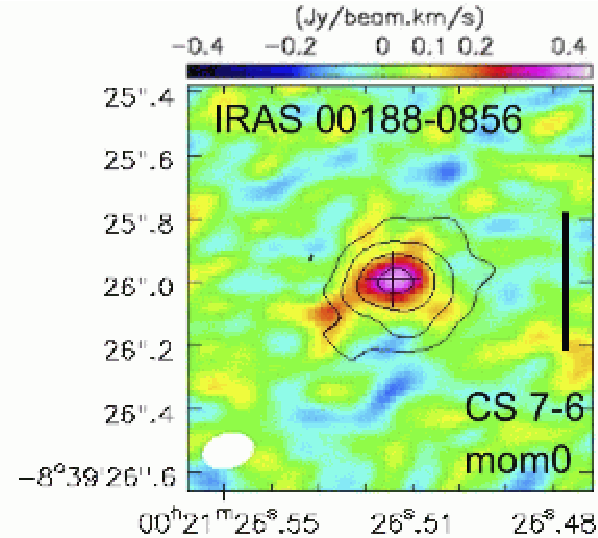} 
\includegraphics[angle=0,scale=.34]{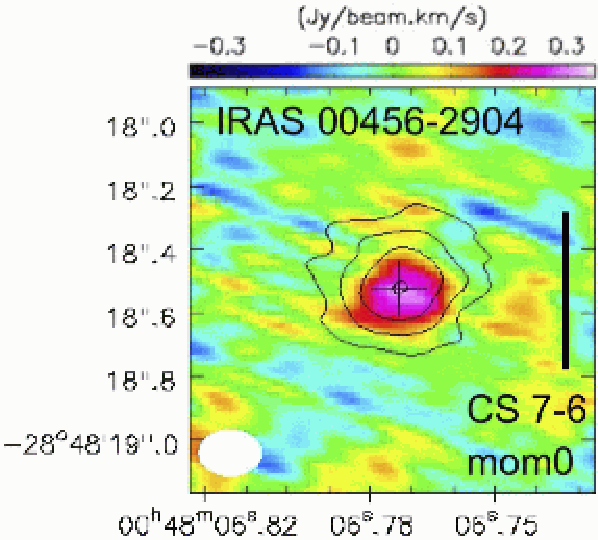} 
\includegraphics[angle=0,scale=.34]{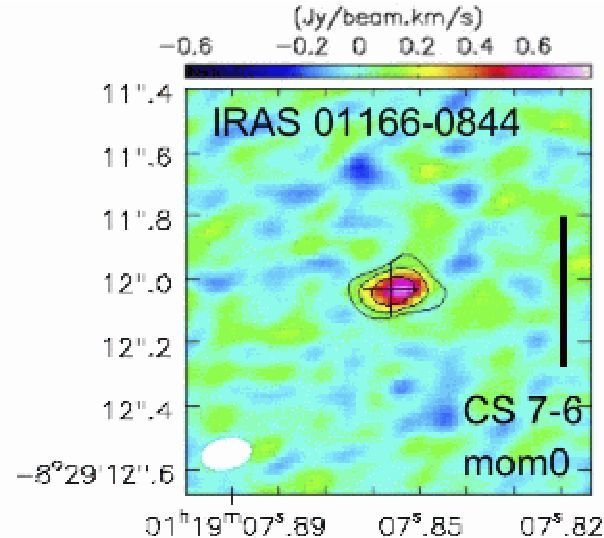} 
\includegraphics[angle=0,scale=.34]{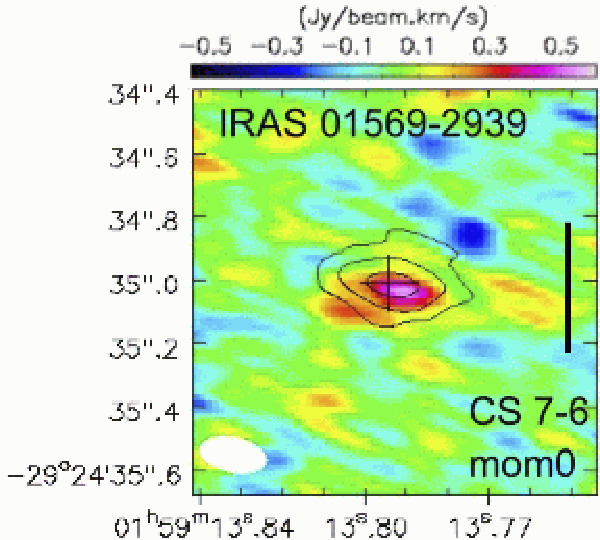}\\ 
\includegraphics[angle=0,scale=.34]{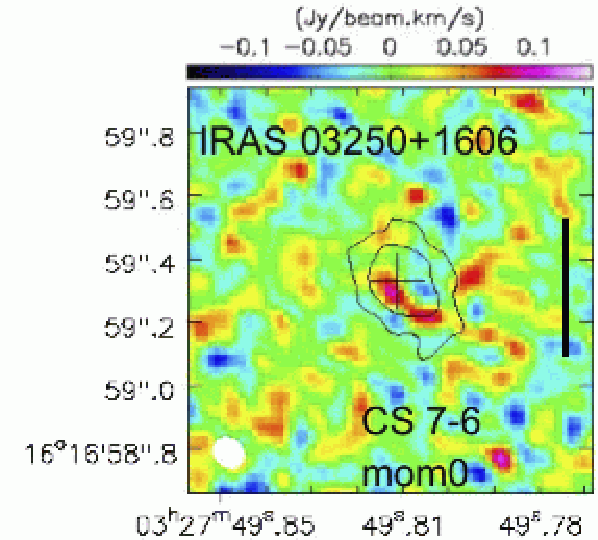} 
\includegraphics[angle=0,scale=.34]{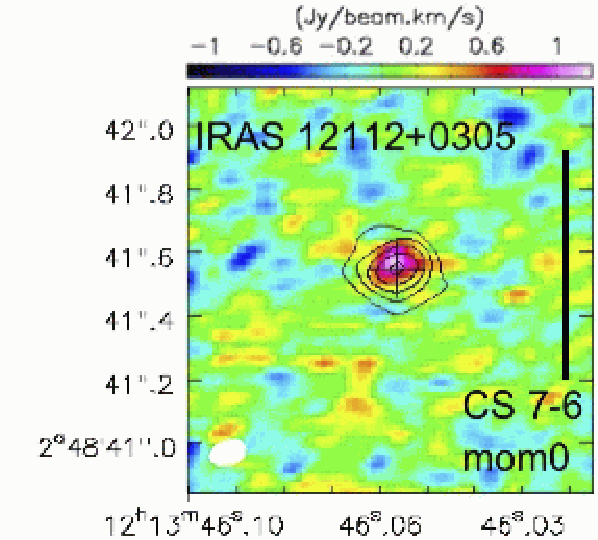} 
\includegraphics[angle=0,scale=.34]{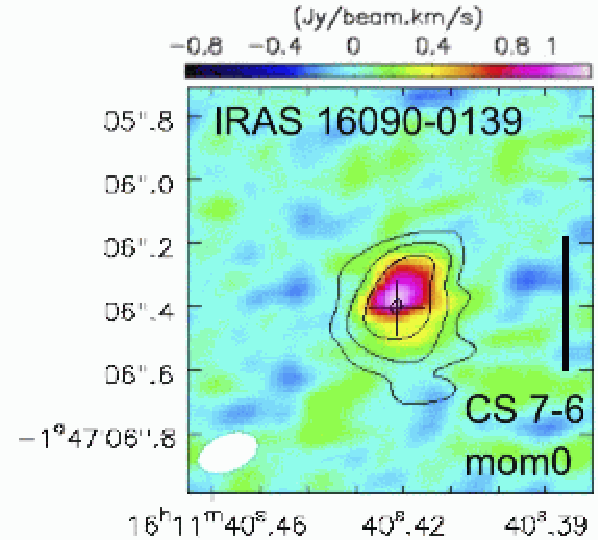} 
\includegraphics[angle=0,scale=.34]{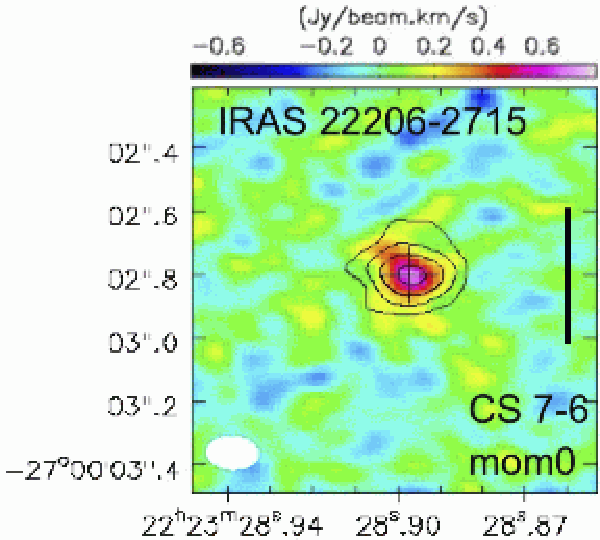}
\includegraphics[angle=0,scale=.34]{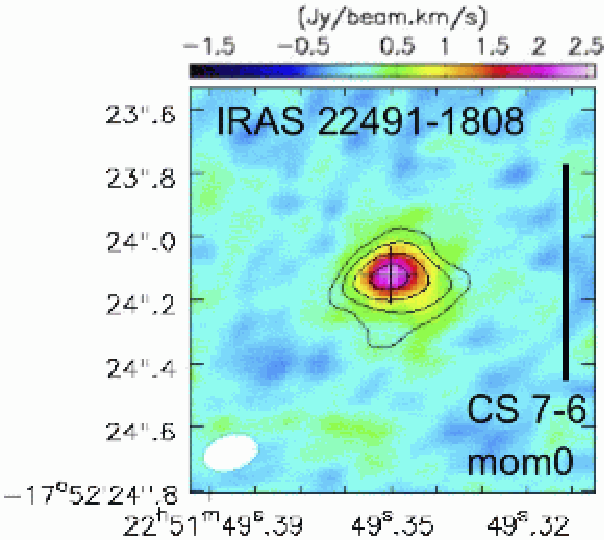}
\end{center}
\caption{Integrated intensity (moment 0) map of the CS J=7--6 line
created from the original-beam-sized data (Table \ref{tab:beam},
column 4).
Continuum emission that was simultaneously obtained (J43) is shown as
contours. 
The contours start from 4$\sigma$ and increase by a factor of 2 
(i.e., 8$\sigma$, 16$\sigma$, 32$\sigma$, and 64$\sigma$) for all
ULIRGs.
Continuum peak position is shown as a cross.
The length of the vertical black solid bar at the right side of each
object corresponds to 1 kpc. 
Beam size for each moment 0 map is shown as a white filled circle 
in the lower-left region.
\label{fig:CS76mom0}
}
\end{figure*}

\begin{figure*}[!hbt]
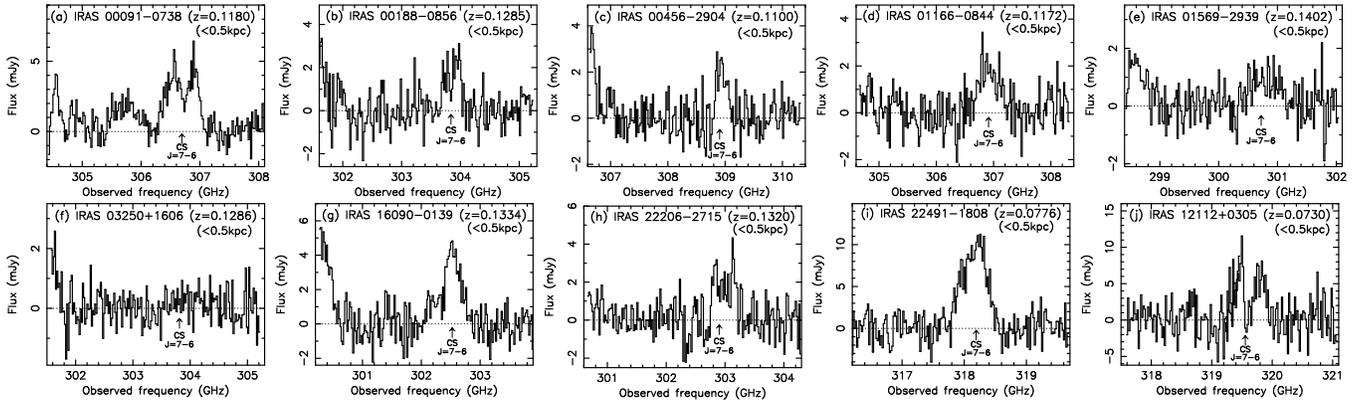

\includegraphics[angle=-90,scale=.155]{fD2a.eps} 
\includegraphics[angle=-90,scale=.155]{fD2b.eps} 
\includegraphics[angle=-90,scale=.155]{fD2c.eps} 
\includegraphics[angle=-90,scale=.155]{fD2d.eps} 
\includegraphics[angle=-90,scale=.155]{fD2e.eps} \\
\includegraphics[angle=-90,scale=.155]{fD2f.eps} 
\includegraphics[angle=-90,scale=.155]{fD2g.eps} 
\includegraphics[angle=-90,scale=.155]{fD2h.eps} 
\includegraphics[angle=-90,scale=.155]{fD2i.eps} 
\includegraphics[angle=-90,scale=.155]{fD2j.eps}
\caption{
0.5 kpc beam-sized spectrum that includes the CS J=7--6 line, taken
during the J43 observation.
The abscissa is observed frequency in GHz and the ordinate 
is flux density in mJy.
An upward arrow is placed at the expected observed frequency of CS
J=7--6 for the adopted redshift (Table \ref{tab:object}, column 2).
The horizontal thin dotted straight line indicates the zero flux
level. 
\label{fig:CS76spec}
}
\end{figure*}

\begin{figure*}[!hbt]
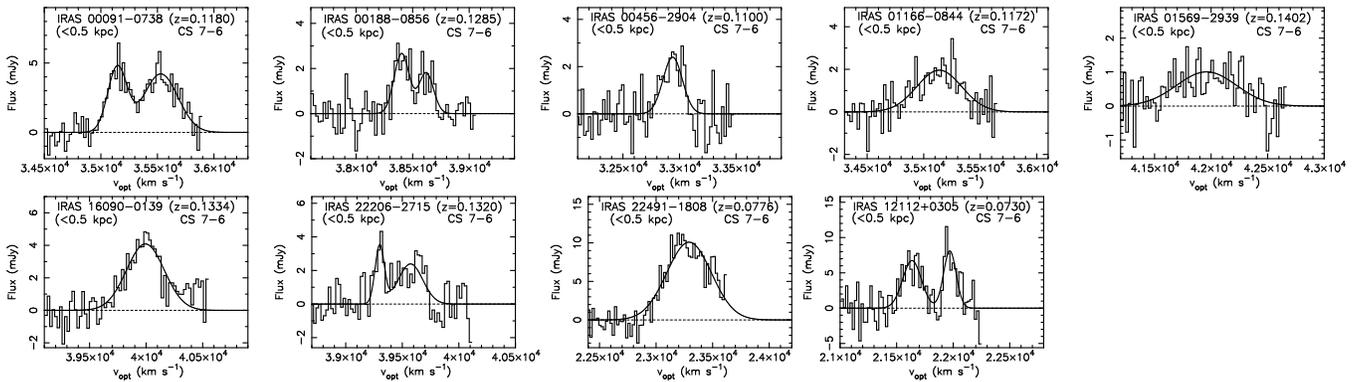

\includegraphics[angle=-90,scale=.145]{fD3a.eps} \hspace{0.1cm}
\includegraphics[angle=-90,scale=.145]{fD3b.eps} \hspace{0.1cm}
\includegraphics[angle=-90,scale=.145]{fD3c.eps} \hspace{0.1cm}
\includegraphics[angle=-90,scale=.145]{fD3d.eps} \hspace{0.1cm}
\includegraphics[angle=-90,scale=.145]{fD3e.eps} \hspace{0.1cm} \\
\includegraphics[angle=-90,scale=.145]{fD3f.eps} \hspace{0.1cm}
\includegraphics[angle=-90,scale=.145]{fD3g.eps} \hspace{-0.1cm}
\includegraphics[angle=-90,scale=.145]{fD3h.eps} 
\includegraphics[angle=-90,scale=.145]{fD3i.eps} \hspace{0.1cm}
\caption{
Adopted Gaussian fit (solid curved line) of the CS J=7--6 emission
line in the 0.5 kpc beam-sized spectra for sources with significant
detection (all ULIRGs except IRAS 03250$+$1606).
The abscissa is optical LSR velocity in km s$^{-1}$ and the ordinate 
is flux density in mJy.
The horizontal thin dotted straight line indicates the zero flux 
level.
\label{fig:CS76Gauss}
}
\end{figure*}

The CS J=7--6 ($\nu_{\rm rest}$=342.883 GHz) emission line was
serendipitously detected in all ULIRGs for which HCN and HCO$^{+}$
J=4--3 observations were made.
Figure \ref{fig:CS76mom0} shows the integrated intensity (moment 0) maps
of CS J=7--6 created from the original-beam-sized data (Table
\ref{tab:beam}; column 4).  
Figure \ref{fig:CS76spec} displays the 0.5 kpc beam-sized spectra that
include the CS J=7--6 line.
Table \ref{tab:CS76Gauss} summarizes the best fit Gaussian parameters
for the CS J=7--6 emission line detected in the 0.5 kpc beam-sized
spectra. 
Figure \ref{fig:CS76Gauss} overplots the best fit Gaussian on the
observed CS J=7--6 emission line in the 0.5 kpc beam-sized spectra. 
The CS J=7--6 emission line was not detected in the LIRG NGC 1614
\citep{ima13a}. 

\begin{deluxetable*}{lcccc}[!hbt]
\tabletypesize{\scriptsize}
\tablecaption{Gaussian Fit of CS J=7--6 Emission Line Extracted from
the 0.5 kpc Beam-sized Spectra \label{tab:CS76Gauss}} 
\tablewidth{0pt}
\tablehead{
\colhead{Object} & \multicolumn{4}{c}{Gaussian fit} \\  
\colhead{} & \colhead{Velocity} & \colhead{Peak} & 
\colhead{FWHM} & \colhead{Flux} \\ 
\colhead{} & \colhead{[km s$^{-1}$]} & \colhead{[mJy]} 
& \colhead{[km s$^{-1}$]} & \colhead{[Jy km s$^{-1}$]} \\  
\colhead{(1)} & \colhead{(2)} & \colhead{(3)} & \colhead{(4)} & 
\colhead{(5)} 
}
\startdata 
IRAS 00091$-$0738 & 35145$\pm$10, 35529$\pm$15 & 4.7$\pm$0.4,
4.2$\pm$0.4 & 199$\pm$23, 358$\pm$41 & 2.3$\pm$0.2 \\
IRAS 00188$-$0856 & 38403$\pm$13, 38620$\pm$26 & 2.7$\pm$0.3,
1.8$\pm$0.4 & 157$\pm$35, 152$\pm$58 & 0.65$\pm$0.15 \\
IRAS 00456$-$2904 & 32940$\pm$15 & 2.4$\pm$0.3 & 201$\pm$36 & 0.46$\pm$0.10 \\
IRAS 01166$-$0844 & 35137$\pm$26 & 2.0$\pm$0.2 & 467$\pm$67 & 0.87$\pm$0.16 \\
IRAS 01569$-$2939 & 41960$\pm$53 & 1.0$\pm$0.2 & 624$\pm$126 & 0.59$\pm$0.15 \\
IRAS 16090$-$0139 & 39994$\pm$15 & 4.1$\pm$0.3 & 393$\pm$52 & 1.5$\pm$0.2 \\
IRAS 22206$-$2715 & 39300$\pm$10, 39573$\pm$34 & 3.5$\pm$0.8,
2.4$\pm$0.4 & 96$\pm$29, 255$\pm$59 & 0.88$\pm$0.20 \\
IRAS 22491$-$1808 & 23292$\pm$12 & 10$\pm$1 & 464$\pm$27 &
4.6$\pm$0.3 \\
IRAS 12112$+$0305 & 21636$\pm$16, 21970$\pm$14 &
6.7$\pm$1.0, 8.1$\pm$1.8 & 195$\pm$42, 131$\pm$31 & 2.4$\pm$0.5 \\ \hline
\enddata

\tablecomments{
Col.(1): Object name.
Cols.(2)--(5): Gaussian fit of the CS J=7--6 emission line detected in
the 0.5 kpc beam-sized spectrum. 
Col.(2): Optical LSR velocity (v$_{\rm opt}$) of emission line peak
in km s$^{-1}$.  
Col.(3): Peak flux in mJy. 
Col.(4): Observed FWHM in km s$^{-1}$.
Col.(5): Gaussian-fit velocity-integrated flux in Jy km s$^{-1}$. 
When fitting results of two Gaussian components are adopted, fluxes of
the two components are added. 
Only Gaussian fitting error (statistical uncertainty) is considered.
}

\end{deluxetable*}

\clearpage

\section{HC$_{3}$N line}

\begin{figure}[!hbt]
\begin{center}
\includegraphics[angle=0,scale=.2]{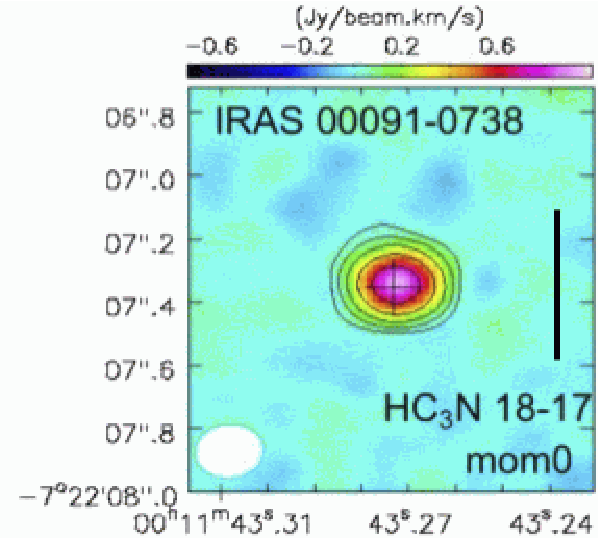} 
\includegraphics[angle=0,scale=.2]{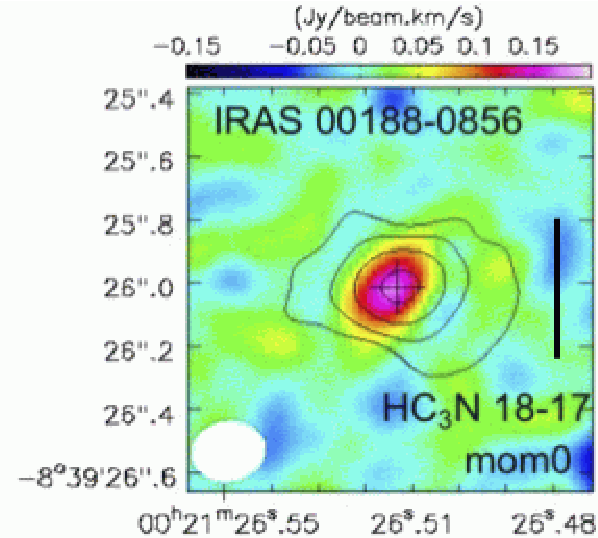} 
\includegraphics[angle=0,scale=.2]{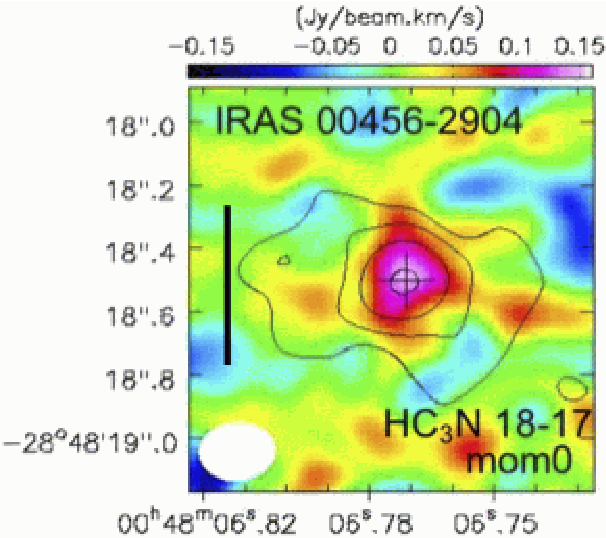} 
\includegraphics[angle=0,scale=.2]{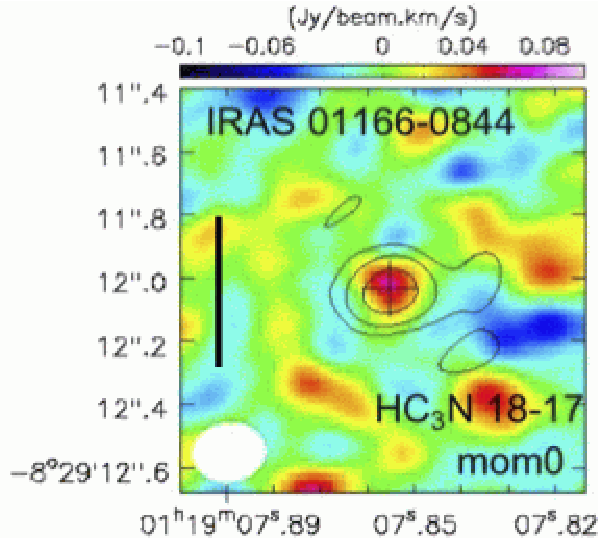} \\
\hspace*{-2.3cm}
\includegraphics[angle=0,scale=.2]{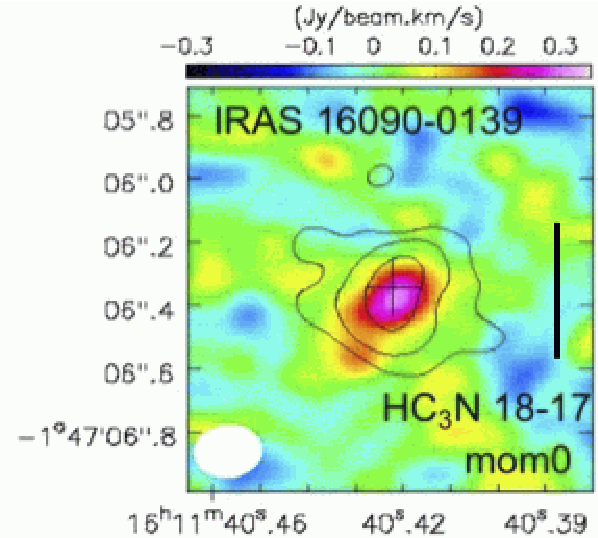} 
\includegraphics[angle=0,scale=.2]{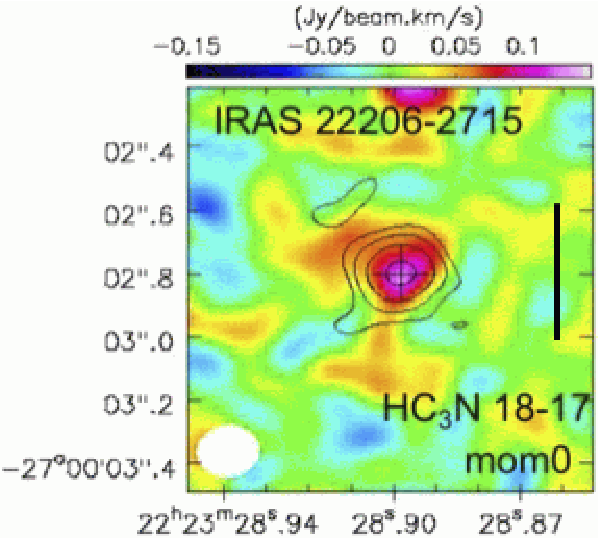} 
\includegraphics[angle=0,scale=.2]{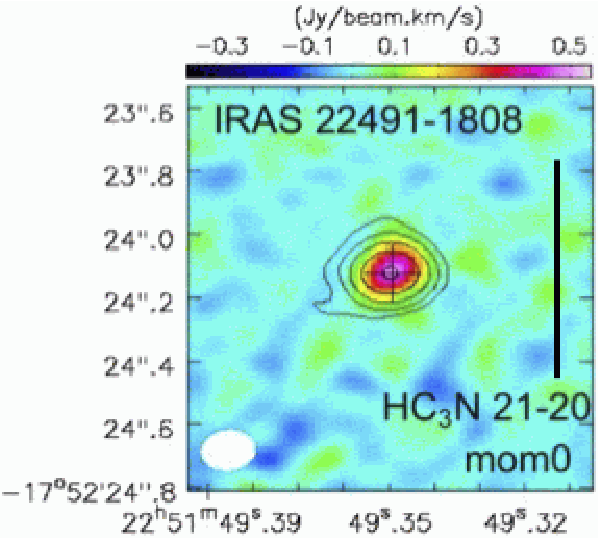} 
\end{center}
\caption{Integrated intensity (moment 0) map of the HC$_{3}$N J=18--17 
(for all ULIRGs but IRAS 22491$-$1808) or 
J=21--20 line (for IRAS 22491$-$1808), created from the
original-beam-sized data (Table \ref{tab:beam}, column 2). 
Only ULIRGs with significant HC$_{3}$N line detection are shown.
Simultaneously obtained continuum emission (J21) is displayed as
contours. 
The contours start from 4$\sigma$ and increase by a factor of 2 
(i.e., 8$\sigma$, 16$\sigma$, 32$\sigma$, and 64$\sigma$) for all
sources.
Continuum peak position is shown as a cross.
The length of the vertical black solid bar for each object corresponds
to 1 kpc.  
Beam size for each moment 0 map is shown as a white filled circle 
in the lower-left region.
\label{fig:HC3Nmom0}
}
\end{figure}

\begin{figure*}[!hbt]
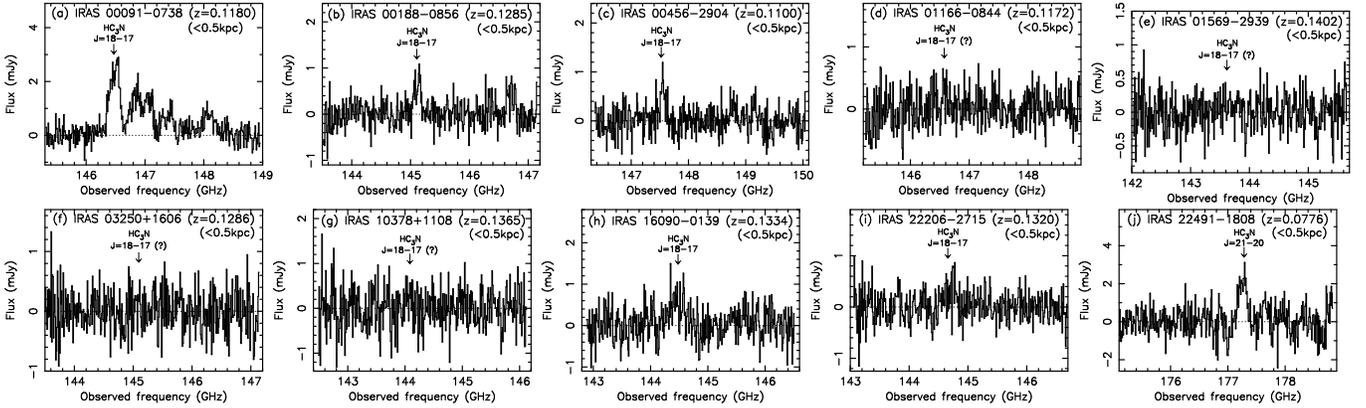

\includegraphics[angle=-90,scale=.155]{fE2a.eps} 
\includegraphics[angle=-90,scale=.155]{fE2b.eps} 
\includegraphics[angle=-90,scale=.155]{fE2c.eps} 
\includegraphics[angle=-90,scale=.155]{fE2d.eps} 
\includegraphics[angle=-90,scale=.155]{fE2e.eps} \\
\includegraphics[angle=-90,scale=.155]{fE2f.eps} 
\includegraphics[angle=-90,scale=.155]{fE2g.eps} 
\includegraphics[angle=-90,scale=.155]{fE2h.eps} 
\includegraphics[angle=-90,scale=.155]{fE2i.eps} 
\includegraphics[angle=-90,scale=.155]{fE2j.eps} 
\caption{
0.5 kpc beam-sized spectrum that includes the HC$_{3}$N J=18--17 or
J=21--20 line, taken during the J21 observation.  
The abscissa is observed frequency in GHz and the ordinate is flux
density in mJy. 
The HC$_{3}$N J=18--17 line ($\nu_{\rm rest}$=163.753 GHz) is covered 
in all ULIRGs, except IRAS 22491$-$1808 for which the HC$_{3}$N J=21--20 
line ($\nu_{\rm rest}$=191.040 GHz) is included.
A downward arrow is shown for the HC$_{3}$N lines at the expected
observed frequency of the adopted redshift (Table \ref{tab:object}, column 2).
The mark ``(?)'' is added when detection is unclear. 
The horizontal thin dotted straight line indicates the zero flux
level. 
\label{fig:HC3Nspec}
}
\end{figure*}

\begin{figure*}[!hbt]
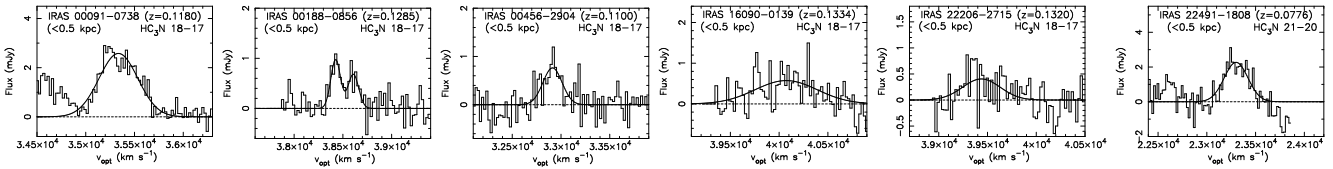

\includegraphics[angle=-90,scale=.125]{fE3a.eps} 
\includegraphics[angle=-90,scale=.125]{fE3b.eps} 
\includegraphics[angle=-90,scale=.125]{fE3c.eps} 
\includegraphics[angle=-90,scale=.125]{fE3d.eps} 
\includegraphics[angle=-90,scale=.125]{fE3e.eps} 
\includegraphics[angle=-90,scale=.125]{fE3f.eps} 
\caption{
Adopted Gaussian fit (solid curved line) of the HC$_{3}$N J=21--20 
(for IRAS 22491$-$1808) or J=18--17 (for other ULIRGs with significant
detection) emission line extracted from the 0.5 kpc beam-sized spectrum.
The abscissa is optical LSR velocity in km s$^{-1}$ and the ordinate 
is flux density in mJy.
The horizontal thin dotted straight line indicates the zero flux 
level.
\label{fig:HC3NGauss}
}
\end{figure*}

HC$_{3}$N J=18--17 ($\nu_{\rm rest}$=163.753 GHz) or 
J=21--20 ($\nu_{\rm rest}$=191.040 GHz) emission lines were
serendipitously detected in some ULIRGs during the observations of 
HCN and HCO$^{+}$ J=2--1.
Figure \ref{fig:HC3Nmom0} shows the integrated intensity (moment 0) maps
of HC$_{3}$N J=18--17 or J=21--20 created from the original-beam-sized
data (Table \ref{tab:beam}; column 2).  
Figure \ref{fig:HC3Nspec} displays the 0.5 kpc beam-sized spectra that
include the HC$_{3}$N J=18--17 or J=21--20 line.
The best fit Gaussian parameters for these HC$_{3}$N emission lines
extracted from the 0.5 kpc beam-sized spectra are summarized in 
Table \ref{tab:HC3NGauss}.
In Figure \ref{fig:HC3NGauss}, the best fit Gaussian is overplotted
to the observed HC$_{3}$N emission line in the 0.5 kpc beam-sized
spectra.   
No HC$_{3}$N line was covered in the spectrum of the LIRG NGC 1614
\citep{ima22}. 

When compared to other dense molecular lines, 
HC$_{3}$N emission can be strong in galaxies with high column density
of obscuring material around energy sources \citep{aal07,lin11},
because HC$_{3}$N is strongly emitted in UV-shielded regions at some
distance from the energy sources \citep{lin11,mei11}.
\citet{lin11} defined HC$_{3}$N-luminous galaxies as sources with 
HC$_{3}$N J=10--9 to HCN J=1--0 flux ratio of $>$0.15, which
constitute less than one-third of observed galaxies, including
those with much less infrared luminosity than (U)LIRGs.
The rest frequency ($\nu_{\rm rest}$) of HC$_{3}$N J=20--19 and HCN
J=2--1 that we observed is approximately twice that of HC$_{3}$N
J=10--9 and HCN J=1--0, respectively.
If emission is thermalized and optically thick, where flux in units of
Jy km s$^{-1}$ increases with $\nu_{\rm rest}$$^{2}$,  then   
the HC$_{3}$N J=20--19 to HCN J=2--1 flux ratios are expected to be 
comparable to the HC$_{3}$N J=10--9 to HCN J=1--0 flux ratios. 
However, the upper excitation energy level for HC$_{3}$N J=20--19 
(E$_{\rm u}$ $\sim$ 92 K) is significantly higher than those of 
HCN J=2--1 (E$_{\rm u}$ $\sim$ 13 K), HC$_{3}$N J=10--9 (24 K), and 
HCN J=1--0 (4 K). 
If only HC$_{3}$N J=20--19 is significantly sub-thermally excited than
other three lines, then the HC$_{3}$N J=20--19 to HCN J=2--1 flux ratio
can be smaller than the HC$_{3}$N J=10--9 to HCN J=1--0 flux ratio.  
In this sense, a source is safely classified as a HC$_{3}$N-luminous
galaxy, if the HC$_{3}$N J=20--19 to HCN J=2--1 flux ratio is 
larger than 0.15.
IRAS 00091$-$0738 corresponds to this case (Table \ref{tab:HC3NGauss},
column 7). 
The strong self-absorption of HCO$^{+}$ and HCN lines at J=2--1,
J=3--2, and J=4--3, detected in IRAS 00091$-$0738 (Figure
\ref{fig:SpectraA}), suggests the presence of large column density of
obscuring material at the foreground side of a background 0.85--2 mm
continuum emitting energy source.   
The bright HC$_{3}$N emission observed in IRAS 00091$-$0738 can come
from highly shielded regions.  

For the remaining ULIRGs, the observed HC$_{3}$N J=20--19 to HCN
J=2--1 flux ratios are comparable, within uncertainty, to the 
threshold of the HC$_{3}$N-luminous galaxies (= 0.15) \citep{lin11}.
However, since the HC$_{3}$N J=10--9 to HCN J=1--0 flux ratio is
expected to be higher than or equal to the HC$_{3}$N J=20--19 to HCN
J=2--1 flux ratio, it is quite possible that other ULIRGs in
Table \ref{tab:HC3NGauss} are categorized as the HC$_{3}$N-luminous
galaxies as well.   
Because ULIRGs' nuclei are usually highly obscured by gas and dust, 
strong HC$_{3}$N emission can naturally arise from UV-shielded regions
within the nuclei. 


\begin{deluxetable*}{ll|cccc|c}[!hbt]
\tabletypesize{\scriptsize}
\tablecaption{Gaussian Fit of HC$_{3}$N Emission Lines Extracted from the
0.5 kpc Beam-sized Spectra \label{tab:HC3NGauss}} 
\tablewidth{0pt}
\tablehead{
\colhead{Object} & \colhead{Line} & \multicolumn{4}{c}{Gaussian fit} 
& $\frac{\rm HC_3N}{\rm HCN J=2-1}$ \\  
\colhead{} & \colhead{} & \colhead{Velocity} & \colhead{Peak} & 
\colhead{FWHM} & \colhead{Flux} & \colhead{} \\ 
\colhead{} & \colhead{} & \colhead{[km s$^{-1}$]} & \colhead{[mJy]} 
& \colhead{[km s$^{-1}$]} & \colhead{[Jy km s$^{-1}$]} & \colhead{} \\  
\colhead{(1)} & \colhead{(2)} & \colhead{(3)} & \colhead{(4)} & 
\colhead{(5)} & \colhead{(6)} & \colhead{(7)} 
}
\startdata 
IRAS 00091$-$0738 & HC$_{3}$N J=18--17 & 35335$\pm$10 & 2.6$\pm$0.1 &
469$\pm$22 & 1.2$\pm$0.1 & 0.55$\pm$0.04 \\
IRAS 00188$-$0856 & HC$_{3}$N J=18--17 & 38426$\pm$12, 38607$\pm$24 &
0.97$\pm$0.17, 0.68$\pm$0.16 & 117$\pm$21, 128$\pm$63 & 0.19$\pm$0.05
& 0.13$\pm$0.04 \\
IRAS 00456$-$2904 & HC$_{3}$N J=18--17 & 32913$\pm$18 & 0.76$\pm$0.15
& 231$\pm$60 & 0.17$\pm$0.056 & 0.13$\pm$0.04 \\
IRAS 16090$-$0139 & HC$_{3}$N J=18--17 & 40072$\pm$69 & 0.57$\pm$0.11
& 753$\pm$148 & 0.41$\pm$0.11 & 0.17$\pm$0.05 \\
IRAS 22206$-$2715 & HC$_{3}$N J=18--17 & 39437$\pm$71 & 0.40$\pm$0.10
& 455$\pm$179 & 0.17$\pm$0.08 ($<$3$\sigma$) & 0.14$\pm$0.06 \\
IRAS 22491$-$1808 & HC$_{3}$N J=21--20 & 23300$\pm$20 & 2.3$\pm$0.4 &
287$\pm$40 & 0.64$\pm$0.14 & 0.13$\pm$0.03 \\ \hline
\enddata

\tablecomments{
Col.(1): Object name.
Col.(1): Line. HC$_{3}$N J=18--17 or J=21--20.
Cols.(3)--(6): Gaussian fit of the HC$_{3}$N emission line in the 0.5
kpc beam-sized spectrum.
Col.(3): Optical LSR velocity (v$_{\rm opt}$) of emission line peak
in km s$^{-1}$.  
Col.(4): Peak flux in mJy. 
Col.(5): Observed FWHM in km s$^{-1}$.
Col.(6): Gaussian-fit velocity-integrated flux in Jy km s$^{-1}$. 
When fitting results of two Gaussian components are adopted, fluxes of
the two components are added. 
Only Gaussian fitting error (statistical uncertainty) is considered.
Col.(7): Flux ratio of HC$_{3}$N J=18--17 to HCN J=2--1 or 
HC$_{3}$N J=21--20 to HCN J=2--1, calculated in units of Jy km
s$^{-1}$.
}

\end{deluxetable*}


\clearpage

\section{Molecular line flux ratios}

The (1) HCN-to-HCO$^{+}$ flux ratios at J=2--1, J=3--2, and J=4--3,
and (2) high-J to low-J flux ratios of HCN and HCO$^{+}$, 
measured in the $\lesssim$0.5 kpc, $\lesssim$1 kpc, $\lesssim$2 kpc,
0.5--1 kpc, and 1--2 kpc spectra, are summarized in Tables
\ref{tab:fluxratio} and \ref{tab:Jratio}, respectively.

\startlongtable
\begin{deluxetable*}{ll|ccc|c}
\tabletypesize{\scriptsize}
\tablecaption{Emission Line Flux Ratio between Different 
Molecules \label{tab:fluxratio}}  
\tablewidth{0pt}
\tablehead{
\colhead{Object} & \colhead{Region} & 
\multicolumn{3}{c}{$\frac{\rm HCN}{\rm HCO^{+}}$} & 
\colhead{$\frac{\rm HCN\ J=4-3}{\rm CS\ J=7-6}$} \\
\colhead{} & \colhead{} & \colhead{J=2--1} &
\colhead{J=3--2} & \colhead{J=4--3} & \colhead{} \\   
\colhead{(1)} & \colhead{(2)} & \colhead{(3)} & \colhead{(4)} &
\colhead{(5)} & \colhead{(6)} 
}
\startdata 
IRAS 00091$-$0738 & $\lesssim$0.5 kpc & 1.3$\pm$0.2 & 1.9$\pm$0.7
 & 1.4$\pm$0.4 & 2.1$\pm$0.3 \\
 & $\lesssim$1 kpc & 1.1$\pm$0.2 & 1.8$\pm$0.8 \tablenotemark{a} &
1.3$\pm$0.3 & --- \\ 
 & $\lesssim$2 kpc & 0.99$\pm$0.28 & 1.6$\pm$0.4 & 0.89$\pm$0.37
\tablenotemark{a} & --- \\
 & 0.5--1 kpc & 0.61$\pm$0.24 & 1.5$\pm$0.6  \tablenotemark{a} & --- &
--- \\ 
 & 1--2 kpc & --- & 0.81$\pm$0.39 \tablenotemark{a} & --- & --- \\ \hline
IRAS 00188$-$0856 & $\lesssim$0.5 kpc & 1.8$\pm$0.1 & 1.8$\pm$0.1 &
1.9$\pm$0.4 & 5.5$\pm$1.3 \\
 & $\lesssim$1 kpc & 1.6$\pm$0.1 & 1.7$\pm$0.1 & 2.1$\pm$0.6 & --- \\
 & $\lesssim$2 kpc & 1.6$\pm$0.2 & 1.7$\pm$0.2 & 2.6$\pm$0.7 & --- \\
 & 0.5--1 kpc & 1.4$\pm$0.2 & 1.6$\pm$0.2 & 2.6$\pm$1.0 & --- \\
 & 1--2 kpc & 1.3$\pm$0.5 & 2.0$\pm$0.6 & --- & --- \\ \hline
IRAS 00456$-$2904 & $\lesssim$0.5 kpc & 1.5$\pm$0.1 & 1.8$\pm$0.2 &
1.6$\pm$0.2 & 7.7$\pm$1.7 \\
 & $\lesssim$1 kpc & 1.4$\pm$0.1 & 1.6$\pm$0.2 & 1.7$\pm$0.2 & --- \\
 & $\lesssim$2 kpc & 1.4$\pm$0.1 & 1.3$\pm$0.2 & 1.7$\pm$0.4 & --- \\
 & 0.5--1 kpc & 1.3$\pm$0.2 & 1.2$\pm$0.2 & 1.8$\pm$0.8
\tablenotemark{a} & --- \\
 & 1--2 kpc & 1.2$\pm$0.3 & 0.66$\pm$0.28 \tablenotemark{a} & --- &
--- \\ \hline 
IRAS 01166$-$0844 & $\lesssim$0.5 kpc & 1.8$\pm$0.2 & 1.3$\pm$0.2 &
1.5$\pm$0.2 & 3.2$\pm$0.6 \\
 & $\lesssim$1 kpc & 1.7$\pm$0.2 & 1.4$\pm$0.3 & 1.4$\pm$0.3 & --- \\
 & $\lesssim$2 kpc & 1.4$\pm$0.3 & 1.7$\pm$0.6 & 1.4$\pm$0.6 & --- \\
 & 0.5--1 kpc & 1.1$\pm$0.5 \tablenotemark{a} & 1.4$\pm$0.6
\tablenotemark{a} & --- & --- \\
 & 1--2 kpc & --- & --- & --- & --- \\ \hline
IRAS 01569$-$2939 & $\lesssim$0.5 kpc & 1.0$\pm$0.2 & 0.85$\pm$0.17 &
0.92$\pm$0.11 & 3.9$\pm$1.1 \\
 & $\lesssim$1 kpc & 0.91$\pm$0.16 & 0.76$\pm$0.19 & 0.94$\pm$0.16 & --- \\
 & $\lesssim$2 kpc & 0.96$\pm$0.14 & 0.78$\pm$0.13 & 0.89$\pm$0.24 & --- \\
 & 0.5--1 kpc & 0.71$\pm$0.26 & 0.67$\pm$0.15 & 1.1$\pm$0.3 & --- \\
 & 1--2 kpc & 1.1$\pm$0.6 \tablenotemark{a} & 0.76$\pm$0.39
\tablenotemark{a} & 0.71$\pm$0.49 \tablenotemark{a} & --- \\ \hline
IRAS 03250$+$1606 & $\lesssim$0.5 kpc & 1.4$\pm$0.3 & 1.4$\pm$0.4 &
1.2$\pm$0.4 & $>$7.6 \\
 & $\lesssim$1 kpc & 1.1$\pm$0.2 & 1.2$\pm$0.4 & 1.2$\pm$0.5 & --- \\
 & $\lesssim$2 kpc & 0.86$\pm$0.19 & 1.0$\pm$0.2 & 1.1$\pm$0.5
\tablenotemark{a} & --- \\
 & 0.5--1 kpc & 0.88$\pm$0.29 & 0.89$\pm$0.34 & 1.3$\pm$0.5 & --- \\
 & 1--2 kpc & --- & 0.82$\pm$0.44 \tablenotemark{a} & --- & --- \\ \hline
IRAS 10378$+$1108 & $\lesssim$0.5 kpc & 1.1$\pm$0.2 & 1.1$\pm$0.1 & --- & ---
\\
 & $\lesssim$1 kpc & 0.95$\pm$0.11 & 1.1$\pm$0.1 & --- & --- \\
 & $\lesssim$2 kpc & 0.87$\pm$0.14 & 1.0$\pm$0.1 & --- & --- \\
 & 0.5--1 kpc & 0.77$\pm$0.19 & 0.96$\pm$0.13 & --- & --- \\
 & 1--2 kpc & 0.69$\pm$0.25 & 1.1$\pm$0.3 & --- & --- \\ \hline
IRAS 16090$-$0139 & $\lesssim$0.5 kpc & 1.3$\pm$0.1 & 1.2$\pm$0.1 &
1.4$\pm$0.1 & 5.5$\pm$0.9 \\
 & $\lesssim$1 kpc & 1.2$\pm$0.1 & 1.1$\pm$0.1 & 1.3$\pm$0.1 & --- \\
 & $\lesssim$2 kpc & 1.1$\pm$0.1 & 0.91$\pm$0.09 & 1.3$\pm$0.1 & --- \\
 & 0.5--1 kpc & 1.1$\pm$0.1 & 0.89$\pm$0.10 & 1.2$\pm$0.1 & --- \\
 & 1--2 kpc & 0.89$\pm$0.16 & 0.68$\pm$0.15 & 1.2$\pm$0.3 & --- \\ \hline
IRAS 22206$-$2715 & $\lesssim$0.5 kpc & 1.5$\pm$0.2 & 1.3$\pm$0.1 &
1.3$\pm$0.2 & 3.2$\pm$0.8 \\
 & $\lesssim$1 kpc & 1.5$\pm$0.1 & 1.4$\pm$0.1 & 1.4$\pm$0.2 & --- \\
 & $\lesssim$2 kpc & 1.8$\pm$0.3 & 1.5$\pm$0.2 & 1.2$\pm$0.3 & --- \\
 & 0.5--1 kpc & 1.4$\pm$0.3 & 1.6$\pm$0.4 & 1.8$\pm$0.7 & --- \\
 & 1--2 kpc & 2.5$\pm$1.3 \tablenotemark{a} & 2.4$\pm$1.4
\tablenotemark{a} & 0.69$\pm$0.52 \tablenotemark{a} & --- \\ \hline
IRAS 22491$-$1808 & $\lesssim$0.5 kpc & 1.5$\pm$0.1 & 1.4$\pm$0.1 &
1.5$\pm$0.1 & 2.7$\pm$0.2 \\
 & $\lesssim$1 kpc & 1.4$\pm$0.1 & 1.4$\pm$0.1 & 1.2$\pm$0.1 & --- \\
 & $\lesssim$2 kpc & 1.4$\pm$0.2 & 1.4$\pm$0.2 & 0.91$\pm$0.21 & --- \\
 & 0.5--1 kpc & 1.2$\pm$0.3 & 1.3$\pm$0.4 & 0.80$\pm$0.23 & --- \\
 & 1--2 kpc & --- & --- & --- & --- \\ \hline
IRAS 12112$+$0305 & $\lesssim$0.5 kpc & --- & --- & 1.7$\pm$0.3 & 4.9$\pm$1.1 \\
 & $\lesssim$1 kpc & ---  & ---  & 1.6$\pm$0.3 &  --- \\
 & $\lesssim$2 kpc & ---  & ---  & 1.5$\pm$0.5 & --- \\
 & 0.5--1 kpc & ---  & ---  & 1.9$\pm$1.4 \tablenotemark{a} & --- \\
 & 1--2 kpc & --- & --- & --- & --- \\ \hline
NGC 1614 & $\lesssim$0.5 kpc & 0.60$\pm$0.04 & 0.44$\pm$0.08 & 0.31$\pm$0.05
& $>$2.4 \\ 
 & $\lesssim$1 kpc & 0.56$\pm$0.04  & 0.38$\pm$0.07  & 0.31$\pm$0.05 &  --- \\
 & 0.5--1 kpc & 0.51$\pm$0.05  & 0.29$\pm$0.14 \tablenotemark{a} & 
0.38$\pm$0.12 & --- \\ \hline
\enddata

\tablenotetext{a}{S/N $<$ 2.5$\sigma$.}

\tablecomments{Col.(1): Object name.
Col.(2): Region.
Central $\lesssim$0.5 kpc, $\lesssim$1 kpc, $\lesssim$2 kpc, 0.5--1
kpc annular, and 1--2 kpc annular regions.  
Cols.(3)--(5): Ratio of HCN-to-HCO$^{+}$ flux calculated in units of 
Jy km s$^{-1}$. 
Col.(3): J=2--1.
Col.(4): J=3--2.
Col.(5): J=4--3.
Col.(6): Ratio of HCN J=4--3 to CS J=7--6 flux measured within the
central $\lesssim$0.5 kpc region in units of Jy km s$^{-1}$.   
In Cols.(3)--(6), only Gaussian fit statistical uncertainty is
considered, because we compare simultaneously taken lines.
No value is shown when the resulting uncertainty of the ratio is too 
large to obtain meaningful information.
}

\end{deluxetable*}


\startlongtable
\begin{deluxetable*}{ll|cc|cc}
\tabletypesize{\scriptsize}
\tablecaption{High-J to Low-J Flux Ratio of HCN and 
HCO$^{+}$ \label{tab:Jratio}}  
\tablewidth{0pt}
\tablehead{
\colhead{Object} & \colhead{Region} & \multicolumn{2}{c}{HCN} & 
\multicolumn{2}{c}{HCO$^{+}$} \\
\colhead{} & \colhead{} & \colhead{$\frac{J=3-2}{J=2-1}$} &
\colhead{$\frac{J=4-3}{J=2-1}$} &
\colhead{$\frac{J=3-2}{J=2-1}$} &
\colhead{$\frac{J=4-3}{J=2-1}$} \\
\colhead{(1)} & \colhead{(2)} & \colhead{(3)} & \colhead{(4)} &
\colhead{(5)} & \colhead{(6)} 
}
\startdata 
IRAS 00091$-$0738 & $\lesssim$0.5 kpc & 1.9$\pm$0.4 & 2.3$\pm$0.2 &
1.3$\pm$0.5 & 2.1$\pm$0.7 \\ 
 & $\lesssim$1 kpc & 2.3$\pm$0.5 & 2.6$\pm$0.4 & 
1.4$\pm$0.5 & 2.3$\pm$0.6 \\ 
 & $\lesssim$2 kpc & 2.4$\pm$0.6 & 2.6$\pm$0.8 
& 1.5$\pm$0.4 & 2.9$\pm$1.2 \tablenotemark{a} \\ \hline
IRAS 00188$-$0856 & $\lesssim$0.5 kpc & 1.6$\pm$0.1 & 2.4$\pm$0.1 &
1.7$\pm$0.1 & 2.3$\pm$0.5 \\ 
 & $\lesssim$1 kpc & 1.8$\pm$0.1 & 2.3$\pm$0.4 & 1.7$\pm$0.2
& 1.8$\pm$0.4 \\  
 & $\lesssim$2 kpc & 1.9$\pm$0.1 & 2.6$\pm$0.3 
& 1.7$\pm$0.2 & 1.6$\pm$0.4 \\
 & 0.5--1 kpc & 2.1$\pm$0.2 & 2.3$\pm$0.5 
& 1.9$\pm$0.3 & 1.2$\pm$0.4 \\\hline 
IRAS 00456$-$2904 & $\lesssim$0.5 kpc & 2.1$\pm$0.1 & 2.6$\pm$0.2  &
1.7$\pm$0.2 & 2.5$\pm$0.3 \\ 
 & $\lesssim$1 kpc & 2.1$\pm$0.1 & 2.5$\pm$0.2 & 1.9$\pm$0.2
& 2.1$\pm$0.3 \\  
 & $\lesssim$2 kpc & 1.9$\pm$0.2 & 2.3$\pm$0.2 
& 2.0$\pm$0.3 & 1.8$\pm$0.4 \\
 & 0.5--1 kpc & 2.0$\pm$0.3 & 2.0$\pm$0.6 &
2.1$\pm$0.4 & 1.4$\pm$0.5 \\ \hline  
IRAS 01166$-$0844 & $\lesssim$0.5 kpc & 1.6$\pm$0.2 & 2.4$\pm$0.2 &
2.2$\pm$0.3 & 2.8$\pm$0.4 \\ 
 & $\lesssim$1 kpc & 2.1$\pm$0.4 & 2.2$\pm$0.3 & 2.6$\pm$0.4
& 2.6$\pm$0.5 \\  
 & $\lesssim$2 kpc & 2.3$\pm$0.7 & 1.9$\pm$0.5 & 1.9$\pm$0.6
& 1.8$\pm$0.7 \\\hline  
IRAS 01569$-$2939 & $\lesssim$0.5 kpc & 1.8$\pm$0.3 & 2.3$\pm$0.3 &
2.1$\pm$0.5 & 2.5$\pm$0.5 \\ 
 & $\lesssim$1 kpc & 2.1$\pm$0.6 & 2.4$\pm$0.4 & 2.5$\pm$0.3
& 2.3$\pm$0.4 \\  
 & $\lesssim$2 kpc & 2.1$\pm$0.4 & 2.1$\pm$0.5 
& 2.6$\pm$0.4 & 2.3$\pm$0.5 \\
 & 0.5--1 kpc & 3.6$\pm$1.3 & 3.0$\pm$1.0 &
3.9$\pm$0.9 & 2.0$\pm$0.7 \\ \hline
IRAS 03250$+$1606 & $\lesssim$0.5 kpc & 1.9$\pm$0.3 & 1.8$\pm$0.4 &
1.9$\pm$0.6 & 2.1$\pm$0.7 \\
 & $\lesssim$1 kpc & 1.9$\pm$0.4 & 1.9$\pm$0.6 & 1.9$\pm$0.5
& 1.8$\pm$0.5 \\  
 & $\lesssim$2 kpc & 2.1$\pm$0.4 & 1.9$\pm$0.5 & 1.8$\pm$0.4
& 1.5$\pm$0.7 \tablenotemark{a} \\ 
 & 0.5--1 kpc & 1.8$\pm$0.6 & 2.2$\pm$0.6 &
1.8$\pm$0.7 & 1.5$\pm$0.6 \tablenotemark{a} \\ \hline 
IRAS 10378$+$1108 & $\lesssim$0.5 kpc & 3.0$\pm$0.3 & --- & 3.0$\pm$0.4
& --- \\ 
 & $\lesssim$1 kpc & 2.7$\pm$0.2 & --- & 2.5$\pm$0.3 &
--- \\ 
 & $\lesssim$2 kpc & 2.3$\pm$0.3 & --- & 1.9$\pm$0.3 &
--- \\
 & 0.5--1 kpc & 2.4$\pm$0.5 & --- & 1.9$\pm$0.4 &
--- \\
 & 1--2 kpc & 1.6$\pm$0.5 & --- & 1.0$\pm$0.4 &
--- \\ \hline 
IRAS 16090$-$0139 & $\lesssim$0.5 kpc & 1.7$\pm$0.1 & 3.5$\pm$0.2 &
1.8$\pm$0.2 & 3.1$\pm$0.2 \\
 & $\lesssim$1 kpc & 1.7$\pm$0.1 & 3.2$\pm$0.1 & 1.9$\pm$0.2
& 2.8$\pm$0.2 \\  
 & $\lesssim$2 kpc & 1.8$\pm$0.1 & 2.9$\pm$0.2 
& 2.1$\pm$0.2 & 2.4$\pm$0.2 \\ 
 & 0.5--1 kpc & 1.8$\pm$0.2 & 2.7$\pm$0.2 &
2.2$\pm$0.2 & 2.5$\pm$0.2 \\ 
 & 1--2 kpc & 1.9$\pm$0.4 & 2.1$\pm$0.5 &
2.4$\pm$0.5 & 1.6$\pm$0.4 \\ \hline 
IRAS 22206$-$2715 & $\lesssim$0.5 kpc & 1.6$\pm$0.1 & 2.2$\pm$0.2 &
1.8$\pm$0.2 & 2.6$\pm$0.3 \\ 
 & $\lesssim$1 kpc & 1.7$\pm$0.1 & 2.2$\pm$0.2 & 1.8$\pm$0.2
& 2.3$\pm$0.3 \\  
 & $\lesssim$2 kpc & 1.7$\pm$0.2 & 2.1$\pm$0.3 
& 2.0$\pm$0.4 & 3.3$\pm$0.9 \\
 & 0.5--1 kpc & 1.9$\pm$0.3 & 2.2$\pm$0.5 &
1.7$\pm$0.5 & 1.8$\pm$0.7 \\ \hline 
IRAS 22491$-$1808 & $\lesssim$0.5 kpc & 1.9$\pm$0.1 & 2.6$\pm$0.1 &
2.0$\pm$0.1 & 2.6$\pm$0.2 \\ 
 & $\lesssim$1 kpc & 1.8$\pm$0.1 & 2.7$\pm$0.2 & 1.8$\pm$0.2
& 3.2$\pm$0.4 \\  
 & $\lesssim$2 kpc & 1.9$\pm$0.2 & 2.4$\pm$0.4 
& 1.8$\pm$0.3 & 3.5$\pm$0.8 \\ 
 & 0.5--1 kpc & 1.4$\pm$0.3 & 3.0$\pm$0.7 &
1.3$\pm$0.4 & 4.4$\pm$1.2 \\ 
\hline
NGC 1614 & $\lesssim$0.5 kpc & 0.79$\pm$0.13 & 0.80$\pm$0.13 &
1.1$\pm$0.1 & 1.5$\pm$0.1 \\ 
 & $\lesssim$1 kpc & 0.62$\pm$0.12 & 0.63$\pm$0.08 & 0.92$\pm$0.08 &
1.1$\pm$0.1 \\ 
 & 0.5--1 kpc & 0.44$\pm$0.21 \tablenotemark{a} &
0.57$\pm$0.14 & 0.78$\pm$0.11 & 0.77$\pm$0.15 \\ \hline 
\enddata

\tablenotetext{a}{S/N $<$ 2.5$\sigma$.}

\tablecomments{Col.(1): Object name.
Col.(2): Region.
Cols.(3)--(6): Ratio of flux measured in units of Jy km s$^{-1}$. 
Col.(3): J=3--2 to J=2--1 for HCN.
Col.(4): J=4--3 to J=2--1 for HCN.
Col.(5): J=3--2 to J=2--1 for HCO$^{+}$.
Col.(6): J=4--3 to J=2--1 for HCO$^{+}$.
In Cols.(3)--(6), only Gaussian fit statistical uncertainty is
considered.
}

\end{deluxetable*}

\clearpage

\section{Comparison of Observed HCN-to-HCO$^{+}$ Flux Ratios 
with RADEX Non-LTE Modeling for Different Parameters}

\begin{figure*}
\begin{center}
\includegraphics[angle=0,scale=.5]{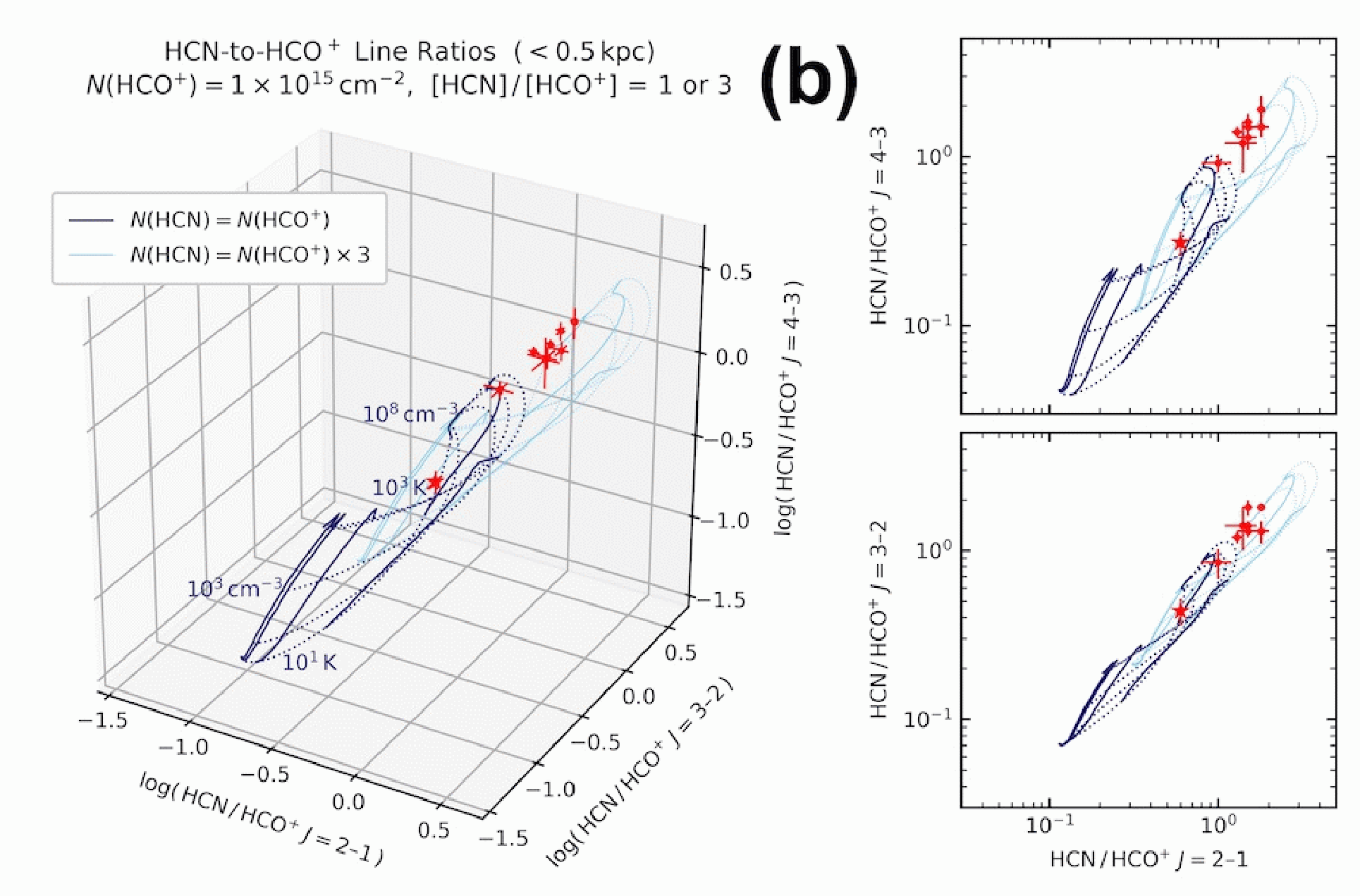} \\
\includegraphics[angle=0,scale=.5]{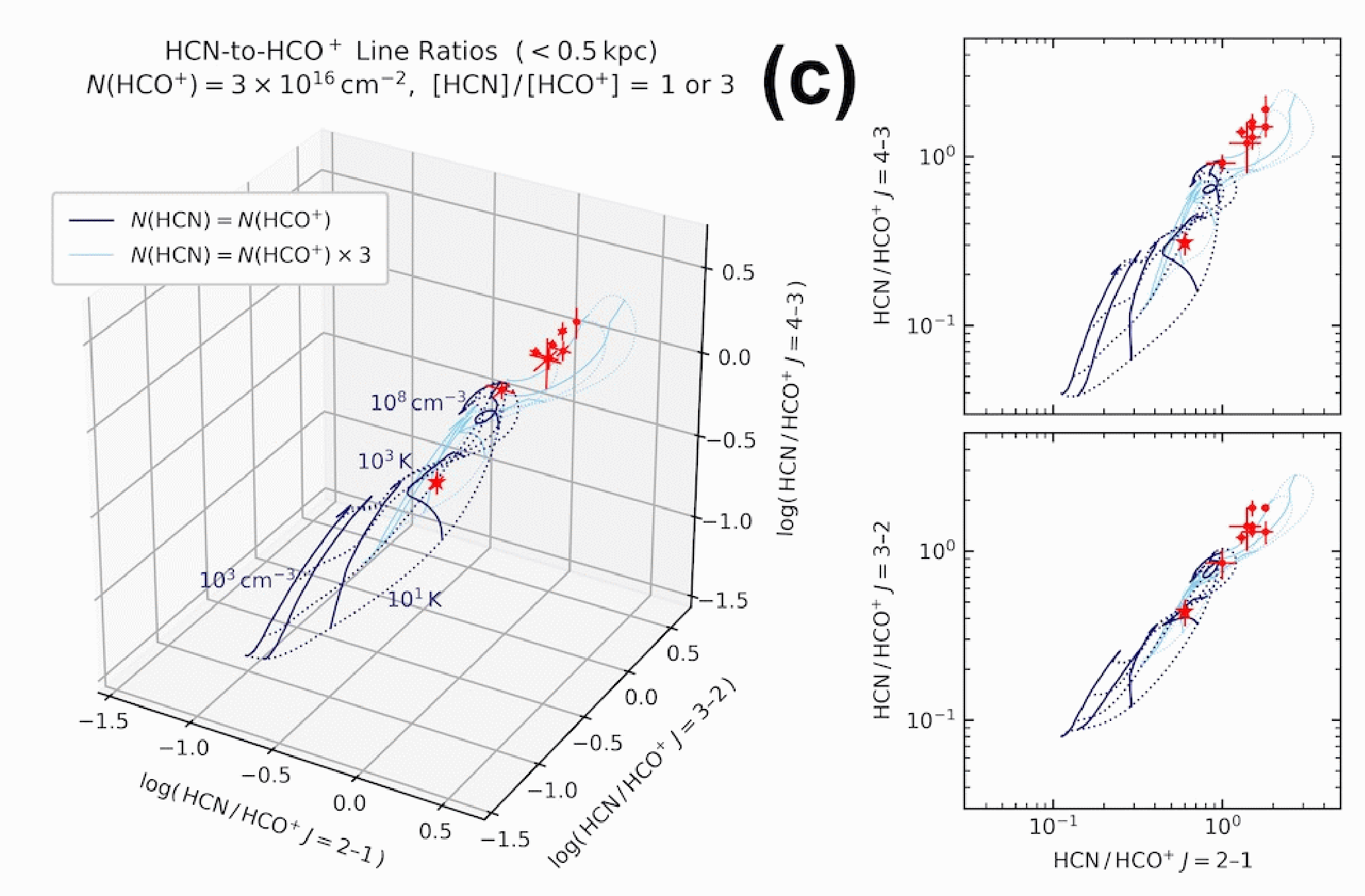} 
\end{center}
\caption{
Same as Figure \ref{fig:RadexComp}, but 
{\it (a)} HCN-to-HCO$^{+}$ abundance ratio of [HCN]/[HCO$^{+}$] = 1
and 7, 
{\it (b)} HCO$^{+}$ column density of N$_{\rm HCO+}$ = 1 $\times$
10$^{15}$ cm$^{-2}$ (a factor of 10 smaller than that adopted in
Figure \ref{fig:RadexComp}), and 
{\it (c)} N$_{\rm HCO+}$ = 3 $\times$ 10$^{16}$ cm$^{-2}$ (a factor of
3 larger).
Red filled circles are the observed HCN-to-HCO$^{+}$ flux ratios of
the same (U)LIRGs as plotted in Figure \ref{fig:RadexComp}.
\label{fig:RadexCompb}
}
\end{figure*}

Figure \ref{fig:RadexCompb} compares the observed HCN-to-HCO$^{+}$
flux ratios of (U)LIRGs' central $\lesssim$0.5 kpc regions with RADEX
non-LTE model calculations, by assuming significantly different value of
HCN-to-HCO$^{+}$ abundance ratio or HCO$^{+}$ column density from that
adopted in Figure \ref{fig:RadexComp} in $\S$5.2.  
Our argument that the overall distribution of the observed
HCN-to-HCO$^{+}$ flux ratios is better explained
with an enhanced HCN-to-HCO$^{+}$ abundance ratio rather than a
comparable ratio, does not change.  

\begin{figure*}[!hbt]
\begin{center}
\includegraphics[angle=0,scale=.55]{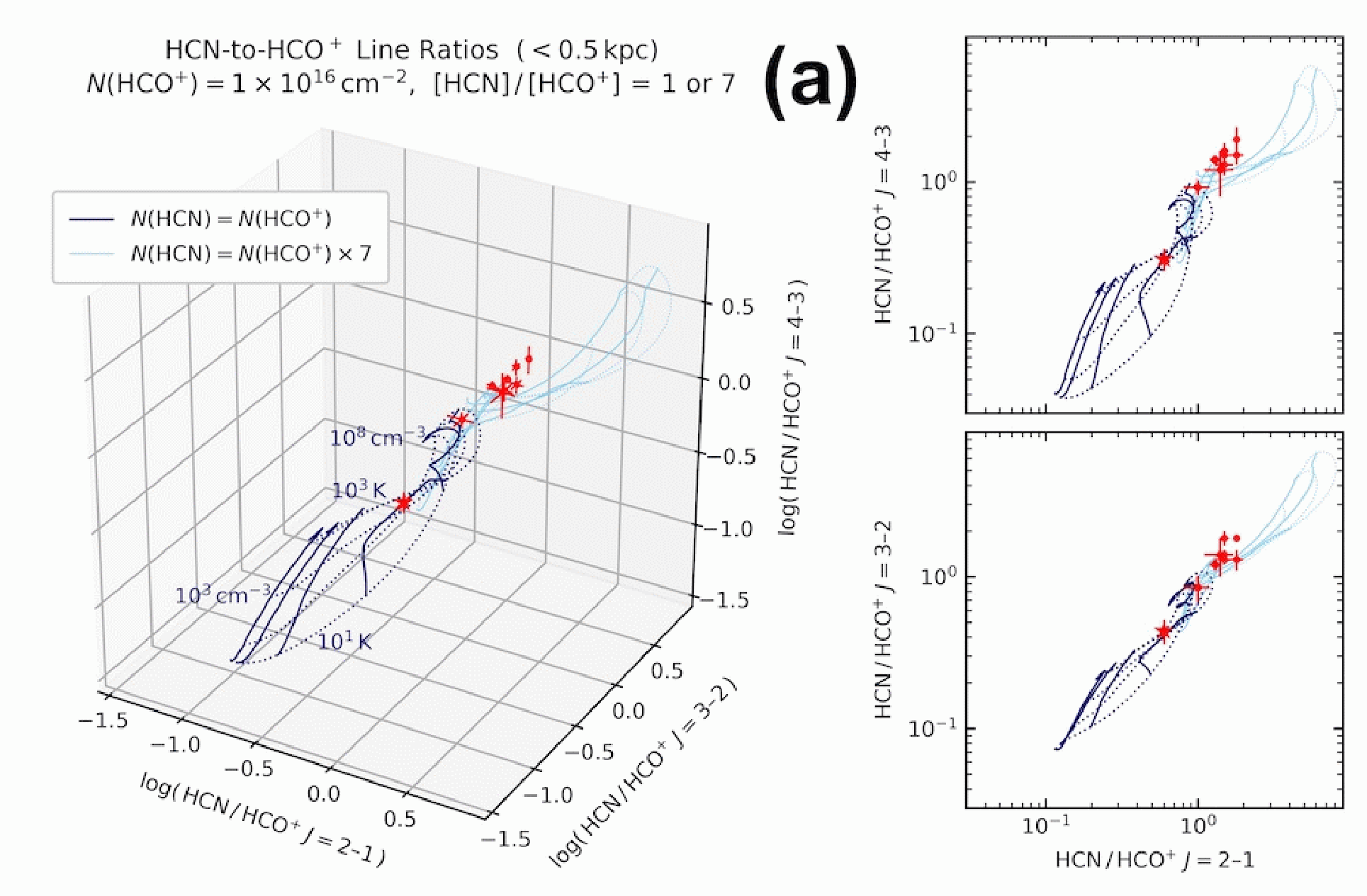} \\
\end{center}
\end{figure*}


\clearpage

\section{Derived Molecular Gas Density and Temperature in the Central 
$\lesssim$1 kpc and $\lesssim$2 kpc Regions} 

\begin{figure*}[!hbt]
\begin{center}
\includegraphics[angle=0,scale=.24]{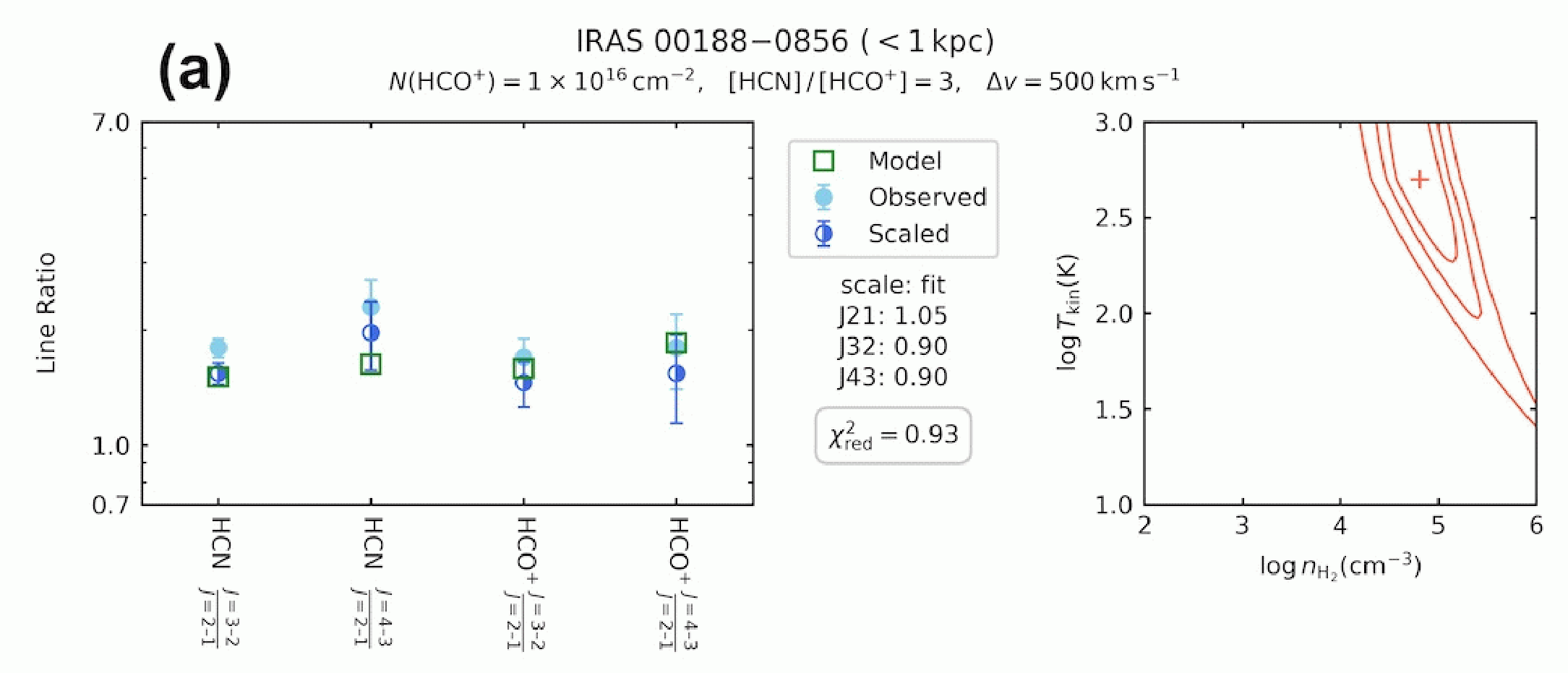} 
\includegraphics[angle=0,scale=.24]{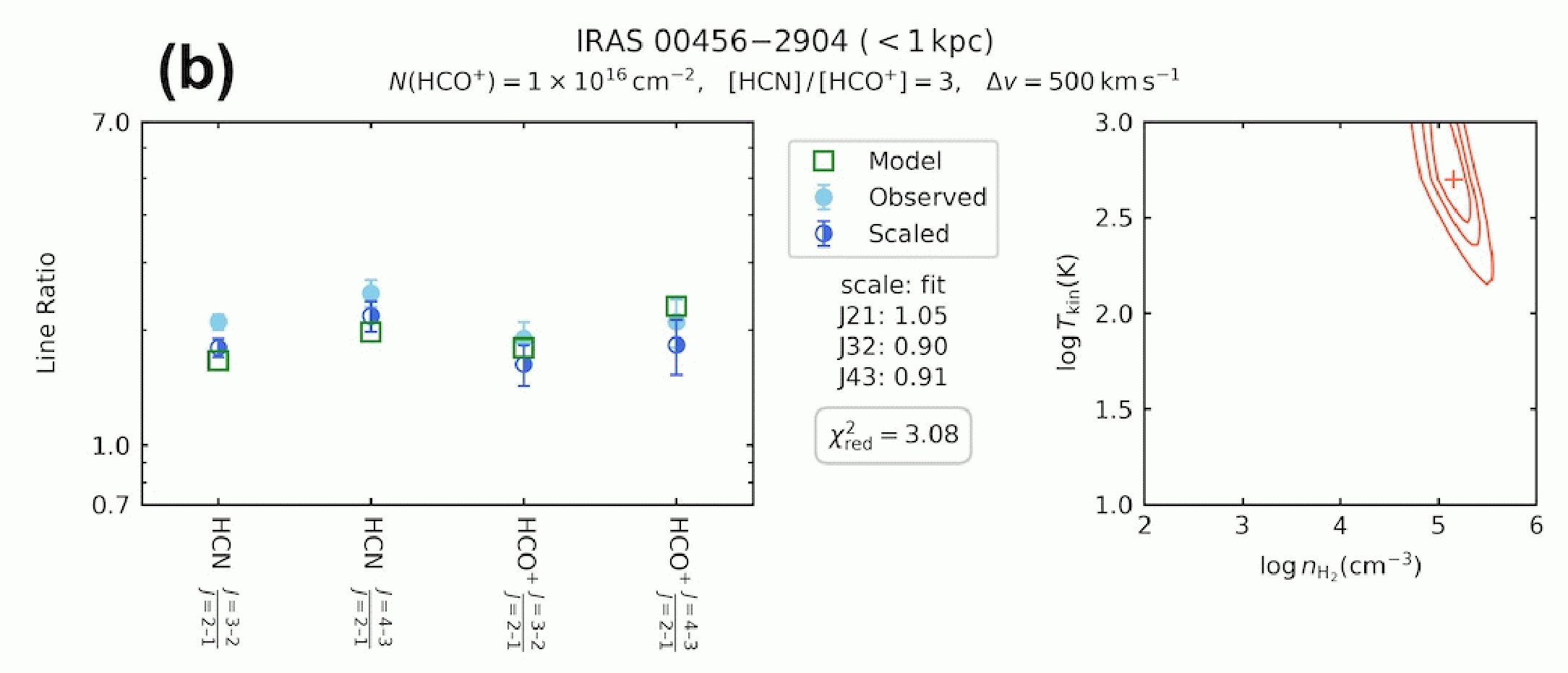} \\
\includegraphics[angle=0,scale=.24]{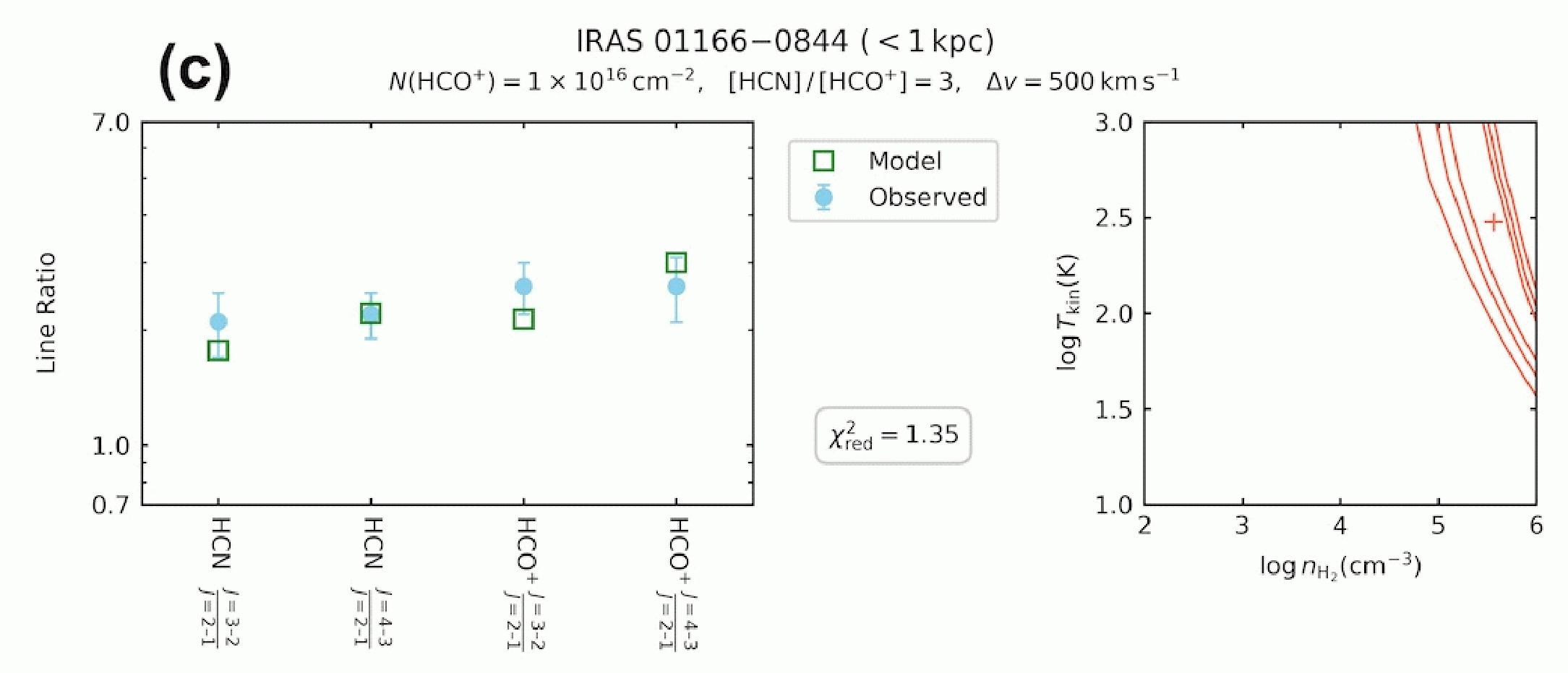} 
\includegraphics[angle=0,scale=.24]{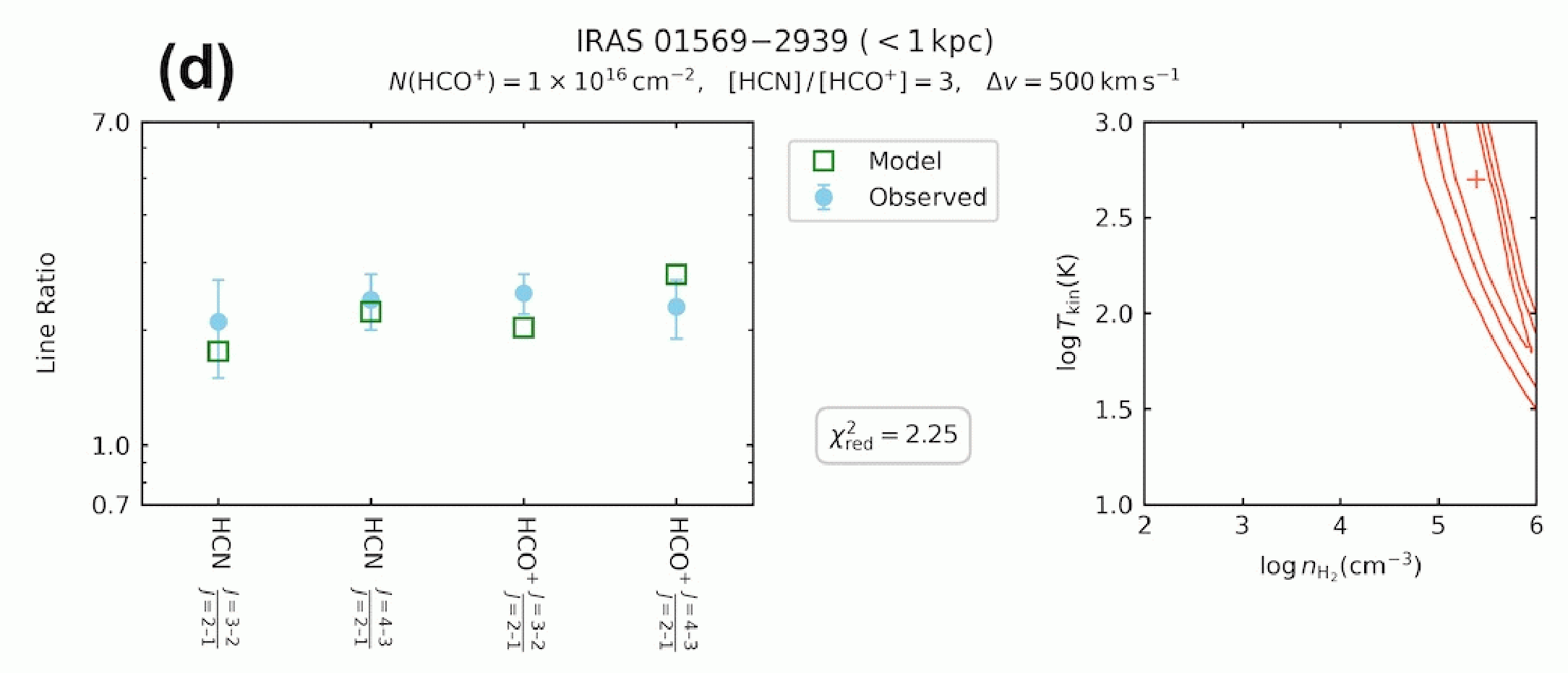} \\
\includegraphics[angle=0,scale=.24]{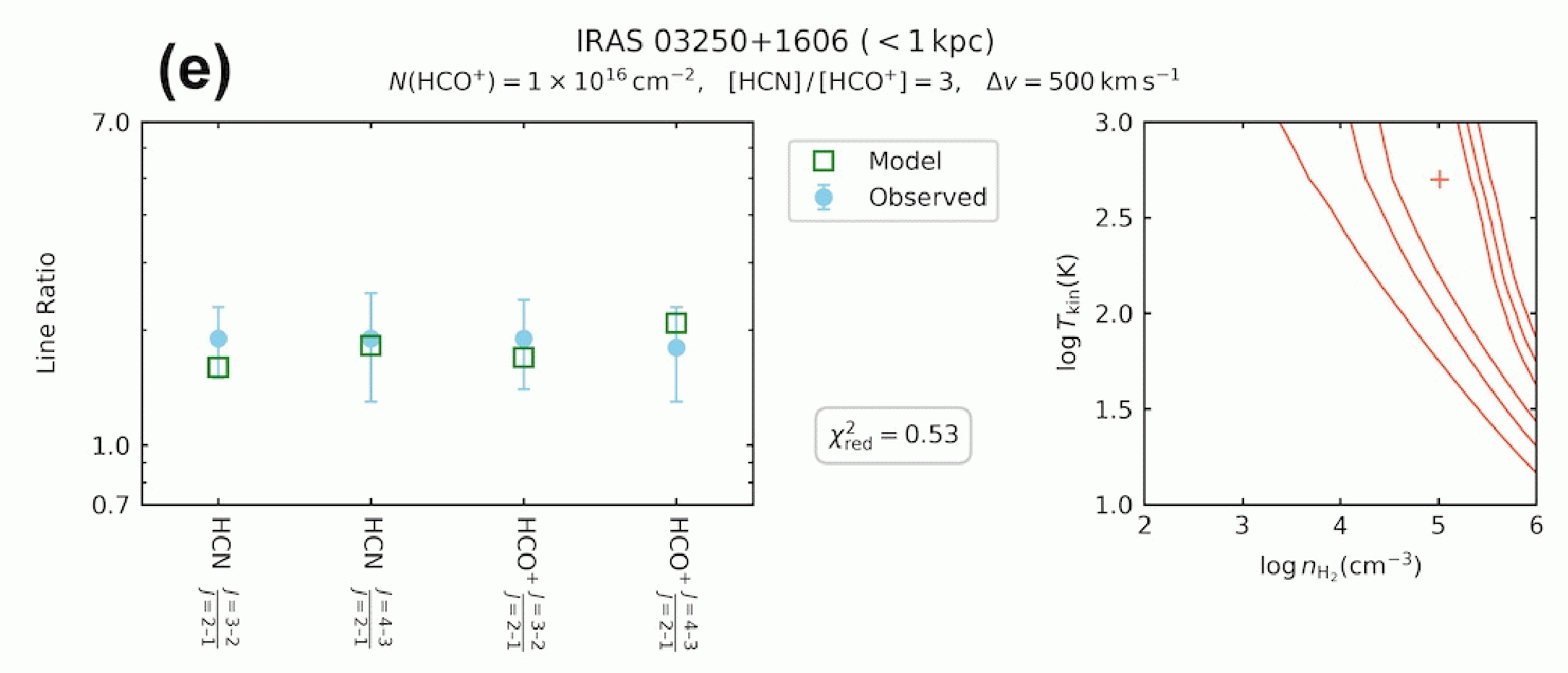} 
\includegraphics[angle=0,scale=.24]{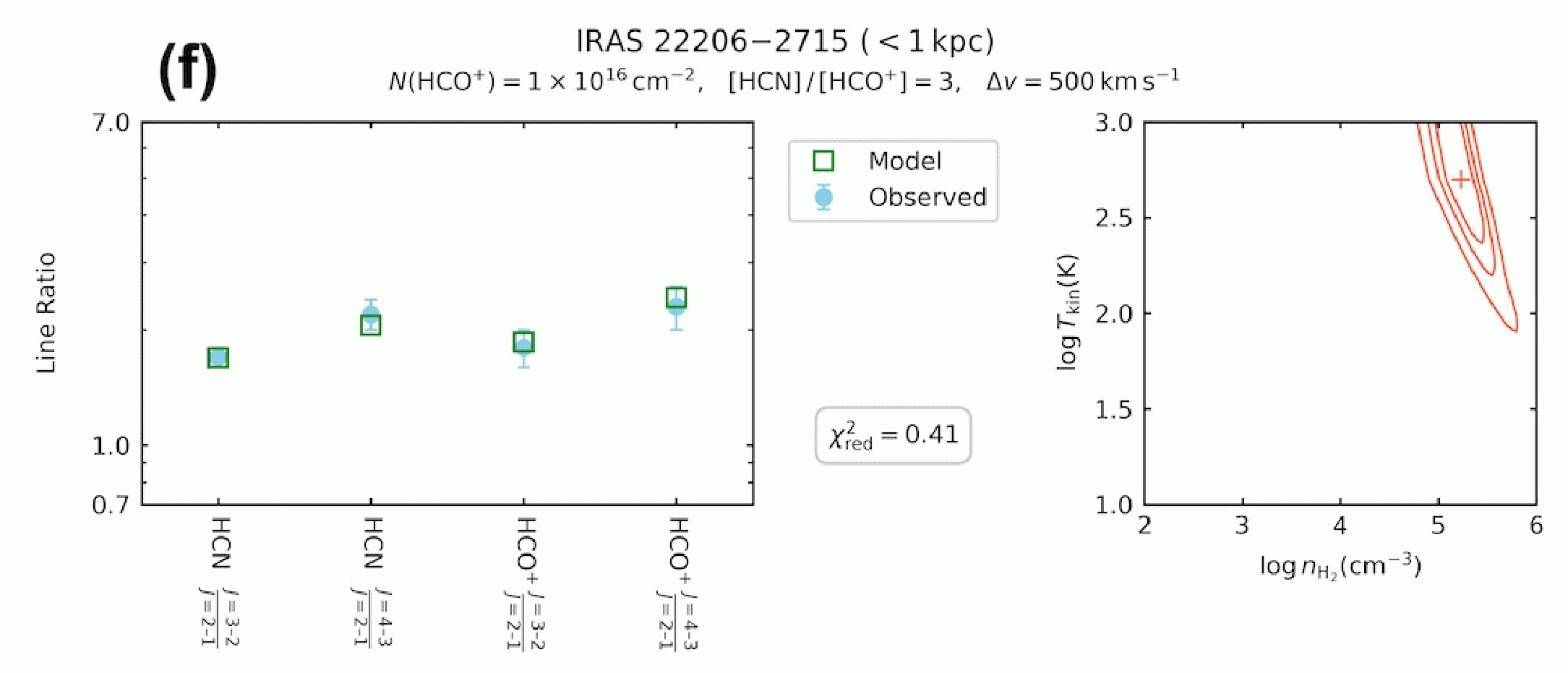} \\
\includegraphics[angle=0,scale=.24]{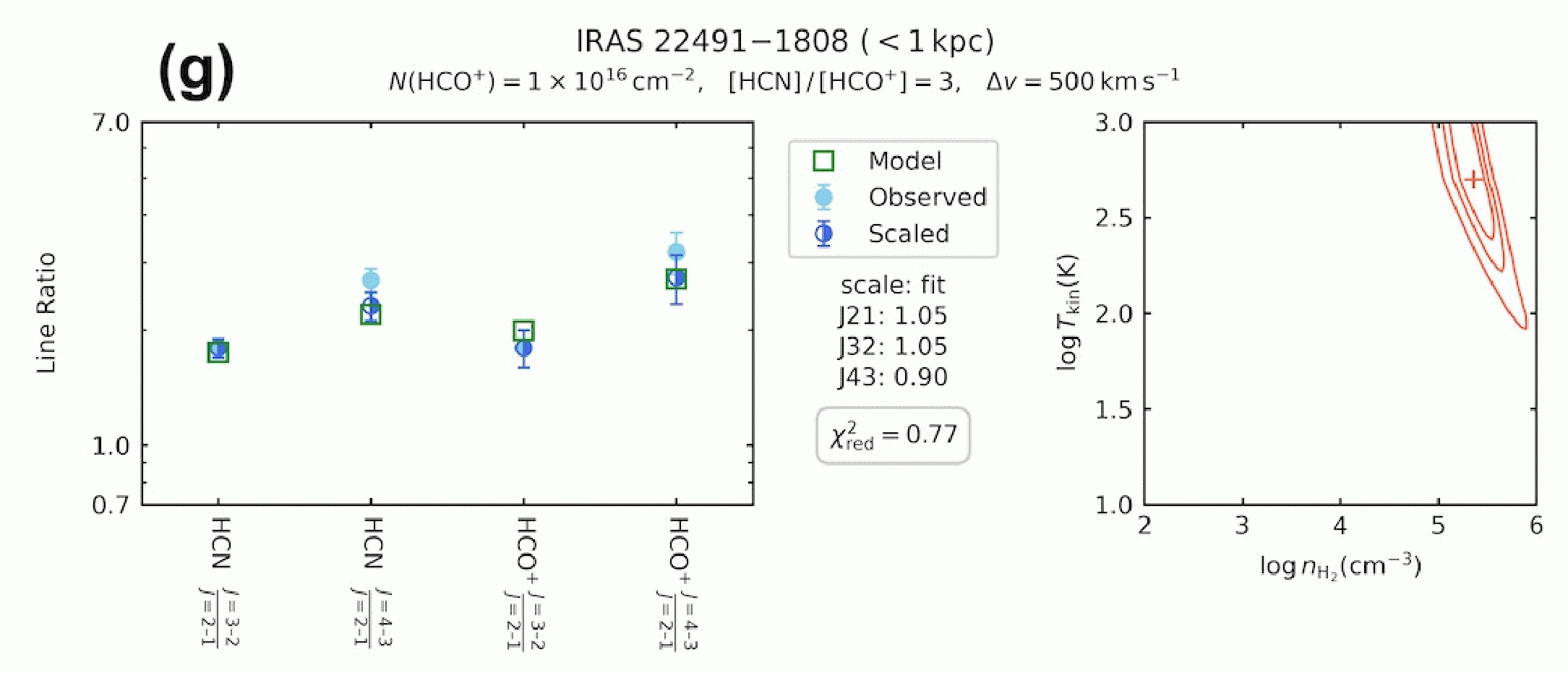} 
\includegraphics[angle=0,scale=.24]{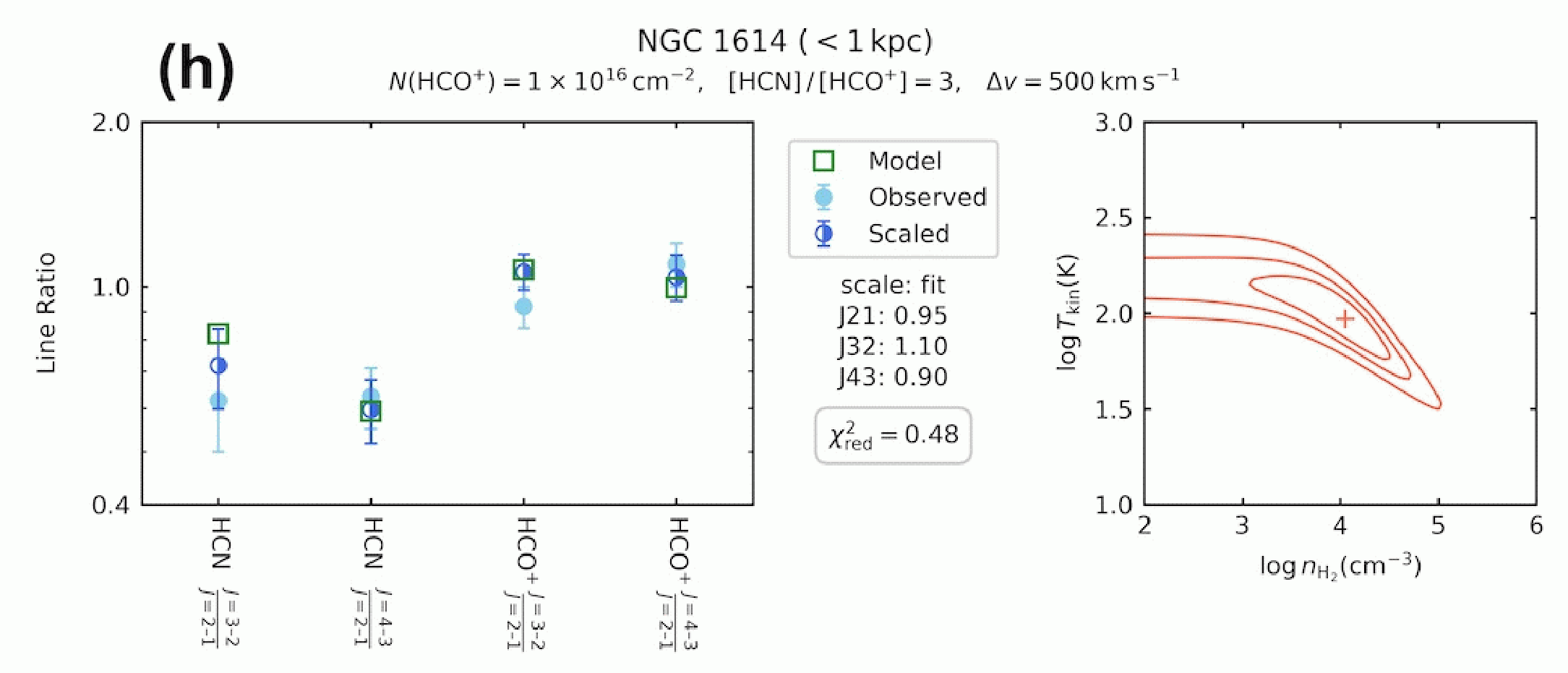} \\
\end{center}
\caption{
Fitting results for the high-J to low-J emission line flux ratios of HCN
and HCO$^+$ derived from the 1 kpc beam-sized spectra, displayed in
the same way as Figure \ref{fig:FitLM}. 
The result of IRAS 16090$-$0139 is shown in Figure \ref{fig:IR16090FitLM}d.
\label{fig:FitLM1kpc}
}
\end{figure*}

\begin{figure*}[!hbt]
\begin{center}
\includegraphics[angle=0,scale=.24]{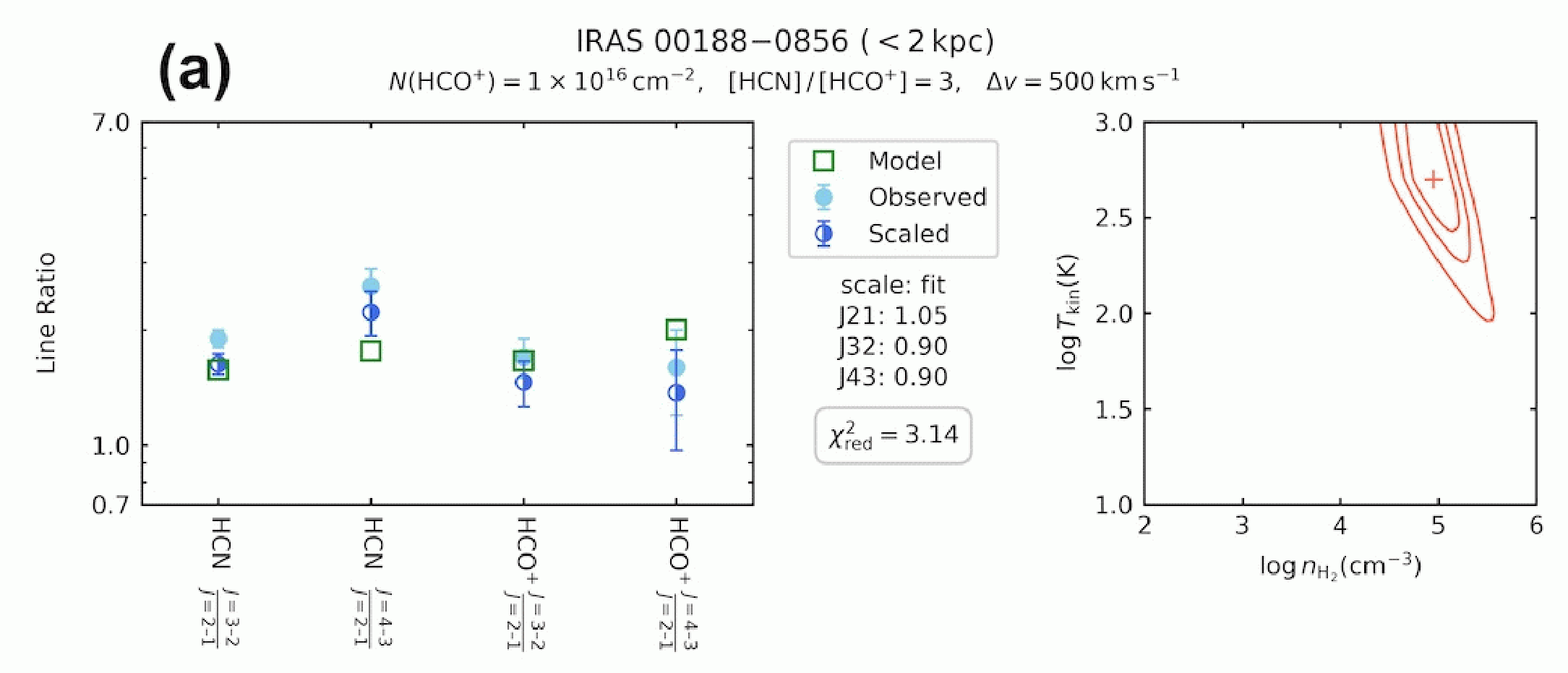} 
\includegraphics[angle=0,scale=.24]{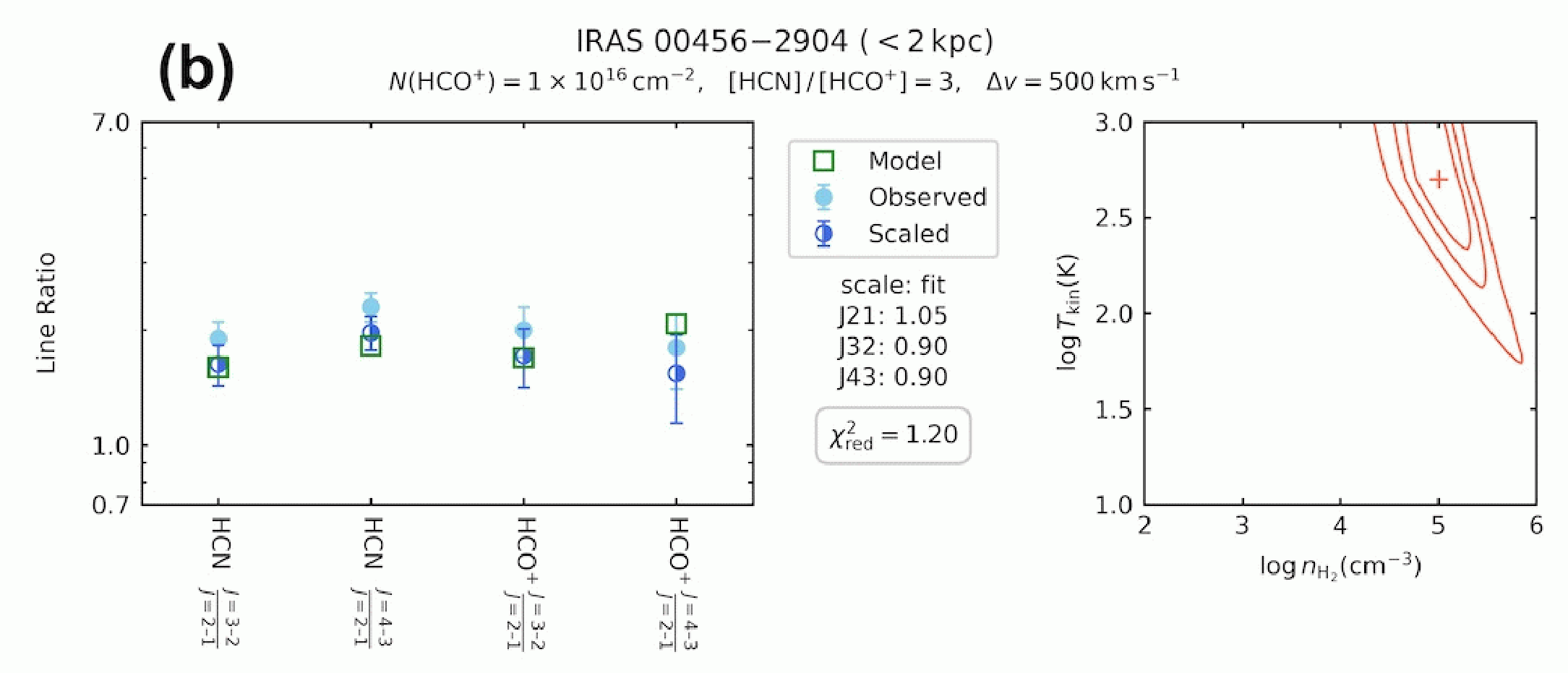} \\
\includegraphics[angle=0,scale=.24]{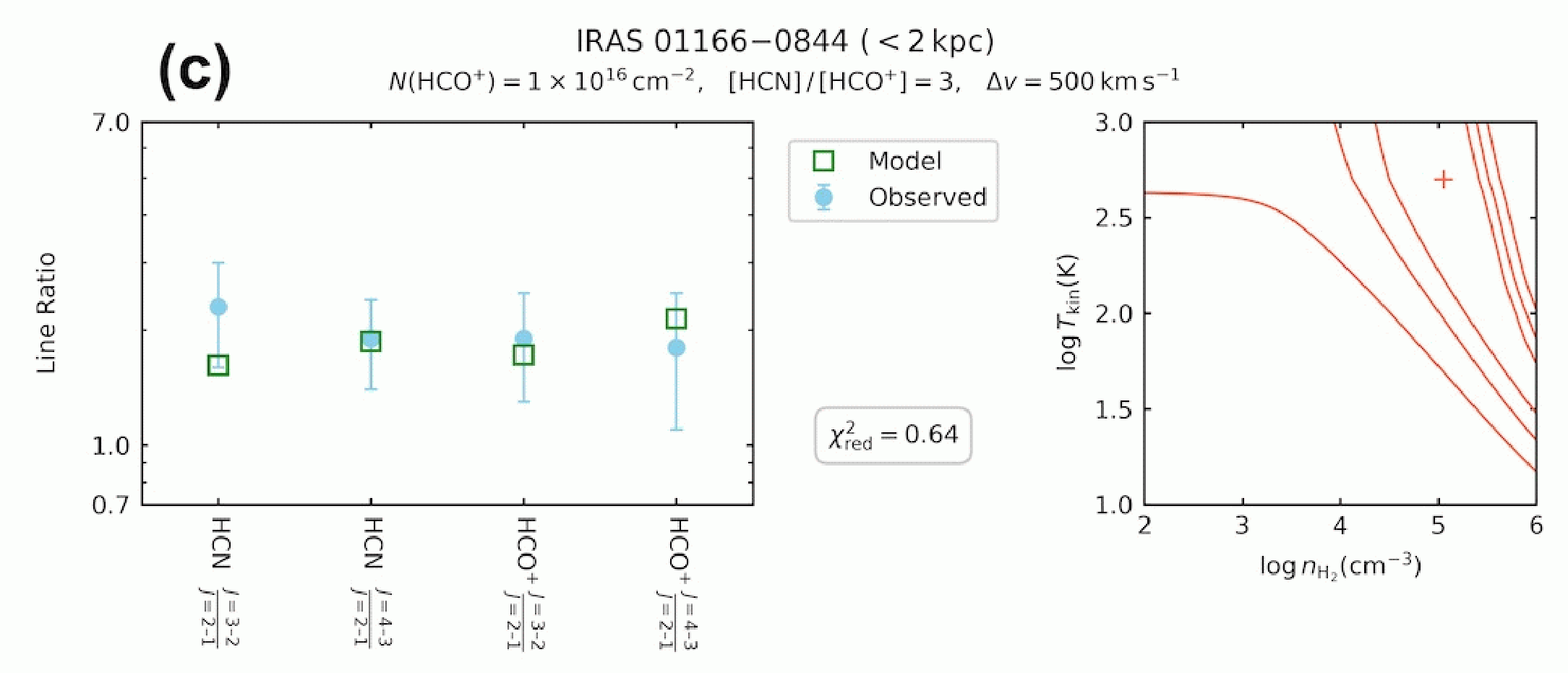} 
\includegraphics[angle=0,scale=.24]{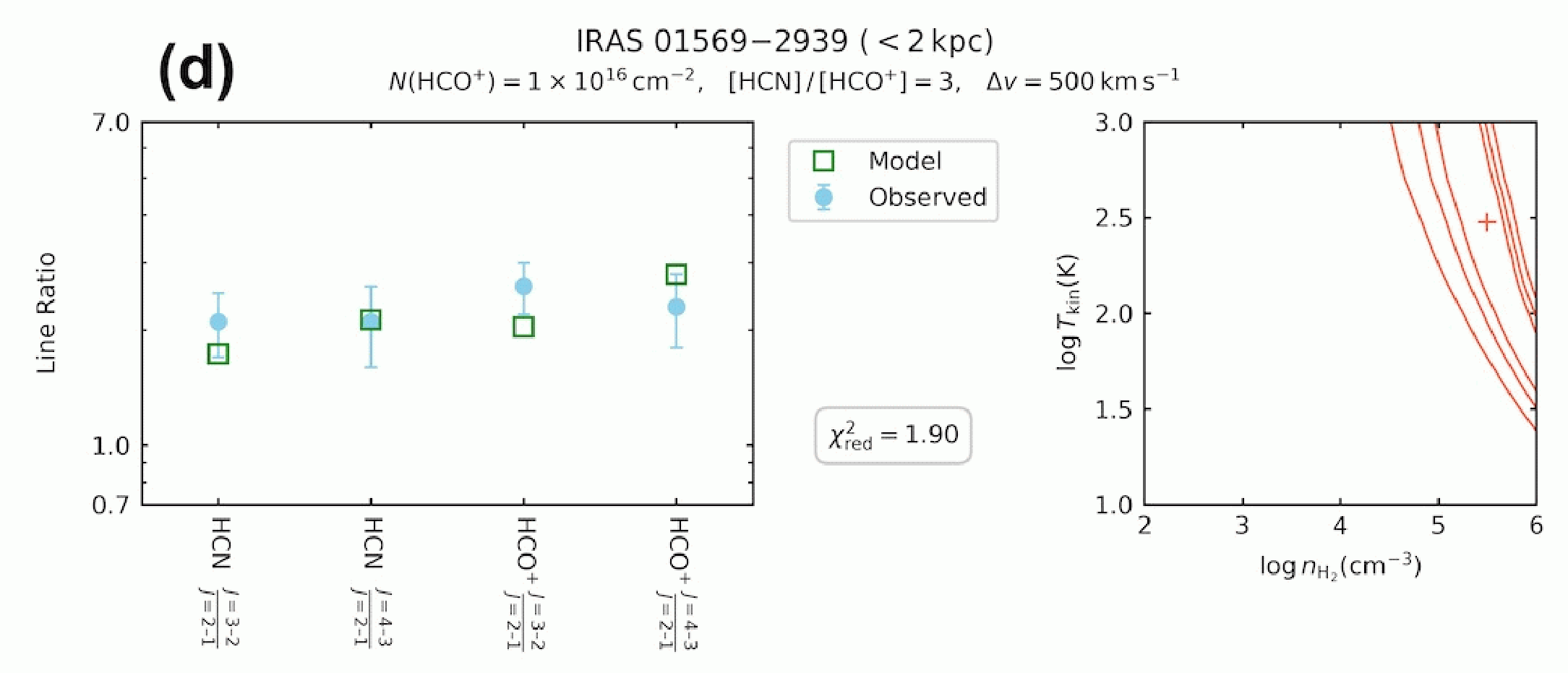} \\
\includegraphics[angle=0,scale=.24]{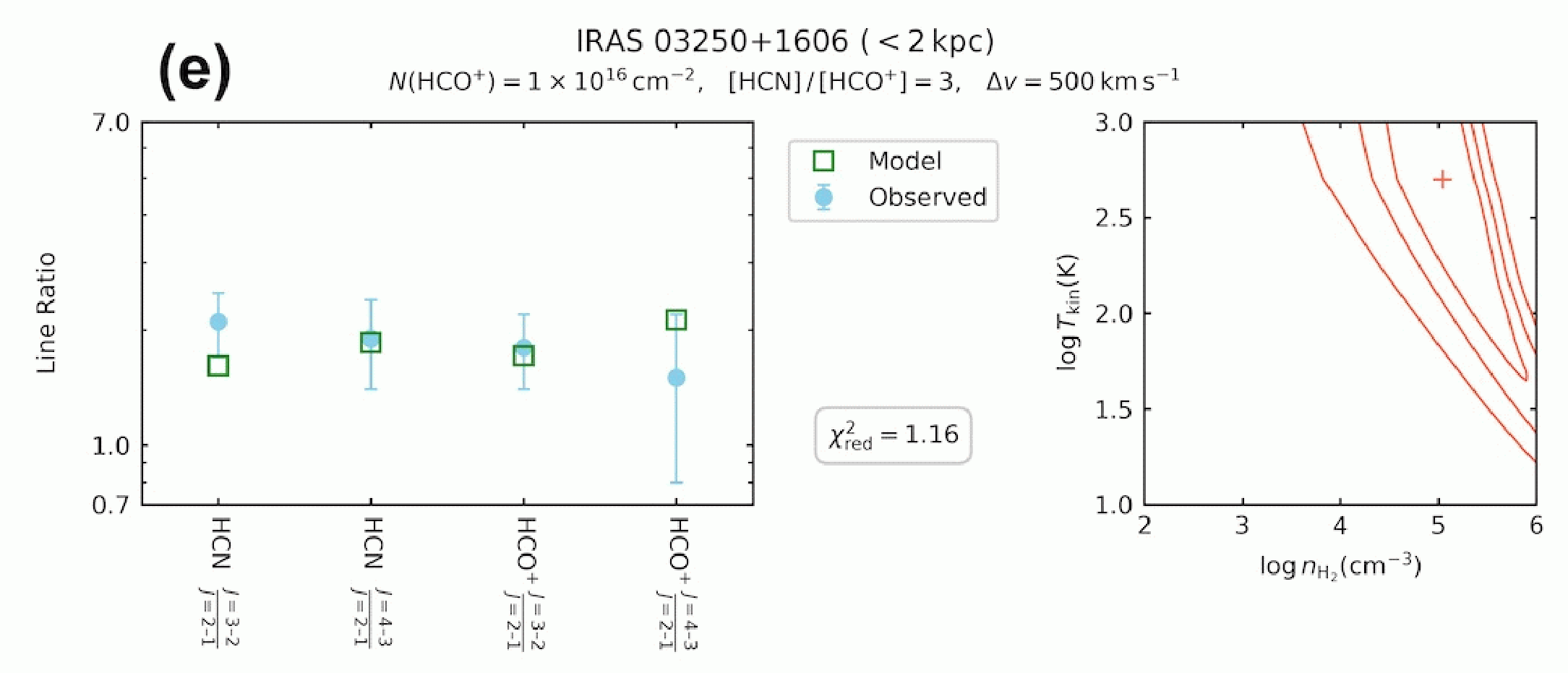} 
\includegraphics[angle=0,scale=.24]{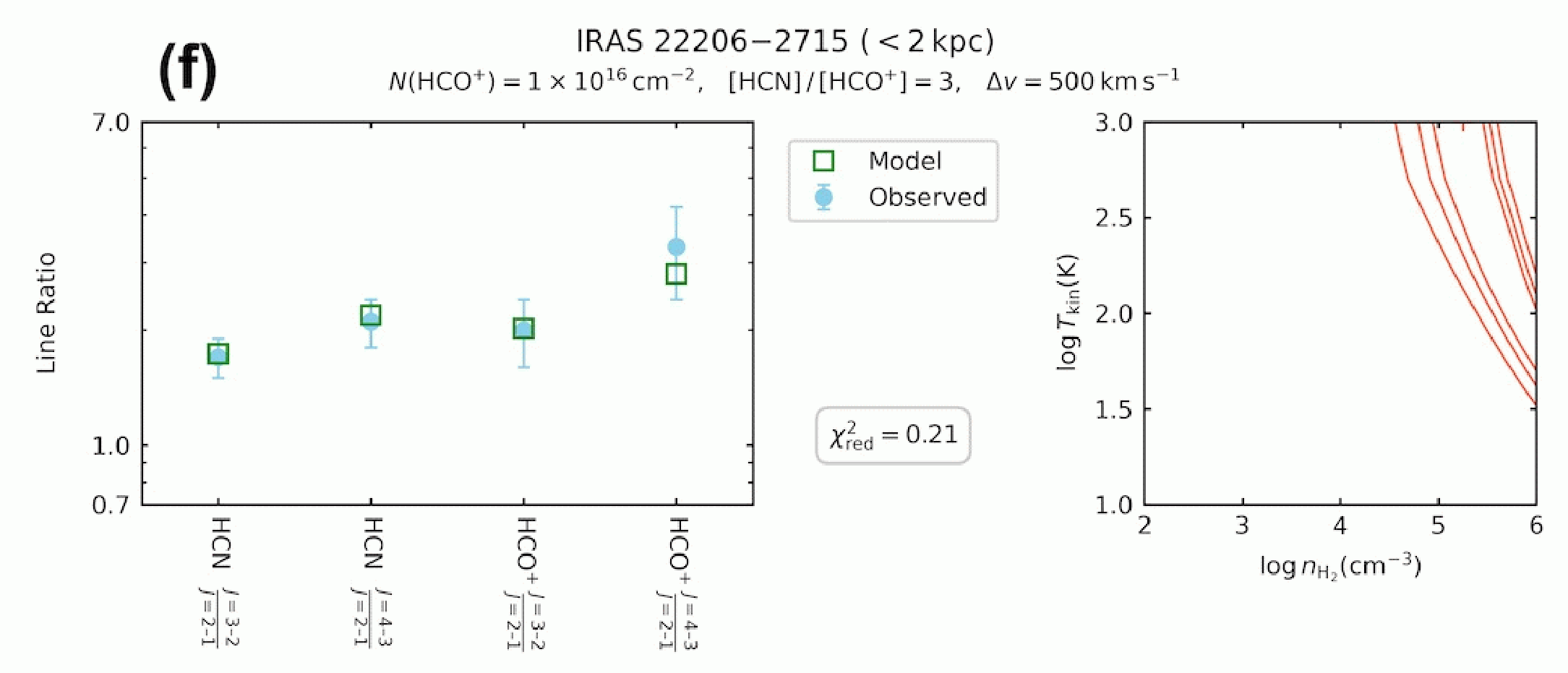} \\
\hspace*{-9cm}
\includegraphics[angle=0,scale=.24]{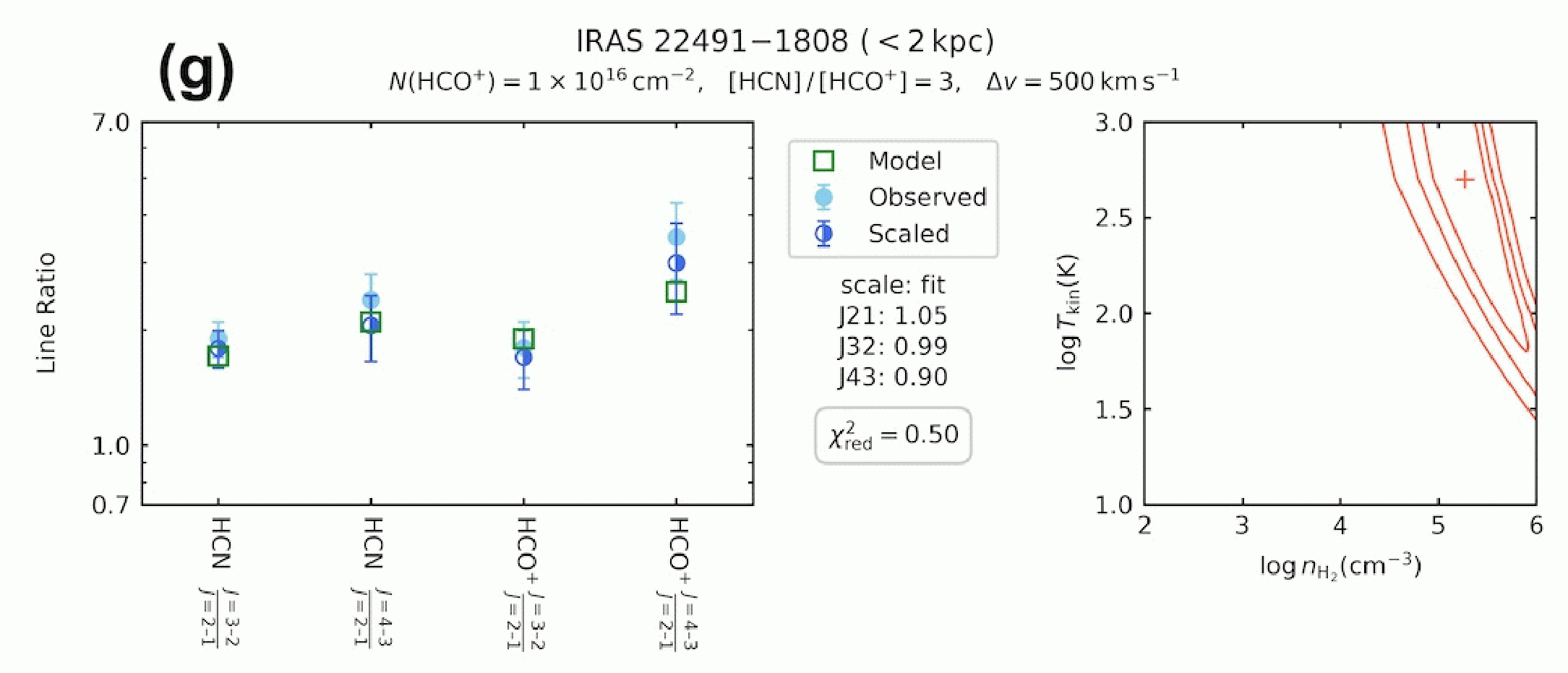} 
\end{center}
\caption{
Same as Figure \ref{fig:FitLM1kpc}, but for the 2 kpc beam-sized
spectra.
The result of IRAS 16090$-$0139 is shown in Figure
\ref{fig:IR16090FitLM}e.
The LIRG NGC 1614 is not shown, for the same reason as explained
in Figure \ref{fig:SpectraA} caption. 
\label{fig:FitLM2kpc}
}
\end{figure*}

Figures \ref{fig:FitLM1kpc} and \ref{fig:FitLM2kpc} display our
fitting results for molecular gas density and temperature based on 
the emission line flux ratios extracted from the 1 kpc and 2 kpc
beam-sized spectra, respectively, using the Levenberg-Marquardt method. 
Table \ref{tab:bestfit2} summarizes these results.
For (U)LIRGs for which we adopt the second fitting results (i.e., flux
scaling on) for the 0.5 kpc beam-sized data (Figure \ref{fig:FitLM}), 
we do for the 1 kpc and 2 kpc beam-sized data as well, for consistent
comparison of molecular gas properties among different regions of each
(U)LIRG. 
We provide these fitting results for the 1--2 kpc beam-sized data because
\citet{ima23} apply the same fitting to different nearby ULIRGs
observed with comparable 1--2 kpc beam sizes. 
The presence of dense ($\gtrsim$10$^{5}$ cm$^{-3}$) and warm
($\gtrsim$300 K) molecular gas in all the observed ULIRGs' nuclei in
this paper is confirmed with the 1--2 kpc beam sizes, as previously
seen for different nearby ULIRGs \citep{ima23}.   
The derived molecular gas density ($\sim$10$^{4}$ cm$^{-3}$) and
temperature ($\sim$100 K) in the only one LIRG NGC 1614 (Figure 
\ref{fig:FitLM1kpc}h) are substantially
smaller than those of the observed ULIRGs in this paper (Figure 
\ref{fig:FitLM1kpc}a--g), which 
also conforms to the result reported by \citet{ima23} for NGC 1614 and
different nearby ULIRGs.

\begin{deluxetable}{lccccc}[!hbt]
\tabletypesize{\small}
\tablecaption{Summary of the Best Fit Values for the 1 kpc and 2 kpc 
Beam-sized Data \label{tab:bestfit2}}
\tablewidth{0pt}
\tablehead{\colhead{Object} & \colhead{Region} & \colhead{Scaling} & 
\colhead{log n$_{\rm H_2}$} & \colhead{log T$_{\rm kin}$} & 
\colhead{Reduced} \\
\colhead{} & \colhead{} & \colhead{} & \colhead{[cm$^{-3}$]} & 
\colhead{[K]} & \colhead{$\chi^{2}$} \\ 
\colhead{(1)} & \colhead{(2)} & \colhead{(3)} & \colhead{(4)} &
\colhead{(5)} & \colhead{(6)} 
} 
\startdata
IRAS 00188$-$0856 & $<$1 kpc & on & 4.8$^{+0.2}_{-0.2}$ &
2.7$^{+\infty}_{-0.3}$ &  0.93 \\ 
 & $<$2 kpc & on & 5.0$^{+0.2}_{-0.2}$ & 2.7$^{+0.4}_{-0.1}$
& 3.1 \\ 
IRAS 00456$-$2904 & $<$1 kpc & on & 5.2$^{+0.1}_{-0.2}$ & 2.7$^{+0.3}_{-0.1}$
& 3.1 \\
 & $<$2 kpc & on & 5.0$^{+0.2}_{-0.2}$ & 2.7$^{+\infty}_{-0.2}$
& 1.2 \\ 
IRAS 01166$-$0844 & $<$1 kpc & off & 5.6$^{+0.3}_{-0.4}$ & 2.5$^{+\infty}_{-0.5}$
& 1.4 \\ 
 & $<$2 kpc & off & 5.1$^{+0.6}_{-0.5}$ & 2.7$^{+\infty}_{-0.7}$
& 0.64 \\ 
IRAS 01569$-$2939 & $<$1 kpc & off & 5.4$^{+0.3}_{-0.2}$ & 2.7$^{+\infty}_{-0.5}$
& 2.3 \\ 
 & $<$2 kpc & off & 5.5$^{+\infty}_{-0.4}$ & 2.5$^{+\infty}_{-0.8}$
& 1.9 \\ 
IRAS 03250$+$1606 & $<$1 kpc & off & 5.0$^{+0.5}_{-0.4}$ & 2.7$^{+\infty}_{-0.7}$
& 0.53 \\ 
 & $<$2 kpc & off & 5.0$^{+0.4}_{-0.4}$ & 2.7$^{+\infty}_{-0.5}$
& 1.2 \\ 
IRAS 22206$-$2715 & $<$1 kpc & off & 5.2$^{+0.2}_{-0.2}$ &
2.7$^{+\infty}_{-0.2}$ & 0.41 \\ 
 & $<$2 kpc & off & 5.3$^{+\infty}_{-0.2}$ & 3.0$^{+\infty}_{-1.2}$
& 0.21 \\ 
IRAS 22491$-$1808 & $<$1 kpc & on & 5.4$^{+0.1}_{-0.2}$ & 2.7$^{+0.6}_{-0.2}$
& 0.77 \\  
 & $<$2 kpc & on & 5.3$^{+0.4}_{-0.3}$ & 2.7$^{+\infty}_{-0.5}$
& 0.50 \\ 
NGC 1614 & $<$1 kpc & on & 4.0$^{+0.3}_{-0.4}$ & 2.0$^{+0.1}_{-0.1}$ & 0.48 \\ 
\enddata

\tablecomments{
Col.(1): Object name. 
Col.(2): Region.
Col.(3): Scaling on or off.
Col.(4): Decimal logarithm of H$_{2}$ gas density in units of cm$^{-3}$. 
Col.(5): Decimal logarithm of gas kinetic temperature in units of K. 
Col.(6): Reduced $\chi^{2}$ value.
The HCO$^{+}$ column density, HCN-to-HCO$^{+}$ abundance ratio, and
molecular line width are fixed at N$_{\rm HCO^+}$ = 1 $\times$
10$^{16}$ cm$^{-2}$, [HCN]/[HCO$^+$] = 3, and $\Delta$v = 500 km
s$^{-1}$, respectively. 
}

\end{deluxetable}



\end{document}